\documentclass[sn-aps, referee]{sn-jnl}%

\usepackage{graphicx}%
\usepackage{multirow}%
\usepackage{amsmath,amssymb,amsfonts}%
\usepackage{amsthm}%
\usepackage{mathrsfs}%
\usepackage[title]{appendix}%
\usepackage{xcolor}%
\usepackage{textcomp}%
\usepackage{manyfoot}%
\usepackage{algorithm}%
\usepackage{algorithmicx}%
\usepackage{algpseudocode}%
\usepackage{listings}%
\usepackage{todonotes}
\usepackage[version=4]{mhchem}
\usepackage{siunitx}
\usepackage{hyperref}
\usepackage[subrefformat=parens]{subcaption}
\usepackage{geometry}
\usepackage{ulem}
\usepackage{threeparttable}
\usepackage{booktabs}%
\usepackage{xr}
\usepackage{physics}
\usepackage{multirow}
\usepackage{orcidlink}

\theoremstyle{thmstyleone}%

\theoremstyle{thmstyletwo}%

\theoremstyle{thmstylethree}%

\begin{document}

\title{MatterSim: A Deep Learning Atomistic Model Across Elements, Temperatures and Pressures}

\author*[1]{\fnm{Han} \sur{Yang}\orcidlink{0000-0002-4531-093X}}\email{hanyang@microsoft.com}
\equalcont{These authors contributed equally to this work.}

\author[1]{\fnm{Chenxi} \sur{Hu}\orcidlink{0009-0006-8486-9230}}
\equalcont{These authors contributed equally to this work.}

\author[1]{\fnm{Yichi} \sur{Zhou}}
\equalcont{These authors contributed equally to this work.}

\author[1]{\fnm{Xixian} \sur{Liu}\orcidlink{0009-0008-9215-3990}}
\equalcont{These authors contributed equally to this work.}

\author[1]{\fnm{Yu} \sur{Shi}\orcidlink{0000-0001-9235-8963}}
\equalcont{These authors contributed equally to this work.}

\author*[1]{\fnm{Jielan} \sur{Li}\orcidlink{0000-0003-4428-2452}}\email{jielanli@microsoft.com}
\equalcont{These authors contributed equally to this work.}

\author[1]{\fnm{Guanzhi} \sur{Li}\orcidlink{0000-0002-4167-6432}}
\equalcont{These authors contributed equally to this work.}

\author[1]{\fnm{Zekun} \sur{Chen}\orcidlink{0000-0002-4183-2941}}
\equalcont{These authors contributed equally to this work.}

\author[1]{\fnm{Shuizhou} \sur{Chen}\orcidlink{0009-0005-2701-5565}}
\equalcont{These authors contributed equally to this work.}

\author[1]{\fnm{Claudio} \sur{Zeni}\orcidlink{0000-0002-6334-2679}}
\author[1]{\fnm{Matthew} \sur{Horton}\orcidlink{0000-0001-7777-8871}}
\author[1]{\fnm{Robert} \sur{Pinsler}\orcidlink{0000-0003-1454-188X}}
\author[1]{\fnm{Andrew} \sur{Fowler}}
\author[1]{\fnm{Daniel} \sur{Z\"ugner}\orcidlink{0000-0003-1626-5065}}
\author[1]{\fnm{Tian} \sur{Xie}\orcidlink{0000-0002-0987-4666}}
\author[1]{\fnm{Jake} \sur{Smith}\orcidlink{0000-0003-0412-1312}}
\author[1]{\fnm{Lixin} \sur{Sun}\orcidlink{0000-0002-7971-5222}}
\author[1]{\fnm{Qian} \sur{Wang}\orcidlink{0009-0007-7680-4514}}
\author[1]{\fnm{Lingyu} \sur{Kong}\orcidlink{0009-0006-2226-5730}}
\author[1]{\fnm{Chang} \sur{Liu}\orcidlink{0000-0001-5207-5440}}
\author*[1]{\fnm{Hongxia} \sur{Hao}\orcidlink{0000-0002-4382-200X}}\email{hongxiahao@microsoft.com}
\author*[1]{\fnm{Ziheng} \sur{Lu}\orcidlink{0000-0003-2239-8526}}\email{zihenglu@microsoft.com}

\affil*[1]{\orgname{Microsoft Research AI for Science}}%

\abstract{
Accurate and fast prediction of materials' properties is central to the digital transformation of materials design. However, the vast design space and diverse operating conditions pose significant challenges for accurately modeling arbitrary material candidates and forecasting their properties.
We present MatterSim, a deep learning model actively learned from large-scale first-principles computations, for efficient atomistic simulations at first-principles level and accurate prediction of broad material properties across the periodic table, spanning temperatures from 0 to \SI{5000}{\kelvin} and pressures up to \SI{1000}{\giga\pascal}.
Out-of-the-box, the model serves as a machine learning force field, and shows remarkable capabilities not only in predicting ground-state material structures and energetics, but also in simulating their behavior under realistic temperatures and pressures, signifying an up to ten-fold enhancement in precision compared to the prior best-in-class. This enables MatterSim to compute materials' lattice dynamics, mechanical and thermodynamic properties, and beyond, to an accuracy comparable with first-principles methods. Specifically, MatterSim predicts Gibbs free energies for a wide range of inorganic solids with near-first-principles accuracy and achieves a \SI{15}{meV\per atom} resolution for temperatures up to \SI{1000}{\kelvin} compared with experiments. This opens an opportunity to predict experimental phase diagrams of materials at minimal computational cost. Moreover, MatterSim also serves as a platform for continuous learning and customization by integrating domain-specific data. The model can be fine-tuned for atomistic simulations at a desired level of theory or for direct structure-to-property predictions, achieving high data efficiency with a reduction in data requirements by up to 97\%.
}

\maketitle

\newpage

\section{Introduction}

Material design stands at the heart of technological advancements in nanoelectronics,\cite{fiori2014electronics,li2007electronic} energy storage,\cite{mizushima1980lixcoo2,ceder1998identification} biomedicine,\cite{tibbitt2015progress} and environmental sustainability.\cite{li2018review,hu2010design} Conventionally, the development of new materials has been a slow and expensive process, dominated by experimental trial and error. Transitioning these efforts \textit{in silico} offers an immense potential to expedite this process.\cite{curtarolo2013high} At the core of this paradigm shift is the ability to accurately and efficiently predict the properties of arbitrary materials under practical synthesis and working conditions.

Advances in deep learning have enabled efficient prediction of materials properties in many domains.\cite{choudhary2022recent,xie2018crystal,merchant2023scaling,chen2024accelerating} A few models based on extensive computational databases can make predictions across many chemical compositions,\cite{lindsey2017chimes,schutt2017schnet,musaelian2023learning,batzner20223} and recent attempts have tried to extend this capability to the entire periodic table.\cite{chen2019graph,choudhary2021atomistic,chen2019graph, chen2022universal, deng2023chgnet, merchant2023scaling, batatia2023foundation,zhang2023dpa, shoghi2023molecules}
One of the most outstanding examples, universal machine learning force field (MLFF), has been proposed based on open-source or proprietary crystalline databases.\cite{chen2019graph, chen2022universal, deng2023chgnet, merchant2023scaling, batatia2023foundation,zhang2023dpa,shoghi2023molecules} These models mark a significant advancement of machine learning towards chemical universality for materials modeling. 
However, the property of a potential candidate material not only depends on its chemical composition and corresponding near-equilibrium atomic structure, but also on thermodynamic conditions including temperature and pressure. This results in a requirement of high predictive accuracy over an enormous configuration space well beyond the ground states or local minima of crystal structures typically captured by current databases and models, which fundamentally limits their applicability for materials design.

\begin{figure}
    \centering
    \includegraphics[width=\textwidth]{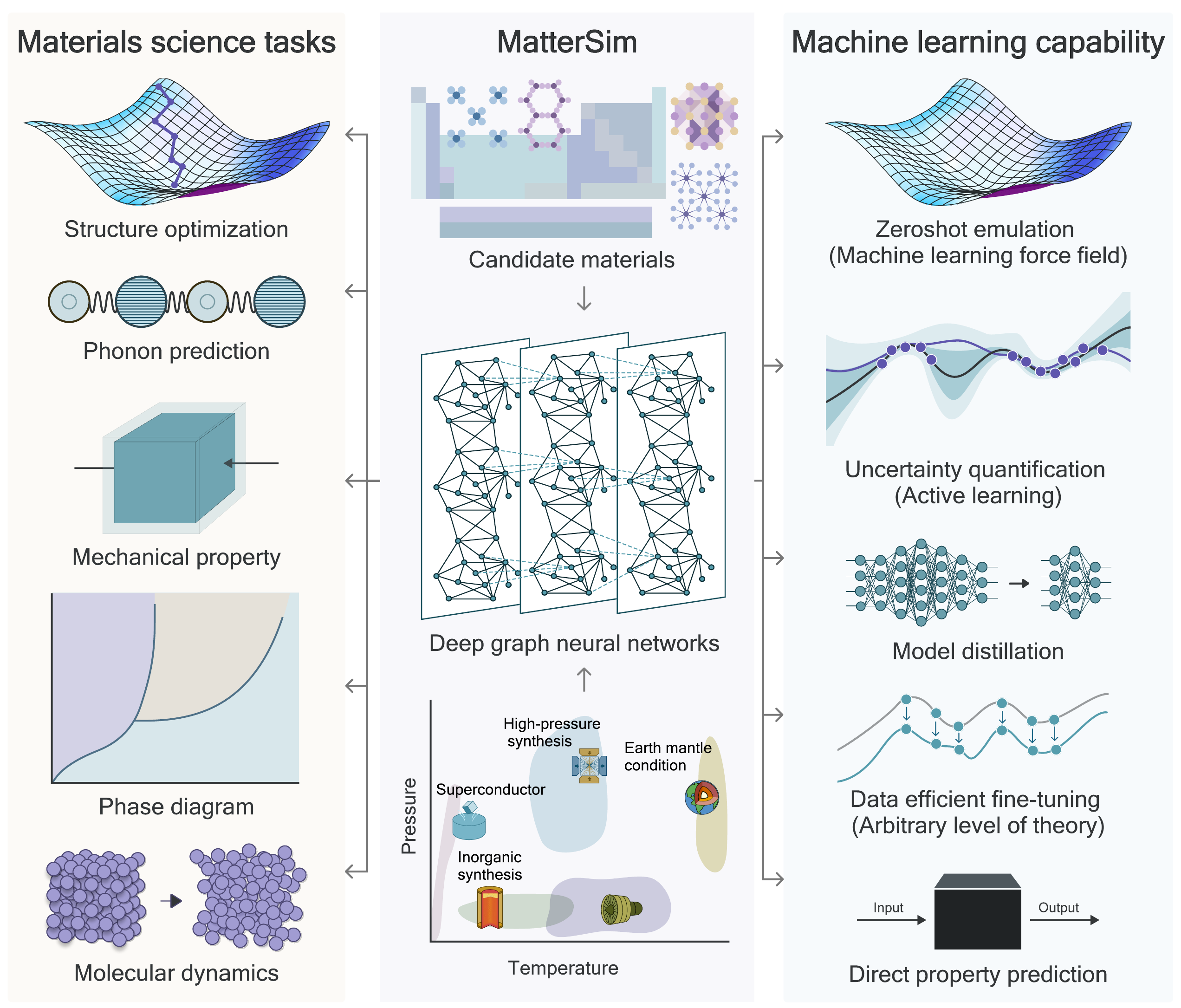}
    \caption{\textbf{MatterSim is a deep learning atomistic model for predicting materials properties  with high predictive accuracy  across chemical elements, temperatures and pressures, enabling a wide range of applicability and functionality.}}
    \label{fig:overview:mattersim-functionality}
\end{figure}

To address this challenge, we introduce MatterSim, a deep learning model designed for emulating materials and predicting their properties under realistic thermodynamic conditions including finite temperatures and pressures, as illustrated in \autoref{fig:overview:mattersim-functionality}.
MatterSim utilizes deep graph neural networks, uncertainty-aware sampling and active learning to explore the vast materials space with first-principles computations as a supervisor for enhanced generalizability.\cite{kresse1996efficient} 
Out-of-the-box, MatterSim operates as a zero-shot MLFF, delivering both efficient and precise predictions of energies and forces, showcasing  proficiency in predicting  energetics near ground states and dynamics under realistic conditions, with a mean absolute error (MAE) of 36 meV/atom (43 meV/atom as chemical accuracy) on \texttt{MPF-TP} (a dataset covering wide ranges of materials structure sampled under finite temperature between 0-5000 K and pressures between 0--1000 GPa), marking a ten-fold increase in accuracy compared with previous efforts.\cite{chen2022universal,deng2023chgnet,batatia2023foundation} Therefore, MatterSim is well-suited for calculating a broad range of properties, including lattice dynamics, mechanical properties, thermodynamics and more. Remarkably, the model is capable of predicting temperature- and pressure-dependent free energies of wide ranges of solid materials comparable with first-principles methods and experimental measurements, thereby opening an opportunity for fast and accurate prediction of phase diagrams of materials. 
Furthermore, MatterSim's extensive coverage of the compositional and configurational space of materials enables it to effectively describe material features in the latent space and to serve as a pre-trained model for continuous learning and further customization, with high data efficiency.
With active learning and fine-tuning, the model can be extended to carry out atomistic simulations of highly complex systems beyond its current data coverage and theory level.  For example, to simulate liquid water, only 3\% of the data is needed to customize MatterSim to obtain the results of a specialized model trained from scratch, and to reproduce the experimental structural and transport properties of water.
Additionally, MatterSim's highly expressive features enable direct structure-to-property prediction of materials, which is also known as end-to-end prediction. After being fine-tuned with a limited amount of data, MatterSim outperforms specialized models trained exclusively with domain specific data on the tasks related to lattice dynamics, electronic and mechanical properties in Matbench.\cite{dunn2020benchmarking}

\section{Results}\label{sec:results}
\subsection{Learning the materials space under first-principles supervision}

\begin{figure}
    \centering
    \includegraphics[width=\textwidth]{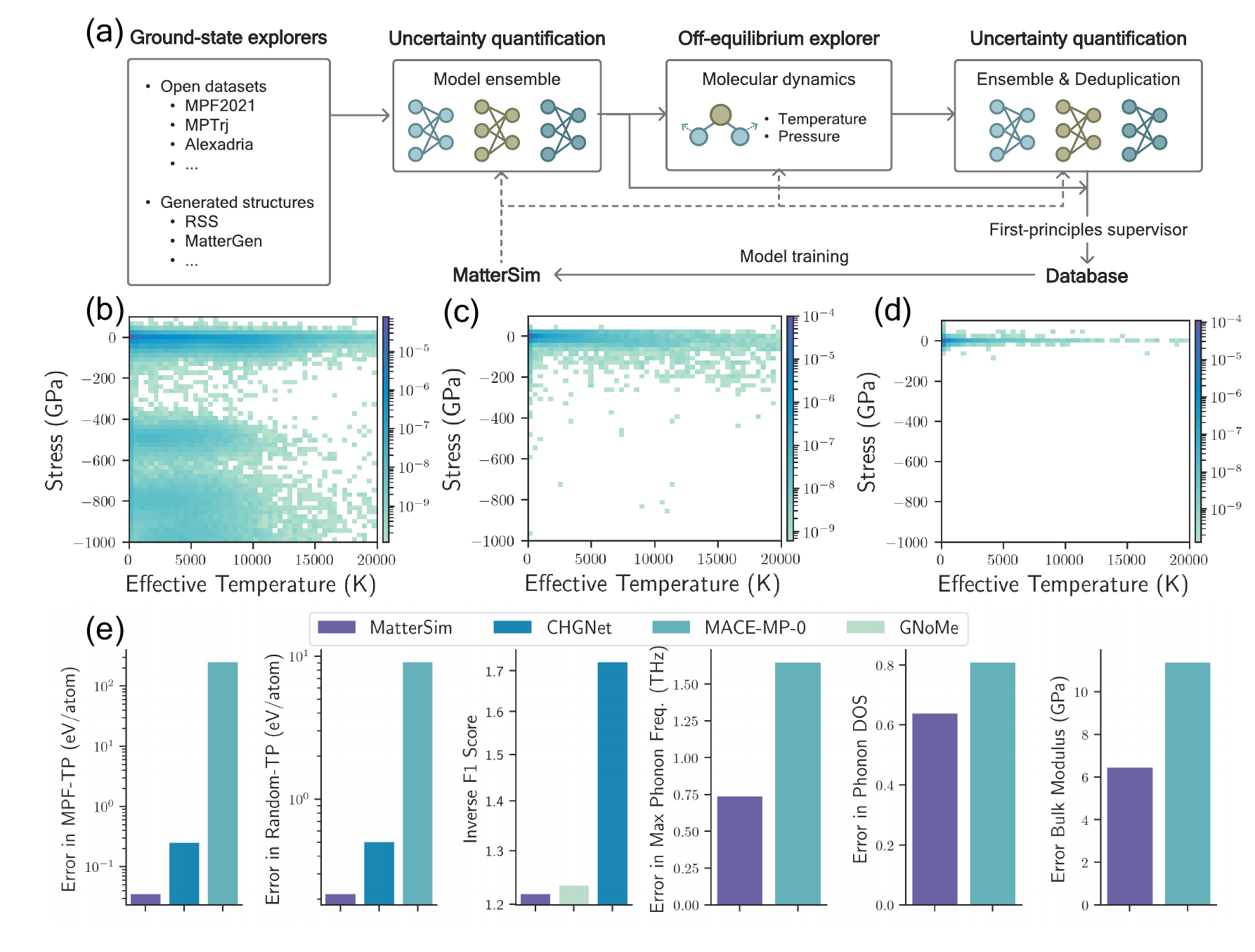}
    \caption{\textbf{MatterSim is developed on an enriched materials space.} \textbf{(a)} A data explorer employed in MatterSim for generating datasets covering wide potential energy surface; Histogram of the stress (GPa) and effective temperature (K) of  \textbf{(b)} the generated materials in this work, \textbf{(c)} the \texttt{MPF2021} dataset and \textbf{(d)} the \texttt{Alexandria} dataset. \textbf{(e)} Comparative performance metrics of MatterSim across six tasks: energy prediction on \texttt{MPF-TP} and \texttt{random-TP} datasets, phonon properties including max frequency and density of states (DOS), Bulk Modulus, and inverse F1 score in MatBench-Discovery leaderboard. Lower scores indicating superior performance for all tasks. Refer to main text and supplementary information for task details.
    }
    \label{fig:overview:data-overview}
\end{figure}

MatterSim employs an active learning approach to explore the extensive materials space, integrating a deep graph neural network, a materials explorer, a first-principles supervisor,\cite{kresse1996efficient,kresse1996efficiency,kohn1965self,hohenberg1964inhomogeneous} and an ensemble uncertainty monitor, see \autoref{fig:overview:data-overview}(a). Starting from an initial dataset curated from existing sources, the first-principles supervisor offers the deep learning model supervision signals relating to energies, forces, and stresses on the given structures at the generalized gradient approximation (GGA) Perdew–Burke–Ernzerhof (PBE)\cite{perdew1996generalized} level of theory with Hubbard U correction\cite{anisimov1991band} for select materials as specified by the Materials Project standard settings\cite{jain2013commentary}. This trained model then functions as an effective surrogate to the first-principles method, guiding the materials explorer to gather more structures, thereby exploring the most uncertain regions of materials space to enrich the samples for model training. These sampled structures will also be labeled by the first-principles supervisor to provide additional training signals to the model following an active learning loop. The MatterSim model, curated from several such iterations, is capable of learning a wide range of materials space with minimal data redundancy.

A key feature of MatterSim is the vast coverage of materials space.
We note that data in current databases has significant chemical and/or structural bias, leading to significant under-sampling of materials space. For example, most open databases are obtained through relaxation of experimental crystal structures with first-principles calculations.\cite{chen2022universal,deng2023chgnet,saal2013materials,kirklin2015open,schmidt2023machine}. As demonstrated in \autoref{fig:overview:data-overview}(c) and \autoref{fig:overview:data-overview}(d), the relaxation trajectory contains highly symmetric structures close to local energy minima with high structural redundancy. Therefore, models trained on such data are deficient in the generalizability and predictive power needed for atomistic simulation of materials under realistic conditions like finite temperatures and pressures. In addition, these databases tend to have a strong bias towards certain elements, which leads to an under representation of many interatomic interactions.\cite{deng2023chgnet} 
Here our designed materials explorers featuring a diverse collection of materials, including ground-state or near-equilibrium structures from public datasets and in-house generated ones by the ground-state materials explorer, as well as off-equilibrium structures (see \autoref{fig:overview:data-overview}(a)) across a wide range of temperatures and pressures by the off-equilibrium materials explorer, signifying a critical expansion of the configurational space.
It is worth noting that active learning is adopted in a batched manner in the sampling process to avoid relabeling structures of high confidence to the model.
With such a scheme, we collected a first-principles labeled dataset with better chemical and structural coverage, with an analysis in \autoref{si-sec:data-distribution}. As of the date of publication, the training dataset contains $\sim$17M structures labeled with first-principles computations.
As shown in \autoref{fig:overview:data-overview}(b), the curated dataset is representative of materials at temperatures and pressures covering 0--\SI{5000}{\kelvin} and 0--\SI{1000}{\giga\pascal}. The element pair distribution (see \autoref{si-fig:pairwise-distribution}) also shows significantly better sampled chemical space with a more uniform distribution. 
More importantly, the dataset contains on average 2 to 3-fold more distinct atomic environments across the entire periodic table compared to previous databases based on DFT relaxation of crystal structures,
and 10-fold or even higher for certain elements especially for noble gas elements. More details on this are provided in Section \autoref{si-fig:latent-space-comparison}. The coverage of the data generated in this work has empowered MatterSim to make accurate and robust predictions for a wide range of applications; in \autoref{fig:overview:data-overview}(e), we list the performance of MatterSim on six tasks, including phonon-related property prediction, materials discovery (MatBench Discovery), and dynamics under realistic conditions (structure benchmark sets sampled from \textit{ab initio} molecular dynamics with wide temperature and pressure ranges). The most noticeable enhancement are observed on the benchmark datasets \texttt{MPF-TP} and \texttt{Random-TP} (sampled from high temperature and pressure), where MatterSim achieves up to 10-fold improvement compared to universal force fields trained on relaxation trajectories. %

The choice of model architecture is of central importance to the performance of MatterSim. It needs to be scalable -- capable of consuming large amount of data by expanding the model size. It also needs to be efficient during inference so that the model can tractably be used to carry out complex simulations for long timescales. Many models have been developed for application as an MLFF as well as to predict other properties. In this work, we opt to use two primary architectures, M3GNet\cite{chen2022universal} and Graphormer\cite{ying2021transformers} as the backbones for MatterSim. M3GNet is an invariant graph neural network model with high data efficiency, which has been used to train models (and ensembles) with data up to 3M. For models with larger data sizes, we turn to Graphormer. Graphormer is a transformer-based model with proven learning capacity and scalability.\cite{shi2022benchmarking,ying2021transformers} In particular, we baked in several additional attributes including invariance to translation and periodic boundary conditions and explicit equivariant features to better accommodate  materials-related tasks, see \autoref{si-sec:model-arch} for more details. Such a model has better accuracy and better generalizability compared with smaller models at the cost of substantially reduced inference speed and higher GPU memory requirements, see \autoref{si-fig:graphormer-m3gnet-resource-comparison} for more details.
Considering that models with different sizes have different accuracy and inference speeds, the choice for which model to use can be made based on the time or accuracy constraints of the relevant task. In this work, we use the M3GNet-based model for all zero-shot simulations due to its fast inference speed, except for the MatBench Discovery task. For MatBench Discovery and end-to-end property prediction, we turn to a Graphormer-based model, which gives better accuracy. A brief comparison of model efficiency and accuracy is provided in \autoref{si-sec:performance-m3gnet-vs-graphormer} of SI.%

\begin{figure}
    \centering
    \includegraphics[width=\textwidth]{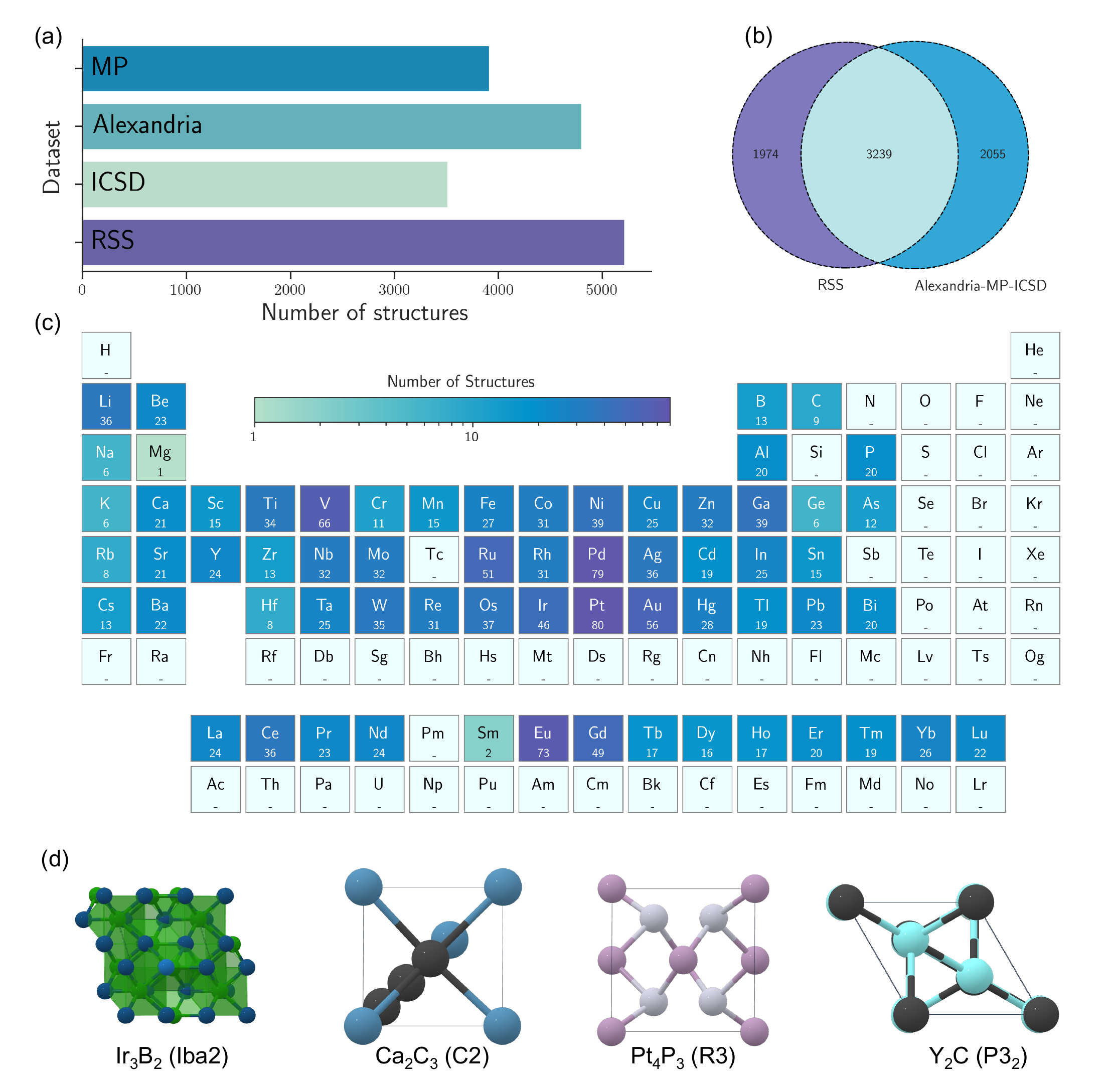}
    \caption{\textbf{MatterSim as a zero-shot emulator empowering materials discovery.} \textbf{(a)} and \textbf{(b)} are the contribution of each dataset to the combined convex hull formed by \texttt{Alexandria-MP-ICSD} dataset (see text) and RSS-generated materials;  \textbf{(c)} Elementwise appearance distribution\cite{riebesell_pymatviz_2022} of the 852 RSS-generated materials found be to on the combined convex hull formed by the \texttt{Alexandria-MP-ICSD}  and  RSS-generated materials.
The materials containing H, Si, N, Sb, O, S, Se, Te, F, Cl, Br, I are removed due to potential issue with how anion corrections are implemented in Materials Project when applied to hypothetical materials\cite{githubstrict_anionsOption}. \textbf{(d)} exhibits examples of materials found to be lower than the \texttt{Alexandria-MP-ICSD} hull, with the corresponding space group in the parentheses.}
    \label{fig:materials-discovery}
\end{figure}

\subsection{MatterSim as a zero-shot atomistic emulator}\label{sec:zero-shot}
MatterSim serves as a universal MLFF to efficiently predict energies, forces, and stresses of structures consisting of any combinations of elements from the periodic table (currently supports the first 89 elements) under simulation conditions of 0-5000 K and 0-1000 GPa, without additional training data. Its universality and accuracy is benchmarked on multiple open datasets as well as three newly created ones with better representation of the model's capability under finite temperatures and pressures. Detailed description of these datasets are provided in \autoref{si-sec:performance-on-benchmark-sets} of SI. As shown in \autoref{si-tab:performance-compare}, MatterSim outperforms force fields trained on open relaxation trajectory databases with a substantial increase in accuracy by an order of magnitude compared to previous best-in-class,  which showcases the model being faithful in reproducing first-principles potential energy surfaces covering wide chemical, temperature, and pressure spaces. Most significant improvements are observed on the \texttt{MPF-TP} and \texttt{Random-TP} datasets, which are sampled from high temperature and pressure simulations; see \autoref{fig:overview:data-overview}(e) for more details. This enables MatterSim to carry out wide ranges of zero-shot simulation tasks including but not limited to materials discovery, phonon prediction, mechanical property prediction, Gibbs free energy prediction, phase diagram construction, and molecular dynamics simulations.

\textbf{Materials Discovery.}
MLFFs combined with high-throughput screening, crystal structure prediction, or generative models, have shown the capability to accelerate materials discovery where the force field is used as an efficient surrogate to first-principles method to compute energy as a measure of materials' stability.\cite{zeni2023mattergen,merchant2023scaling,xie2021crystal} At the core lies the accuracy in measuring ground-state energies of materials of wide elemental combinations. MatterSim shows a strong capability in predicting stability of new materials, with an F1 score of 0.83 and a mean absolute energy error of the formation energy of \SI{25}{meV/atom} benchmarked on MatBench Discovery, details to follow in \autoref{si-tab:matbench}, marking the SOTA capability to relax the initial structures as well as to accurately label the energies of the relaxed ones. To further demonstrate its potential at scale, we carried out an exhaustive search on all binary chemical systems using random structure search (RSS),\cite{pickard2011ab}
 and the computational details are listed in \autoref{si-sec:rss-details} of SI. 
RSS has the advantage of baring a theoretical guarantee of being exhaustive, but its applicability has been constrained due to the prohibitive computational cost of relaxations with first-principles methods and the lack of an MLFF that is capable of predicting materials near and far from their equilibrium positions. (The initial structures of RSS are far from equilibrium.) Using MatterSim, we carried out materials screening on all 4,005 unary and binary chemical systems of 89 elements with 45 chemical compositions for binary chemical systems, up to 12 atoms in the unit cell. For each chemical system, we generate 20,000 candidate materials, resulting in about 80 million structures in total. By taking the most stable three structures from each chemical composition according to MatterSim's energy prediction,  and using first-principles computations for verification, we identified 16,399 structures to be on or below the energy convex hull defined by the \texttt{Alexandria-MP-ICSD} structures\cite{jain2013commentary,schmidt2022dataset,schmidt2022large,bergerhoff1987crystallographic} (See Ref.~\citenum{zeni2023mattergen} for more details). Importantly, on the combined energy convex hull formed by \texttt{Alexandria-MP-ICSD} and the \texttt{RSS} datasets, the current \texttt{RSS} constitutes the largest contribution of 5,213 out of 7,268 materials, representing the best coverage of 71\%, compared with any previous efforts as shown in \autoref{fig:materials-discovery}(a). Among the 5,213 stable structures, 1,974 of them are newly discovered, i.e., not present in the \texttt{Alexandria-MP-ICSD} dataset (see \autoref{fig:materials-discovery}(b)). \autoref{fig:materials-discovery}(c) presents the element-wise appearance of the 852 materials out of the 5,213 structures on the combined hull, excluding materials that would be \textit{potentially} impacted by anion correction implementation in Materials Projects. Our findings underscore the vast potential for discovering diverse new materials, even within binary chemical systems.

\begin{figure}
\centering
    \includegraphics[width=0.8\textwidth]{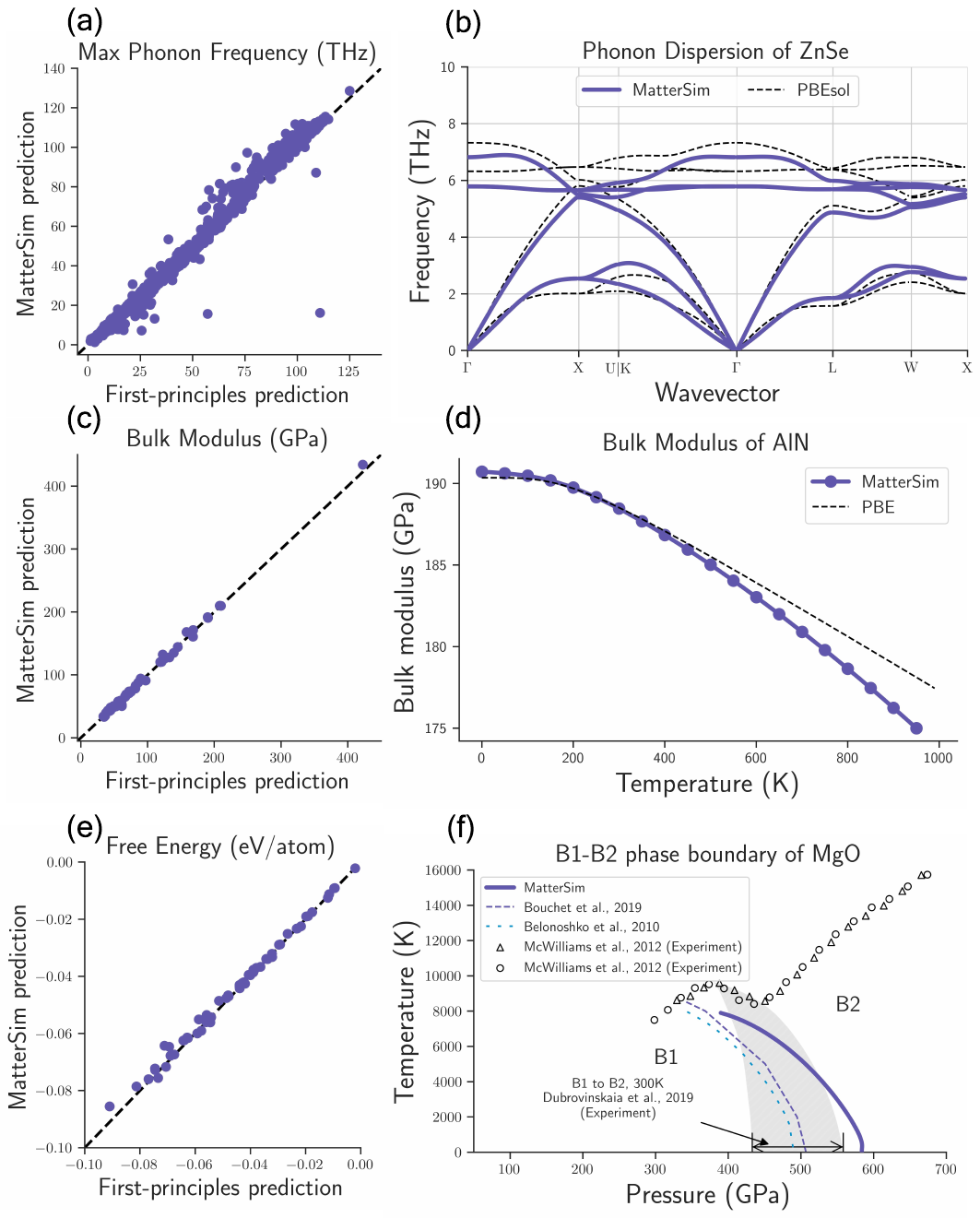}
    \caption{\textbf{MatterSim as a zero-shot emulator for predicting lattice dynamics and thermodynamic properties.} \textbf{(a)}, \textbf{(c)} and \textbf{(e)} are parity plots of maximum phonon frequency, bulk modulus and computed free energy difference between 0 and \SI{300}{\kelvin}, respectively; \textbf{(b)} is the phonon dispersion of \ce{ZnSe}, \textbf{(d)} is the temperature dependent bulk modulus of \ce{AlN} and \textbf{(f)} is the predicted B1-B2 phase boundary of \ce{MgO} with a comparison to first-principles studies and experimental measurements.}\label{fig:phonon-related-properties}
\end{figure}

\textbf{Phonons.} 
Phonons are pivotal in solid-state physics and materials science,\cite{bloch1929quantenmechanik,baroni2001phonons} acting as key indicators of dynamic stability and the paramount foundation for predicting mechanical properties and free energies, but it is computationally expensive to compute phonons using first-principles methods.\cite{baroni1987green,giannozzi1991ab,kresse1995ab,yang2021combined,yang2022computational} 
While machine learning can in principle accelerate this process\cite{batatia2023foundation, chen2022universal}, enhanced quantitative predictive power is needed.\cite{fang2024phonon}
MatterSim achieves high accuracy in predicting phonon spectra of materials thanks to its faithful and robust reproduction of the potential energy surface close to local minima.  \autoref{fig:phonon-related-properties}(a) shows the benchmark results on the materials from the PhononDB,\cite{PhononDB} using maximum phonon frequency as an indicator, and a good agreement is achieved with a mean absolute error (MAE) of \SI{0.87}{\tera\hertz}. An example phonon dispersion of \ce{ZnSe} is shown in \autoref{fig:phonon-related-properties}(b). Compared with first-principles references, not only is the highest frequency reproduced but also the entire spectra. More phonon dispersions of example materials and their comparison with first-principles calculations can be found in \autoref{si-sec:phonon_prediction} in the SI. 

\textbf{Mechanical Properties.}
Understanding mechanical properties is crucial in materials design and engineering to ensure safety and reliability, especially when taking into consideration temperature and pressure dependence. We showcase the capability of MatterSim to predict mechanical properties by computing the bulk modulus of a wide range of ordered inorganic crystals gathered from previous studies (see \autoref{si-sec:mechanical-properties} for details) and their temperature dependence under quasi-harmonic approximation (QHA) with computational details list in \autoref{si-sec:mechanical-properties}.
\autoref{fig:phonon-related-properties}(c) shows the parity plot of the \SI{0}{\kelvin}-bulk modulus predicted from MatterSim and their first-principles references. A remarkable agreement is achieved with an MAE of only \SI{2.47}{\giga\pascal}. In addition, as a model that predicts materials under finite temperature, we also predict the temperature dependence of the bulk modulus of materials. As an example, \autoref{fig:phonon-related-properties}(d) exhibits the temperature dependence of the bulk modulus of \ce{AlN} predicted by MatterSim, with the MAE being \SI{0.97}{\giga\pascal} over the temperature range and with a percentage error less than 5\% up to \SI{1000}{\kelvin} compared to first-principles references.
A detailed comparison on other materials are shown in \autoref{si-sec:mechanical-properties} of SI. These results demonstrate MatterSim's robustness and accuracy in predicting the effects of materials' properties under a wide range of temperatures. In addition to temperature dependence, we demonstrate that MatterSim is capable of predicting the pressure-dependent behavior in \autoref{si-fig:enthalpy-parity-plot} up to \SI{1000}{\giga\pascal}. Such capability further signifies the importance of data coverage, especially under realistic temperatures and pressures.

\textbf{Free Energy and Phase Diagrams.}
Over the last two decades, high-throughput computations driven by first-principles methods\cite{jain2013commentary} and more recently by large-scale machine learning\cite{merchant2023scaling,chen2022universal} have been proposed in hope to accelerate the discovery of new materials. However, such methods heavily relied on the \textit{energy above hull} metric to determine the stability of proposed candidates, suffering from the \textit{zero-Kelvin curse}\cite{tolborg2022free} --- the stability of materials is measured by their ground-state energies without considering  temperature effect. Formally, the thermodynamic stability of a material is determined by its Gibbs free energy under synthesizing and operating conditions. While such quantity can be computed using first-principles methods in principle, the cost to transit from `energy above hull' to `free energy above hull' is prohibitive. We benchmark MatterSim for its efficiency and accuracy on free energy prediction on wide ranges of ordered inorganic solids by comparing with both first-principles calculations using the PBE functional,\cite{perdew1996generalized} and experimental measurements.\cite{bartel2018physical} %
As shown in \autoref{fig:phonon-related-properties}(e)-(f), and \autoref{si-fig:free-energy-examples-and-histogram}, MatterSim achieves a sub-\SI{10}{meV\per atom} error for temperatures up to \SI{1000}{\kelvin} when compared with QHA computations at PBE level of theory, signifying a near-first-principles predictive power. More importantly, when compared with experimental measurements on over 200 materials as shown in \autoref{si-fig:free-energy-comparison-vs-expt-overview}, it achieves an MAE of \SI{15}{meV\per atom}(see \autoref{si-sec:free-energy-prediction}), lower than dedicated models directly trained on experimental data.\cite{bartel2018physical} 
With such a result, we demonstrate as a proof-of-concept construction of temperature- and pressure-dependent phase diagrams using MatterSim. Under QHA (See \autoref{si-sec:mechanical-properties} for details), we computed the phase boundaries (\autoref{fig:phonon-related-properties}(f)) of \ce{MgO} and Si (discussed in \autoref{si-sec:phase_diagram_computation}). In \autoref{fig:phonon-related-properties}(f), MatterSim predicts the transition pressure of MgO from B1 to B2 at \SI{300}{\kelvin} to be \SI{584}{\giga\pascal}, which is very close the recent experimental measurement 429--\SI{562}{\giga\pascal}\cite{dubrovinskaia2019b1} and a recent first-principles prediction \SI{520}{\giga\pascal}\cite{zhang2023toward}. In addition, \autoref{fig:phonon-related-properties}(f) plots the B1--B2 boundary over temperatures up to \SI{16000}{\kelvin} and pressures up to \SI{700}{\giga\pascal}. MatterSim not only computes phase transition pressures in good agreement with experiments for ambient temperature, but also predicts the phase stability under extreme temperatures and pressures well --- the predicted phase boundary falls into the shaded region connecting experiments reported in literature and is very close to the experimentally measured boundary for temperatures higher than \SI{4000}{\kelvin}.\cite{dubrovinskaia2019b1,mcwilliams2012phase} Notably, such a prediction not only requires good description of free energy with the temperature dependence, but also its pressure dependence, signifying the importance of generalizability hardly achieved by directly fitting to limited experimental data.\cite{bartel2018physical}

\begin{figure}
    \centering
    \includegraphics[width=0.85\textwidth]{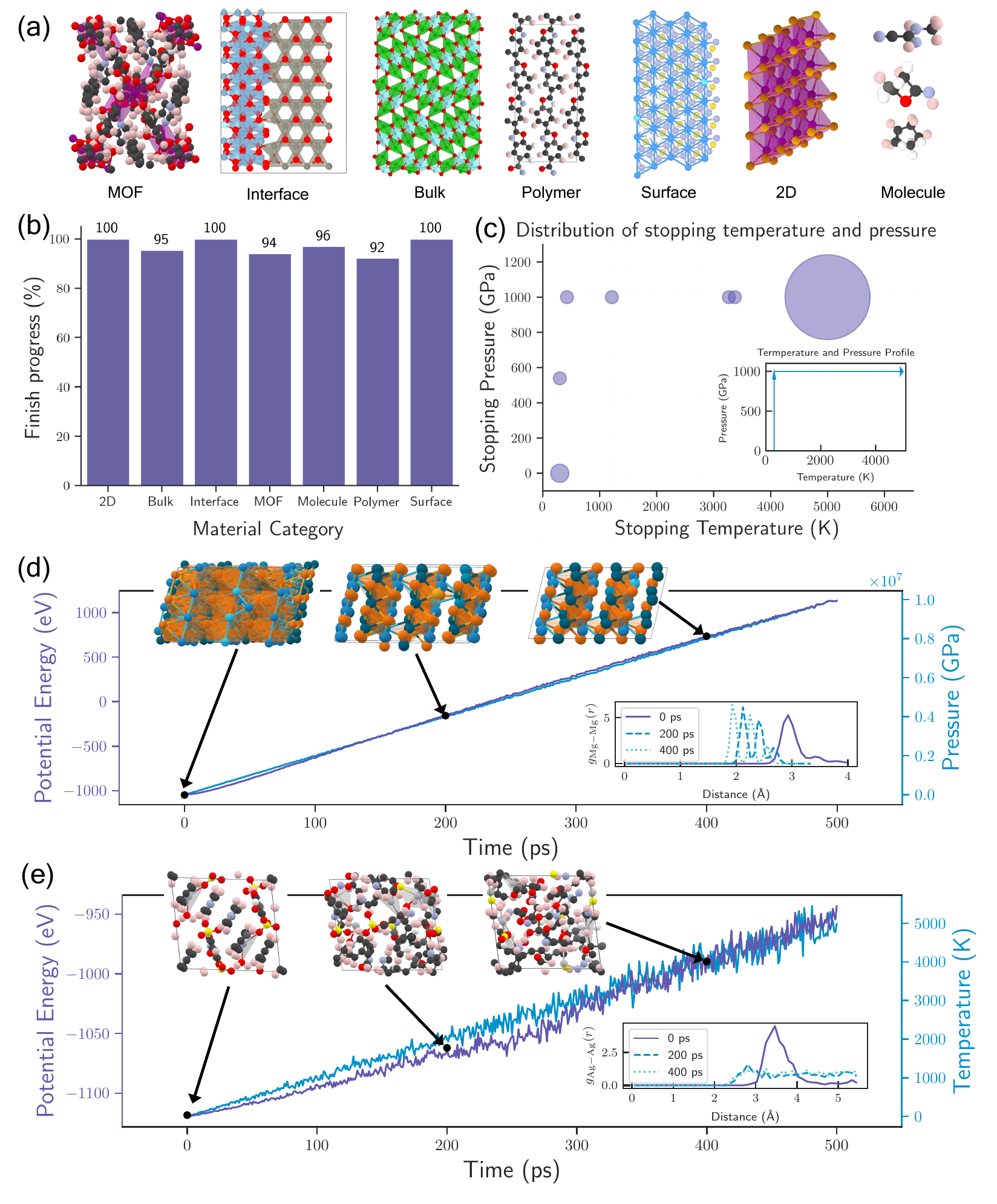}
    \caption{\textbf{MatterSim as a zero-shot molecular dynamics (MD) engine.} \textbf{(a)} Examples materials selected for running molecular dynamics; \textbf{(b)} the success rate of molecular dynamics with increasing temperature and pressure for various categories of materials; \textbf{(c)} analysis of the stopping temperature and pressure of the molecular dynamics trajectories, with the temperature and pressure profile of the trajectory shown in the inset; \textbf{(d)} the potential energy of a MOF material under increasing temperature and NVT ensemble, with the inset being the radial distribution function of the \ce{Ag}-\ce{Ag} atoms $g_\mathrm{Ag-Ag}(r)$ at 0, 200 and \SI{400}{\pico\second} of the trajectory; and \textbf{(e)} the potential energy of a bulk inorganic material under increasing pressure and NPT ensemble, with the inset being the radial distribution function of the \ce{Mg}-\ce{Mg} atoms $g_\mathrm{Mg-Mg}(r)$ at 0, 200 and \SI{400}{\pico\second} of the trajectory.}\label{fig:md}
\end{figure}

\textbf{Molecular Dynamics Simulations.}
MLFFs have shown significant acceleration in molecular dynamics simulations and high accuracy compared with first-principles methods, if trained properly.\cite{unke2021machine}
The universality and robustness of MLFFs heavily rely on the coverage of underlying training data. A lack of chemical, configurational, or compositional coverage inevitably leads to erroneous or even diverging simulations.\cite{fu2022forces} This issue becomes particularly pronounced when the simulation temperatures and pressures are high.
Benefiting from the data collection and training pipeline, MatterSim serves as a surrogate model to first-principles methods to carry out robust, efficient, and accurate molecular dynamics for complex materials under finite temperature and pressure conditions. 
To validate MatterSim's robustness under arbitrary simulation conditions (especially on finite temperature and pressure tasks), we randomly selected 118 systems including bulk inorganic materials, metal organic frameworks, two-dimensional materials, interfaces, molecular crystals, polymers and surfaces, with details of the selected materials discussed in \autoref{sec:si_md} of SI.
All of these systems are subject to heating from 0 to \SI{5000}{\kelvin} in a relatively short time frame to benchmark the emulator's robustness to deal with both the crystalline and the liquid or disordered structures, as well as the phase transition between them. A success rate is defined as the ratio of the actual runtime to the preset total time in MD simulations. As shown in \autoref{fig:md}(b),  MatterSim achieved more than 90\% success rate for all the material families tested, exhibiting robustness over wide temperature ranges.
In addition, all bulk systems are subject to additional compression from 0 to \SI{1000}{\giga\pascal} (followed by heating from 300 to \SI{5000}{\kelvin}) to further benchmark MatterSim's pressure response, see the inset in \autoref{fig:md}(c) for details. As \autoref{fig:md}(c) shows, the vast majority of systems have completed the entire simulation process and the average finished rate is above 90\% as well. 
Beyond robustness, MatterSim also achieves high accuracy. As an indicator, MatterSim has an up-to-10-fold lower prediction error compared with previous universal MLFFs on the \texttt{Random-TP} and \texttt{MPF-TP} datasets that are created under wide temperature and pressure ranges, see \autoref{fig:overview:data-overview}(e) and \autoref{si-tab:performance-compare} for more details. Interestingly, MatterSim demonstrates good generalizability to material systems that is not trained on. We depicted two example MD trajectories, including a metal-organic framework (MOF) compound under NPT ensemble and a bulk inorganic material under NVT ensemble to show the accuracy of MatterSim in \autoref{fig:md}(c) and \autoref{fig:md}(d) -- for the six snapshots, MatterSim predicts a mean error of energy lower than \SI{50}{meV\per atom}, within wide temperature and pressure ranges.

\subsection{MatterSim as an active learner}

Uncertainty quantification and continuous learning are critical to the successful application of a machine learning model to predict material properties or carry out meaningful simulations. This is especially the case for molecular dynamics because making prediction on out-of-distribution (OOD) configurations can lead to erroneous energies and forces, which in turn results in unphysical simulation trajectories or even simulation failure. 
Considering that the pretrained MatterSim model covers wide ranges of atomic configurations, the idea is that only a small amount of new data is needed to supplement the model to capture the OOD  configurations. We show that with the help of a model ensemble, MatterSim provides confidence estimates in simulating any system without performing actual first-principles computations. More importantly, whenever the pretrained model is deemed unconfident, MatterSim only requires a small fraction of the trajectory being labeled by first-principles computations as additional training context to reach the same level of accuracy compared with training from scratch.

\begin{figure}
    \centering
    \includegraphics[width=\textwidth]{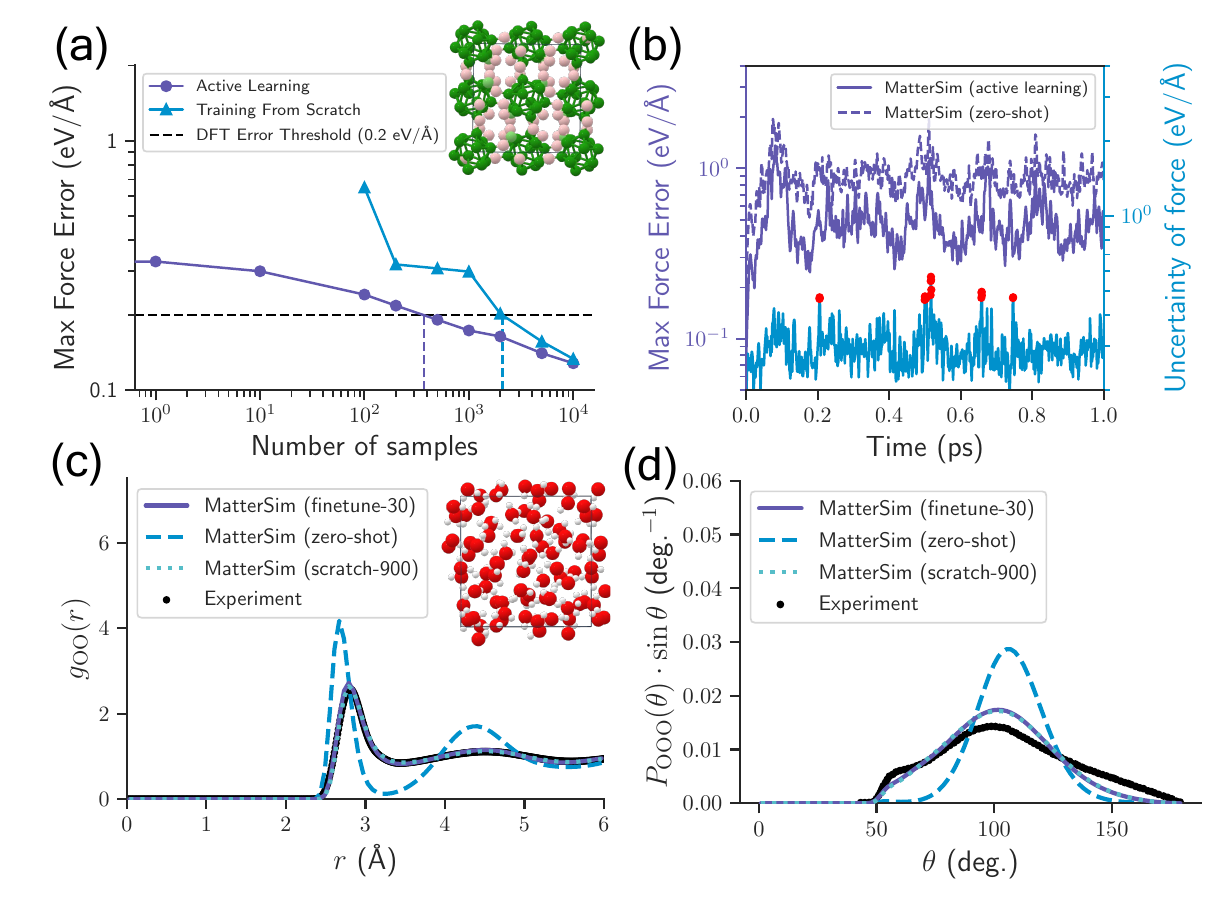}
    \caption{\textbf{MatterSim's efficiency for complex simulation tasks using active learning and fine-tuning.} 
    \textbf{(a)} Max force error against first-principles results in the simulation trajectory of \ce{Li2B12H12} from the actively learned model and the model trained from scratch  and its crystal structure, with increasing number of data samples. \textbf{(b)} The inference error and uncertainty of force along the AIMD trajectory for \ce{Li2B12H12}. The red points represent the structures with above-threshold uncertainty taken for active learning.
    \textbf{(c)} Oxygen-oxygen radial distribution functions of water obtained from MD simulations performed by the zero-shot (blue thick dotted dash), scratch-900 (blue thin dotted dash) and finetune-30 (dark purple solid line) models. Black dots represent experimental references.\cite{skinner2014structure,chen2016ab} \textbf{(d)} Oxygen-oxygen-oxygen-angular distribution functions,$P_{OOO}(\Theta)$, obtained from MD simulations performed by the zero-shot (blue thick dotted dash), scratch-900 (blue thin dotted dash) and finetune-30 (dark purple solid line) models. Black dots symbolize the empirical potential structural refinement (EPSR) of joint Xray-Neutron measurements for bulk water at \SI{298}{K}.\cite{soper2008quantum} Further details of each model can be found in \autoref{si-sec:finetuning-and-molecular-dynamics-on-liquid-water} in the SI.}\label{fig:al-and-ft}
\end{figure}

Building on the strengths of MatterSim and its active learning capabilities, we applied it to a few intricate systems to showcase its efficacy, including molten phosphorus, molten boron, and an ionic superconductor (lithium dodecahydro-closo-dodecaborate, \ce{Li2B12H12}), whose structures are depicted in \autoref{fig:al_structures} and the inset of \autoref{fig:al-and-ft}(a).
Such systems demonstrate complex interatomic interactions and intrinsically require heavy effort in data generation and model training.\cite{deringer2020general, zhou2023structure} In our study, MatterSim selected the structures for active learning based on an ensemble criterion described in \autoref{sec:si_active_learning} in the SI, and it only requires including a small fraction of the structures in the simulation trajectory as additional training data to recover a high prediction accuracy. As shown in \autoref{fig:al-and-ft}(a), the model reproduces similar level of accuracy for \ce{Li2B12H12}, while including only 15\% of the data if it were trained from scratch. Similar performance was also observed for phosphorus and boron shown in \autoref{fig:si_B_al} and \autoref{fig:si_P_al}. In addition, we show in \autoref{fig:al-and-ft}(b) that by incorporating additional first-principles supervision signal on the data points of high uncertainties in the active learning process, we notably reduced the maximum error compared to the zero-shot prediction of MatterSim, which clearly demonstrated the efficacy of MatterSim's capability as an active learner.

\subsection{MatterSim with arbitrary level of theory}
The potency of predictions generated by the machine learning emulator is inherently constrained by the theoretical level of the training data. Here, MatterSim capitalizes on the supervisory signal derived from GGA-PBE\cite{perdew1996generalized} functional (and Hubbard U correction\cite{anisimov1991band} for qualified materials; the detailed information is listed in \autoref{si-sec:computation-details}).
This underlying constraint imposes a limitation on the accuracy of MatterSim when applied to a more expansive range of systems and applications.
Here as a demonstration, we conjugate MatterSim with the fine-tuning technique to recover the structural and dynamical property of liquid water at the rev-PBE0-D3 level of theory and more remarkably show its high data efficiency. Details of the fine-tuning setup can be found in \autoref{si-sec:finetuning-and-molecular-dynamics-on-liquid-water}. %

When simulated with MatterSim in the zero-shot mode at PBE level of theory, water displays an overstructured feature demonstrated by both the radial distribution function including $g_\mathrm{OO}(r)$, $g_\mathrm{OH}(r)$, $g_\mathrm{HH}(r)$ (see \autoref{fig:al-and-ft}(c) and \autoref{fig:rdf-finetune-oh-and-hh}), and angular distribution function of the oxygen-oxygen-oxygen triplets (see \autoref{fig:al-and-ft}(d)) in contrast to experimental measurements. Such overstructured behavior for water systems has been reported in literature, indicating the deficiency of the PBE functional in correctly simulating water properties.\cite{skinner2014structure,chen2016ab,distasio2014individual} However, with only 30 configurations sampled from an \textit{ab initio} molecular dynamics (AIMD) trajectory with rev-PBE0-D3 level of theory\cite{cheng2019ab,monserrat2020liquid}, MatterSim can be efficiently fine-tuned and the resultant model (indicated as finetune-30 model) reaches the same performance as the model trained from scratch with all 900 configurations from the AIMD trajectory (denote as scratch-900) for water structural properties (see  \autoref{fig:al-and-ft}(c) and \autoref{fig:al-and-ft}(d)). In addition to structural properties of liquid water, we found that the dynamical properties, here exemplified by the self diffusion coefficient $\mathcal{D}$, can also be greatly improved. The finetune-30 model predicts $\mathcal{D}$=\SI{1.862d-5}{\centi\meter^2\per\second} for liquid water, with an error below 20\% w.r.t. the experimental measurements (2.3--\SI{2.4d-5}{\centi\meter^2\per\second})\cite{chen2017ab}. The detailed discussion of $\mathcal{D}$ prediction can be found in \autoref{si-sec:finetuning-and-molecular-dynamics-on-liquid-water}. Therefore, by finetuning MatterSim, we were able to obtain the similar accuracy of the scratch-900 model on this task while using only 1/30 of the training data, which clearly underscores the effectiveness of supervised pretraining on large-scale high-fidelity first-principles data with wide coverage.

\subsection{MatterSim as a direct  property predictor}\label{sec:e2e-model}
Direct prediction of materials' properties from their structures is critical to large-scale virtual screening due to its low computational cost when compared to first-principles methods. This becomes particularly noticeable when it comes to properties that are computationally intractable to calculate or experimentally challenging to measure. Machine learning models are promising in this area due to their capability of capturing non-linear mappings in high dimensional spaces. However, the prediction error is typically too large to be of practical use due to the lack of data in specific domains. MatterSim adopts an architecture with high transferability (graphormer) and is pretrained on high-quality first-principles data with a wide coverage. Such data contains the rich dynamics of off-equilibrium systems, presenting a wealth of complex information for deep learning models to learn from. This pre-training enables the model to extract expressive features of materials to accommodate domain-specific property data, facilitating the application to a wide range of downstream tasks beyond atomistic simulations. 

MatterSim's capability to directly predict materials properties from structures is benchmarked on a few regression tasks from MatBench,\cite{dunn2020benchmarking} including prediction of computed band gaps, shear moduli, dielectric constants, max phonon frequency of bulk crystalline materials, and the exfoliation energies of two-dimensional materials. Detailed training scheme is provided in \autoref{si-sec:end-to-end-model} of SI. As shown in \autoref{tab:end-to-end-prediction}, leveraging the expressive features extracted from the vast materials data, the model is able to learn with minimal number of data points and reaches the highest accuracy in predicting all of these properties. Interestingly, the improvement in predictive power is regardless of the model architecture when the model is pre-trained on the large-scale materials database presented in this work, as shown in \autoref{si-tab:comparison-e2e-models}. This signifies the importance of data coverage in extraction of representative materials features when training deep learning models for robust domain-specific property predictions.

\begin{table}[htbp]
    \centering
    \caption{\textbf{MatterSim as a direct property predictor.} The performance of MatterSim's performance of the predictions of the properties in MatBench leaderboard\cite{dunn2020benchmarking} with a comparison with previous models trained exclusively with domain specific data.
    \label{tab:end-to-end-prediction}}
    \begin{tabular}{cccc}
    \toprule
    Property & Specialized model & Training from scratch & MatterSim \\
    \hline
    MP Gap (eV) & 0.1559\cite{ruff2023connectivity}  &  0.3031 & \textbf{0.1290} \\
    $\log G_\mathrm{VRH}$ (GPa) & 0.0670\cite{ruff2023connectivity} & 0.0895  & \textbf{0.0608} \\
    $\log K_\mathrm{VRH}$ (GPa) & 0.0491\cite{ruff2023connectivity} & 0.0687 & \textbf{0.0488} \\
    Dielectric (unitless) & 0.2711\cite{de2021robust}  & 0.3823 & \textbf{0.2516} \\
    Phonons (\si{\per\centi\meter}) & 28.7606\cite{chen2019graph} & 65.8220 & \textbf{26.0220} \\
    jdft2d (\si{meV\per atom}) & 33.1918\cite{de2021robust} &  47.8040  & \textbf{32.7620} \\
    \bottomrule
    \end{tabular}
\end{table}

\section{Discussion}
The accurate prediction of material properties and the simulation of their behaviors without constraints on chemical elements, compositions and configurations are crucial to the digital transformation of materials design. While deep learning has already shown promise in making such predictions, their practical use is still constrained due to the limited generalizability across the vast materials space. This challenge is particularly pronounced when factoring in temperature and pressure, as the configurational space becomes exceedingly large. MatterSim addresses this by combining deep graph neural networks, active learning, and large-scale first-principles computations. The model achieves up to 10-fold increase compared with previous best-in-class, in the prediction accuracy of energies, forces, and stresses for off-equilibrium material structures sampled from an extensive chemical space under finite temperature and pressure, benchmarked against first-principles computations. This enables robust and accurate zero-shot emulation of materials' ground-state energetics, as well as their dynamical behaviors under arbitrary temperatures and pressures. 
 Remarkably, the free energies computed using the model agree well with experimental results; this opens the possibility to efficiently predict experimental phase diagrams of candidate materials. 

More importantly, MatterSim provides a platform for adaptive learning and customization based on the specific materials design request. Starting from the model pretrained on diverse first-principles results, only a small amount of new data needs to be brought to the model to refine the pretrained potential energy surface, thanks to the good coverage of the initial dataset. In addition, the level of theory of the emulator can then be customized by incorporating a small amount of expensive data with beyond the PBE level of theory when necessary. For example, MatterSim allows fine-tuning to achieve the hybrid functional level of theory with only 3\% of the data needed to train from scratch. Such two-step adaptivity, i.e., fine-griding the sampling, and fine-tuning the level of theory, enables extreme data efficiency thanks to the pretrained model. Finally, the model also allows direct connection with real-world experiments without complex ground-up simulations by building end-to-end property predictors, thanks to the expressive feature extracted from the pretrained model.

Despite these advancements, MatterSim could be improved in several areas. From the perspective of model development, it currently utilizes a semi-local description of atomic interactions, where long-range interactions majorly leverages message passing or updates on attention weights through graph nodes. Even though this model demonstrates superior performance compared to models that rely solely on local environments,\cite{chen2022universal,chmiela2023accurate}, it doesn't perform effectively in scenarios where long-range interactions dominate the properties, such as polymeric and heterogeneous systems\cite{ko2023accurate}. As for data coverage, the current model is trained only on homogeneous bulk systems, without explicit inclusion of surface and interface data that are crucial for applications such as catalysis. Additionally, the model currently only naively supports inferences with DFT-PBE level of theory, limiting its use for systems involving complex interactions, such as polymers and organic liquids. Inclusion of additional data with different theory levels by multi-task pretraining could aid in this respect. Further improvement on data efficiency and the model's prediction accuracy is possible with semi-supervised pretraining. Currently, the model only supports the native prediction of energy, forces, and stresses. Including more data modalities, such as charge, spin, magnetic moments, and even more complex electronic structure features, could further enhance the model's accuracy and applicability.

\backmatter

\section*{Acknowledgements}
We thank Chris Bishop, Tie-Yan Liu, Haiguang Liu, Tao Qin, Bin Shao, Jia Zhang, and Karin Strauss for their invaluable discussions and expert input which helped to shape this work. We also recognize the valuable input from Chi Chen which helped training of the M3GNet-based model. Special thanks are due to Prof. Davide Donadio for his comments concerning the application of MatterSim. Our manuscript has been enhanced by the thorough reviews and constructive feedback from Bichlien Nguyen, Yu Xie and Jonas K\"ohler. We appreciate Ryota Tomioka for the implementation of DFT computation pipelines, and Deniz Gunceler and Maik Riechert for their support with our computational infrastructure. Shoko Ueda and Peggy Dai have been instrumental in managing the project, and we are indebted to Lina Lu and Yang Ou for their artistic contributions that have vividly brought our work to life. Finally, we extend our gratitude to the entire Microsoft Research AI for Science team for the enriching daily discussions and collective wisdom that have been integral to our progress. HY extends a personal note of gratitude to Zifan Ye and Cunzhi Zhang for their feedback on water simulation and free energy prediction. ZL wishes to express gratitude to Pascal Salzbrenner and Prof. Chris Pickard for discussions on crystal structures under high pressure. 

\section*{Contributions}
HY, YZ, JL, HH, and ZL conceived the study, HY, CH, YZ, XL, YS, JL, GL, ZC, SC, CL, HH, ZL implemented the methods, HY, CH, YZ, XL, YS, JL, GL, ZC, SC, CZ, HH, ZL performed experiments, CZ, MH, RP, AF, DZ, TX, JS, LS helped with the implementation of the methods and all authors wrote this manuscript. ZL led the research.

\clearpage

\setcounter{section}{0}
\setcounter{figure}{0}  
\setcounter{table}{0}  
\renewcommand{\theHsection}{Supplement.\thesection} %
\renewcommand{\theHsubsection}{Supplement.\thesubsection}  
\renewcommand{\theHsubsubsection}{Supplement.\thesubsubsection}  
\renewcommand{\thefigure}{S\arabic{figure}}  
\renewcommand{\thetable}{S\arabic{table}}  
\renewcommand{\thesection}{S\arabic{section}}  
\renewcommand{\thesubsection}{S\arabic{section}.\arabic{subsection}}  
\renewcommand{\thesubsubsection}{S\arabic{section}.\arabic{subsection}.\arabic{subsubsection}}

\phantomsection %
\addcontentsline{toc}{part}{Supplementary Information}

\begin{center}  
    \Large \textbf{Supplementary Information} \\[0.5cm] %
    \normalsize  
    Han Yang\textsuperscript{1*\textdagger},  
    Chenxi Hu\textsuperscript{1\textdagger},  
    Yichi Zhou\textsuperscript{1\textdagger},  
    Xixian Liu\textsuperscript{1\textdagger},  
    Yu Shi\textsuperscript{1\textdagger},  \\
    Jielan Li\textsuperscript{1*\textdagger},
    Guanzhi Li\textsuperscript{1\textdagger},
    Zekun Chen\textsuperscript{1\textdagger},
    Shuizhou Chen\textsuperscript{1\textdagger},
    Claudio Zeni\textsuperscript{1}, \\
    Matthew Horton\textsuperscript{1},
    Robert Pinsler\textsuperscript{1},
    Andrew Fowler\textsuperscript{1},
    Daniel Z\"ugner\textsuperscript{1},\\
    Tian Xie\textsuperscript{1},
    Jake Smith\textsuperscript{1},
    Lixin Sun\textsuperscript{1},
    Qian Wang\textsuperscript{1},
    Lingyu Kong\textsuperscript{1},\\
    Chang Liu\textsuperscript{1},
    Hongxia Hao\textsuperscript{1*},
    Ziheng Lu\textsuperscript{1*} \\[1em]  
      
    \textsuperscript{1}{Microsoft Research AI for Science} \\[1em]
      
    \textsuperscript{*}Corresponding author: \href{mailto:hanyang@microsoft.com.com}{hanyang@microsoft.com}; \href{mailto:jielanli@microsoft.com}{jielanli@microsoft.com};\\  
    \href{mailto:hongxiahao@microsoft.com}{hongxiahao@microsoft.com}; \href{mailto:zihenglu@microsoft.com}{zihenglu@microsoft.com} \\
    \textsuperscript{\textdagger}These authors contributed equally to this work.  
\end{center}

\clearpage
\section{Model architecture and training details}\label{si-sec:model-arch}
\subsection{Materials Graphs}
\begin{figure}
    \centering
    \includegraphics[width=1\linewidth]{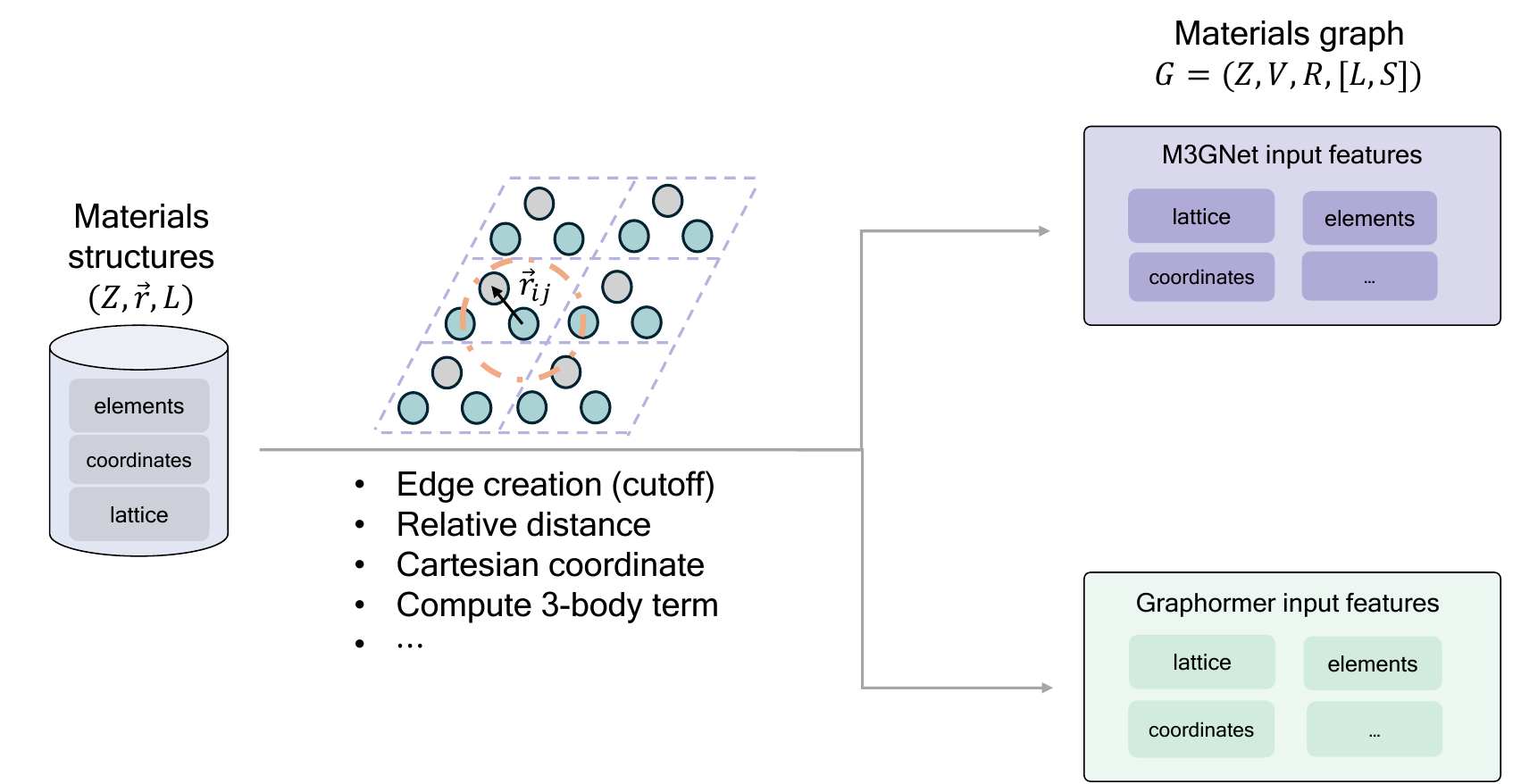}
    \caption{MatterSim leverages materials graphs built upon point clouds to represent atomic interactions and geometric features in Euclidean space.  }%
    \label{si-fig:mattersim-materials-graph}
\end{figure}

The input data for the MatterSim model are constructed from material graphs built upon the underlying point clouds in the three-dimensional Euclidean space with periodic boundary conditions. Each point represents an atom with an associated element from the periodic table. We define a materials graph $\mathcal{G} = (\boldsymbol{Z},\boldsymbol{V},\boldsymbol{R},[\boldsymbol{L}, \boldsymbol{S}])$ (see \autoref{si-fig:mattersim-materials-graph}) with the following components: $\boldsymbol{Z}$ denotes the atomic number $z_i$ and additional features. The geometric features are encapsulated by $\boldsymbol{V}$ and atomic coordinates $\boldsymbol{R}$, with each atomic position $\boldsymbol{r}$ in Euclidean space $\mathbb{R}^{3}$. $\boldsymbol{V}$ represents the relative vectors, such as the bond information between two atoms. $\boldsymbol{S}$ and $\boldsymbol{L}$ are additional optional information, where $\boldsymbol{S}$ is the global scalar state information, such as temperature, pressure, and other conditions, and $\boldsymbol{L}$ is the $3\times 3$ lattice matrix in crystals. Within material graphs, nodes correspond to individual atoms and edges are formed based on a predefined rule. Here, a radial cutoff distance $r_c$ is used to construct edges. For any two atoms $\boldsymbol{r}_i$ and $\boldsymbol{r}_j$, there exists an edge if the Euclidean distance between them is less than or equal to $\leq r_c$. It should be noted that if the coordinates are fractional, we scale them to Cartesian coordinates. As a form of geometric graph, materials graphs exhibit roto-translational symmetry in Euclidean space; specifically, MatterSim maintains roto-translational invariance for scalar properties, such as total energy of materials, and equivariance for vectorial properties like forces. Given a material graph, MatterSim adapts different input representations and crystalline features compatible with the underlying architectures, M3GNet and Graphormer, which will be discussed in more details in the following sections.

\subsection{M3GNet}
M3GNet is a graph neural network that explicitly incorporates two- and three-body interactions,\cite{chen2022universal} enabling high-accuracy predictions of material properties. This architecture is summarized briefly below; for further information, we refer readers to the original M3GNet publication\cite{chen2022universal}.
The major innovation of M3GNet relies on incorporation of three-body interaction into its message-passing framework, thereby enriching the updated atomic and bond features with three-body information. This is achieved through the following formulation:
\begin{equation}
\begin{aligned}
\tilde{\boldsymbol{e}}_{ij} &= \sum_k j_l\left(z_{ln}\,\frac{\norm{\boldsymbol{r}_{ik}}}{r_c}\right)Y^0_l(\theta_{jik})\odot \xi(\boldsymbol{W}_v \boldsymbol{v}_k + \boldsymbol{b}_v)f_c(\norm{\boldsymbol{r}_{ij}})f_c(\norm{\boldsymbol{r}_{ik}}), \\
\boldsymbol{e}_{ij}^{\prime} &= \boldsymbol{e}_{ij} + g(\tilde{\boldsymbol{W}}_2 \tilde{\boldsymbol{e}}_{ij}+\tilde{\boldsymbol{b}}_2)\odot \xi(\tilde{\boldsymbol{W}}_1 \tilde{\boldsymbol{e}}_{ij}+\tilde{\boldsymbol{b}}_1),
\end{aligned}
\end{equation}
where $\boldsymbol{e}_{ij}$ is the input edge feature on the bond connecting atoms $i$ and $j$, $\boldsymbol{e}_{ij}^\prime$ is the edge update message containing three-body information, and $\boldsymbol{x}_i$ is the feature of atom $i$.
Here, $\boldsymbol{r}_{ij}$ represents the relative positions of atoms $i$, $j$; $\theta_{jik}$ represents the angle between bonds $\boldsymbol{e}_{ij}$ and $\boldsymbol{e}_{ik}$; $\tilde{\boldsymbol{W}}$ and $\tilde{\boldsymbol{b}}$ are learnable parameters of the neural network,;$j_l$ is the spherical Bessel function with roots $z_{ln}$, $Y^0_l$ is the spherical harmonics function with $m=0$ and $r_c$ is the cutoff radius. In addition, $f_c(r) = 1- 6(r/r_c)^5+15(r/r_c)^4-10(r/r_c)^3$ is a smooth cutoff function, $\xi(\cdot)$ is the sigmoid activation function, $g(x) = x \xi(x)$, and $\odot$ represents the element-wise product. It is worth noting that edge feature $\tilde{\boldsymbol{e}}_{ij}$ is a vector of $n_\mathrm{max}l_\mathrm{max}$ elements, with $n=0,\cdots,n_\mathrm{max}-1$ and $l = 0, \cdots, l_\mathrm{max}-1$, and $n_\mathrm{max}$ and $l_\mathrm{max}$ are user-defined model hyperparameters. The edge update message $\boldsymbol{e}_{ij}^\prime$ are then passed to several graph convolution steps to update both the atom and bond information $\boldsymbol{x}_i$ and $\boldsymbol{e}_{ij}$.

In M3GNet, message passing described above are conducted multiple times, and the resulting atom and bond features $\boldsymbol{v}_i$ and $\boldsymbol{e}_{ij}$ are passed to a gated multi-layer perceptron (MLP) to obtain the prediction of energies $E$. Forces and stresses are predicted by taking gradient of energy with respect to atomic positions $\boldsymbol{f} = -\partial E/\partial \boldsymbol{r}$ and lattice strain $\boldsymbol{\sigma} = V^{-1}\partial E/\partial \boldsymbol{\varepsilon}$ via auto-differentiation, where $\boldsymbol{r}$ are the atomic positions, $V$ is the lattice volume and $\boldsymbol{\varepsilon}$ are lattice strains. In this work, we used a re-implemented version of M3GNet with PyTorch\cite{Ansel_PyTorch_2_Faster_2024} based on original TensorFlow implementation. We note that during the development of MatterSim, another PyTorch version of the M3GNet emerged in the MatGL library.\cite{Ko_Materials_Graph_Library_2021}

\subsection{Graphormer}
\begin{figure}
    \centering
    \includegraphics[width=0.5\linewidth]{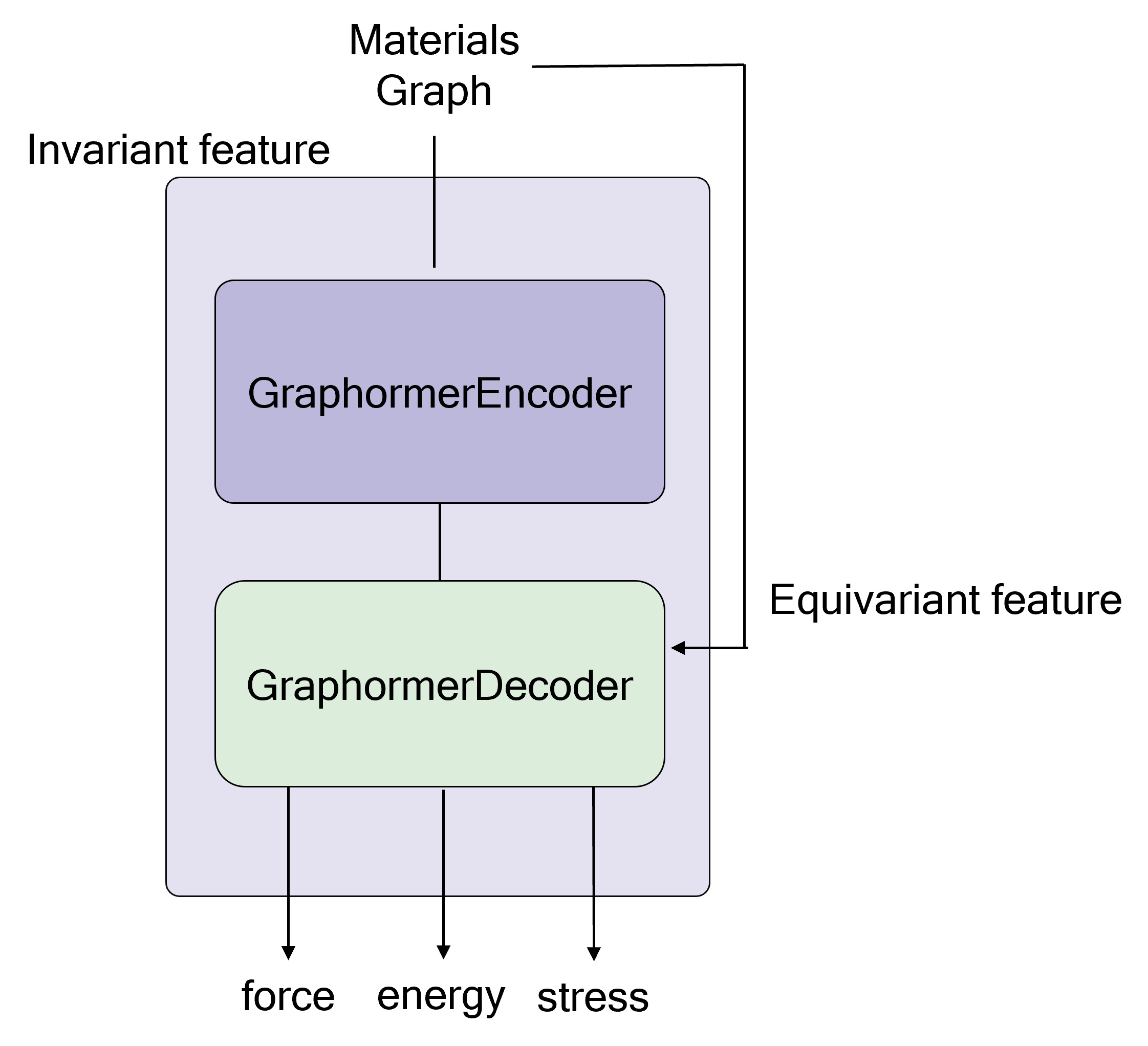}
    \caption{The overview of the Graphormer model.}
    \label{si-fig:graphormer-overview}
\end{figure}

In this section, we provide a short introduction to the design and implementation of Graphormer, with a focus on the modification made to the original architecture for better adaptation to materials structures. Graphormer model consists of two major parts: the structural encoder and property decoder, as illustrated in \autoref{si-fig:graphormer-overview}.

\begin{figure}
    \centering
    \includegraphics[width=0.8\linewidth]{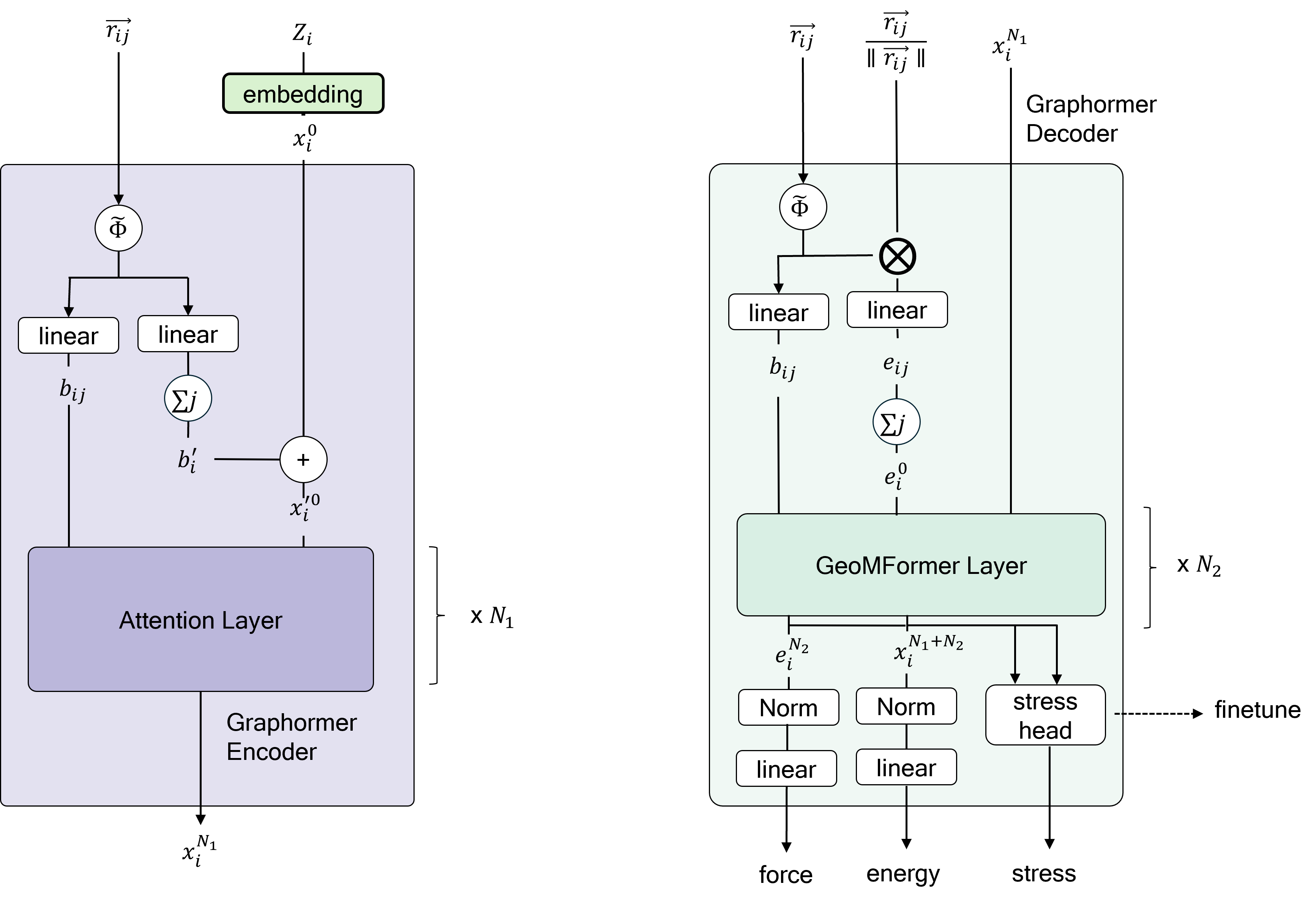}
    \caption{The structural encoder and property decoder architecture of the Graphormer model.}
    \label{si-fig:graphormer-encode-decoder}
\end{figure}

\textbf{Structural Encoder.} %
The structural encoder is depicted in the left panel of \autoref{si-fig:graphormer-encode-decoder}. Here, we provide a concise overview of the encoder's components, which facilitate the mapping of atomic species $Z_i$ and positions $\boldsymbol{r}_i$ into an embedding feature $\boldsymbol{x}_i$. Similar to existing studies, the embedding features are initialized with an embedding function, $\boldsymbol{x}^0_i = \mathrm{Embedding}(Z_i)$. 

Before going through the attention layer to learn the interaction among atoms, we first model the spatial relationships between atoms to obtain an attention bias, which will be added to the later self-attention layer. With $\boldsymbol{r}_{ij} = \boldsymbol{r}_i - \boldsymbol{r}_j$ being the relative positions between atoms $i$ and $j$, and $\tilde{\Phi}(\cdot)$ being the Gaussian basis kernel function,  the attention bias writes
\begin{equation}
    \boldsymbol{b}_{ij} = \mathrm{Linear}(\tilde{\Phi}(\norm{\boldsymbol{r}_{ij}})).
\end{equation}
Then, Graphormer captures the significance of different atoms through centrality encoding by aggregating the spatial encodings and passing them to a linear layer %
\begin{equation}
\boldsymbol{b}^\prime_i = \sum_{j\in\mathcal{N}(i)}\mathrm{Linear}\left({m_{ij} \cdot \tilde{\Phi}(\norm{\boldsymbol{r}_{ij}})}\right),
\end{equation}
where $\mathcal{N}(i)$ is the neighbor list of the $i$-th atom within a predefined cutoff radius $r_c$, and the local mask is given by $m_{ij} = 1 - 6 \left({\norm{r_{ij}}}/{r_c}\right)^5 + 15 \left({\norm{r_{ij}}}/{r_c}\right)^4 - 10 \left(\norm{r_{ij}}/{r_c}\right)^3$ for two atoms $i$ and $j$. The centrality encoding is used to update the initial embeding $\boldsymbol{x}_i^{\prime 0} = \boldsymbol{x}_i^0 + \boldsymbol{b}_i^\prime$, which will be passed to the attention module.

In the multi-head self-attention module of the $h$-th layer, $\boldsymbol{Q}$, $\boldsymbol{K}$, and $\boldsymbol{V}$ are obtained through linear mappings from the features $\boldsymbol{x}^h_i$ and the features are updated as follows 
\begin{equation}
\boldsymbol{x}_i^{h+1} = \sum_j \mathrm{Softmax}\left[\left(\frac{\boldsymbol{Q}\boldsymbol{K}^T}{\sqrt{d}}\right)_{i,j} + b_{ij}\right] \cdot m_{ij} \cdot \boldsymbol{V}_{j}.
\end{equation}
Here $d$ is the hidden dimension and $\sqrt{d}$ is used to normalize the product. It should be noted that $\boldsymbol{K}$ and $\boldsymbol{V}$ for the expanded atoms are directly copied from the initial atoms to ensure the same representation is shared between an atom within the unit cell and its images outside the unit cell. 

To account for the periodic boundary conditions (PBC) inherent in crystal structures, we have adapted the original Graphormer by incorporating the multi-graph construction introduced in Ref.~\citenum{xie2018crystal}. This approach enables us to represent atoms within the unit cell as a series of periodic graphs. In these graphs, image atoms from neighboring lattices are included up to a pre-specified cutoff distance. Similar to exisiting studies, the interaction of information between different atoms is influenced by a smooth cutoff function, which is based on the interatomic distance and a predefined cutoff threshold, allowing us to smoothly decrease the influence of long-distance atomic pairs.

\textbf{Property Decoder}
For the Decoder part of Graphormer (see \autoref{si-fig:graphormer-encode-decoder}), we adopted the GeoMFormer\cite{chen2023geomformer} module, which uses Transformer modules accommodated for SO(3)-equivariant vectors.\cite{vaswani2017attention}
It consists of two separate streams to maintain and learn invariant and equivariant representations. Meanwhile, it also includes a cross-attention module that connects the two streams, enabling information fusion between the two steams and enhancing geometric modeling in each stream. 

To accommodate periodic boundary condition in the decoder,  we made the following modifications. In the initialization part, we no longer use the original vector $\boldsymbol{e}_i^0 = {\boldsymbol{r}_i}/{\norm{\boldsymbol{r}_i}} \tilde{\Phi}(\norm{\boldsymbol{r}_i})$, as it does not maintain invariance to translation in periodic systems. Instead, we adopt the following initialization: 
\begin{equation}
\boldsymbol{e}_i^0 = \sum_j m_{ij} \cdot \frac{\boldsymbol{r}_{ij}}{\norm{\boldsymbol{r}_{ij}}} \tilde{\Phi}(\norm{\boldsymbol{r}_{ij}}),
\end{equation}
Furthermore, we adopted the multi-graph technique similar to that in the structural encoder for the atoms in the unit cell. %
Finally, for the output features $\boldsymbol{e}_i^{N_2}$ of the $N_2$-th layer, a linear layer is used to obtain the force:  
\begin{equation}
\boldsymbol{f}_i  = \mathrm{Linear}(\norm{\boldsymbol{e}_i^{N_2}})
\end{equation}

\begin{figure}
    \centering
    \includegraphics[width=0.5\linewidth]{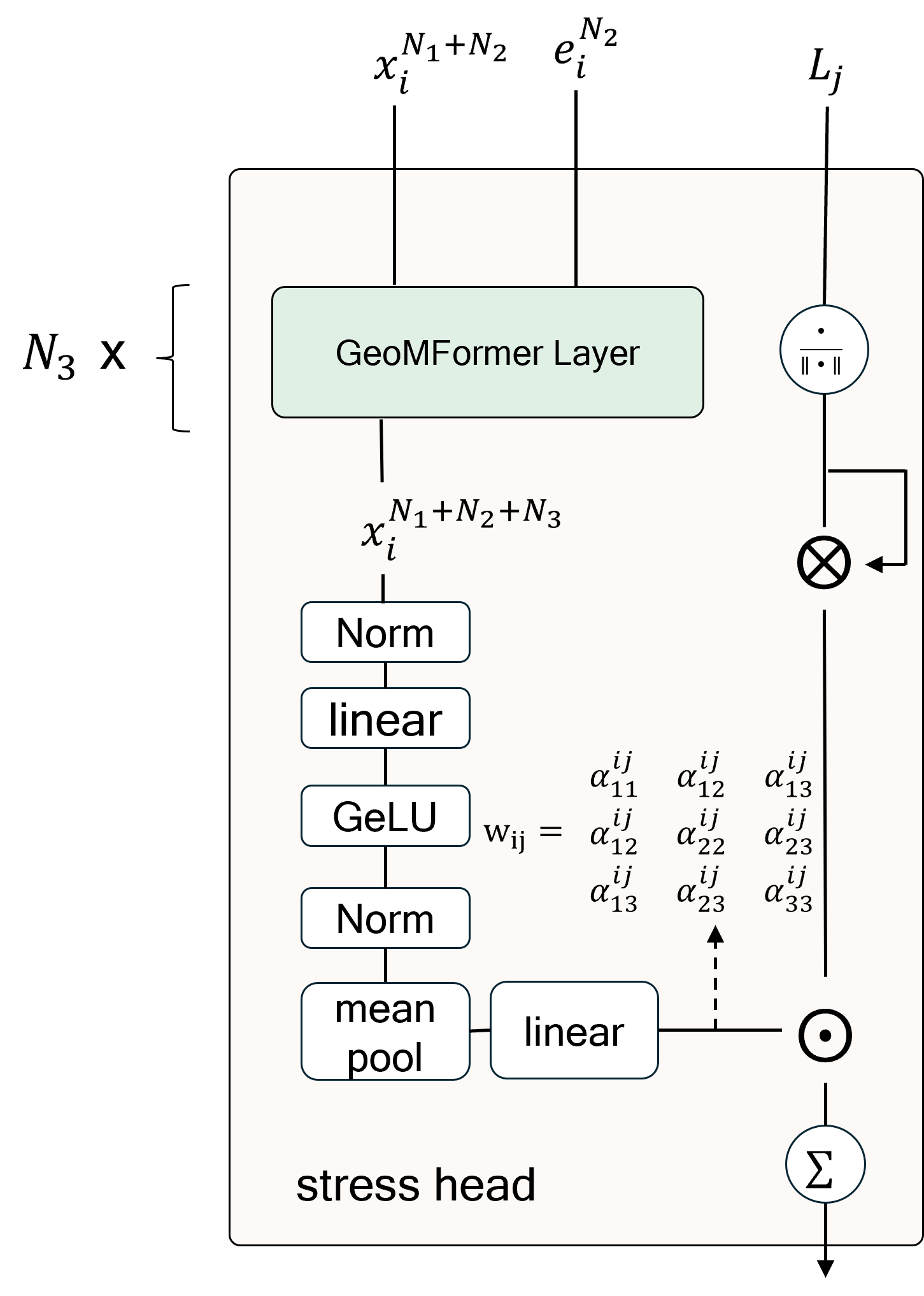}
    \caption{Illustration of the newly designed stress head for the Graphormer model, utilizing a symmetrized weight matrix $\boldsymbol{w_{ij}}$ and normalized lattice vector outer products to predict the stress tensor $\sigma$.}
    \label{si-fig:graphormer-new-stress-head}
\end{figure}

The stress prediction was not supported in the original implementation of Graphormer. To enable the model to predict stress, we designed a new stress head to evaluate stress as follows:
\begin{equation}
\boldsymbol{\sigma} = \sum _{ij} \boldsymbol{w}_{ij} \frac{\boldsymbol{L}_i}{\norm{\boldsymbol{L}_i} } \otimes \frac{\boldsymbol{L}_j}{\norm{\boldsymbol{L}_j}},
\end{equation}
where $\boldsymbol{L}$ represents lattice vectors.
Here, $\boldsymbol{w_{ij}}$ is derived from the features from the decoder followed by the transformations defined in \autoref{si-fig:graphormer-new-stress-head}.

\subsection{Training details}
For the training of the M3GNet, we referred to the training setups in the original implementation.\cite{chen2022universal}. To be specific, the loss function 
\begin{equation}
    L = l(e,e_\mathrm{DFT}) + \omega_f l(\boldsymbol{f},\boldsymbol{f}_\mathrm{DFT}) + \omega_\sigma l(\boldsymbol{\sigma}, \boldsymbol{\sigma}_\mathrm{DFT})
\end{equation}
was used, where $l(\cdot, \cdot)$ is the Huber loss function, $e$ is the energy per atom, $\boldsymbol{f}$ is the force vector on each atom, $\boldsymbol{\sigma}$ is the stress, and $\omega_f$ and $\omega_\sigma$ are the weights of forces and stress, respectively. For the models used in this work, $\omega_f = 1$ and $\omega_\sigma = 0.1$ are used. The initial learning rate was set to be 0.001 for the Adam optimizer which decays in a cosine manner to 1\% to the original values in 100 epochs, and the training process stops after running for 200 epochs with a batch size of 128 on 8 NVIDIA A100 GPUs.
As the training data size increases up to 3M, the the total number of parameters in M3GNet increase accordingly from 880K to 4.5M. 
Without  modifying model architecture and training scheme, further increase of model parameters led to instability under current settings.

For the training of Graphormer, we have configured the structural encoder with 24 attention layers and the property decoder with 10 layers of GeoMFormer (2 layers in stress head). All attention heads are set to 32, and the dimension of hidden layers and feed-forward layers is set to 768. The number of Gaussian basis kernels is set to 128. In the structural encoder, all dropout rates are set to 0.0, while in the property decoder, the activation dropout is set at 0.1, with all other dropout rates at 0.0. The cutoff for expansion is set to \SI{20}{\angstrom}, the smooth function cutoff is set to \SI{5}{\angstrom}, the maximum number of expanded atoms is capped at 256, and the offsets for expansion range from -5 to 5 in each direction. We use AdamW\cite{loshchilov2017decoupled} as the optimizer with the hyper-parameter $\epsilon$ set to 1e-8, and $\beta _1$ and $\beta _2$ set to 0.9 and 0.999, respectively. The peak learning rate is set to 2e-4, and weight decay is set to 0.0. The model is trained for a total of 1,562,500 steps with a warm-up period of 93,750 steps. After the warm-up, the learning rate linearly decreases to 0. The batch size for training is set to 256, and all labels use the mean absolute error (MAE) as the loss function. The energy loss factor, force loss factor, and stress loss factor are all set to 1.0. The model training is conducted on 64 NVIDIA A100 GPUs. The total parameters of Graphormer is 182M. In the first round of training, only energies and forces are trained. After training is complete, the relevant parameters are frozen, and the stress head is further trained.

\subsection{Model comparison}\label{si-sec:performance-m3gnet-vs-graphormer}
\begin{figure}
    \centering
    \includegraphics[width=1.0\linewidth]{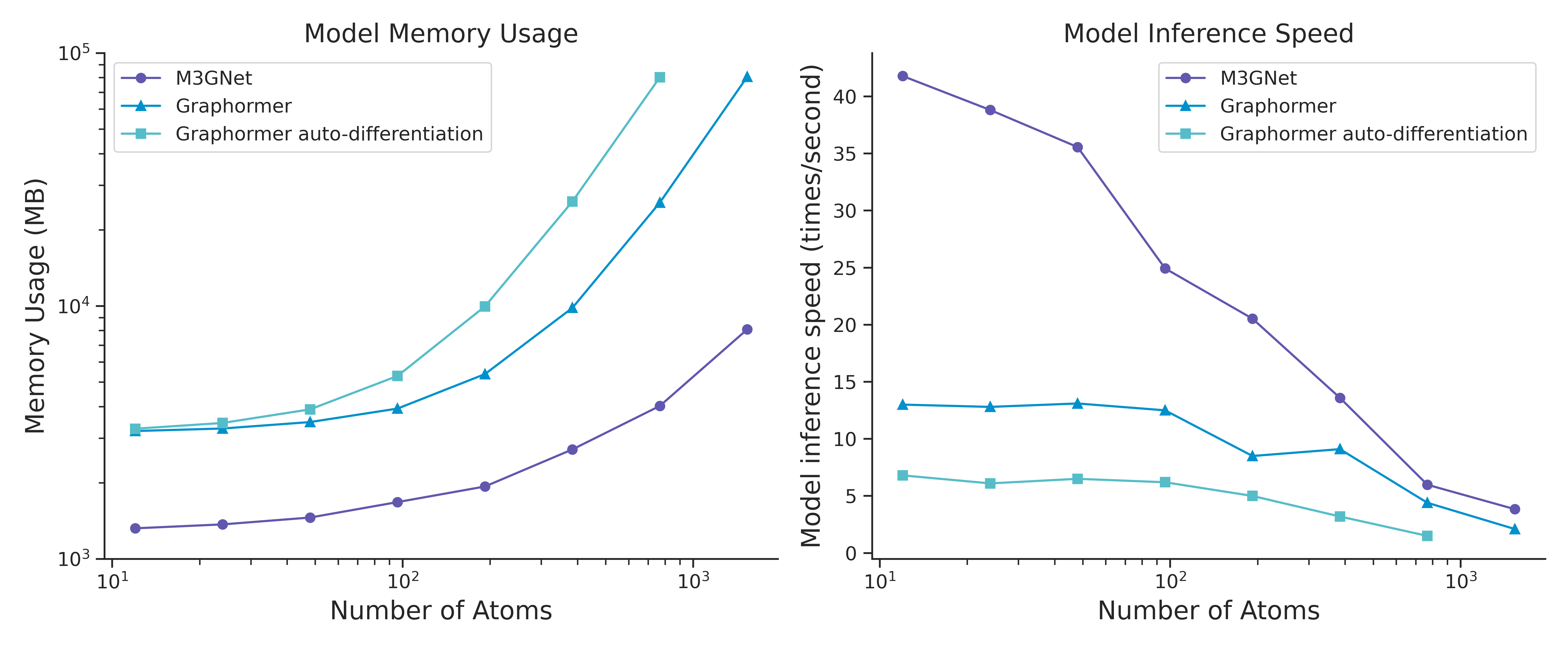}
    \caption{Model Inference Speed and GPU Memory Usage}
    \label{si-fig:graphormer-m3gnet-resource-comparison}
\end{figure}

Graphormer offers a higher degree of model complexity and the potential for increased predictive accuracy with its larger model parameters to consume the large training dataset. However, this architectural design inherently leads to slower computational speed compared to M3GNet, especially when automatic differentiation is employed to predict forces and stresses. Furthermore, the memory demands of Graphormer are significantly higher. This means that for larger atomic systems, Graphormer's need for GPU memory often reaches the upper limit of what is available even on high-end GPUs such as the A100. %
On the other hand, M3GNet serves as a more resource efficient alternative, especially in environments with constrained GPU resources. Its computational framework is optimized for speed, making it a more practical choice for processing a majority of the tasks encountered in our research. M3GNet's design balances performance and resource utilization, allowing for the analysis of large datasets and complex systems without the memory limitations faced by Graphormer. 

We further demonstrate the different level of needs of computing resources, we benchmarked the performance of M3GNet and Graphormer on a set of materials with different number of atoms in the unit cell in \autoref{si-fig:graphormer-m3gnet-resource-comparison}. For example, for a material with 100 atoms, Graphormer requires around 10-fold more GPU memories compared to our implementation of M3GNet. The accuracy of the models based on M3GNet and Graphormer is discussed in \autoref{si-sec:performance-on-benchmark-sets} and shown in \autoref{si-fig:performance-legecy-checkpoints}.

\section{Materials explorer}\label{si-sec:materials-explorer}
MatterSim's predictive capabilities are underpinned by a two-part materials structure explorer, as shown in \autoref{fig:overview:data-overview}(a), that enhances the training datasets through both equilibrium and off-equilibrium structural data.

\begin{figure}
    \centering
    \includegraphics[width=0.6\textwidth]{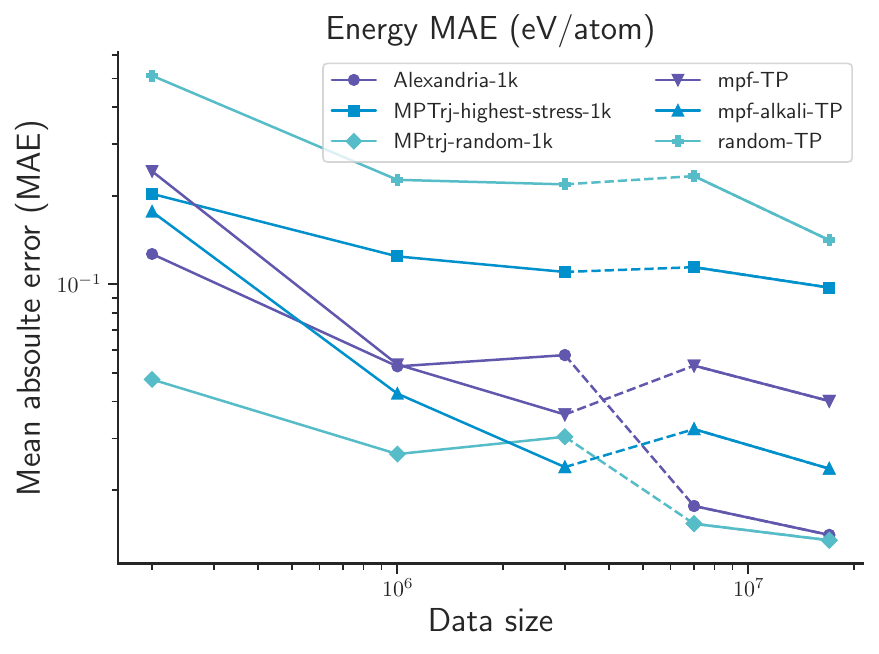}
    \caption{Performance of the intermediate checkpoints of MatterSim obtained on the iteratively generated structures on three test datasets: \texttt{MPF-alkali-TP}, \texttt{MPF-TP} and \texttt{random-TP}. The details of the generation of the test datasets can be found in \autoref{si-sec:performance-on-benchmark-sets}.}
    \label{si-fig:performance-legecy-checkpoints}
\end{figure}

The \textbf{ground-state explorer} focuses on materials at or near atomistic equilibrium positions. It primarily utilizes an uncertainty-based method with ensemble models to selectively incorporate data from both public repositories and internally generated datasets into the database. This approach ensures the inclusion of the most informative structures for model training, enhancing the accuracy of property predictions at equilibrium. The \textbf{off-equilibrium explorer}, targets materials with  off-equilibrium atomistic positions. It conducts molecular dynamics (MD) simulations under a wide range of pressures, including 0, 500, 800, and \SI{1000}{\giga\pascal}. Under each pressure, we incrementally increase the temperature from 0 to \SI{5000}{\kelvin} within \SI{200}{\pico\second}.
These simulations are crucial for sampling a wide range of atomic configurations, allowing MatterSim to learn and predict material properties under high-pressure and high-temperature scenarios that deviate significantly from equilibrium states.
Together, these explorers provide a dataset that spans a vast configurational space over the entire periodic table, ensuring that MatterSim is equipped with the necessary information to predict material properties across a full spectrum of conditions. In addition to the data explorers, a sub-sampling procedure is applied according to the uncertain evaluated on an ensemble of models. The details of the uncertainty evaluation can be found in \autoref{si-sec:uncertainty}. As an example to showcase the validity of this data generation strategy, we tested the performance of intermediate model checkpoints up to 3M structures used in model training in \autoref{si-fig:performance-legecy-checkpoints}. As the the dataset increases, the model performance of force, energy and stress prediction is improving on the test datasets.

We note that the dataset fueling MatterSim is part of a dynamic and continually evolving scheme, ensuring that the model's predictive power is constantly refined and updated. As of the time of the release of this manuscript, 17 million data points have been compiled in this dataset, encompassing materials sampled from publicly available databases, for example, Materials Project or Alexandria, internally generated datasets, and molecular dynamics trajectories under ambient to extreme conditions.

\section{Data distribution}\label{si-sec:data-distribution}
MatterSim relies on the materials explorer defined in \autoref{fig:overview:data-overview}(a) to expand coverage of compositional space configurational space so as to describe materials under a wide range of temperature and pressure. 
In this section, we analyze the distribution of the dataset generated in this work and compare with public datasets, including \texttt{MPF2021},\cite{chen2022universal} \texttt{MPtrj},\cite{deng2023chgnet} and \texttt{Alexandria}\cite{schmidt2023machine} datasets. We first compare the elementwise and pair-element appearance of the elements in the periodic table through different datasets, then we compare the number of atoms distribution. Finally, we illustrate the definition of the effective temperature used in \autoref{fig:overview:data-overview}(b), and highlight again the comparison of the distribution in the effective temperature and pressure space across different database.

\subsection{Elemental and elemental pairwise distribution}\label{si-sec:elemental-and-elemental-pairwise-distribution}
\begin{figure}
    \centering
    \includegraphics[width=\textwidth]{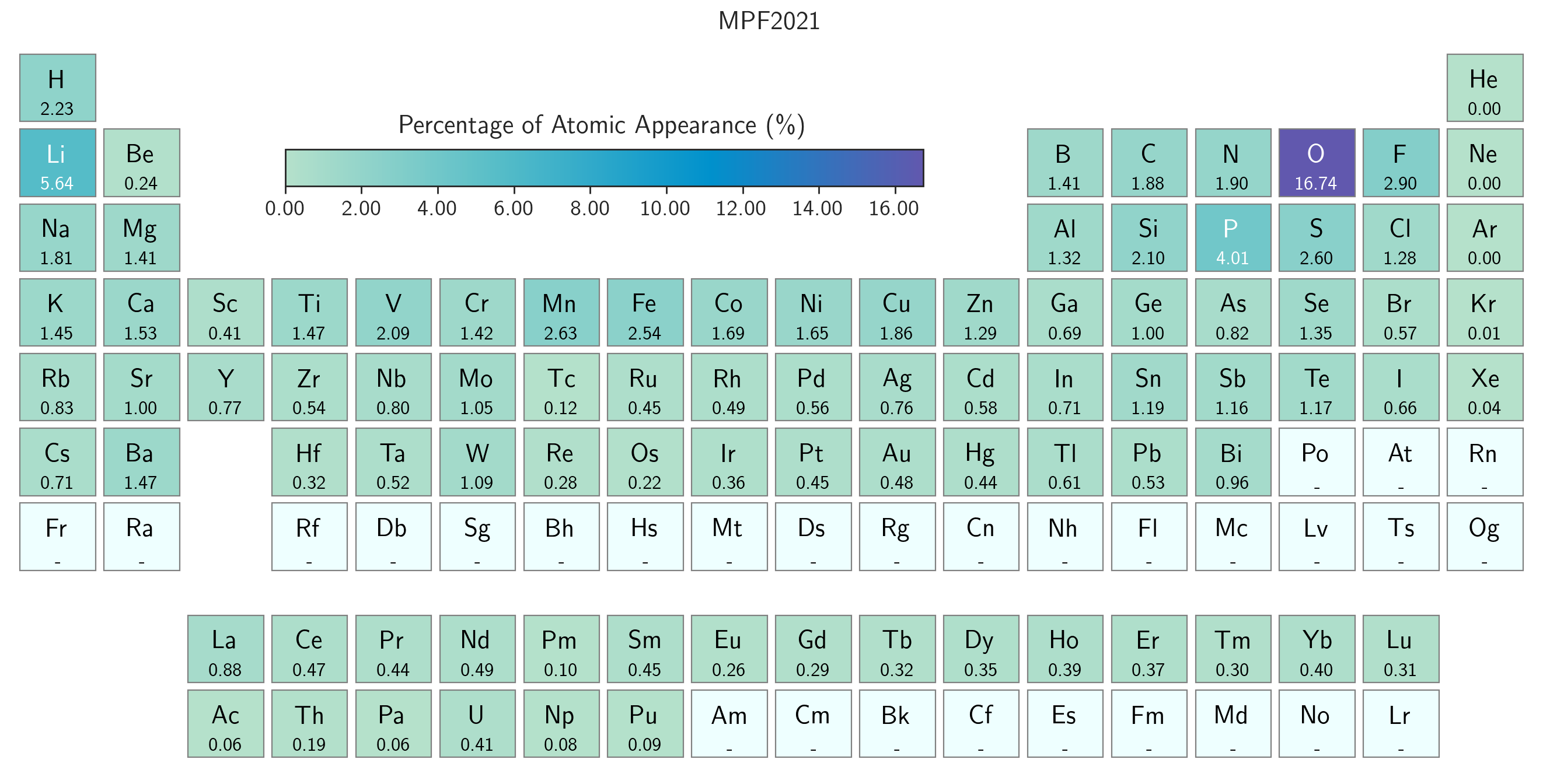}
    \caption{Elementwise percentage of atomic appearance in \texttt{MPF2021} dataset.}
    \label{si-fig:elementwise-mpf2021-distribution}
\end{figure}

\begin{figure}
    \centering
    \includegraphics[width=\textwidth]{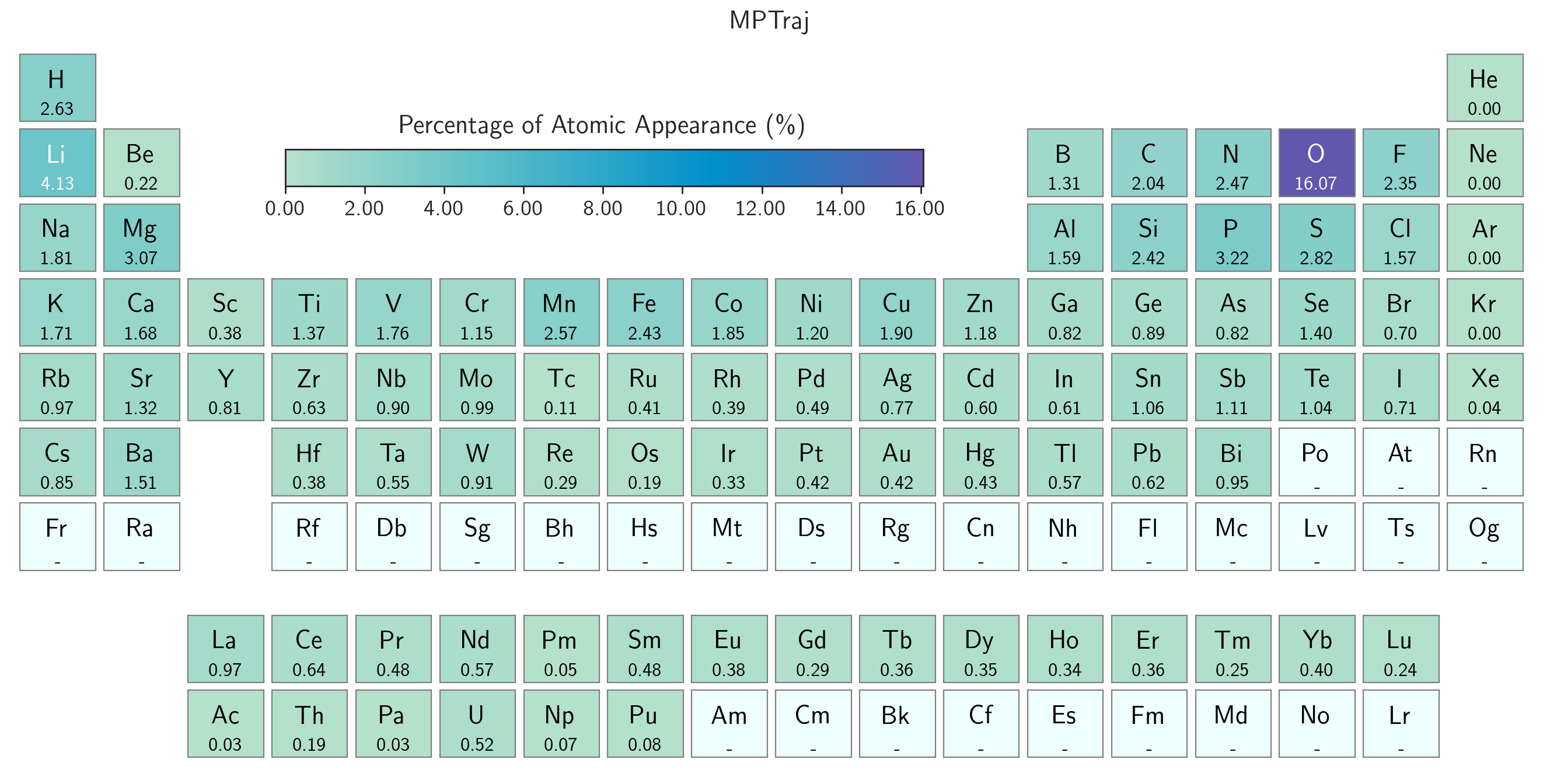}
    \caption{Elementwise percetage of atomic appearance of 1M structures randomly sampled from \texttt{MPtrj} dataset.}
    \label{si-fig:elementwise-mptrj-distribution}
\end{figure}

\begin{figure}
    \centering
    \includegraphics[width=\textwidth]{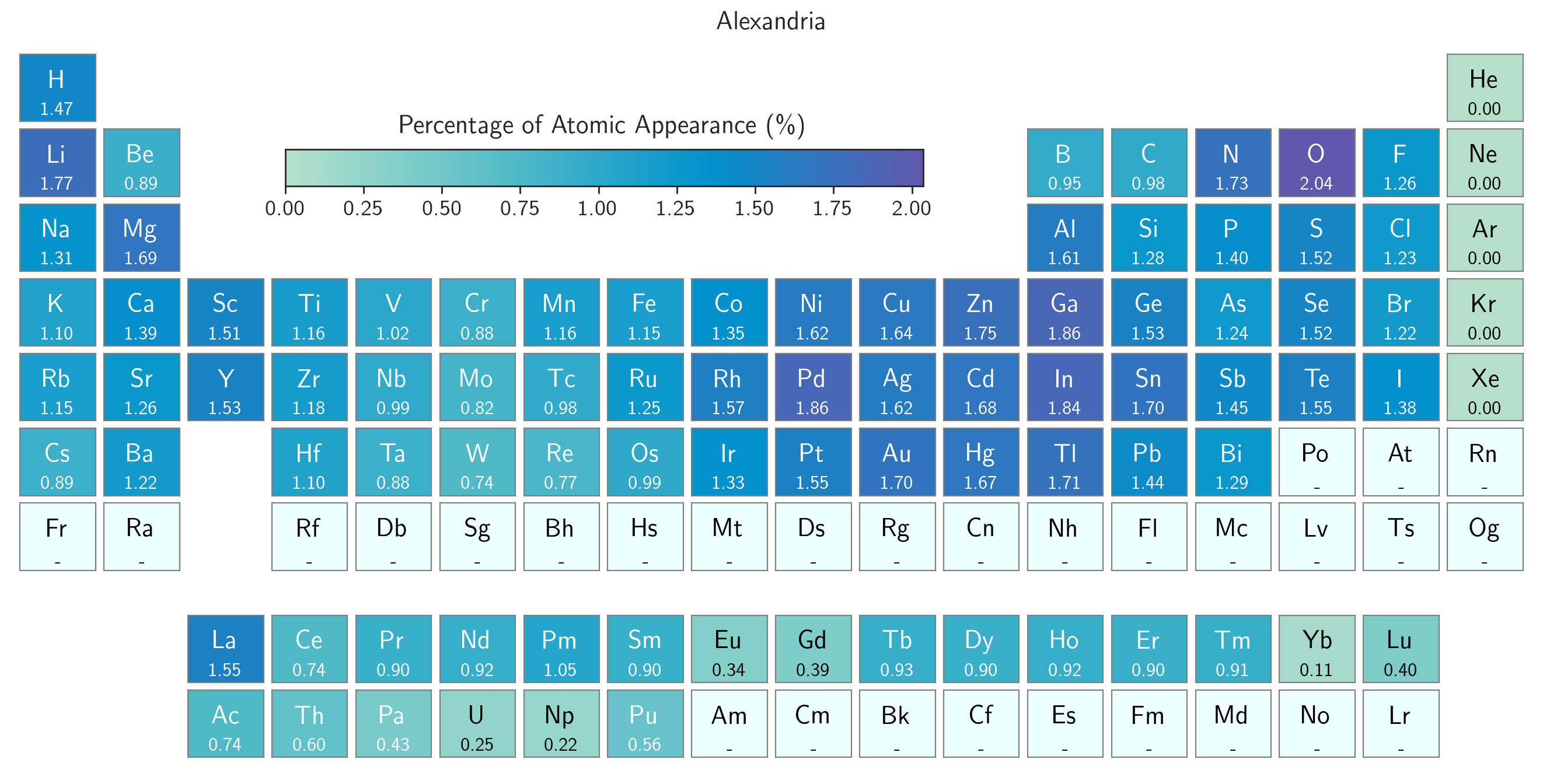}
    \caption{Elementwise percentage of atomic appearance of 1M structures randomly sampled from \texttt{Alexandria} dataset.}
    \label{si-fig:elementwise-alexandria-distribution}
\end{figure}

\begin{figure}
    \centering
    \includegraphics[width=\textwidth]{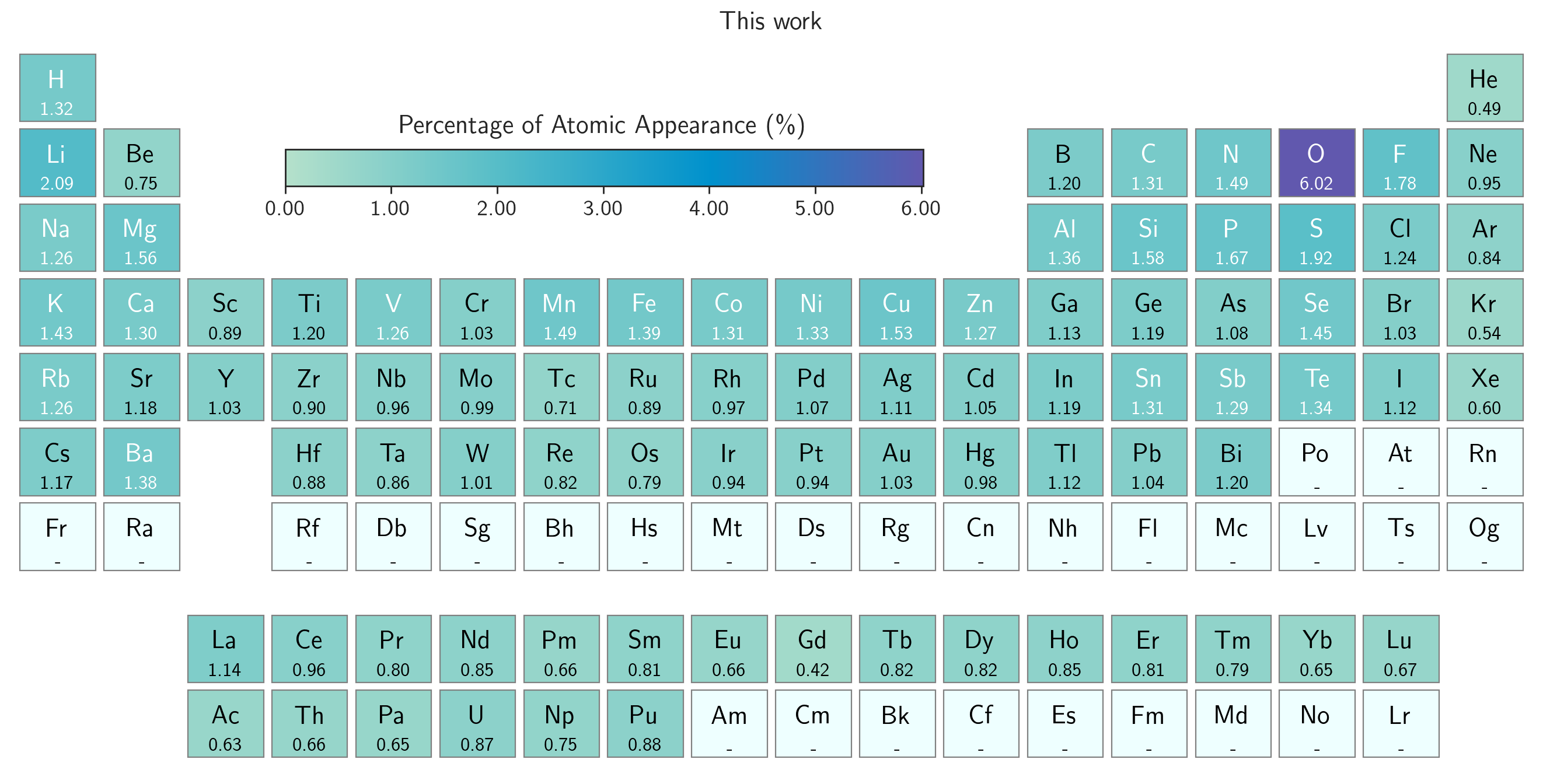}
    \caption{Elementwise percentage of atomic appearance of 1M structures randomly sampled from the dataset in this work.}
    \label{si-fig:elementwise-this-work-distribution}
\end{figure}

\begin{figure}
    \centering
    \begin{subfigure}[b]{0.45\textwidth}
    \includegraphics[width=\textwidth]{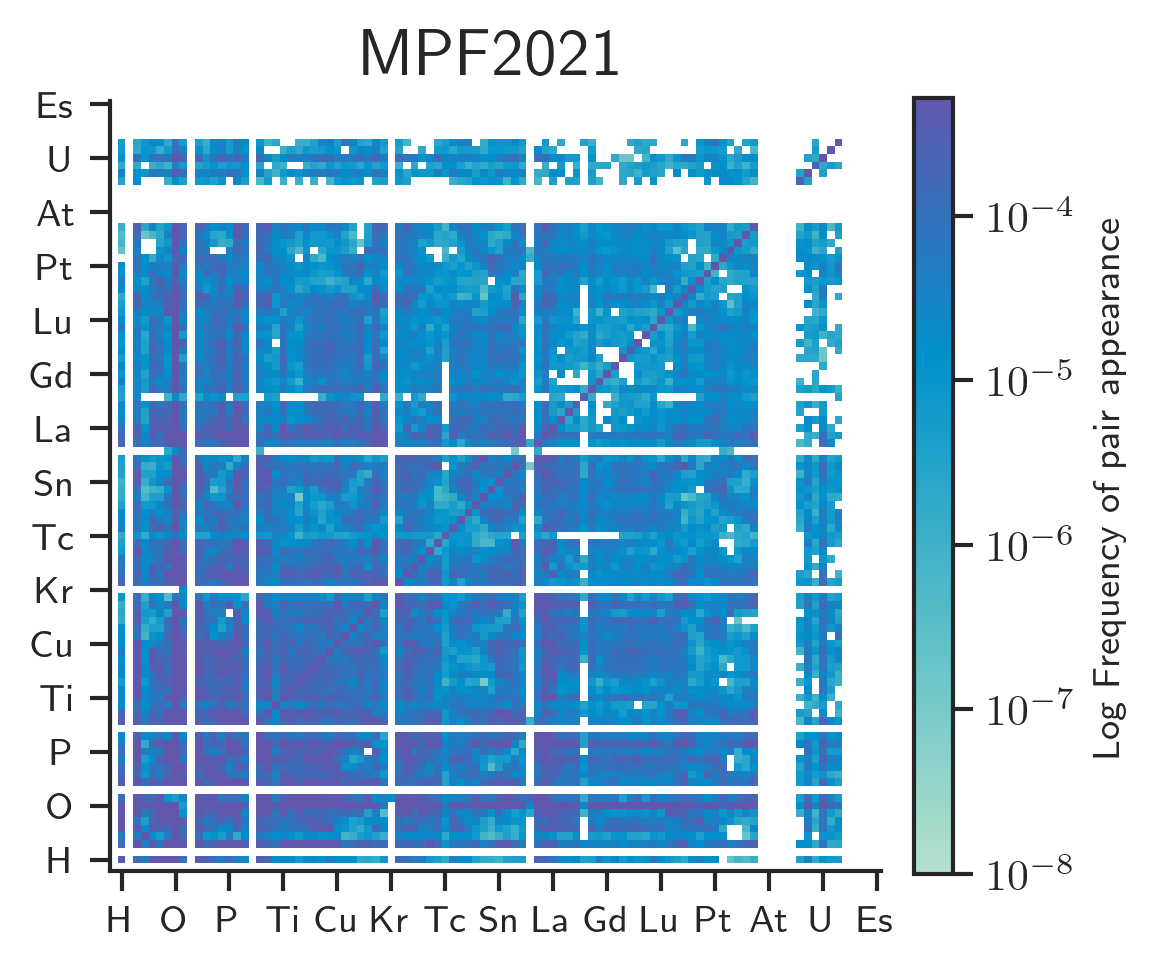}
    \end{subfigure}
    \begin{subfigure}[b]{0.45\textwidth}
    \includegraphics[width=\textwidth]{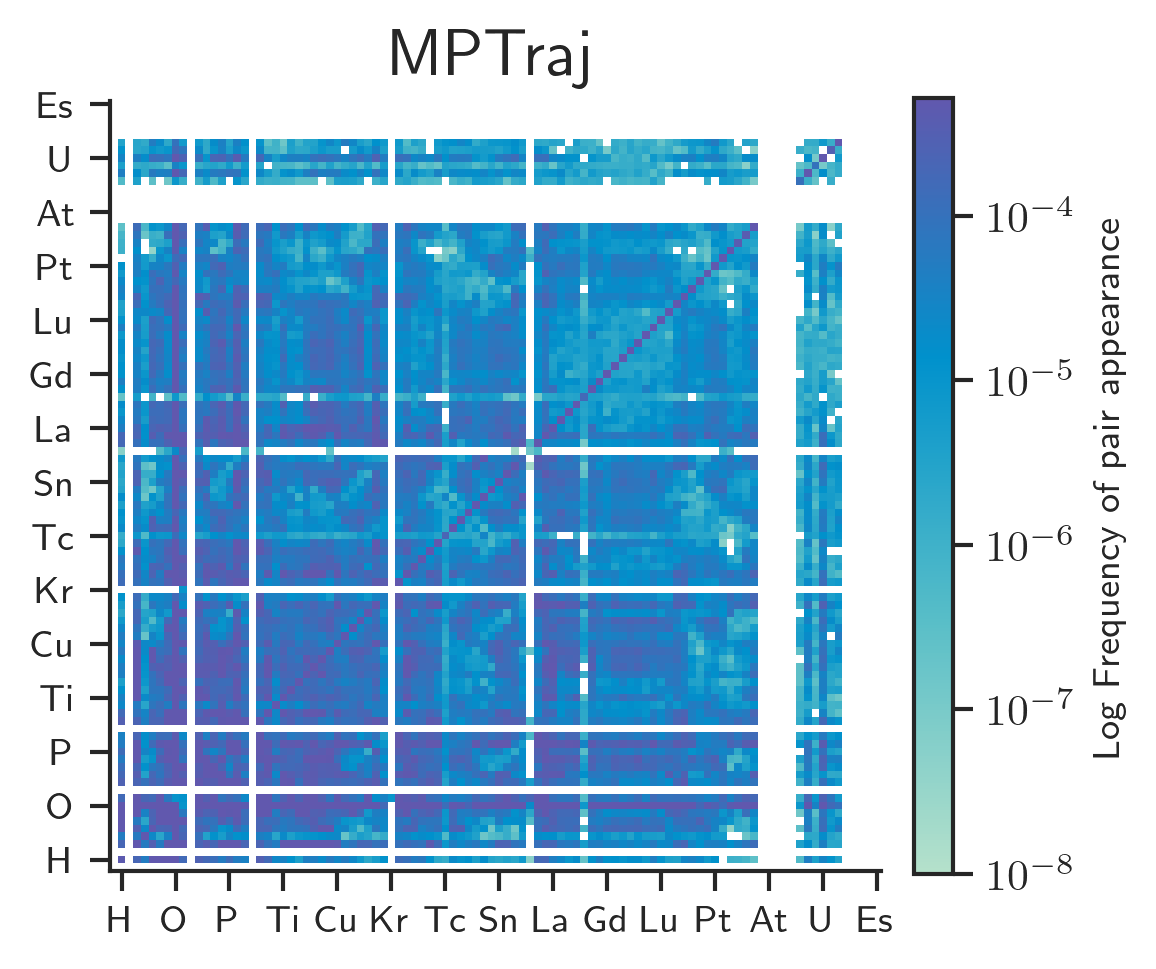}
    \end{subfigure}
    \begin{subfigure}[b]{0.45\textwidth}
    \includegraphics[width=\textwidth]{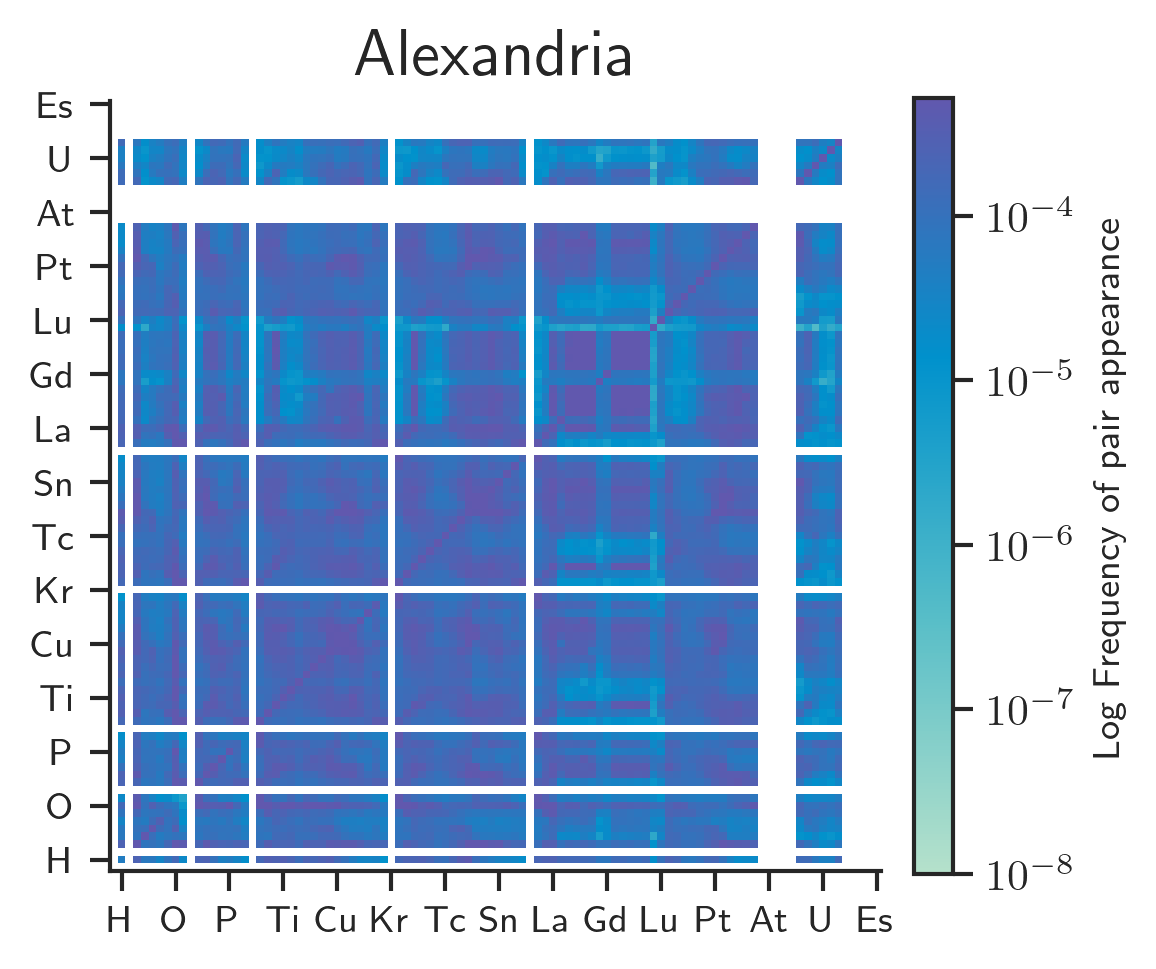}
    \end{subfigure}
    \begin{subfigure}[b]{0.45\textwidth}
    \includegraphics[width=\textwidth]{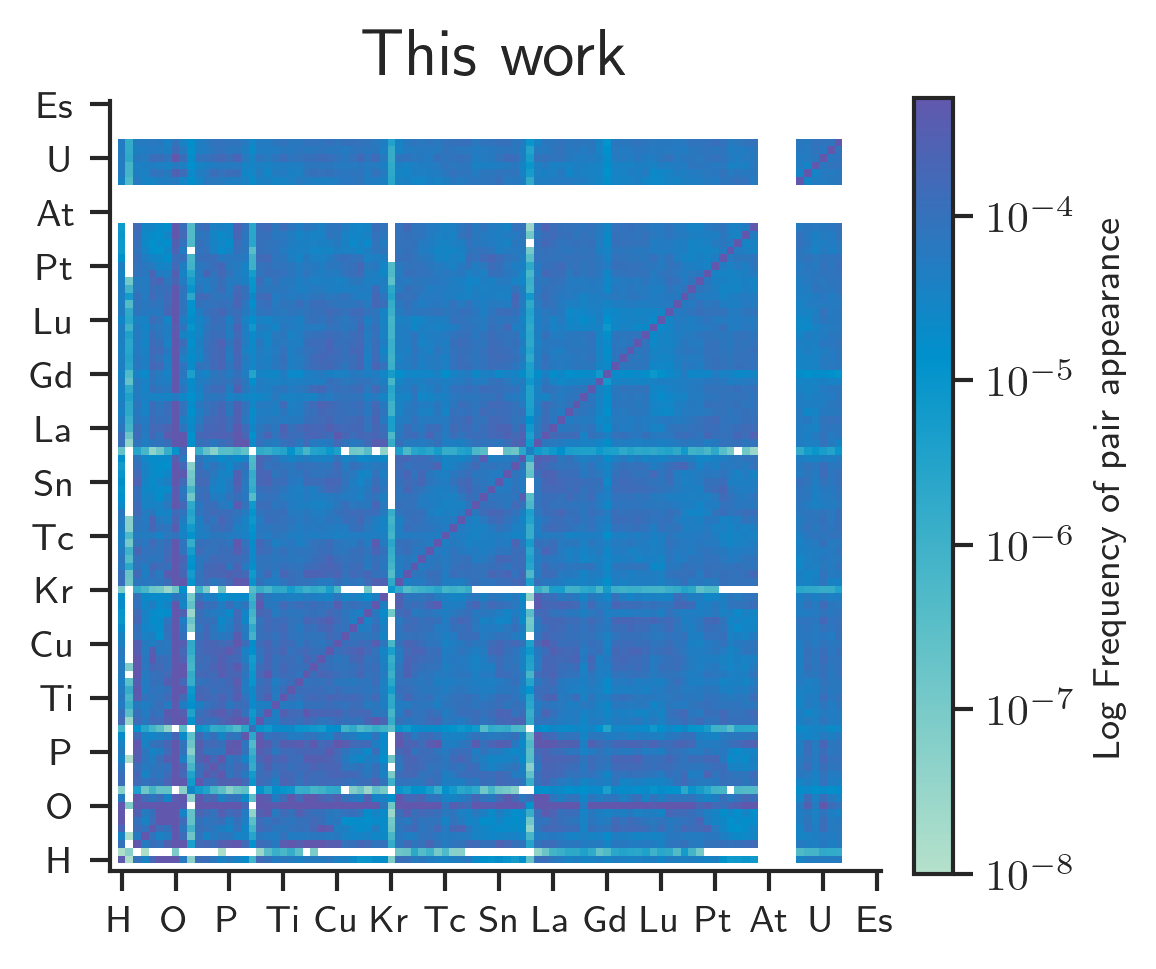}
    \end{subfigure}
    \caption{Pairwise elemental distribution of entire \texttt{MPF2021} dataset, 1M structures randomly sampled from \texttt{MPtrj} dataset, 1M structures randomly sampled from \texttt{Alexandria} dataset and 1M structures randomly sampled from the dataset in this work.}
    \label{si-fig:pairwise-distribution}
\end{figure}

 \autoref{si-fig:elementwise-mpf2021-distribution}, \autoref{si-fig:elementwise-mptrj-distribution}, \autoref{si-fig:elementwise-alexandria-distribution} and \autoref{si-fig:elementwise-this-work-distribution} plot the elementwise percentage atomic appearance of the entire \texttt{MPF2021} dataset, 1M structures uniformly sampled from \texttt{MPtrj} dataset, 1M structures uniformly sampled from \texttt{Alexandria} dataset and 1M structures uniformly sampled from the dataset generated and used in this work, respectively. In the distribution of \texttt{MPF2021} and \texttt{MPtrj}, a significant bias to oxygen is observed -- oxygen has 8-fold more percentage of appearance compared to most elements in the periodic table. The distribution of \texttt{Alexandria}, on the other hand, has almost uniform distribution over the periodic table, however we have shown in \autoref{fig:overview:data-overview}(d) that \texttt{Alexandria} only has a peaked distribution around the pressure of 0 GPa. The dataset generated in this work, as shown in \autoref{fig:overview:data-overview}(b), \autoref{si-fig:elementwise-this-work-distribution} and \autoref{si-fig:pairwise-distribution}, not only ameliorates the biased distribution to oxides, but it also effectively explored the configurational space and covers a much wider domain in the effective temperature and pressure space. 

 In addition to elementwise distribution, in \autoref{si-fig:pairwise-distribution}, we also compare the element pairwise distribution in the four datasets. One pairwise appearance is counted when a pair of atoms exists with a separation distance lower than \SI{5}{\angstrom}. \texttt{MPF2021} and \texttt{MPtrj} are both derivatives of the Materials Project, and thus it is no surprise that we notice a common distribution of them for elements with atomic number larger than the Lanthanum element -- there are noticeable missing or close to negligible element pairs. The \texttt{Alexandria} database again has a uniform distribution for most of the elements, however, there are still missing columns or rows involving the noble gas elements. In contrast to public ones, the dataset generated in this work has a relatively uniform distribution and also a almost full coverage of all the combination of element pairs. 


\subsection{Number of atoms}
\begin{figure}
    \centering
    \includegraphics[width=0.75\textwidth]{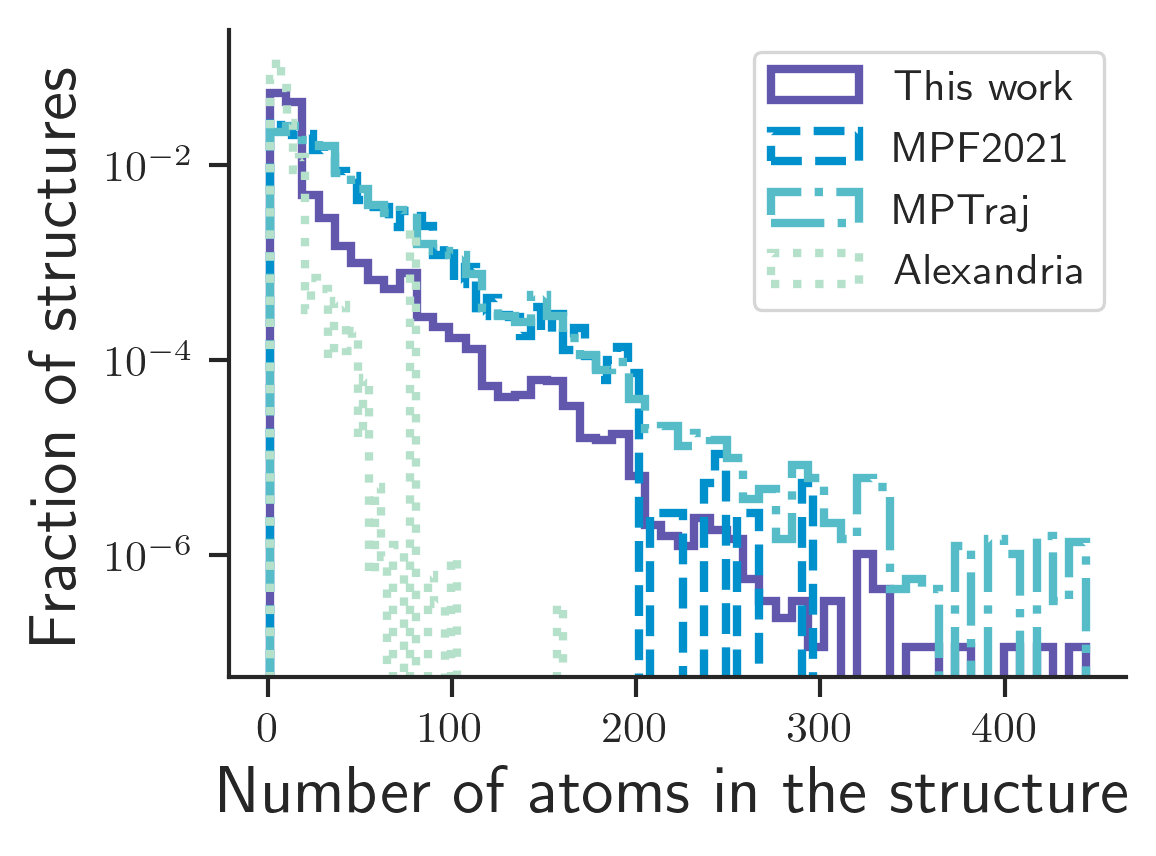}
    \caption{Distribution of number of atoms in the structures in \texttt{MPF2021}, \texttt{MPtrj} (1M randomly sampled structures), \texttt{Alexandria} (1M randomly sampled structures) and the dataset generated in this work (1M randomly sampled structures).}
    \label{si-fig:hist-num-atoms-distribution}
\end{figure}

In addition to the elemental and element pair distributions, we also compare the number of atoms in the datasets. \autoref{si-fig:hist-num-atoms-distribution} shows the histogram of the number of atoms in the structures from the \texttt{MPF2021}, \texttt{MPtrj}, \texttt{Alexandria} datasets and this work. Again, we use the entire \texttt{MPF2021} datasets, and 1M randomly sampled structures from the other three datasets. In \autoref{si-fig:hist-num-atoms-distribution}, we clearly see that the \texttt{Alexandria} has a biased distribution over structures with less than 100 atoms, while \texttt{MPF2021} and \texttt{MPtrj} datasets have more dense distributions over materials with larger than 100 atoms. We also notice that \texttt{MPF2021} has much less distribution over 200 atoms. The dataset generated in this work has a less biased distribution compared with \texttt{Alexandria} and has a more smooth decreasing in the distribution curve from structures with lower than 100 atoms to those with more 300 atoms, empowering the model to handle materials ranging from simplest diamond structures to very large complicated ones.

\subsection{Latent space coverage}
\begin{figure}
    \centering
    \begin{subfigure}[b]{0.45\textwidth}
    \includegraphics[width=\textwidth]{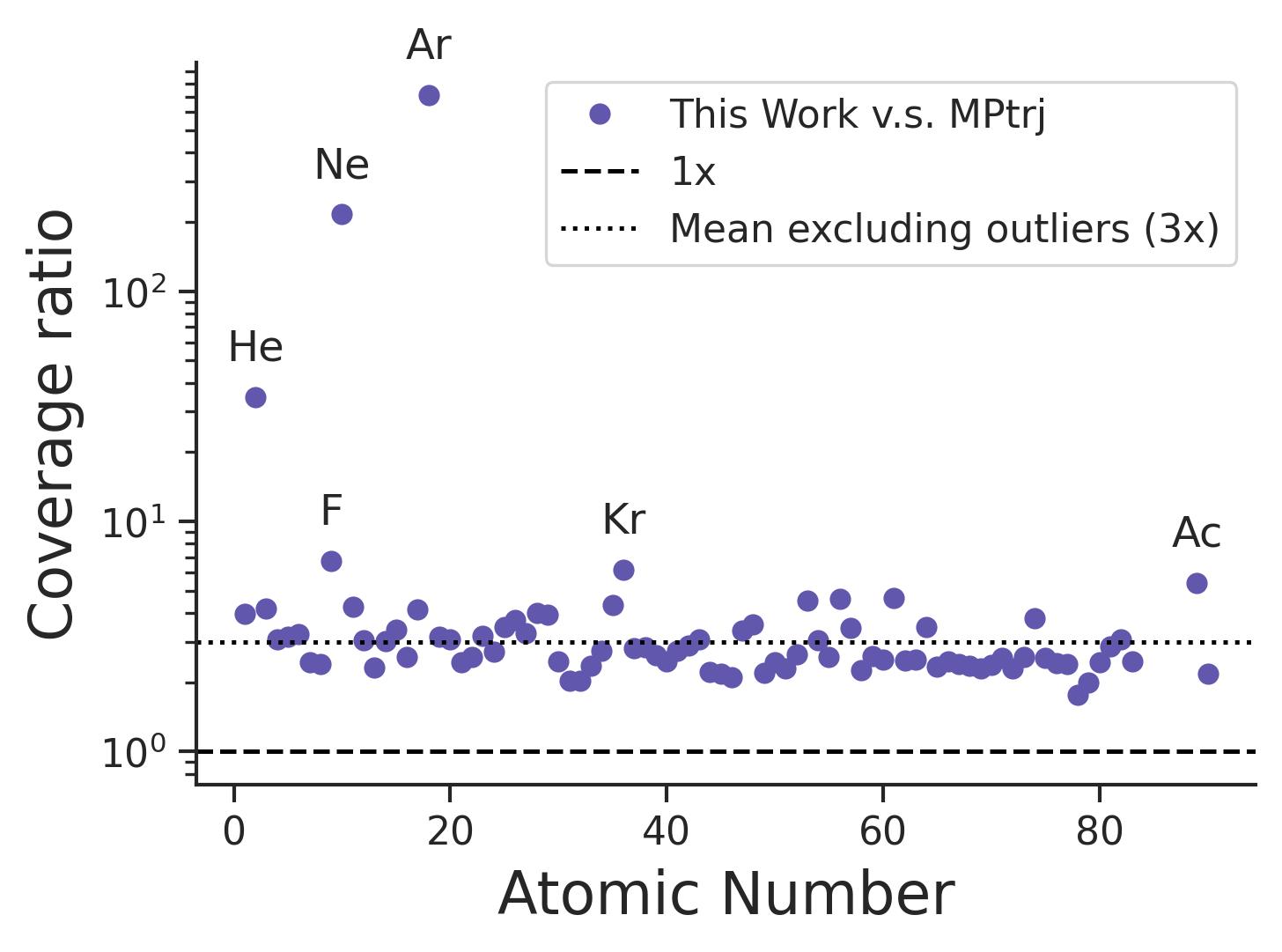}
    \caption{}
    \label{si-fig:latent-space-this-work-vs-mptrj}
    \end{subfigure}
    \begin{subfigure}[b]{0.45\textwidth}
    \includegraphics[width=\textwidth]{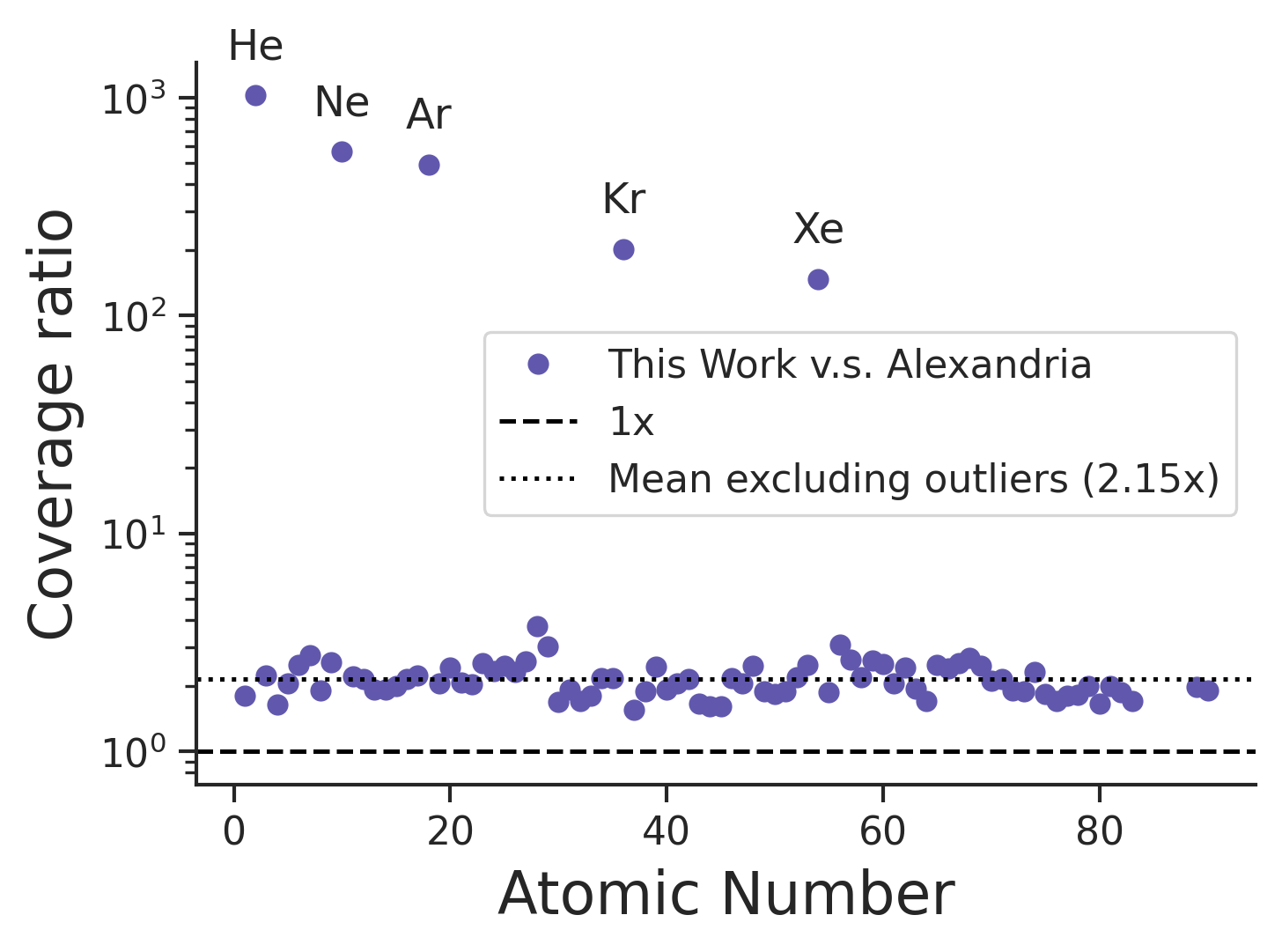}
    \caption{}
    \label{si-fig:latent-space-this-work-vs-alex}
    \end{subfigure}
    \begin{subfigure}[b]{0.8\textwidth}
    \includegraphics[width=\textwidth]{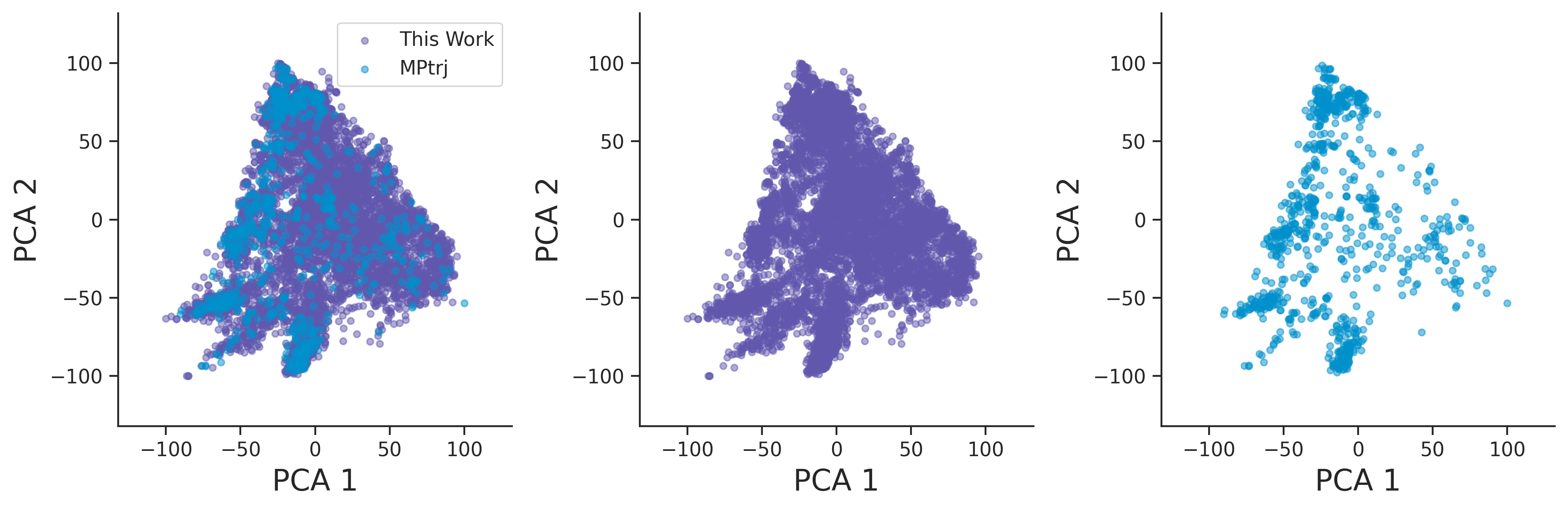}
    \caption{}
    \label{si-fig:latent-space-carbon-this-work-vs-mptrj}
    \end{subfigure}
    \begin{subfigure}[b]{0.8\textwidth}
    \includegraphics[width=\textwidth]{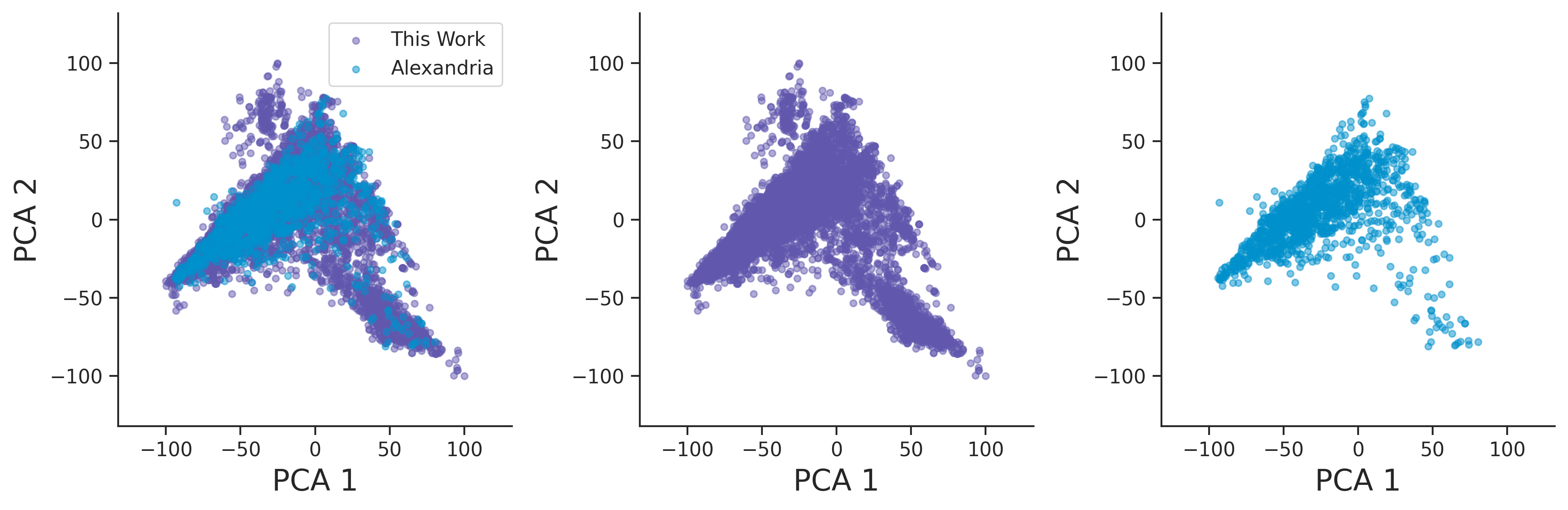}
    \caption{}
    \label{si-fig:latent-space-zirconium-this-work-vs-alexandria}
    \end{subfigure}
    \begin{subfigure}[b]{0.8\textwidth}
    \includegraphics[width=\textwidth]{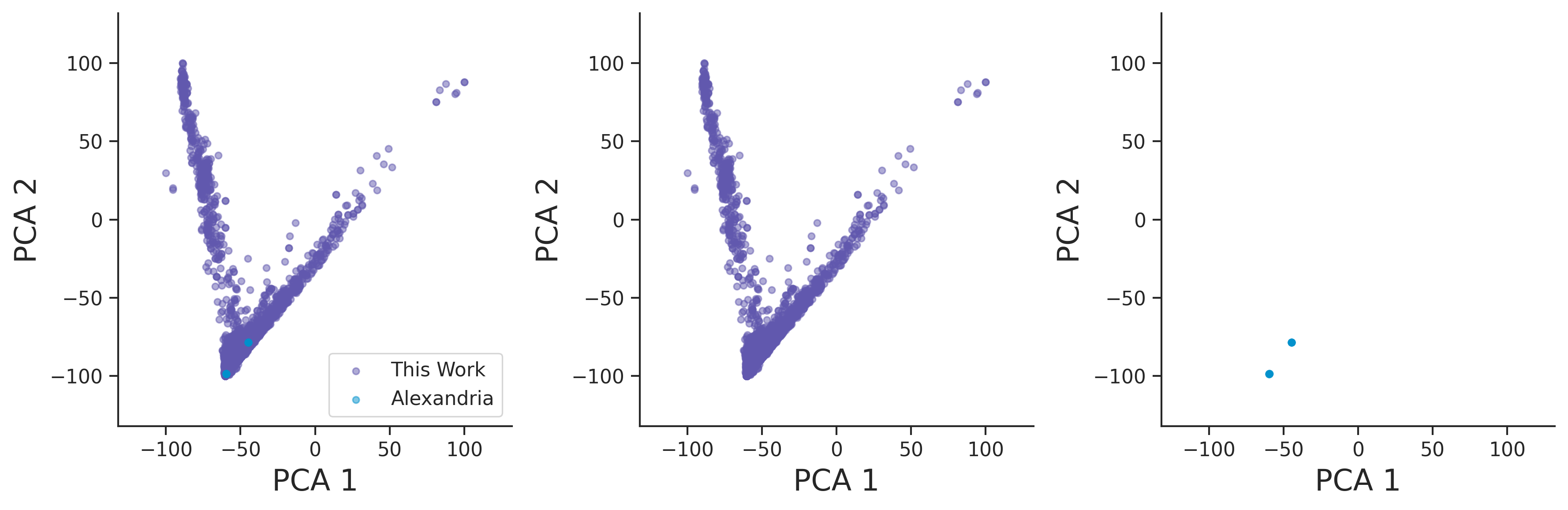}
    \caption{}
    \label{si-fig:latent-space-xenon-this-work-vs-alexandria}
    \end{subfigure}
    \caption{(a) The coverage ratio between the latent space of this work and that  of the \texttt{MPtrj} dataset for the entire periodic table; (b) the coverage ratio between the latent space of this work and that of the \texttt{Alexandria} dataset for the entire periodic table; (c) Principal component analysis (PCA) of the latent space of carbon atoms sampled from this work and \texttt{MPtrj} dataset; (d) Principal component analysis (PCA) of the latent space of Zirconium atoms sampled from this work and \texttt{Alexandria} dataset; (e) Principal compotent analysis (PCA) of the latent space of Xenon atoms sampled from this work and \texttt{Alexandria} dataset. Overlap and separate plots of PCAs are shown for clarity.}
    \label{si-fig:latent-space-comparison}
\end{figure}

In \autoref{si-fig:latent-space-comparison}, we compare the atomic embeddings in \texttt{MPtrj}, \texttt{Alexandria} and the dataset generated in this work. To make a fair comparison, we sampled 10,000 atomic embeddings from the 1M subset of our dataset, 1,000 atomic embeddings from the 1M subset of \texttt{MPtrj} and 1,500 atomic embeddings from the 1M subset of \texttt{Alexandria} for each element -- this choice is based on the ratio of number of structures in each dataset. The principal component analysis (PCA) is done for the embeddings to reduce the embeddings to a 2-dimensional space. The principals are scaled to the range of -100 to 100 range and a circle of radius of 2 is assigned to each data point, and finally the coverage is computed as the total areas of all circles excluding all the overlaps. \autoref{si-fig:latent-space-this-work-vs-mptrj} and \autoref{si-fig:latent-space-this-work-vs-alex} show the coverage ratio between this work and \texttt{MPtrj} dataset, and between this work and \texttt{Alexandria} dataset, exhibiting 3-fold and 2.15-fold larger coverage on average through the entire periodic table with two examples of Carbon and Zirconium elements shown in \autoref{si-fig:latent-space-carbon-this-work-vs-mptrj} and \autoref{si-fig:latent-space-zirconium-this-work-vs-alexandria}, respectively. Interestingly, we also noticed that the noble gas elements are extremely scared in the \texttt{MPtrj} and \texttt{Alexandria} datasets that we can barely sample enough atomic embeddings from the 1M subsets, leading to much higher coverage of our dataset, see \autoref{si-fig:latent-space-xenon-this-work-vs-alexandria}, and they are excluded from the computation of mean coverage ratio in \autoref{si-fig:latent-space-this-work-vs-mptrj} and \autoref{si-fig:latent-space-this-work-vs-alex}.

\subsection{Temperature and pressure distribution}
MatterSim is an emulator designed for modeling materials under real-world temperature and pressure conditions and the workflow designed in Fig.~\ref{fig:overview:data-overview}(a) is capable of generating an enriched dataset that covers a broad range of these conditions. To straightforwardly illustrate the distribution, we have included a two-dimensional histogram of the effective temperature and stress of our generated dataset in Fig.~\ref{fig:overview:data-overview}(b), where the effective temperature is defined as follows,
\begin{enumerate}
    \item For each material from a given dataset, we evaluate its total energy per atom ($\varepsilon$) with MatterSim;
    \item Then, we optimize the atomic positions with fixed lattice parameters for at most 500 steps until the max forces converge to \SI{0.01}{eV\per \angstrom}, and evaluate the relaxed total energy per atom ($\varepsilon_0$);
    \item Finally, the effective temperature of this given material is evaluated by 
    \begin{equation*}
        T_\mathrm{eff} = \frac{\varepsilon-\varepsilon_0}{k_\mathrm{B}},
    \end{equation*}
    where $k_\mathrm{B}$ is the Boltzmann's constant.
\end{enumerate}
With the effective temperature, we compare the distribution of the \texttt{MPF2021} dataset and 1M randomly sampled structures from \texttt{Alexandria}, as shown in \autoref{fig:overview:data-overview}(a), \autoref{fig:overview:data-overview}(c) and \autoref{fig:overview:data-overview}(d). Since the structures are relaxed to their corresponding local minima, it is not surprising to find that they are densely packed around the \SI{0}{\giga\pascal} in stress, with very scattered  data points of high effective temperature and high pressure. The dataset generated in this work, however, has a much wider coverage over the effective temperature space (0 -- \SI{2e4} {\kelvin}) and stress space (0 -- \SI{1000}{\giga\pascal} in magnitude). We note that the effective temperature should not be direcly interpreted as the physical temperature or the temperature employed in the simulations, instead it is an intuitive metric to measure the energy distribution of the dataset.

\section{Uncertainty quantification}\label{si-sec:uncertainty}
Uncertainty quantification plays a crucial role in the predictive modeling of materials properties and simulations, such as those involving MLFFs. Accurately assessing the uncertainty in predictions is essential because it provides insight into the reliability of the models' outputs and informs decision-making processes. In the context of materials science, where the potential for innovation is vast, but the costs of errors are high, being able to trust the predictions of computational models is paramount. Current methods of uncertainty quantification often involve statistical approaches that estimate the confidence intervals or prediction errors, such as Bayesian methods, bootstrapping, and ensemble techniques.\cite{tan2024enhanced,krajewski2022extensible,thomas2023calibration,thaler2023scalable} These methods help to understand the limits of model predictions and to identify areas where the model may require further training or refinement.

\begin{figure}
    \centering
    \includegraphics[width=\textwidth]{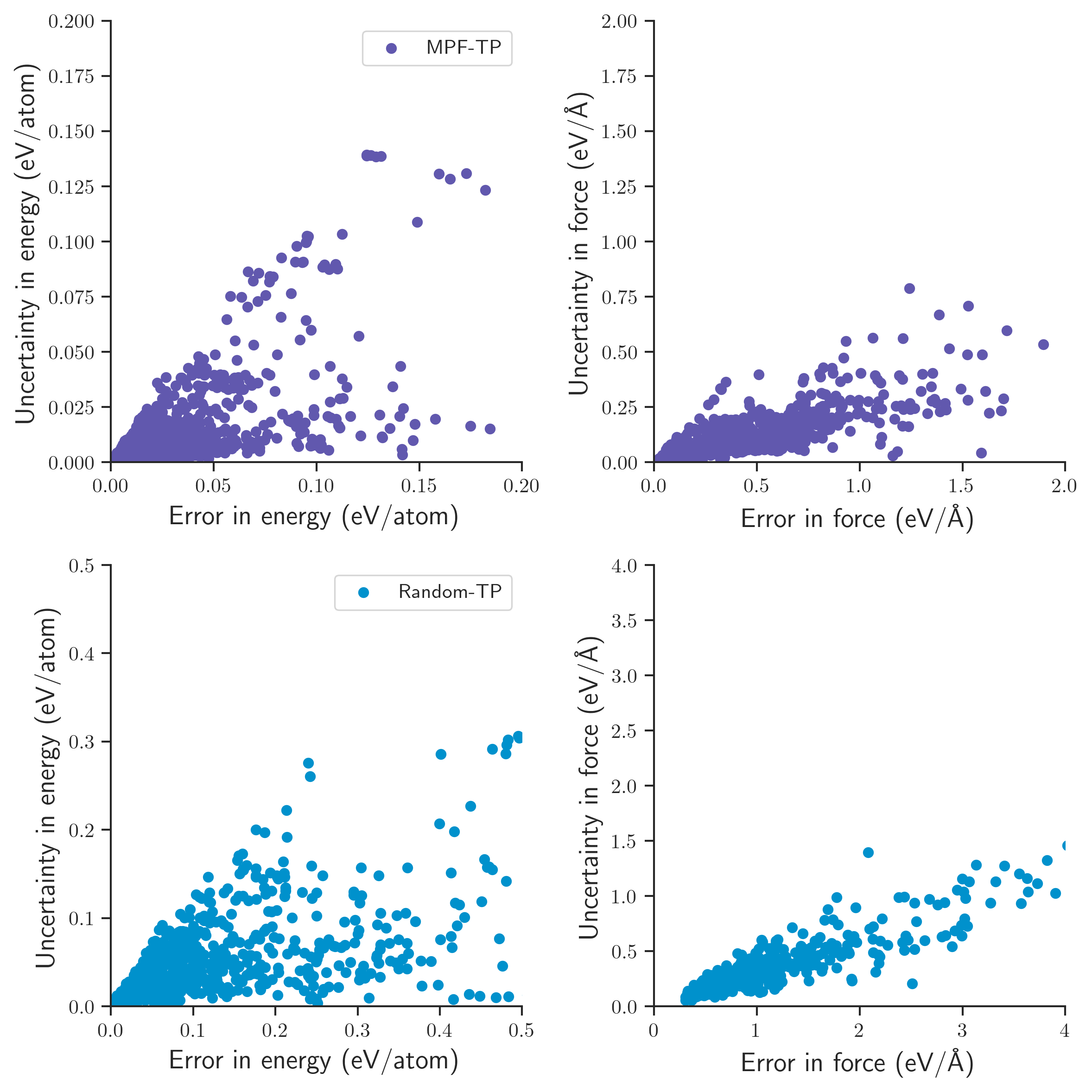}
    \caption{Parity plots of the prediction errors and the prediction uncertainties of energy per atom and forces computed for the \texttt{MPF-TP} and Random-TP datasets.}
    \label{si-fig:uncertainty-plot}
\end{figure}

In the case of MatterSim, uncertainty quantification is addressed through an ensemble approach. By training a set of five distinct models with different random initialization, the ensemble of models gives an estimation of the uncertainty on both the energies and forces. The forces offer insight into the dynamical behavior of atoms and can be particularly revealing in scenarios where the inferences of only a small fraction of the atoms within the simulation cell is deemed unreliable. Conversely, energy predictions are often more informative in cases for crystals with small number atoms in the cell. By integrating both energies and forces into the uncertainty analysis, MatterSim ensures a robust and reliable assessment of uncertainties, enhancing the confidence in its predictive capabilities for material properties and simulations. As shown in \autoref{si-fig:uncertainty-plot}, the uncertainty is measured by the standard deviation of both energies and forces on a set of randomly selected structures and is plotted in contrast to the error with respect to ground truth. While the well-known underestimation of uncertainty is present,\cite{caldeira2020deeply} the ensemble-based uncertainty is still capable of  distinguishing materials with high error from the rest.

\section{First-principles computation details}\label{si-sec:computation-details}
The DFT parameters employed in this work are generated with the \texttt{MPRelaxSet} class defined in the \texttt{pymatgen} library\cite{ong2013python} and the calculations are conducted with Vienna Ab-initio Simulation Package (VASP) (version 6.3.0)\cite{kresse1996efficiency,kresse1996efficient} using the projector augmented wave (PAW) method\cite{blochl1994projector} and Perdew-Burke-Ernzerhof (PBE)\cite{perdew1996generalized} exchange-correlation functional with Hubbard U parameter for \ce{Co}, \ce{Cr}, \ce{Fe}, \ce{Mn}, \ce{Mo}, \ce{Ni}, \ce{V}, \ce{W} elements in oxides and fluorides chosen to be 3.32, 3.7, 5.3, 3.9, 4.38, 6.2, 3.25, and \SI{6.2}{eV}, respectively, to compensate on-site electronic repulsion. 
The cutoff of plane-wave basis set is \SI{520}{eV} and the convergence threshold for total energy is \SI{5e-5}{eV\per atom}.
For each material, the total energy, forces on each atom and the stress are computed, stored and used for training.
We encounter convergence difficulties with elements such as \ce{Gd} and \ce{Eu}, particularly in off-equilibrium structures where self-consistent cycles fail to converge, or energies vary significantly for two structures with nearly identical atomic positions. Such calculations are consequently excluded from the study.

To construct the energy hull from random structure search results, a double relaxation defined by \texttt{DoubleRelaxMaker} and \texttt{MPRelaxSet}, followed by a static VASP calculation defined in \texttt{StaticMaker} is carried out on each selected structure. For more detailed information about our construction of the \texttt{Alexandria-MP-ICSD} dataset, the energy hull of the \texttt{Alexandria-MP-ICSD} dataset and the new combined hull formed by the \texttt{Alexandria-MP-ICSD} hull and our RSS-generated, one may refer to the supplementary information in Ref.~\citenum{zeni2023mattergen}.

\section{Benchmark datasets and results}\label{si-sec:performance-on-benchmark-sets}
The zero-shot performance of MatterSim is benchmarked on a few datasets computed using the same level of DFT as the training data of MatterSim.
\texttt{MPtrj-1k} and \texttt{Alexandria-1k} are collected by sampling randomly one thousand materials from the \texttt{MPtrj} and \texttt{Alexandria} dataset, respectively. These two datasets contain structures that are close to local energy minima and reflect the capability of models on predicting near-equilibrium-position properties, which is useful in evaluating materials' chemical stability. The \texttt{MPtrj-highest-stress-1k} contains the 1,000 materials with the highest stress in magnitude computed using DFT from the \texttt{MPtrj} datasets. This benchmark set evaluates the models performance in the high pressure domain. 

\begin{figure}
    \centering
    \begin{subfigure}[b]{0.32\textwidth}
    \includegraphics[width=\textwidth]{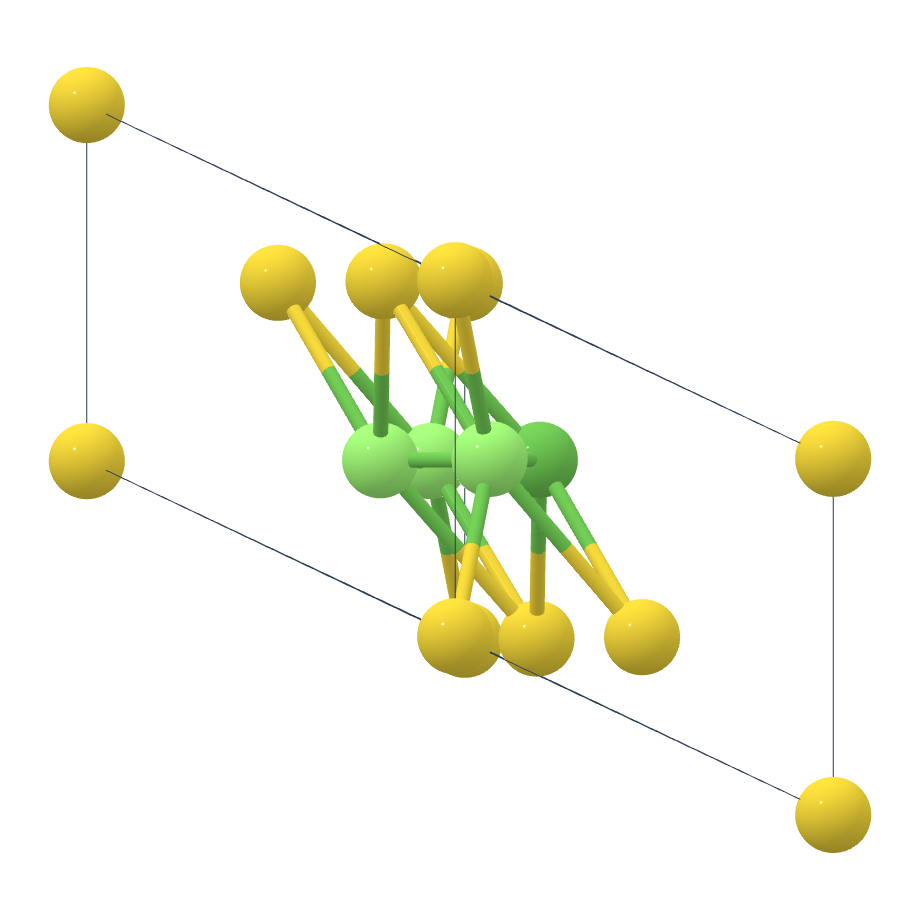}
    \caption{}
    \end{subfigure}
    \begin{subfigure}[b]{0.32\textwidth}
    \includegraphics[width=\textwidth]{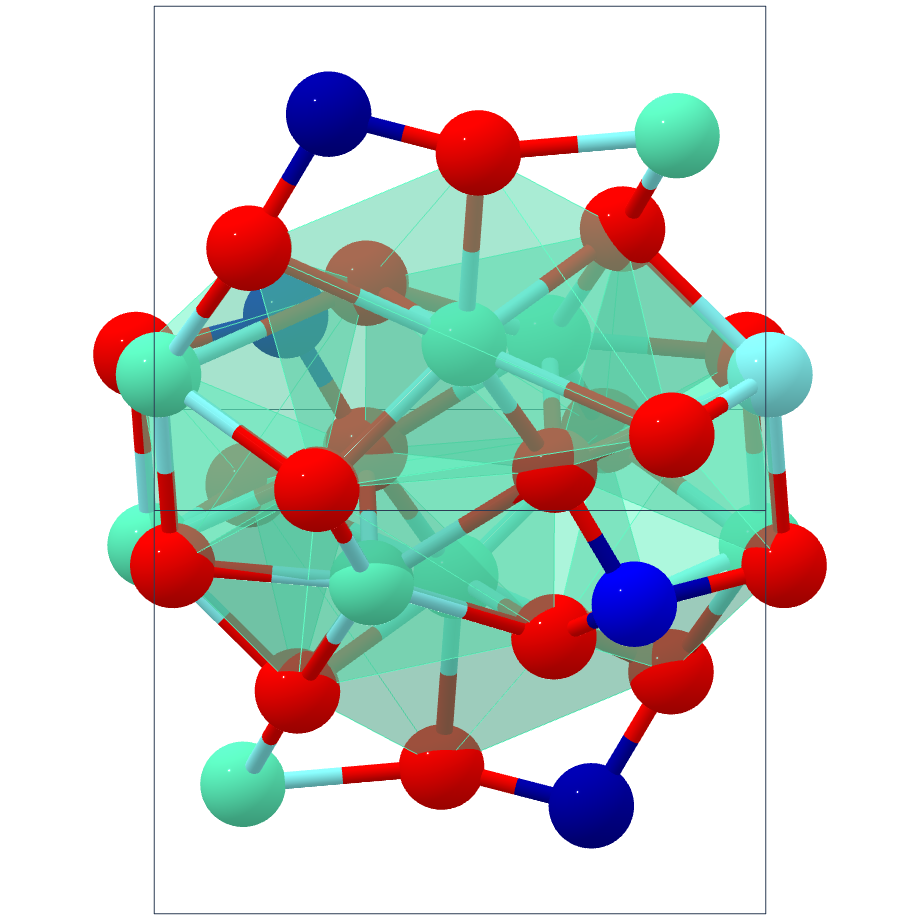}
    \caption{}
    \end{subfigure}
    \begin{subfigure}[b]{0.32\textwidth}
    \includegraphics[width=\textwidth]{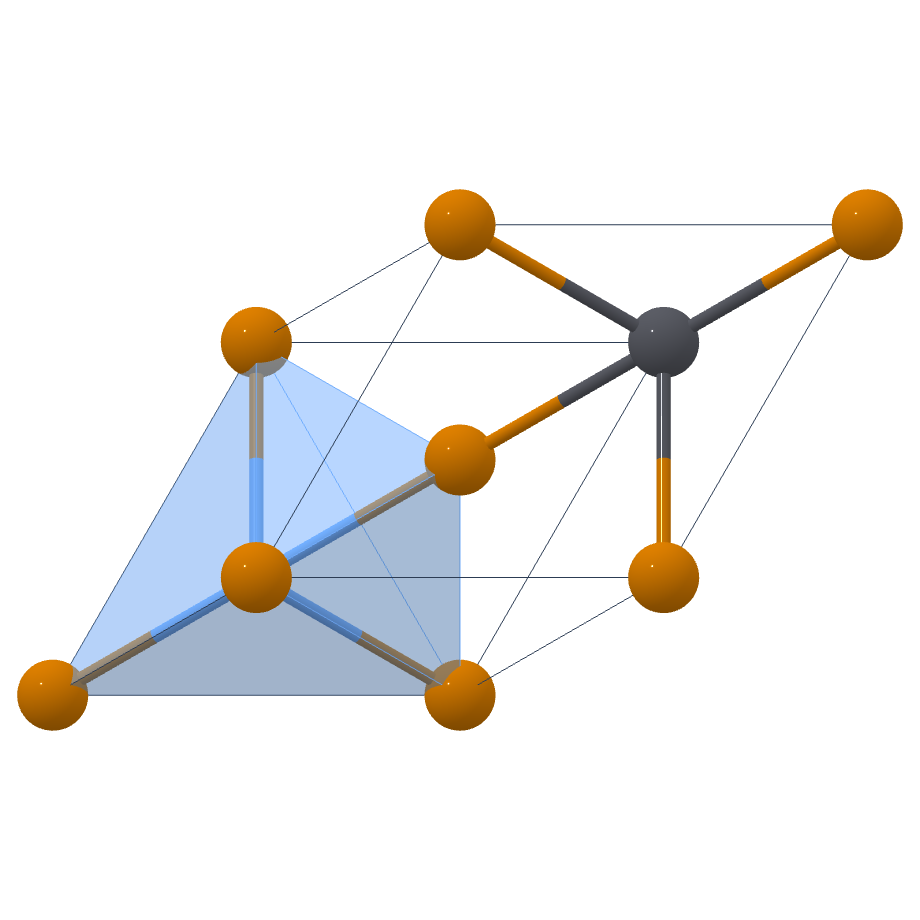}
    \caption{}
    \end{subfigure}
    \begin{subfigure}[b]{0.32\textwidth}
    \includegraphics[width=\textwidth]{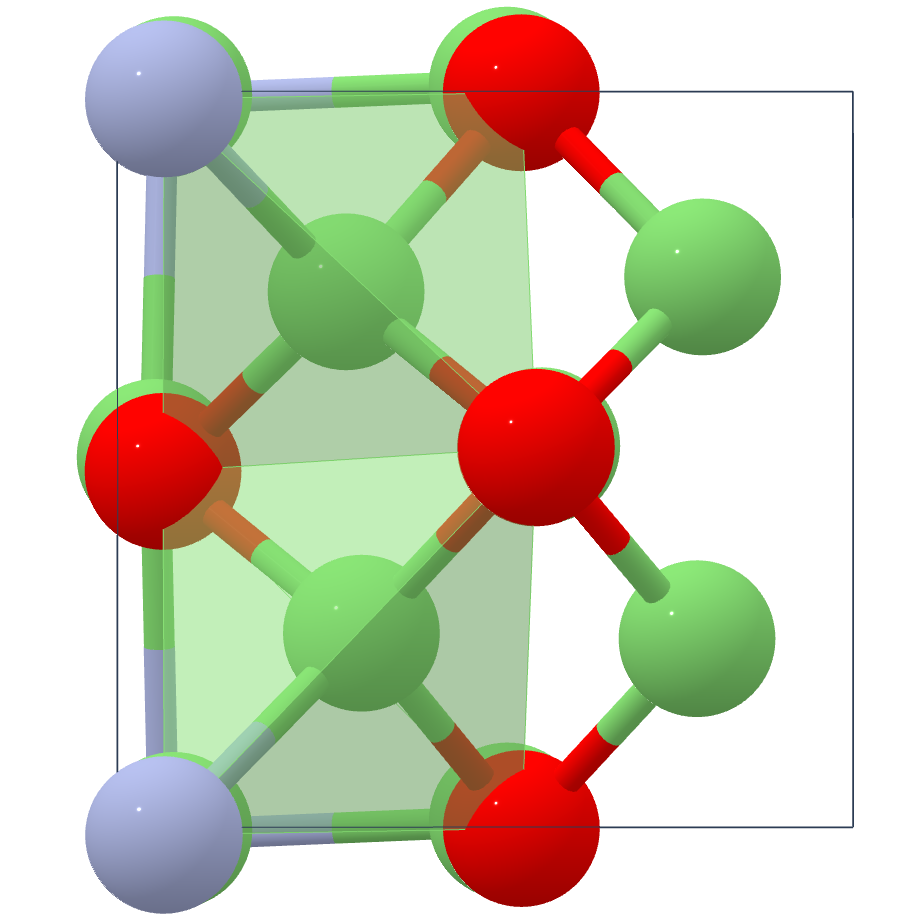}
    \caption{}
    \end{subfigure}
    \begin{subfigure}[b]{0.32\textwidth}
    \includegraphics[width=\textwidth]{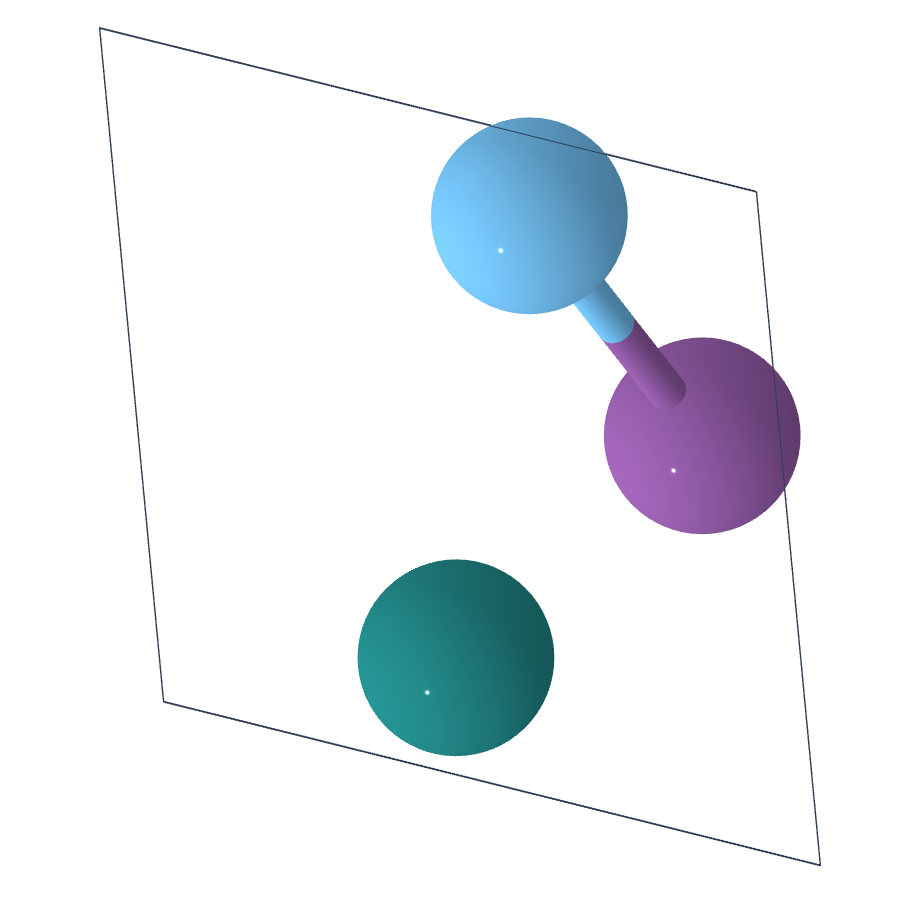}
    \caption{}
    \end{subfigure}
    \begin{subfigure}[b]{0.32\textwidth}
    \includegraphics[width=\textwidth]{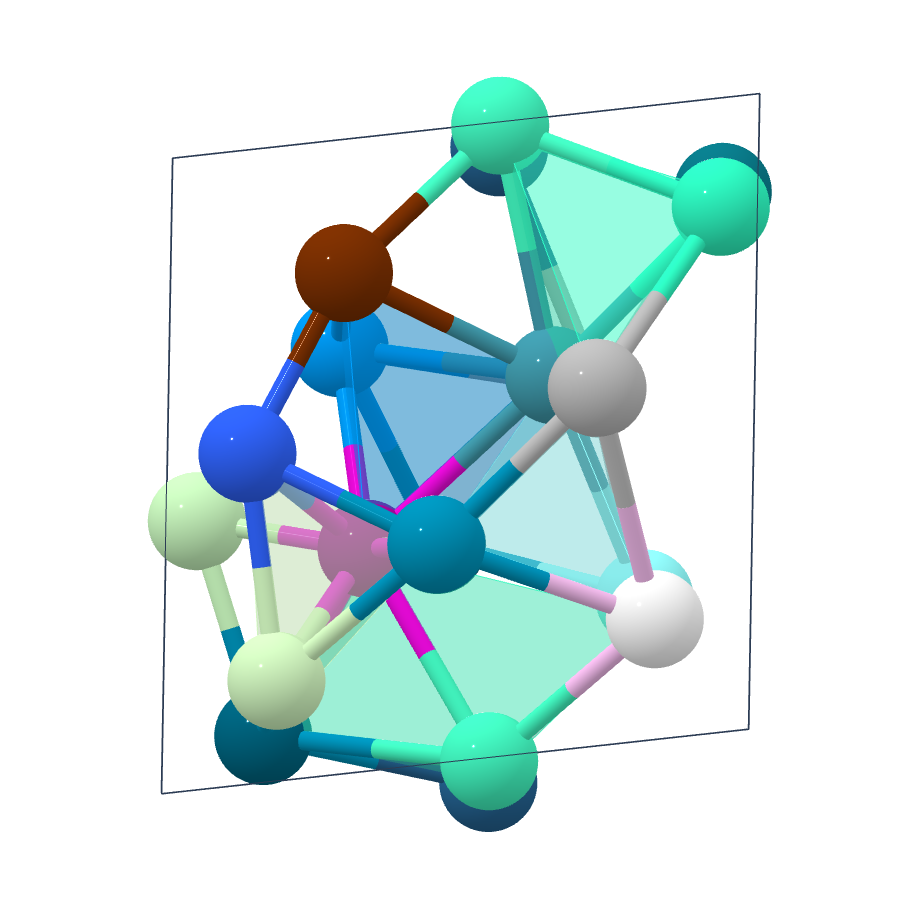}
    \caption{}
    \end{subfigure}
    \caption{Visualization of example materials in the benchmarks sets. (a) \ce{Na5As4} from \texttt{MPtrj-random-1k}; (b) \ce{Eu2CoO4} from \texttt{MPtrj-highest-stress-1k}; (c) \ce{AcTe2Pb} from \texttt{Alexandria-1k}; (d) \ce{Li8NO3} from \texttt{MPF-Alkali-TP}; (e) \ce{TiSbRu} from \texttt{MPF-TP}; and (f) \ce{ArTbPrGdYPaMnCuAgOsPd2RhXeBrKr} from \texttt{Random-TP}. }
    \label{si-fig:visualize-benchmark-materials}
\end{figure}

The \texttt{MPF-Alkali-TP}, \texttt{MPF-TP}, \texttt{Random-TP} benchmark sets are created with increasing complexity to evaluate the models' performance on materials under finite temperature and pressure conditions with far from equilibrium atomic positions. All of these benchmark sets are created with first-principles molecular dynamics trajectories initialized with corresponding structures. The \texttt{MPF-Alkali-TP} dataset is sampled from AIMD trajectories of materials that contains alkali metals in Materials Project and this dataset serves to assess the performance of the model for predicting ionic conductors. The selection rule of elements is that the compound should contain at least one alkali metal and at least one elements from N, O, P, S, Se, F, Cl, Br, I. In total, 50 compounds are selected randomly from Materials Project following the selection rule. Similarly, the \texttt{MPF-TP} contains molecular dynamics trajectories on 50 randomly selected compounds from \texttt{MPF2021} dataset without elemental constraints. For \texttt{Random-TP}, the initial structures are created by randomly placing 20 atoms with random elements in a simulation box with. Again, 50 random structures are used for later molecular dynamics simulation. During collection of the molecular dynamics simulation trajectories, each starting compound is first relaxed followed by running an NPT simulation using VASP, in which the cutoff energy of plane wave is controlled to be 520 eV and only gamma point was sampled in the reciprocal space to ensure the speed of sampling. The simulation was carried out for 100 ps for each material and the during the last 20 ps, 5 frames were uniformly collected for VASP calculation under MPRelaxSet setting, which were used in the final benchmark set. The temperatures and pressures are all random sampled. For the temperature, it is uniformly sampled between 0 to 5000K. For the pressure, we used a log scale when carrying out the sampling. The pressure range is from 0 to 1000 GPa. By such a way of creation, these three datasets reflects the power of the emulators for finite-temperature and pressure simulations with increasing difficulty in generalizability from simple ionic compounds to complex random hypothetical structures. Since the temperature and pressure ranges are wide, these benchmark sets are also reflective of model performance on crystalline materials, amorphous materials, liquids, and pressured materials. Typical structures from these datasets are all shown in \autoref{si-fig:visualize-benchmark-materials}.

To evaluate the performances, the per-atom mean absolute energy error, the mean error on forces, and mean error on stress are computed for each benchmark set. The results are shown in \autoref{si-tab:performance-compare} where the comparison is carried out between MatterSim and a few open-source universal MLFFs, including M3GNet\cite{chen2022universal}, CHGNet\cite{deng2023chgnet}, MACE-MP-0\cite{batatia2023foundation}. For M3GNet, the \texttt{M3GNet-MP-2021.2.8-PES} checkpoint defined in the MatGL library is used. The CHGNet model is accessed via the GitHub repository. For MACE-MP-0, the large version of the model defined in the commit \texttt{4d2d1c4} in the repo (https://github.com/ACEsuit/mace) is used. To evaluate the MAE of CHGNet the Materials Project 2020 Compatibility corrections are applied to the benchmark sets, while others are not. 
The results are shown in  \autoref{si-tab:performance-compare}.

\begin{table}[htpb]
    \footnotesize
    \setlength\tabcolsep{2pt}
    \centerline{
    \begin{tabular}{ccccccc}
    \toprule
    Test Set &  MAE    & M3GNet &  CHGNet & MACE-MP-0  & MatterSim(M3GNet) & MatterSim(Graphormer)\\
    \midrule
    \multirow{3}{*}\texttt{MPTrj-random-1k}           & Energy [eV/atom] &  0.032     & 0.027        &  0.015      & 0.030    & \textbf{0.012}\\
                                               & Force  [eV/\AA]  &  0.189     & 0.120        &  0.117      & 0.149    & \textbf{0.077}\\
                                               & Stress [GPa]     &  0.268     & 0.290        &  0.468      & 0.241    & \textbf{0.164}\\
    \midrule
    \multirow{3}{*}\texttt{MPTrj-highest-stress-1k}   & Energy [eV/atom] &  0.214     & 0.142        &  0.124      &  0.110   & \textbf{0.100}\\
                                               & Force  [eV/\AA]  &  0.875     & 0.689        &  0.534      &  0.417   & \textbf{0.314}\\
                                               & Stress [GPa]     & 12.288     & 8.085        & 43.284      &  6.230   & \textbf{5.921}\\
    \midrule
    \multirow{3}{*}\texttt{Alexandria-1k}             & Energy [eV/atom] &  0.119     &  0.150                     &  0.092                    & 0.058     & \textbf{0.0131}\\
                                               & Force  [eV/\AA]  &  0.112     &  0.108                     &  0.095                    & 0.086    & \textbf{0.006}\\
                                               & Stress [GPa]     &  1.431     &  1.643                     &  0.160           & 0.761     & \textbf{0.049}\\
    \midrule
    \multirow{3}{*}\texttt{MPF-Alkali-TP}              & Energy [eV/atom] &  0.165     &  0.250                     &  1.351                    &  \textbf{0.024}   &  \textbf{0.024}\\
                                               & Force  [eV/\AA]  &  1.139     &  1.636                     & 15.819                    &  0.332            & \textbf{0.326}\\
                                               & Stress [GPa]     &  4.911     & 12.625                     & 25.723                    &  \textbf{0.851}   & 1.072\\
    \midrule 
    \multirow{3}{*}\texttt{MPF-TP}                    & Energy [eV/atom] &  0.207     &  0.254                     &  256.340                  & \textbf{0.036}    &  0.0400\\
                                               & Force  [eV/\AA]  &  1.224     &  3.313                     &  1506.854                 & \textbf{0.431}    & \textbf{0.421}\\
                                               & Stress [GPa]     &  5.575     & 25.208                     &  202.093                  & \textbf{1.318}    & 1.917\\
    \midrule
    \multirow{3}{*}\texttt{Random-TP}                 & Energy [eV/atom] &  0.537     &  0.506                     &  9.184                    & 0.219     & \textbf{0.141}\\
                                               & Force  [eV/\AA]  &  1.789     &  3.950                     & 88.327                    & 0.937    & \textbf{0.813}\\
                                               & Stress [GPa]     &  3.216     &  7.230                     & 19.224                    & \textbf{2.518} & 2.696\\
    \bottomrule
    \end{tabular}
    }
    \caption{Performance of  CHGNet, MACE-MP-0 and MatterSim on benchmark datasets. \label{si-tab:performance-compare}}
\end{table}

\section{Benchmark on Matbench Discovery}
Materials discovery is an innovative field that recently starts to harnesses the power of data-driven approaches to revolutionize the way new materials are found and developed. This rapidly evolving domain leverages machine learning models to predict and analyze the properties of materials before they are physically synthesized, thereby significantly reducing research time and cost. The MatBench Discovery task, in particular, are designed to test the effectiveness of these machine learning models in predicting the stability of new materials based on a set of structural substitutions derived from the Wang-Botti-Marques (WBM) dataset.\cite{wang2021predicting} In particular, we focus on the \texttt{IS2RE} task which challenge models to predict relaxed energy from the input structures. After comparing the results with materials project, these structures classified as stable or unstable by constructing energy hulls. Finally, a few metrics are gathered including binary classification F1 score, precision/recall rates, and MAE are gathered to evaluate models' performance.

MatterSim is applied to tackle the \texttt{IS2RE} task of MatBench Discovery.
The initial structures in the WBM dataset are used as input of the model and a \texttt{FIRE} optimizer was used to relax the structures. The lattice is also relaxed using the \texttt{ExpCellFilter} function of Atomic Simulation Environment\cite{larsen2017atomic}. The force convergence criteria is set to be \SI{0.01}{eV\per\angstrom}.
The final results are shown in \autoref{si-tab:matbench}. MatterSim achieves the highest performance in all metrics compared with all opensource and commercial model. An F1 score of 0.83 and an mean absolute energy error of the formation energy  \SI{0.026} {eV\per atom}, demonstrating better success rate in finding new materials, despite that the model is trained on only a smaller amount of the data, signifying the importance of less data redundancy.

\begin{table}[htpb]
    \centerline{
\begin{tabular}{c|ccccccccc}
\toprule
Model      &  F1  & DAF  & Precision &  Accuracy & TPR & TNR & MAE & RMSE & $R^2$ \\
\midrule
MatterSim & \textbf{0.83} &  4.84 &  \textbf{0.83}  &   \textbf{0.96}  & \textbf{0.82} &  \textbf{0.97} &  \textbf{0.03} & \textbf{0.08} & \textbf{0.81} \\
GNoMe             & 0.81 & \textbf{4.86} & \textbf{0.83} & 0.94 & 0.80 & \textbf{0.97} & \textbf{0.03} & \textbf{0.08} &  0.78     \\
CHGNet            & 0.58 & 3.06 & 0.52 & 0.84 & 0.66 & 0.88 & 0.07 & 0.11 &  0.61   \\
M3GNet            & 0.57 & 2.67 & 0.45 & 0.80 & 0.77 & 0.81 & 0.07 & 0.11 &  0.60    \\
MACE              & 0.57 & 2.78 & 0.47 & 0.81 & 0.72 & 0.83 & 0.07 & 0.11 &  0.63   \\
ALIGNN            & 0.56 & 2.92 & 0.50 & 0.83 & 0.65 & 0.87 & 0.09 & 0.15 &  0.27    \\
MEGNet            & 0.51 & 2.70 & 0.46 & 0.81 & 0.57 & 0.86 & 0.13 & 0.20 & -0.28    \\
CGCNN             & 0.51 & 2.63 & 0.45 & 0.81 & 0.59 & 0.85 & 0.14 & 0.23 & -0.62    \\
CGCNN+P           & 0.51 & 2.40 & 0.41 & 0.78 & 0.67 & 0.80 & 0.11 & 0.18 &  0.03    \\
Wrenformer        & 0.48 & 2.13 & 0.36 & 0.74 & 0.69 & 0.75 & 0.10 & 0.18 & -0.04    \\
BOWSR             & 0.44 & 1.91 & 0.32 & 0.68 & 0.74 & 0.67 & 0.12 & 0.16 &  0.14    \\
Voronoi RF        & 0.34 & 1.51 & 0.26 & 0.67 & 0.51 & 0.70 & 0.14 & 0.21 & -0.31    \\
Dummy             & 0.19 & 1.00 & 0.17 & 0.68 & 0.23 & 0.77 & 0.12 & 0.18 &  0.00  \\
\bottomrule
\end{tabular}
}
\caption{Matbench discovery results using the potential enabled by MatterSim. For results, we relax the input structures, relax for 500 steps until the max magnitude of forces is lower than \SI{0.01}{eV\per\angstrom}, and evaluate the outputted energies. The results showcase that the interatomic potentials trained as part of this work showcase SOTA performance on downstream tasks. Results of GNoMe were taken from Ref.~\citenum{merchant2023scaling} and all other models from Ref.~\citenum{riebesell2023matbench}. Bolded numbers indicate the models with best performance on each metric. \label{si-tab:matbench}}
\end{table}

\section{Random structure search}\label{si-sec:rss-details}
\subsection{Search setup and computation details}
Random structure search (RSS) is carried out using a python-interfaced-version of AIRSS package \cite{pickard2006high,pickard2011ab}. Searches are carried out on all possible 4005 unary and binary chemical systems of the first 89 elements.
For each chemical system, two-consecutive round of searches are carried out. In the first round, we sample 10,000 structures in each binary system. The number of atoms in the unit cell is randomly sampled to be between 2 to 12. A uniform elemental-wise minimum separation between atoms in \r{A} was set by \texttt{MINSEP} = 0.7-3. The number of symmetry operations of the initially generated structures is set to be 2 to 4, i,e., \texttt{SYMMOPS} = 2-4. All proposed structures are relaxed using MatterSim with the lattice being optimized as well. Then in the second round of search, the same amount of structures is generated. During this round of generation, we use parameters extracted from the lowest energy structure during the first round of search to confine the search space. In particular, the \texttt{MINSEP} and the per-atom-volume \texttt{VARVOL} are extracted from the lowest energy structure in each reduced composition and then used for the generation.  Relaxations are again carried out on these structures using MatterSim. Such two-round search is a standard routine to carry out RSS as the first batch tries to cover a large volume and interatomic distance range while the second batch focuses on the most likely setup and does a thorough search. After the two round of searches, the resulting structures are collected together and deduplication is performed using \texttt{pymatgen}'s \texttt{StructureMatcher}\cite{ong2013python}. Finally, the top three structures with the lowest energies estimated by MatterSim is sent to first-principles computation following the double relaxation and static calculation protocol as discussed in \autoref{si-sec:computation-details}. These final DFT results are used for the construction of the final hull by combining with the \texttt{Alexandria-MP-ICSD} dataset \cite{zeni2023mattergen}.

\subsection{Search results and discussions}
RSS is uniquely comprehensive due to its exhaustive nature, although it is traditionally limited by its high computational demands. This search is facilitated by the MatterSim, which assesses energy inferences across an extensive set of unary and binary chemical combinations. The screening encompassed 4005 unary and binary chemical systems between 89 elements, with each pairing examined across 45 varying compositions. This leads to an astronomical number of energy inferences, totaling over 30 billion, assuming 400 relaxation steps needed for each structure. Remarkably, the use of MatterSim enabled the completion of this vast screening process within a week—a task that would otherwise span an estimated 100 years if approached with DFT methods using 1,000 CPU cores. (The estimation is based on that a single-point DFT energy computation on a 12-core CPU node takes around 10 seconds.)

\begin{figure}
    \centering
    \includegraphics[width=0.8\textwidth]{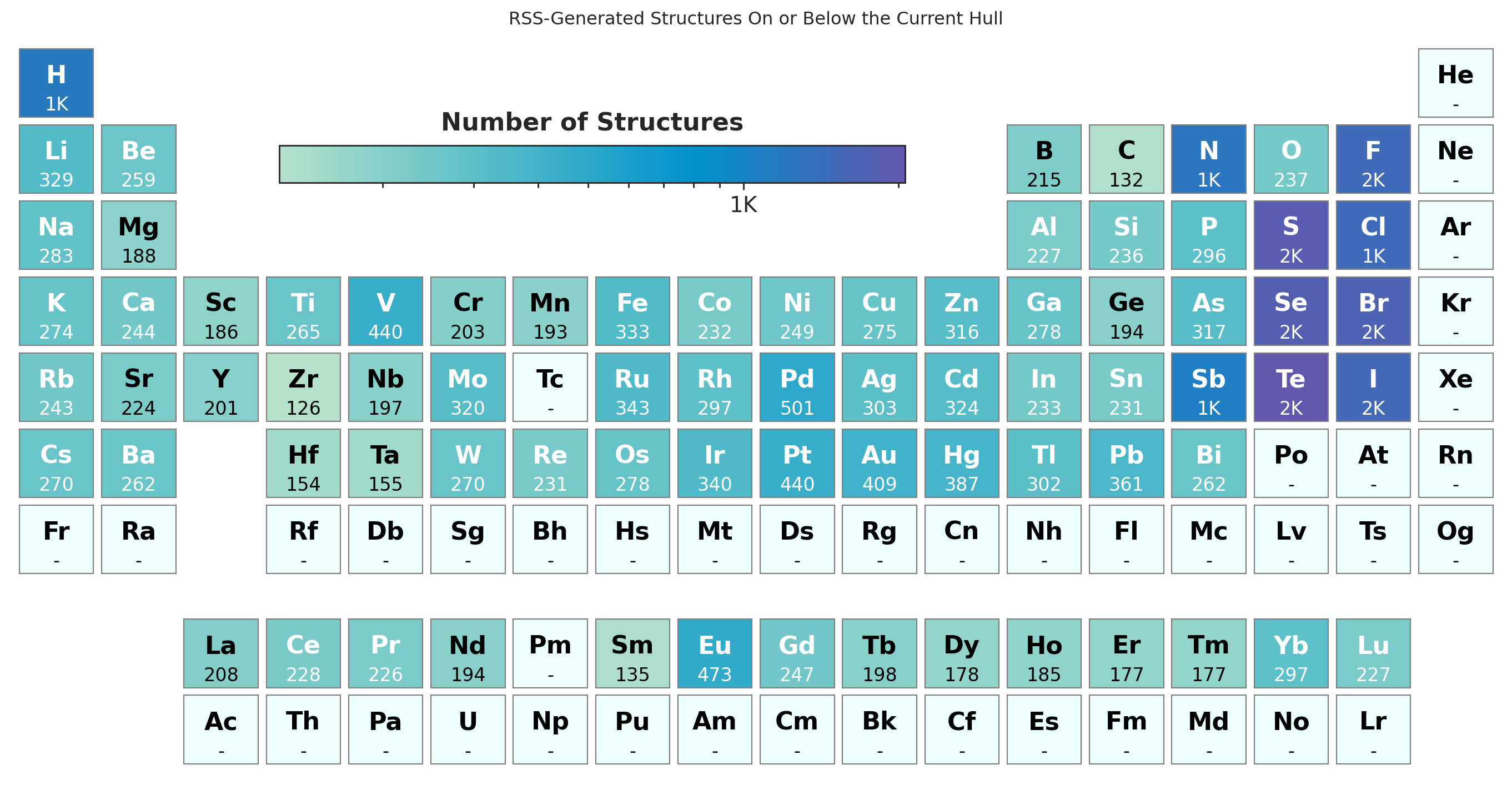}
    \caption{Elementwise appearance distribution of the 16,399 RSS-generated materials found to be on or below the current convex hull.}
    \label{si-fig:rss-below-current-hull}
\end{figure}

The final structures out of RSS covers around 90 percent of the elementary and binary structures within 12 atoms in Materials Project, demonstrating the exhaustive nature of this search. Among them, we carried out DFT calculations on the most stable 1\% structures of each composition according to the energies predicted by MatterSim. This leads to around 500,000 structures computed using DFT in total. Within these structures, we identified 16,399 structures to be lower than or on the current energy hull defined by the \texttt{Alexandria-MP-ICSD} dataset, as illustrated in \autoref{si-fig:rss-below-current-hull}. In this plot, we observed a bias towards anion-rich compounds consisting of O, S,  F,  Cl,  Br,  I,  N,  H,  Se,  Si,  Sb, and Te elements, which are potentially affected by Materials Project's anion correction. 
While this correction works fine for compounds with usual oxidation states, the off-stoichiometric nature of many candidates in RSS search leads to over-estimation of their stability in anion-rich compounds. 
Therefore, when analyzing the RSS-generated results, we excluded all the materials containing these elements, and even after this removal, we still find 852 materials on the new hull defined by the combination of the RSS-generated structures and the \texttt{Alexandria-MP-ICSD} dataset, as illustrated in \autoref{fig:materials-discovery}(c). Considering the fact that we only included the RSS-generated candidates with the lowest 1\% energy of each chemical composition, we expect more stable materials to be confirmed with first-principles verifications.
Such results further reveal that the current known materials space only covers a small percentage of the entire space, far from exhaustive, even for simple binary systems.

\begin{figure}
    \centering
    \includegraphics[width=0.8\textwidth]{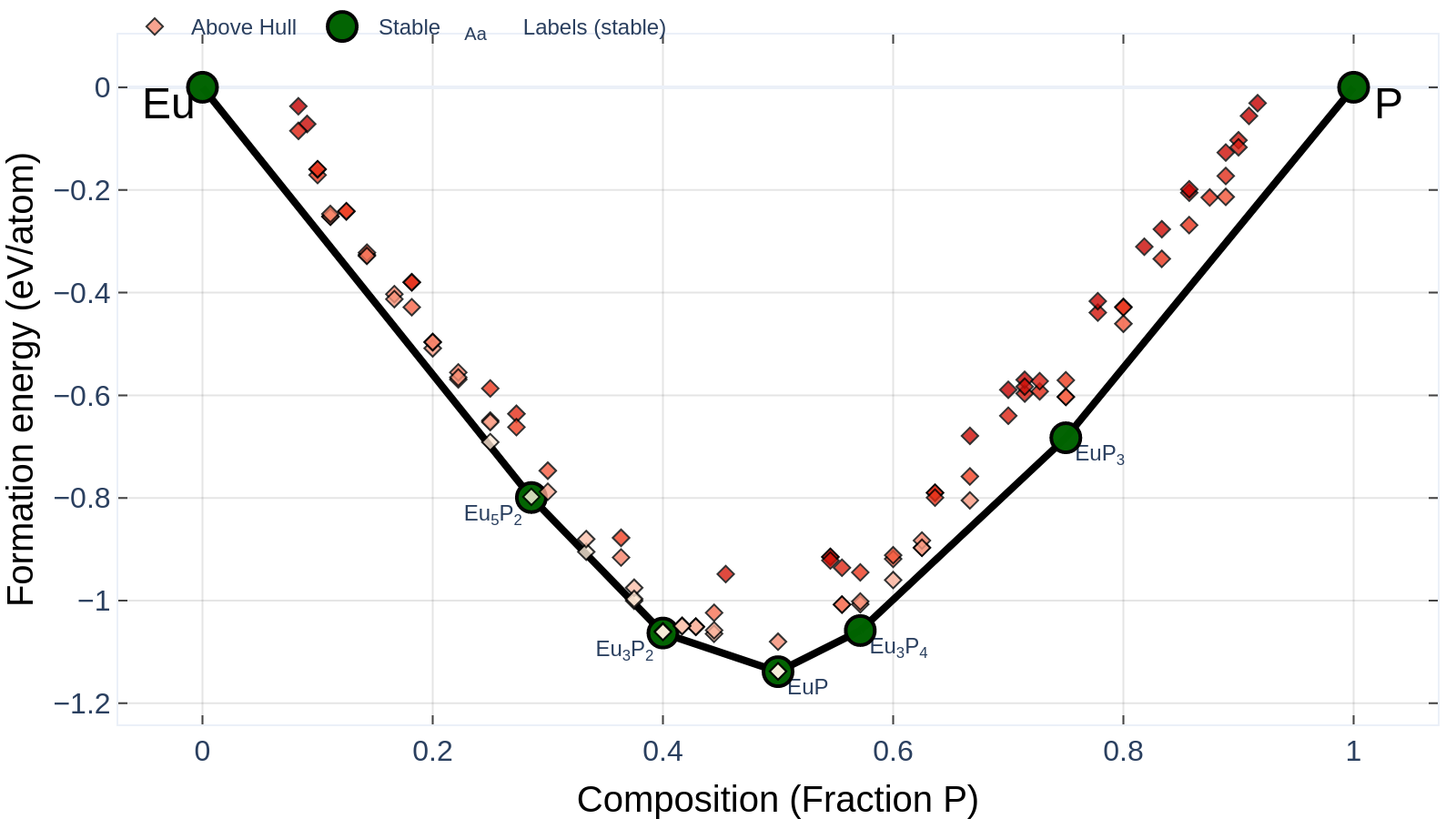}
    \caption{The formation energy of the RSS-generated materials for the \ce{Eu}-\ce{P} chemical system, with the black segments being the combined convex hull defined by the \texttt{Alexandria-MP-ICSD} dataset and our RSS-generated materials, and the green dots being the on-hull materials. }
    \label{fig:Eu-P-chemical-system}
\end{figure}

\section{Phonon prediction}\label{si-sec:phonon_prediction}

\subsection{Benchmark Dataset}
We benchmark against Materials Data Repository (MDR) phonon calculation database (also known as PhononDB),\cite{PhononDB} a database of phonon properties derived from first-principles calculations. PhononDB encompasses various materials, each characterized by phonon properties computed using the finite displacement method via the Phonopy software package.\cite{phonopy-phono3py-JPCM, phonopy-phono3py-JPSJ} The force constants for these calculations are obtained through the VASP\cite{kresse1996efficiency,kresse1996efficient}. Furthermore, the Perdew-Burke-Ernzerhof for solids (PBEsol) exchange-correlation functional\cite{perdew2008restoring,Togo2024private} is utilized within the DFT framework. MatterSim's performance is rigorously assessed against the entire PhononDB database.

\subsection{Method}
Here we continue to utilize Phonopy software package to interface with MatterSim. The phonon dispersion curves and density of states (DOS) are computed using the finite displacement method. For each material, a supercell is constructed from the primitive cell. Due to the large number of materials, an algorithm is designed to automatically choose the supercell size in the following. For each material, we set the maximum number of atoms in the supercell ($N_\mathrm{max}$) used in the phonon calculations to be 300, except Fd\={3}m, Fm\={3}m, F\={4}3m and $\mathrm{P6_3mc}$ space graps for which 216, 216, 216 and 450 are used, respectively. 
For a primitive cell containing $N_p$ atoms with lattice vector length $a$, $b$, $c$, and maximum $N_{max}$ atoms in the supercell, the supercell size is $n_x \times n_y \times n_z$ are computed 
\begin{equation}
    \begin{aligned}
    n_x &= \max\left(\left\lfloor \left( \frac{N_{max}}{N_p} \frac{bc}{a^2} \right)^{\frac{1}{3}} + 0.5 \right\rfloor , \ 1\right) \\
    n_y &= \left\lfloor n_x \frac{a}{b} + 0.5\right\rfloor, \\
    n_z &= \left\lfloor n_x \frac{a}{c} + 0.5\right\rfloor,
    \end{aligned}
\end{equation}
assuming $a$ is the longest side of the primitive cell.
To generate force constant matrices, displacements compatible with the space group are introduced to atomic positions within the supercell as implemented in \texttt{Phonopy}.
With a magnitude of \SI{0.03}{\angstrom}, consistent with settings in PhononDB, the forces acting on each displaced atom are then predicted using MatterSim or other MLFFs. The resulting forces serve as input for Phonopy, which computes the dynamical matrices, phonon frequencies and dispersions.

\subsection{Results}

\begin{figure}
    \centering
    \begin{subfigure}[b]{0.45\textwidth}
    \includegraphics[width=\textwidth]{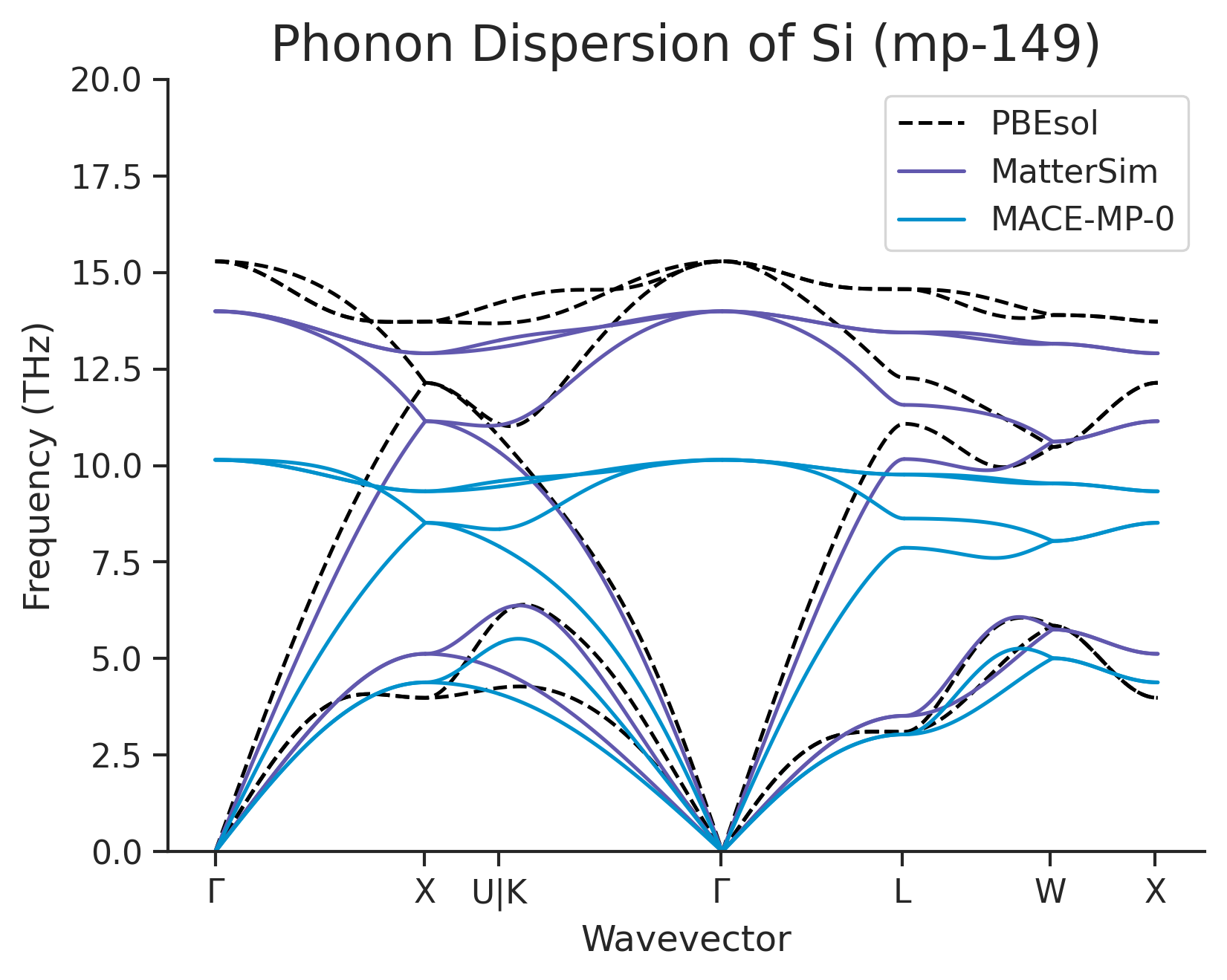}
    \caption{}
    \label{si-fig:phonon-Si}
    \end{subfigure}
    \begin{subfigure}[b]{0.45\textwidth}
    \includegraphics[width=\textwidth]{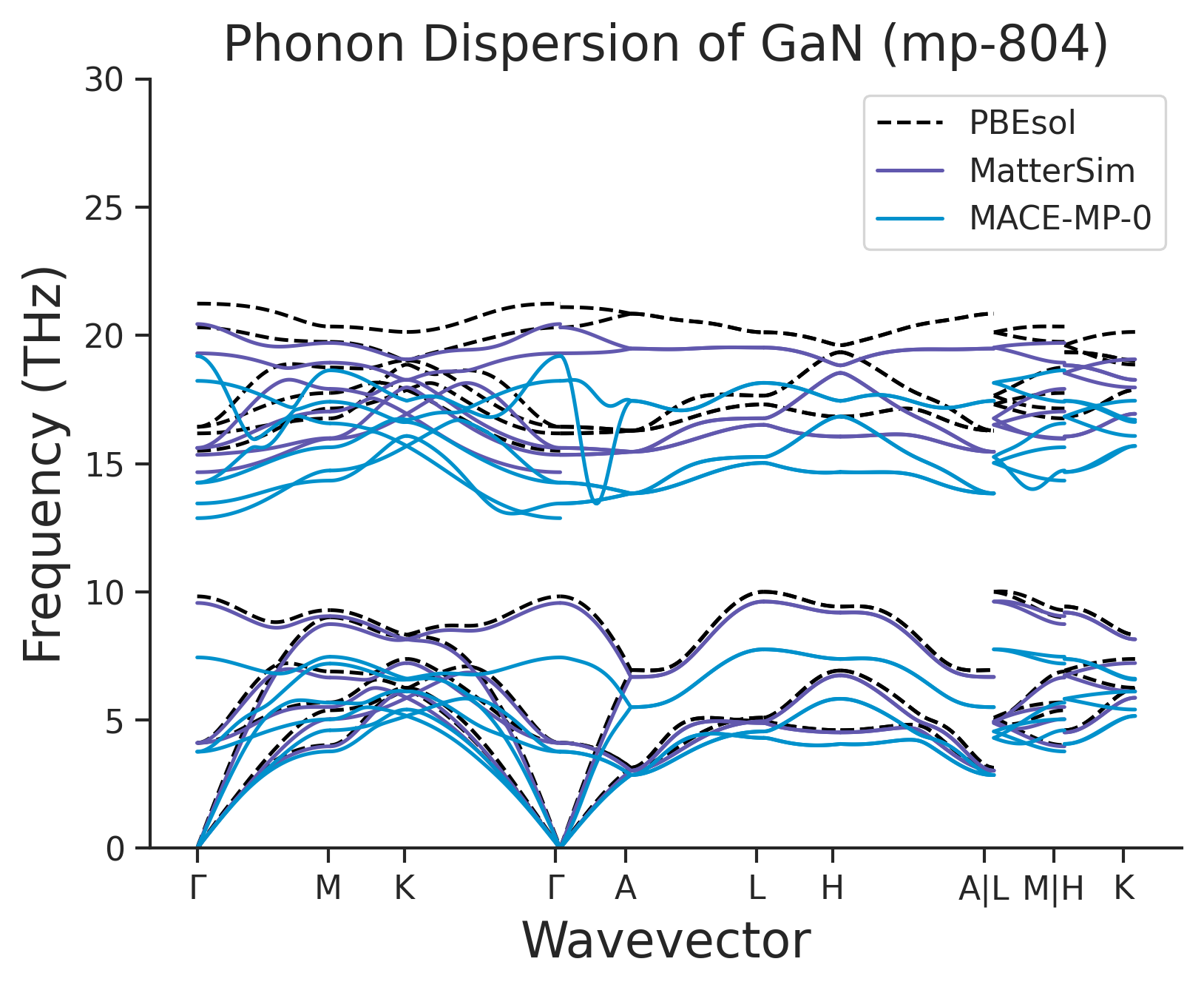}
    \caption{}
    \label{si-fig:phonon-GaN}
    \end{subfigure}
    \begin{subfigure}[b]{0.45\textwidth}
    \includegraphics[width=\textwidth]{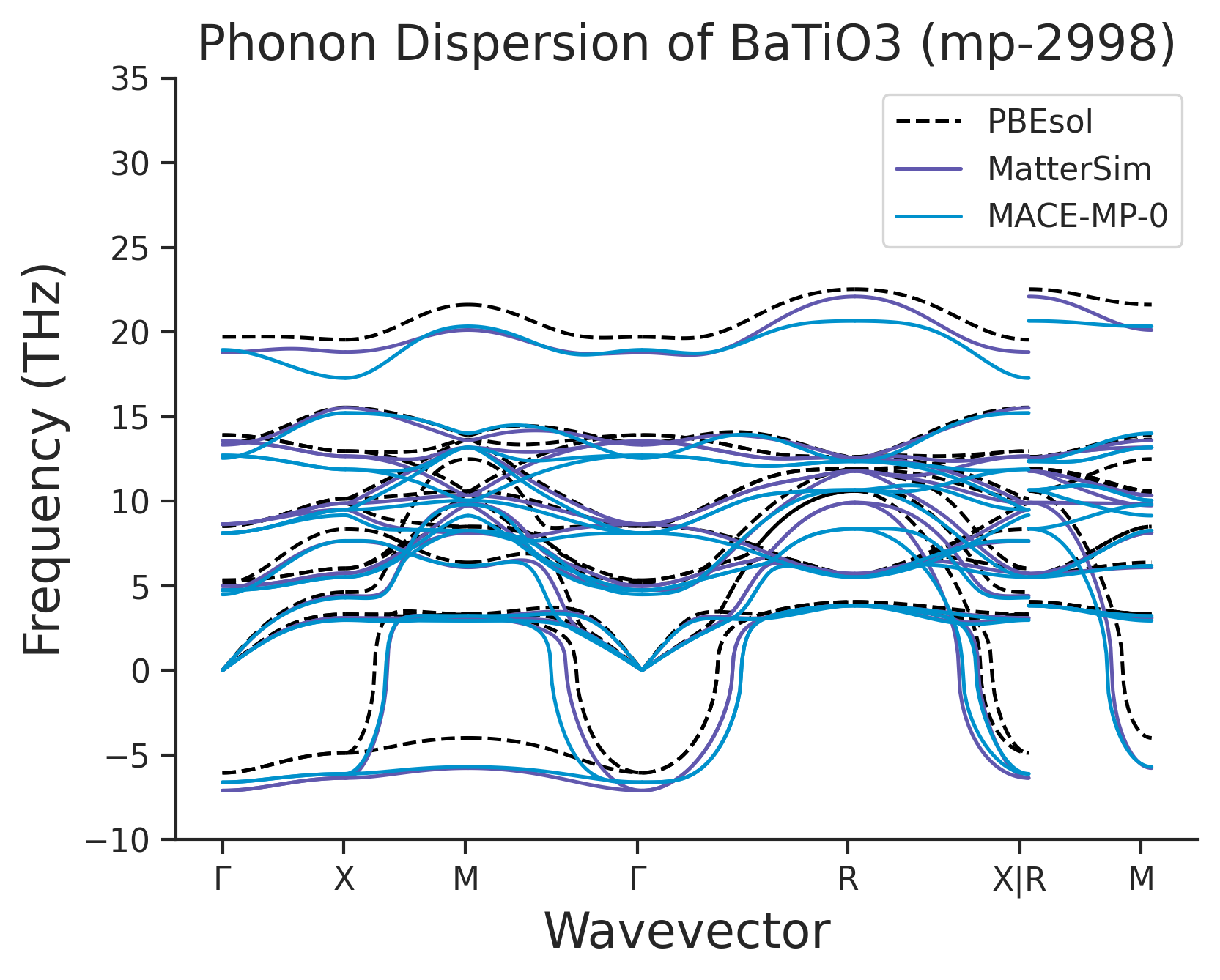}
    \caption{}
    \label{si-fig:phonon-BaTiO3}
    \end{subfigure}
    \begin{subfigure}[b]{0.45\textwidth}
    \includegraphics[width=\textwidth]{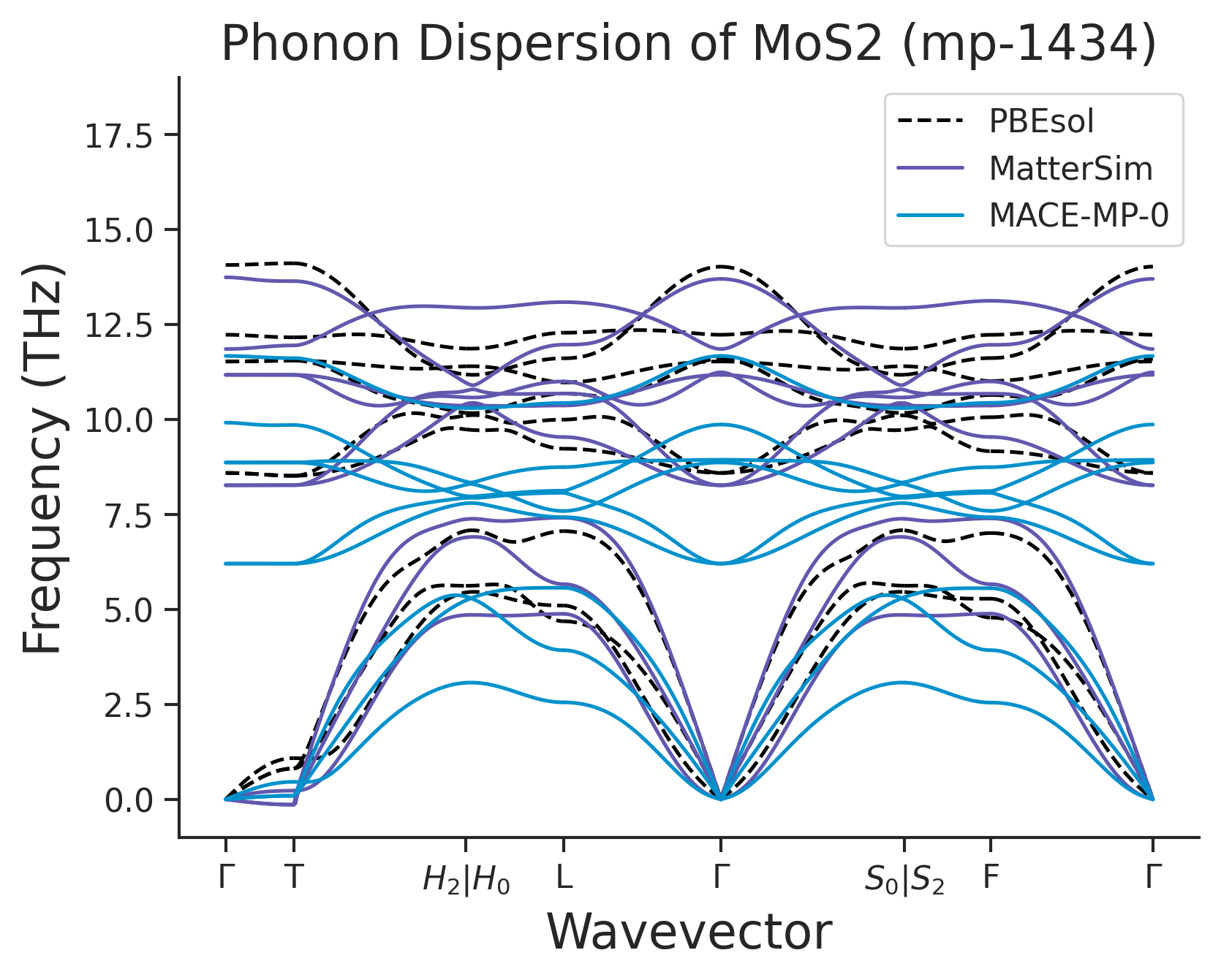}
    \caption{}
    \label{si-fig:phonon-MoS2}
    \end{subfigure}    
    \caption{Comparative analysis of phonon dispersion and DOS of (a)\ce{Si}, (b)\ce{GaN}, (c)\ce{BaTiO3} and (d) \ce{MoS2}: Predictions from MatterSim, MACE-MP-0, and PBEsol calculations sourced from PhononDB.}
    \label{si-fig:phonon-examples}
\end{figure}

We demonstrate the accuracy of MatterSim as an efficient alternative to traditional first-principles approaches for predicting phonon dispersion in bulk materials and we do a benchmark on the entire PhononDB database. Four example phonon dispersions of silicon (\ce{Si}), the binary compound (\ce{GaN}), a perovskite (\ce{BaTiO3}) and a layered structure (\ce{MoS2}) are shown in Figs.~\ref{si-fig:phonon-Si}, \ref{si-fig:phonon-GaN}, \ref{si-fig:phonon-BaTiO3} and \autoref{si-fig:phonon-MoS2}, respectively. The results are compared with those obtained using the recently proposed  model MACE-MP-0\cite{batatia2023foundation} and PBEsol calculations taken from PhononDB. As presented in the figures, MatterSim accurately predicted the phonon dispersion and DOS in all of the four materials, with slight underestimate of the highest frequency. In the case of \ce{MoS2}, MatterSim has a remarkable overall prediction with a particular good agreement for the frequency of the highest optical phonon at $\Gamma$ point.  Although MACE-MP-0 predicted \ce{BaTiO3} very well, it significantly underestimated the phonon dispersion in the other three cases, which is observed with M3GNet as well.\cite{chen2022universal}
In addition, we observed a non-physical abrupt change of phonon frequencies along the $\Gamma$--A direction in \ce{GaN} predicted by MACE-MP-0. This points to the importance of the underlying training data on phonon prediction.

\begin{figure}
    \centering
    \begin{subfigure}[b]{0.45\textwidth}
    \includegraphics[width=\textwidth]{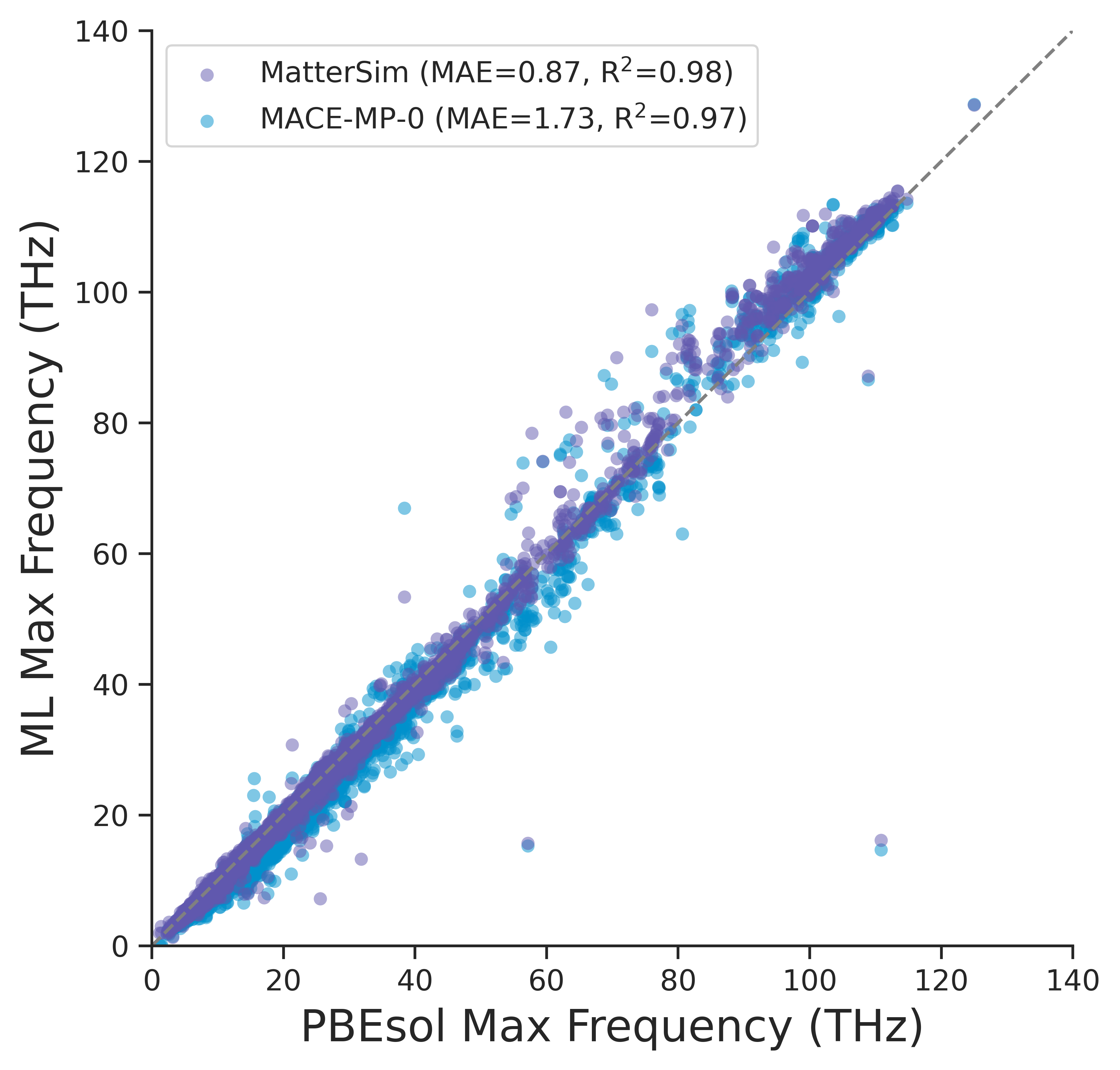}
    \caption{}
    \label{si-fig:phonon-max-freq}
    \end{subfigure}
    \begin{subfigure}[b]{0.45\textwidth}
    \includegraphics[width=\textwidth]{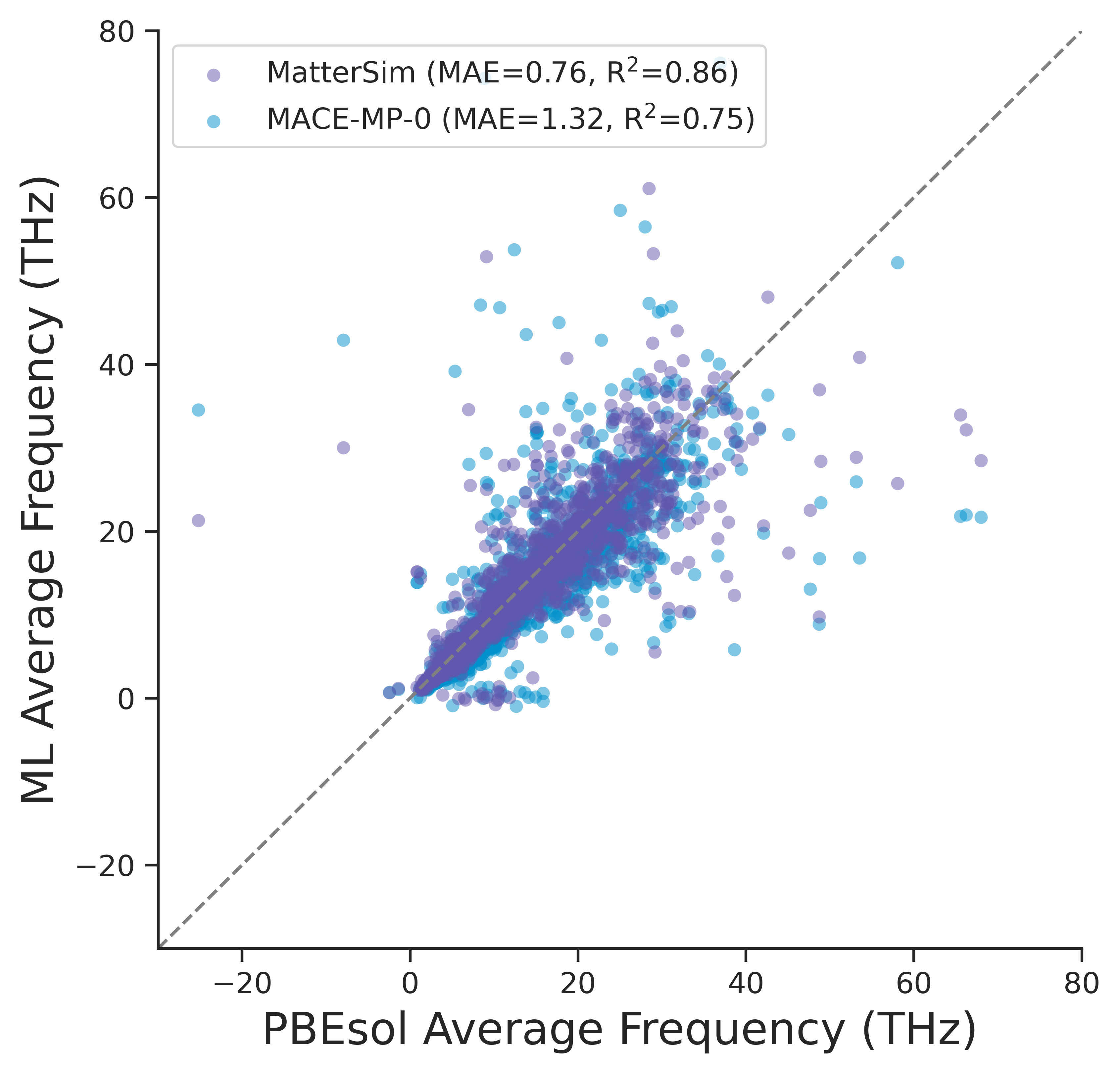}
    \caption{}
    \label{si-fig:phonon-average-freq}
    \end{subfigure}
    \begin{subfigure}[b]{0.45\textwidth}
    \includegraphics[width=\textwidth]{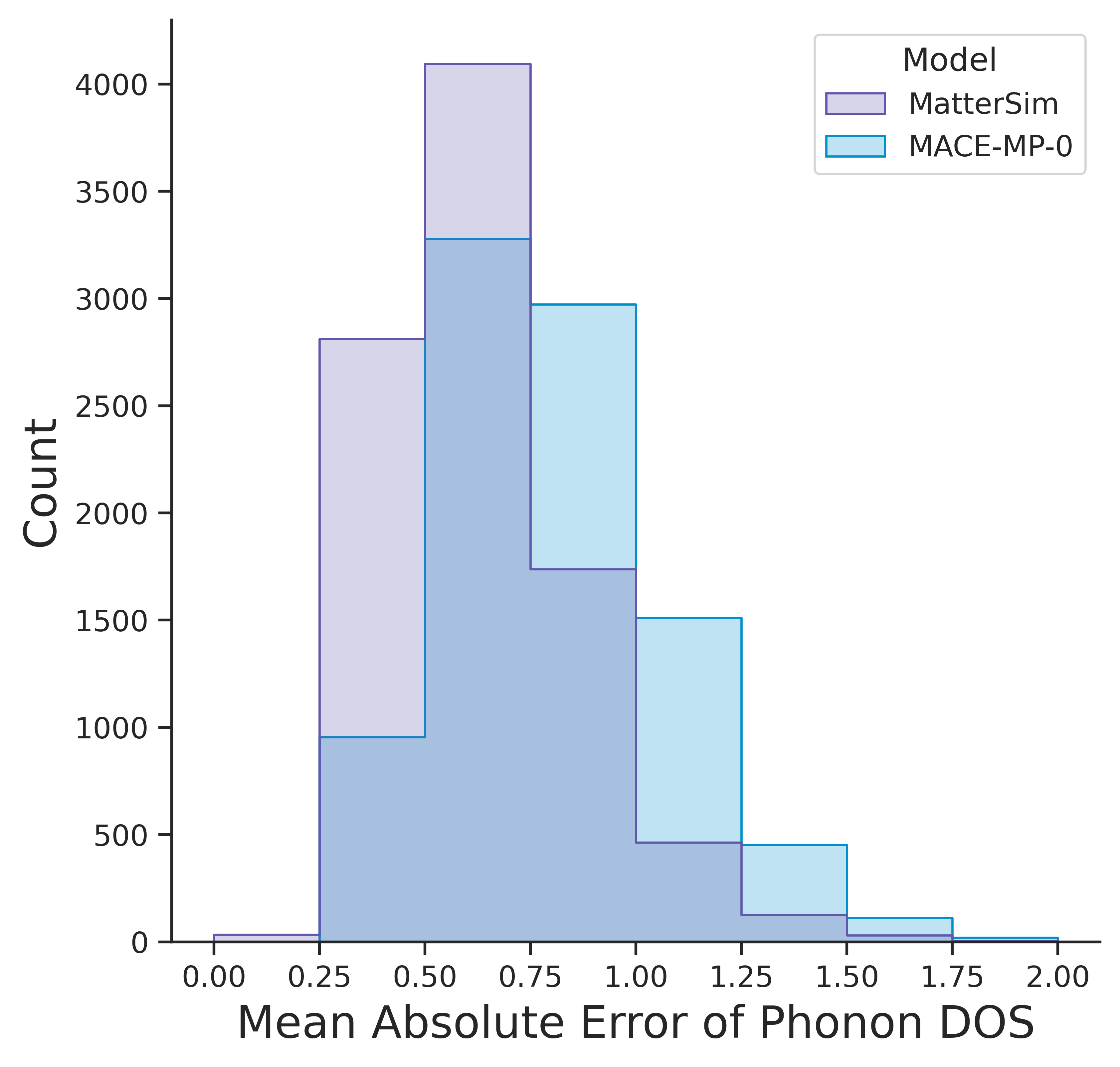}
    \caption{}
    \label{si-fig:phonon-abs-err}
    \end{subfigure}
    \begin{subfigure}[b]{0.45\textwidth}
    \includegraphics[width=\textwidth]{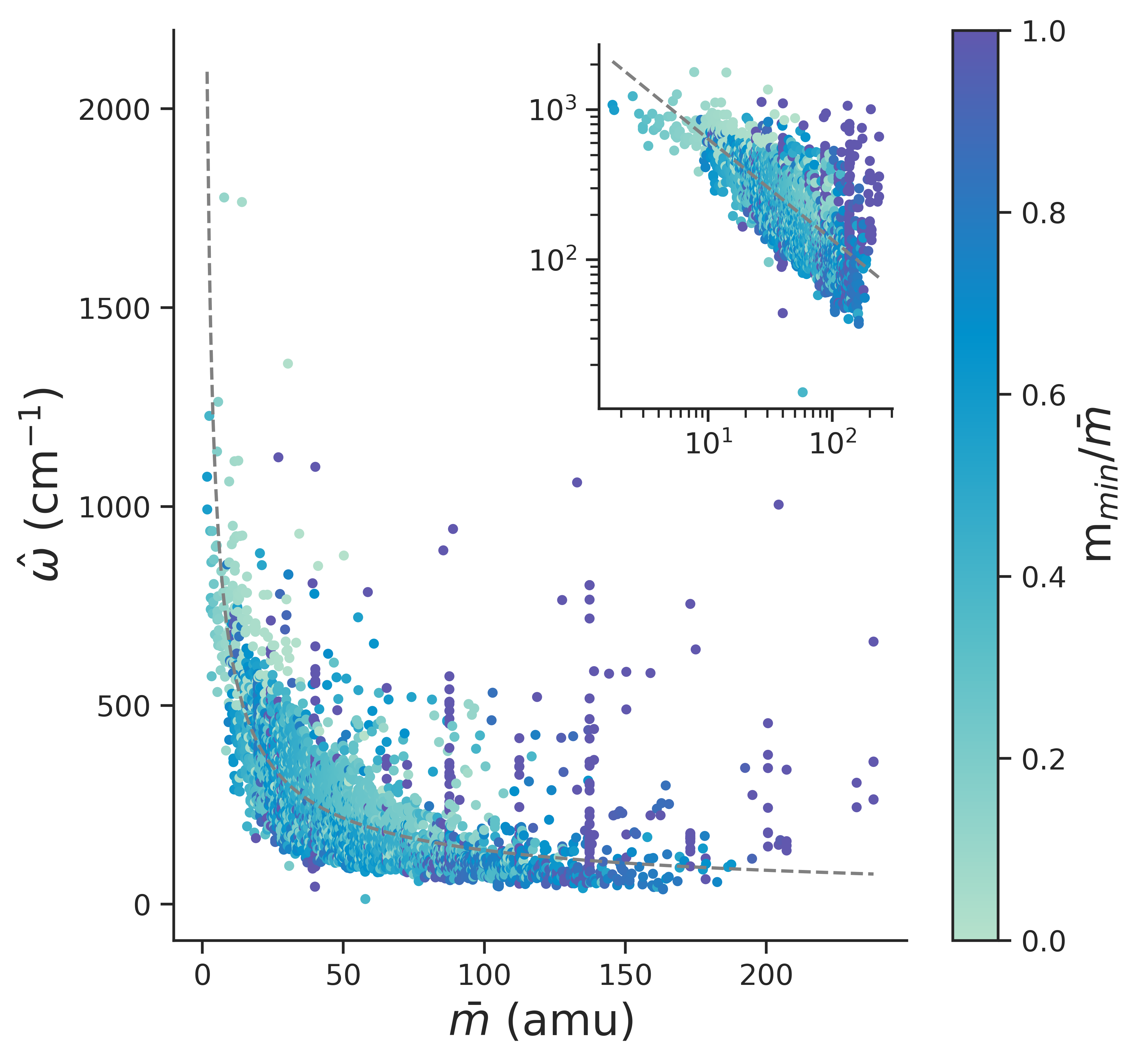}
    \caption{}
    \label{si-fig:phonon-freq-vs-mass}
    \end{subfigure}
    \caption{Performance evaluation of phonon predictions: (a) Phonon maximum frequency. (b) Phonon average frequency. (c) MAE of phonon DOS. (d) The correlation between the phonon average frequency and average atomic mass.}
    \label{si-fig:phonon-metrics}
\end{figure}

To quantitatively evaluate the performance of MatterSim's prediction of phonons, we computed phonon maximum frequency and average frequency of all computed dispersion, square difference between PBEsol-calculated and ML-predicted phonon DOS, and the phonon average frequency versus the average atomic mass, as illustrated in Fig.~\ref{si-fig:phonon-metrics}. In our comparative analysis of phonon maximum frequency and average frequency, MatterSim exhibited superior performance to previous models based on crystal relaxation trajectories when evaluated using MAE and R-squared (R$^2$) metrics, as visualized in Figs.~\ref{si-fig:phonon-max-freq} and \ref{si-fig:phonon-average-freq}. In the prediction of phonon maximum frequency, MatterSim demonstrated a lower MAE of \SI{0.87}{THz} compared to MACE-MP-0's MAE of \SI{1.73}{THz}. Similarly, in the prediction of phonon average frequency, which is defined using phonon frequency $\omega$ and DOS $g(\omega)$
\begin{equation}
    \bar{\omega} = \frac{\int \omega g(\omega)\, \mathrm{d}\omega}{\int g(\omega)\,\mathrm{d}\omega},
\end{equation}
MatterSim maintained its superior performance with an MAE of \SI{0.76}{THz} relative to MACE-MP-0's \SI{1.32}{THz}, and an R$^2$ score of $0.86$ compared to $0.75$. MatterSim maintained a consistently high level of performance across the prediction of both phonon maximum frequency and phonon average frequency, whereas MACE-MP-0 exhibited a marked decline in R$^2$ score when faced with the prediction of phonon average frequency, which is a more challenging task because it requires an accurate description of the full phonon DOS. To evaluate the two models' performance in the prediction of phonon DOS, we calculated the MAE of the DOS,
\begin{equation}
    \mathrm{MAE}_{\mathrm{{DOS}}} = \int \left\vert g_\mathrm{PBEsol}(\omega) - g_\mathrm{ML}(\omega) \right\vert\,\mathrm{d}\omega,
\end{equation}
where $g_\mathrm{PBEsol}(\omega)$ and $g_\mathrm{ML}(\omega)$ are PBEsol-calculated and ML-predicted phonon DOS, respectively. The distribution of the MAE of calculated materials is presented in \autoref{si-fig:phonon-abs-err}. Upon examining the histogram, it is evident that in the three bins representing the lower MAE values, the count for MatterSim is significantly higher than the count for MACE-MP-0. This suggests that MatterSim has a larger number of materials with lower MAE, demonstrating that it performs better in terms of accuracy for the DOS prediction when compared to models based on crystal relaxation trajectories. The average MAE over calculated materials for phonon DOS predicted by MatterSim and MACE-MP-0 was $0.64$ and $0.81$, respectively. The correlation between the average phonon frequency $\bar{\omega}$ and the average atomic mass of the material $\bar{m}$, which is defined by the atomic mass of each atom $M_\kappa$ and the number of atoms $n$ in the material as
\begin{equation}
    \bar{m} = \left( \frac{1}{n} \sum_{\kappa} \sqrt{M_\kappa} \right)^2,
\end{equation}
is presented in \autoref{si-fig:phonon-freq-vs-mass}. Only materials that have no negative frequencies were considered in the figure. The plot is in agreement with the work by Ref.~\citenum{petretto2018high} and Ref.~\citenum{chen2022universal}. The data was fit to the following form,
\begin{equation}
    \log\bar{\omega} = k \log\bar{m} + b.
\end{equation}
MatterSim's fitting parameters yielded a slope of $k = -0.67$ and an intercept of $b = 8.00$. These results exhibit a remarkable agreement with those obtained from the PhononDB dataset, where the fitted parameters were $k = -0.64$ for the slope and $b = 7.95$ for the intercept. This close agreement suggests that MatterSim is robust and reliably captures the trends in diverse materials.


\section{Mechanical properties} \label{si-sec:mechanical-properties}
\subsection{Quasi-Harmonic Approximation} \label{si-sec:mechanical-properties-qha}
The harmonic approximation (HA) assumes that atoms in a crystal vibrate about their equilibrium positions and the potential energy can be approximated by a quadratic function of the atomic displacements. This model is accurate at low temperature where anharmonic effects are negligible. However, as the temperature increases, anharmonic contributions become significant, and the harmonic approximation fails to predict the correct thermodynamic behavior. To this end, the quasi-harmonic approximation (QHA) is introduced as an extension to the HA, and it takes into account the anharmonicity by computing the volume dependence of the phonons. While the shape of the potential energy surface may change with the volumes, QHA assumes that the HA is applicable for each volume. In this way, QHA is capable of describing anharmonicity and thermal expansion effects. In this work, QHA is employed to predict mechanical properties, enthalpies and free energies of ordered crystals.

Under QHA, the Helmholtz free energy $F$ at a given temperature ($T$) and volume ($V$) can be expressed as:
\begin{equation}
    F(T, V) = U_\mathrm{el}(V) + F_\mathrm{ph}(T, V),
\end{equation}
where $U_\mathrm{el}$ is electronic total energy and $F_\mathrm{ph}$ is phonon Helmholtz free energy. $F_\mathrm{ph}$ is obtained by
\begin{equation}
    F_\mathrm{ph}(T, V) = \frac{1}{2} \sum_{\mathbf{q}, i} \hbar \omega_{\mathbf{q}, i}(V) + k_B T \sum_{\mathbf{q}, i} \ln \left[ 1 - \exp \left( -\hbar \omega_{\mathbf{q}, i}(V) / k_B T \right) \right],
\end{equation}
where $\mathbf{q}$ is the wave vector, $i$ is the band index, $\omega$ is the phonon frequency, $k_B$ is the Boltzmann constant and $\hbar$ is the reduced Planck constant.

The Gibbs free energy $G$ is obtained by
\begin{equation}
    G(T, p) = \min_{V} \left[ F(T, V) + p V \right],
\end{equation}
where $p$ is the pressure.

The bulk modulus of the system $K$ can also be obtained as
\begin{equation}
    K(T) = V(T) \left. \frac{\partial^2 F(T, V)}{\partial V^2} \right\vert_T
\end{equation}
\subsection{Bulk Modulus Prediction}
To benchmark the prediction accuracy of bulk modulus and other thermodynamic properties (See \autoref{si-sec:enthalpy-prediction} and \autoref{si-sec:free-energy-prediction}) against first-principles results, we collected a wide range of ordered inorganic solids including inorganic elementary substances,  oxides, nitrides, carbides and a few half-Heusler compounds, whose phononic, mechanical, and transport properties have been studied using either experimental or first-principles methods.\cite{slack1958thermal,labotz1963thermal,martin1972thermal,takahashi1980porosity,gerlich1982temperature,moore1985thermal,morelli1995low,hohl1999efficient,young2000thermoelectric,inbook,popov2010thermal,mann2010hydrothermal,book,toberer2011phonon,lindsay2013phonon,xiao2013cubic,seko2015prediction,togo2015distributions, van2016high, campi2017first, skelton2017lattice,qian2019thermal,xia2020high,rakesh2021anomalous,zhu2021charting,ju2021exploring,tranaas2022lattice,cao2023anomalous}. To achieve consistency during benchmark, all of these materials are recomputed with first-principles method using the same setups under which we obtained the training set of MatterSim. The material whose first-principles QHA computations are converged are curated as a list and their Materials Project ids are shown in \autoref{si-tab:benchmark-materials}. Bulk moduli are computed at zero pressure and over a temperature range from \SI{0}{K} to \SI{1000}{K} with QHA implemented in Phonopy\cite{phonopy-phono3py-JPCM, phonopy-phono3py-JPSJ}, among which 59 materials were finished without error with PBE functional, MatterSim and MACE-MP-0 models. 
To perform a comparative analysis of the prediction and quantify the predictive accuracy of the bulk modulus, we employed the MAE of a bulk modulus curve as a metric,  compared with reference values obtained with PBE calculations. The MAE of a material for a model is defined as,
\begin{equation}
    \mathrm{MAE}_{\mathrm{K}} = \frac{1}{T_{\mathrm{max}}} \int_{0}^{T_{\mathrm{max}}} \left\vert \mathrm{K}_\mathrm{PBE}(T) - \mathrm{K}_\mathrm{ML}(T) \right\vert\,\mathrm{d}T,
\end{equation}
where $K$ is bulk modulus and $T_{\mathrm{max}}$ is \SI{1000}{K}.
As shown in \autoref{si-fig:bulk-modulus-histogram}, we present the distribution of the MAEs of MatterSim and MACE-MP-0 with respect to PBE calculations.
The average MAEs over 59 materials predicted by MatterSim and MACE-MP-0 are \SI{4.11}{\giga\pascal} and \SI{11.35}{\giga\pascal}, respectively. This suggests that under finite temperature conditions, MatterSim provides a more precise prediction of the bulk modulus, demonstrating its potential as a reliable tool in the prediction of mechanical properties under varying thermal environments.

\subsection{Enthalpy Prediction}\label{si-sec:enthalpy-prediction}
\begin{figure}
    \centering
    \begin{subfigure}[b]{0.45\textwidth}
    \includegraphics[width=\textwidth]{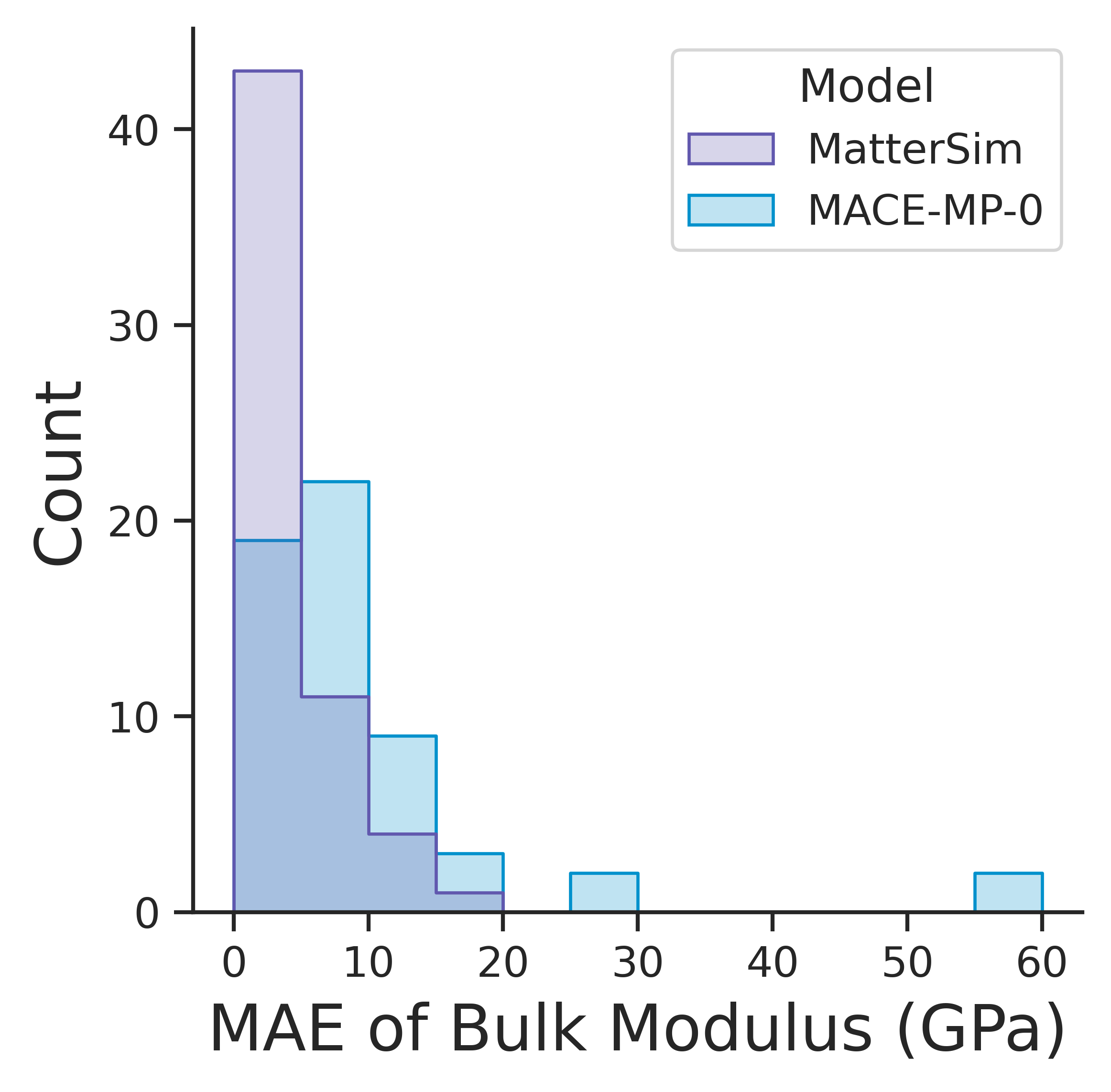}
    \caption{}
    \label{si-fig:bulk-modulus-histogram}
    \end{subfigure}
    \begin{subfigure}[b]{0.45\textwidth}
    \includegraphics[width=\textwidth]{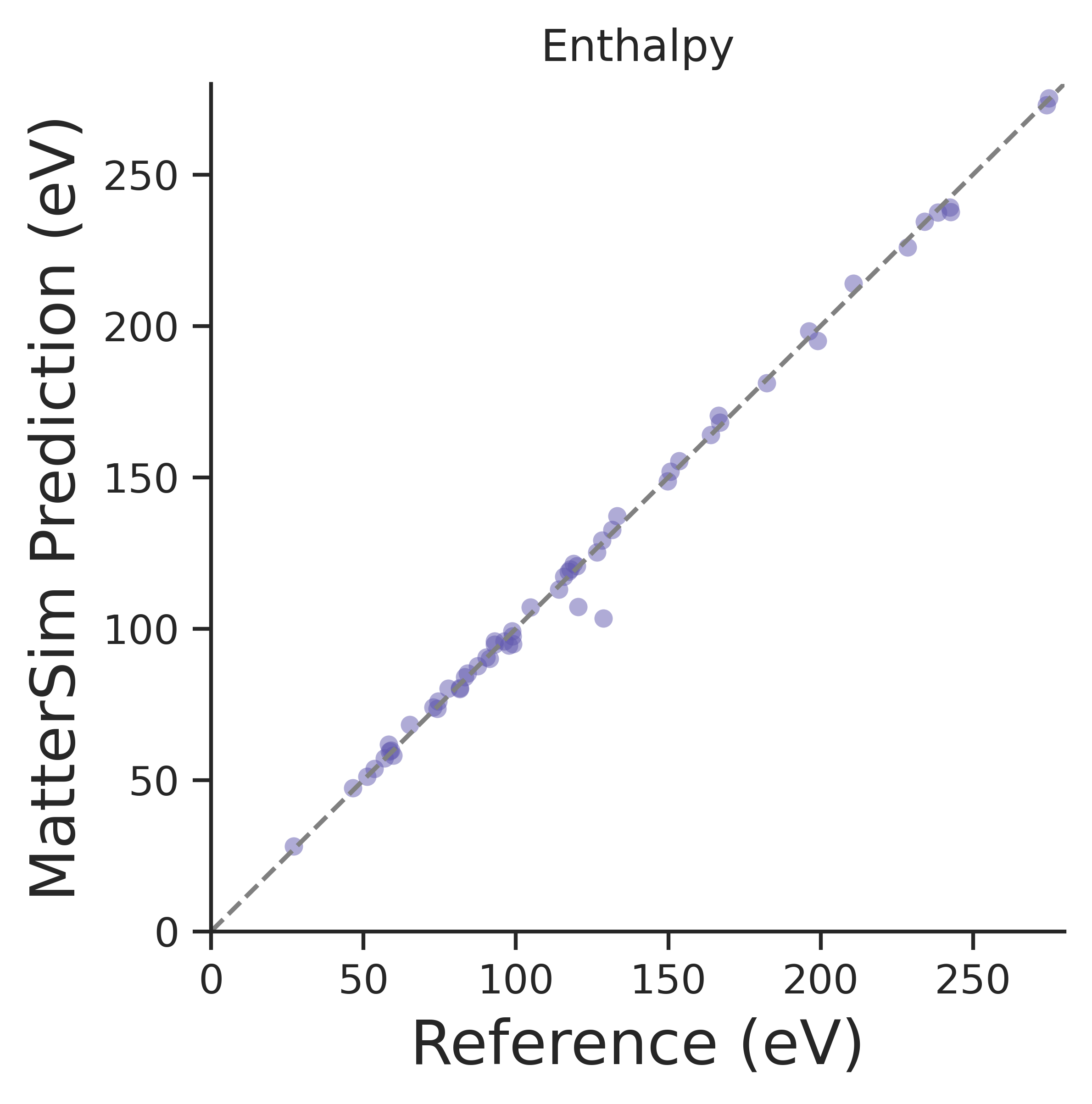}
    \caption{}
    \label{si-fig:enthalpy-parity-plot}
    \end{subfigure}
    \caption{(a) Distribution of bulk modulus's MAPE . (b) Parity plots of enthalpy at \SI{1000}{\giga\pascal}.}
\end{figure}

The enthalpy ($H$) under pressure $p$ is expressed as
\begin{equation}
    H(P) = U + p V,
\end{equation}
where $U$ is the internal energy and $V$ is the volume. The primitive cell size at a certain pressure is determined by MatterSim using volume changing relaxation in which the enthalpy of the system is minimized instead of the internal energy. The relaxed primitive cell was used to compute enthalpy using both MatterSim and PBE calculations. The pressure dependence of enthalpy of 59 materials are computed by MatterSim and compared with PBE results as shown in \autoref{si-fig:enthalpy-parity-plot}. The accuracy of our model was rigorously evaluated by comparing the MatterSim-computed enthalpies with the PBE-computed enthalpies at \SI{1000}{\giga\pascal}. This comparison yielded an MAE of \SI{2.23}{eV}, indicating the average deviation of our model's predictions from the PBE values was minimal. Furthermore, our model demonstrated excellent predictive capabilities, as evidenced by achieving an R$^2$ score of $1.00$. As evidenced by the low MAE and the perfect R$^2$ score, MatterSim's predictive performance highlight its capability to accurately predict the stability of materials under high-pressure conditions, underscoring the potential of MatterSim as a robust tool for investigating the thermodynamic stability of materials.

\begin{table}
\centering
\caption{Summary of material candidates and their corresponding ID in Materials Project used for the prediction of bulk modulus, enthalpy and Gibbs free energy with MatterSim.}
\begin{tabular}{cccc}
\toprule
Materials & mp-id & Materials & mp-id \\
\hline
\ce{C} & \href{https://next-gen.materialsproject.org/materials/mp-66/}{mp-66} & \ce{ZrNiSn} & \href{https://next-gen.materialsproject.org/materials/mp-924129}{mp-924129} \\
\ce{h-AlN} & \href{https://next-gen.materialsproject.org/materials/mp-661}{mp-661} & \ce{CaSe} & \href{https://next-gen.materialsproject.org/materials/mp-1415/}{mp-1415} \\
\ce{BAs} & \href{https://next-gen.materialsproject.org/materials/mp-10044/}{mp-10044} & \ce{SrSe} & \href{https://next-gen.materialsproject.org/materials/mp-2758/}{mp-2758} \\
\ce{h-ZnTe} & \href{https://next-gen.materialsproject.org/materials/mp-8884/}{mp-8884} & \ce{CdS} & \href{https://next-gen.materialsproject.org/materials/mp-2469/}{mp-2469} \\
\ce{GeC} & \href{https://next-gen.materialsproject.org/materials/mp-1002164/}{mp-1002164} & \ce{h-AlAs} & \href{https://next-gen.materialsproject.org/materials/mp-8881/}{mp-8881} \\
\ce{BSb} & \href{https://next-gen.materialsproject.org/materials/mp-997618/}{mp-997618} & \ce{SiC} & \href{https://next-gen.materialsproject.org/materials/mp-8062/}{mp-8062} \\
\ce{h-CdSe} & \href{https://next-gen.materialsproject.org/materials/mp-1070/}{mp-1070} & \ce{h-SiC} & \href{https://next-gen.materialsproject.org/materials/mp-7140/}{mp-7140} \\
\ce{MgS} & \href{https://next-gen.materialsproject.org/materials/mp-1315/}{mp-1315} & \ce{Mg2Si} & \href{https://next-gen.materialsproject.org/materials/mp-1367/}{mp-1367} \\
\ce{BP} & \href{https://next-gen.materialsproject.org/materials/mp-1479/}{mp-1479} & \ce{GaN} & \href{https://next-gen.materialsproject.org/materials/mp-830/}{mp-830} \\
\ce{h-GaAs} & \href{https://next-gen.materialsproject.org/materials/mp-8883/}{mp-8883} & \ce{SrO} & \href{https://next-gen.materialsproject.org/materials/mp-2472/}{mp-2472} \\
\ce{AlN} & \href{https://next-gen.materialsproject.org/materials/mp-1700/}{mp-1700} & \ce{CaTe} & \href{https://next-gen.materialsproject.org/materials/mp-1519/}{mp-1519} \\
\ce{h-GaN} & \href{https://next-gen.materialsproject.org/materials/mp-804/}{mp-804} & \ce{MgSe} & \href{https://next-gen.materialsproject.org/materials/mp-10760/}{mp-10760} \\
\ce{GaP} & \href{https://next-gen.materialsproject.org/materials/mp-2490/}{mp-2490} & \ce{BeTe} & \href{https://next-gen.materialsproject.org/materials/mp-252/}{mp-252} \\
\ce{BeSe} & \href{https://next-gen.materialsproject.org/materials/mp-1541/}{mp-1541} & \ce{SrS} & \href{https://next-gen.materialsproject.org/materials/mp-1087/}{mp-1087} \\
\ce{Si} & \href{https://next-gen.materialsproject.org/materials/mp-149/}{mp-149} & \ce{CaS} & \href{https://next-gen.materialsproject.org/materials/mp-1672/}{mp-1672} \\
\ce{InN} & \href{https://next-gen.materialsproject.org/materials/mp-20411/}{mp-20411} & \ce{ZnS} & \href{https://next-gen.materialsproject.org/materials/mp-10695/}{mp-10695} \\
\ce{AlSb} & \href{https://next-gen.materialsproject.org/materials/mp-2624/}{mp-2624} & \ce{h-CdTe} & \href{https://next-gen.materialsproject.org/materials/mp-12779/}{mp-12779} \\
\ce{h-GaP} & \href{https://next-gen.materialsproject.org/materials/mp-8882/}{mp-8882} & \ce{Mg2Ge} & \href{https://next-gen.materialsproject.org/materials/mp-408/}{mp-408} \\
\ce{AlP} & \href{https://next-gen.materialsproject.org/materials/mp-1550/}{mp-1550} & \ce{ZnO} & \href{https://next-gen.materialsproject.org/materials/mp-1986/}{mp-1986} \\
\ce{h-AlP} & \href{https://next-gen.materialsproject.org/materials/mp-8880/}{mp-8880} & \ce{MgO} & \href{https://next-gen.materialsproject.org/materials/mp-1265/}{mp-1265} \\
\ce{h-ZnO} & \href{https://next-gen.materialsproject.org/materials/mp-2133/}{mp-2133} & \ce{MgTe} & \href{https://next-gen.materialsproject.org/materials/mp-13033/}{mp-13033} \\
\ce{ZnSe} & \href{https://next-gen.materialsproject.org/materials/mp-1190/}{mp-1190} & \ce{MgSe} & \href{https://next-gen.materialsproject.org/materials/mp-13031/}{mp-13031} \\
\ce{GaAs} & \href{https://next-gen.materialsproject.org/materials/mp-2534/}{mp-2534} & \ce{BaS} & \href{https://next-gen.materialsproject.org/materials/mp-1500/}{mp-1500} \\
\ce{h-ZnSe} & \href{https://next-gen.materialsproject.org/materials/mp-380/}{mp-380} & \ce{TiCoSb} & \href{https://next-gen.materialsproject.org/materials/mp-5967/}{mp-5967} \\
\ce{h-CdS} & \href{https://next-gen.materialsproject.org/materials/mp-672/}{mp-672} & \ce{TiNiSn} & \href{https://next-gen.materialsproject.org/materials/mp-924130/}{mp-924130} \\
\ce{h-MgTe} & \href{https://next-gen.materialsproject.org/materials/mp-1039/}{mp-1039} & \ce{AlAs} & \href{https://next-gen.materialsproject.org/materials/mp-2172}{mp-2172} \\
\ce{h-InSb} & \href{https://next-gen.materialsproject.org/materials/mp-1007661}{mp-1007661} & \ce{h-GaSb} & \href{https://next-gen.materialsproject.org/materials/mp-1018059}{mp-1018059} \\
\ce{h-AlSb} & \href{https://next-gen.materialsproject.org/materials/mp-1018100}{mp-1018100} & \ce{h-InN} & \href{https://next-gen.materialsproject.org/materials/mp-22205}{mp-22205} \\
\ce{BeS} & \href{https://next-gen.materialsproject.org/materials/mp-422}{mp-422} & \ce{h-ZnS} & \href{https://next-gen.materialsproject.org/materials/mp-560588}{mp-560588} \\
\ce{GaSb} & \href{https://next-gen.materialsproject.org/materials/mp-1156/}{mp-1156} & & \\
\bottomrule
\end{tabular}\label{si-tab:benchmark-materials}
\end{table}

\section{Free energy and phase diagram computation}\label{si-sec:phase_diagram_computation}

\subsection{Gibbs free energy prediction}\label{si-sec:free-energy-prediction}
The Gibbs free energy of ordered crystalline materials are computed using MatterSim via quasi-harmonic approximation (QHA) implemented in Phonopy as described in \autoref{si-sec:mechanical-properties}.
We benchmark the free energy predictions made by MatterSim to both first-principles calculations of the dataset collected in \autoref{si-sec:mechanical-properties} and experimental measurements from \texttt{FactSage} released in Ref.~\citenum{bartel2018physical}. 

\begin{figure}
    \centering
    \begin{subfigure}[b]{0.45\textwidth}
    \includegraphics[width=\textwidth]{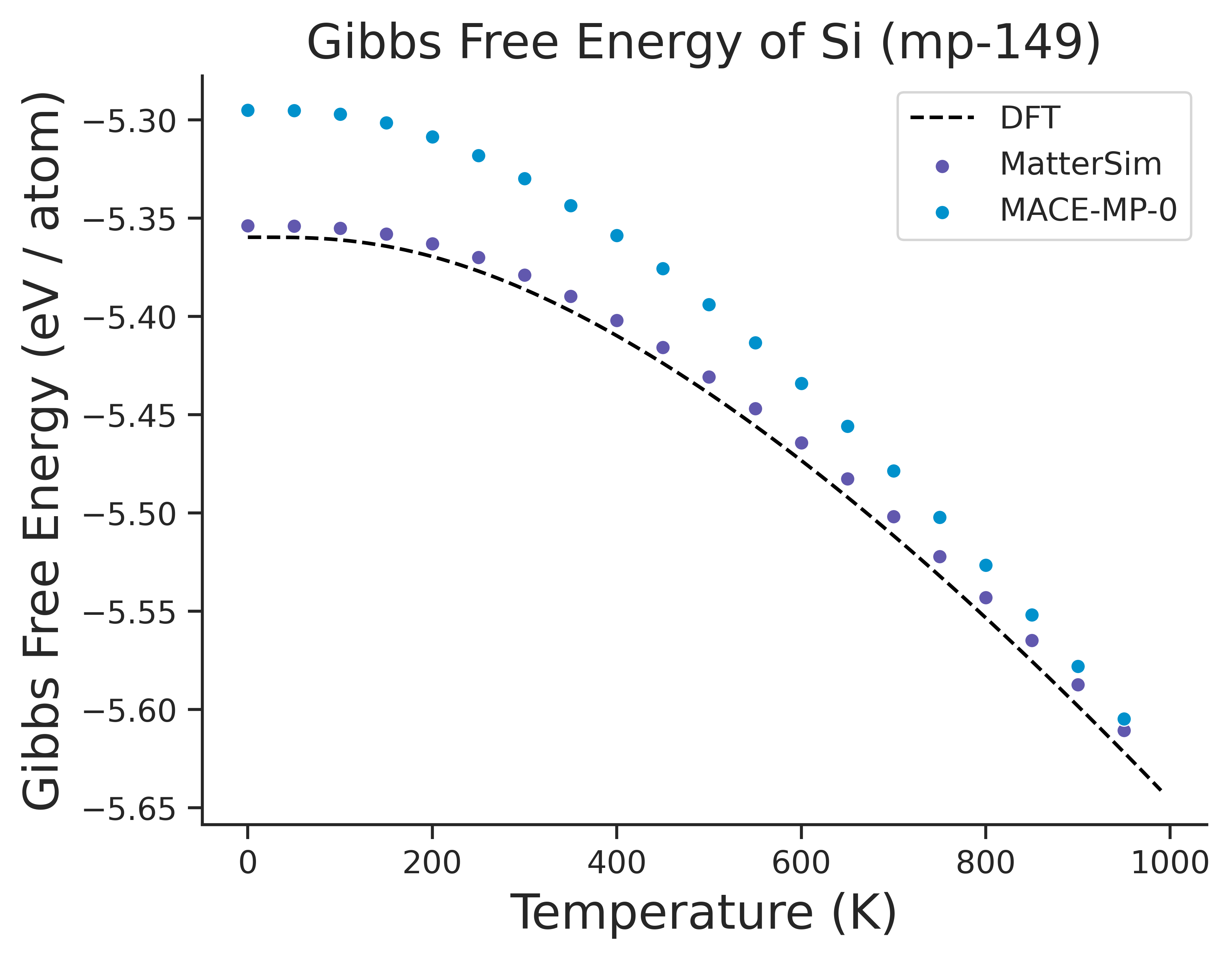}
    \caption{}
    \label{si-fig:free-energy-Si}
    \end{subfigure}
    \begin{subfigure}[b]{0.45\textwidth}
    \includegraphics[width=\textwidth]{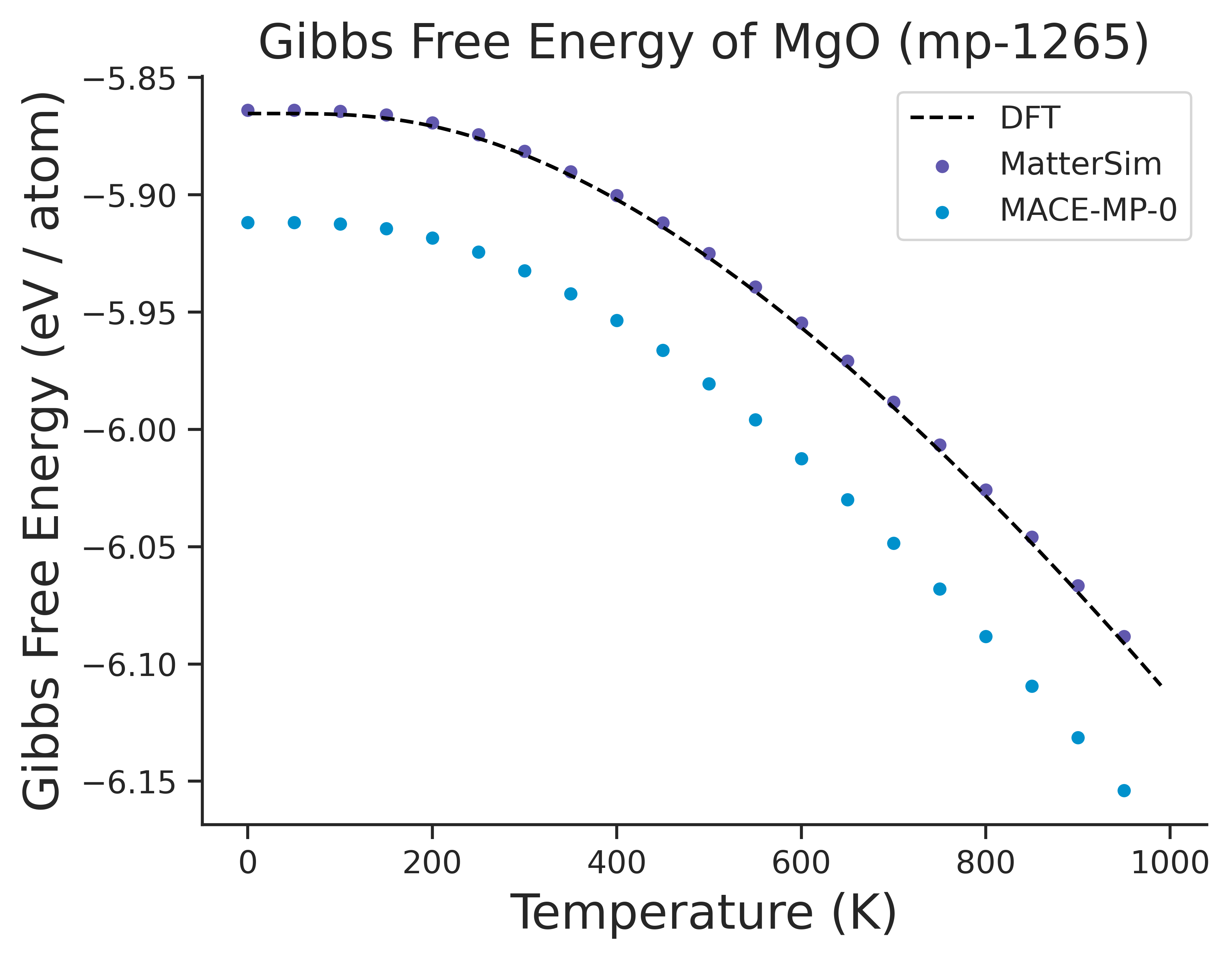}
    \caption{}
    \label{si-fig:free-energy-MgO}
    \end{subfigure}
    \begin{subfigure}[b]{0.45\textwidth}
    \includegraphics[width=\textwidth]{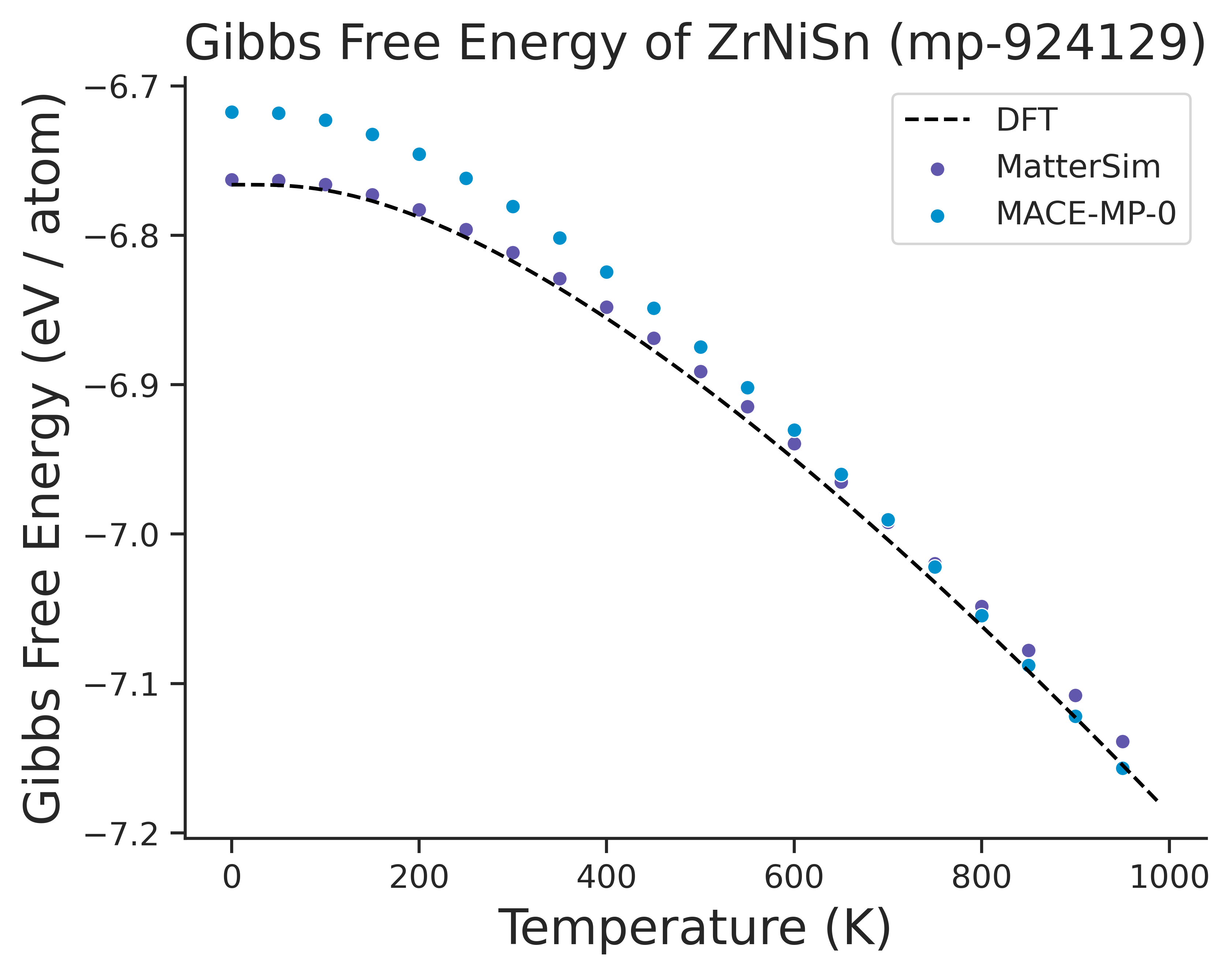}
    \caption{}
    \label{si-fig:free-energy-ZrNiSn}
    \end{subfigure}
    \begin{subfigure}[b]{0.45\textwidth}
    \includegraphics[width=\textwidth]{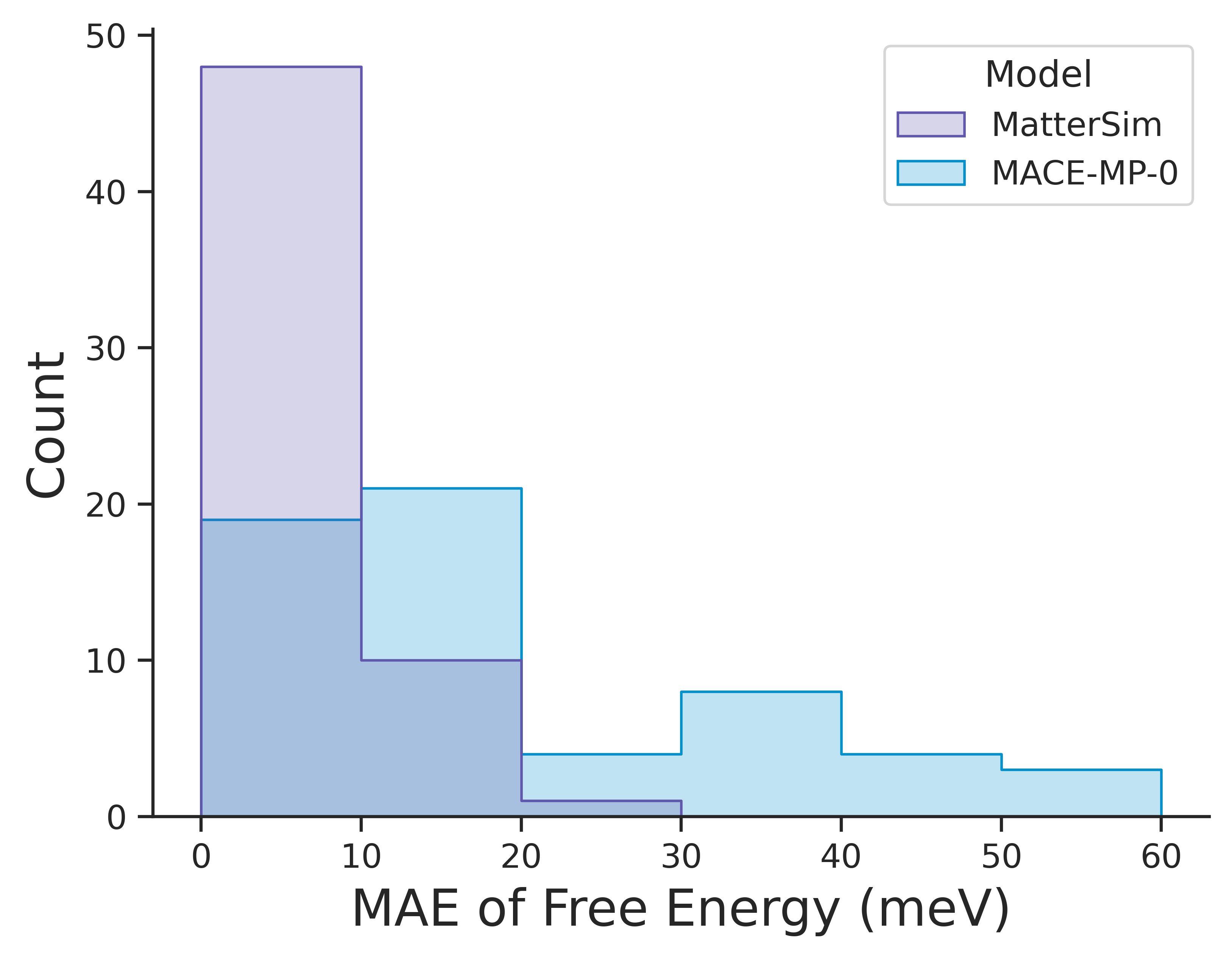}
    \caption{}
    \label{si-fig:free-energy-MAE-hist}
    \end{subfigure}    
    \caption{Comparative analysis of free energy of (a)\ce{Si}, (b)\ce{MgO} and (c)\ce{ZrNiSn} from MatterSim, MACE-MP-0, and PBE calculations. (d) Distribution of Gibbs free energy's MAE}
    \label{si-fig:free-energy-examples-and-histogram}
\end{figure}

The free energies for the set of 59 materials over a temperature range from 0 K to 1000 K at \SI{0}{\giga\pascal} are calculated with MatterSim and are compared with the PBE calculations. We present the examples of \ce{Si}, \ce{MgO} and \ce{ZrNiSn} in \autoref{si-fig:free-energy-examples-and-histogram} and a parity plot of the prediction for the 59 materials in \autoref{fig:phonon-related-properties}(e).
The overall performance is quantified with  mean absolute error of the Gibbs free energy over the 0--\SI{1000}{\kelvin} temperature range, which is defined as
\begin{equation}
    \mathrm{MAE}_{\mathrm{G}} = \frac{1}{t_{\mathrm{max}}} \int_{0}^{t_{\mathrm{max}}} \left\vert \mathrm{G}_\mathrm{PBE}(t) - \mathrm{G}_\mathrm{ML}(t) \right\vert\,\mathrm{d}t,
\end{equation}
where $G$ is Gibbs free energy and $t_{\mathrm{max}}$ is \SI{1000}{K}. We report the distribution of the MAE of Gibbs free energy in \autoref{si-fig:free-energy-MAE-hist} with the results from MACE-MP-0, a model trained on relaxation trajectories of crystals. While MACE-MP-0 already achieves remarkable robustness and universality, MatterSim's predictions are in quantitative consistency with the PBE reference data with an MAE of Gibbs free energy for all the 59 materials being \SI{6.51}{meV}, underscoring the model's accuracy and reliability.

\begin{figure}
    \centering
    \begin{subfigure}{0.45\textwidth}
    \includegraphics[width=\textwidth]{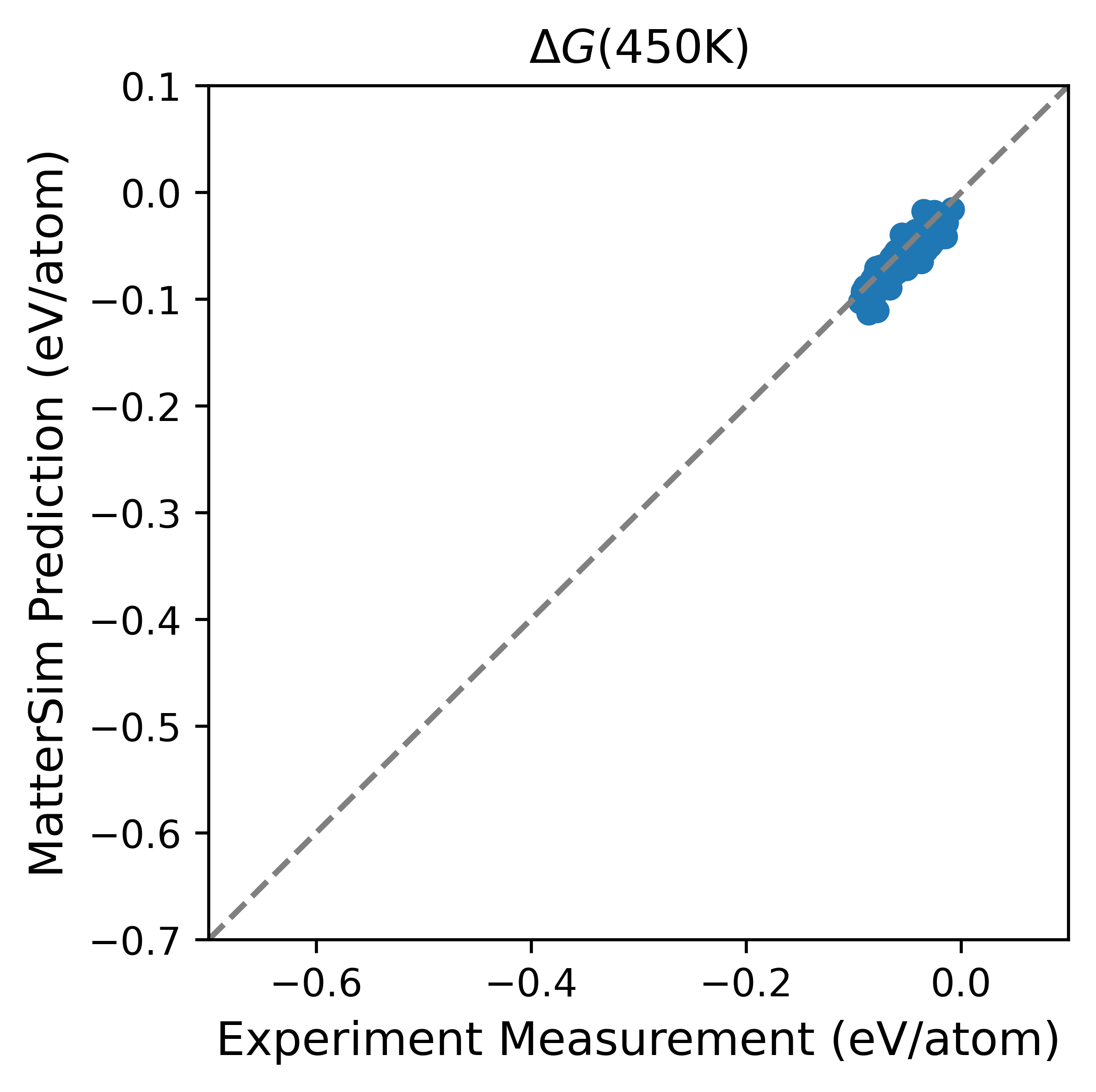}
    \caption{}
    \label{si-fig:free-energy-comparison-vs-expt-450}
    \end{subfigure}
    \begin{subfigure}{0.45\textwidth}
    \includegraphics[width=\textwidth]{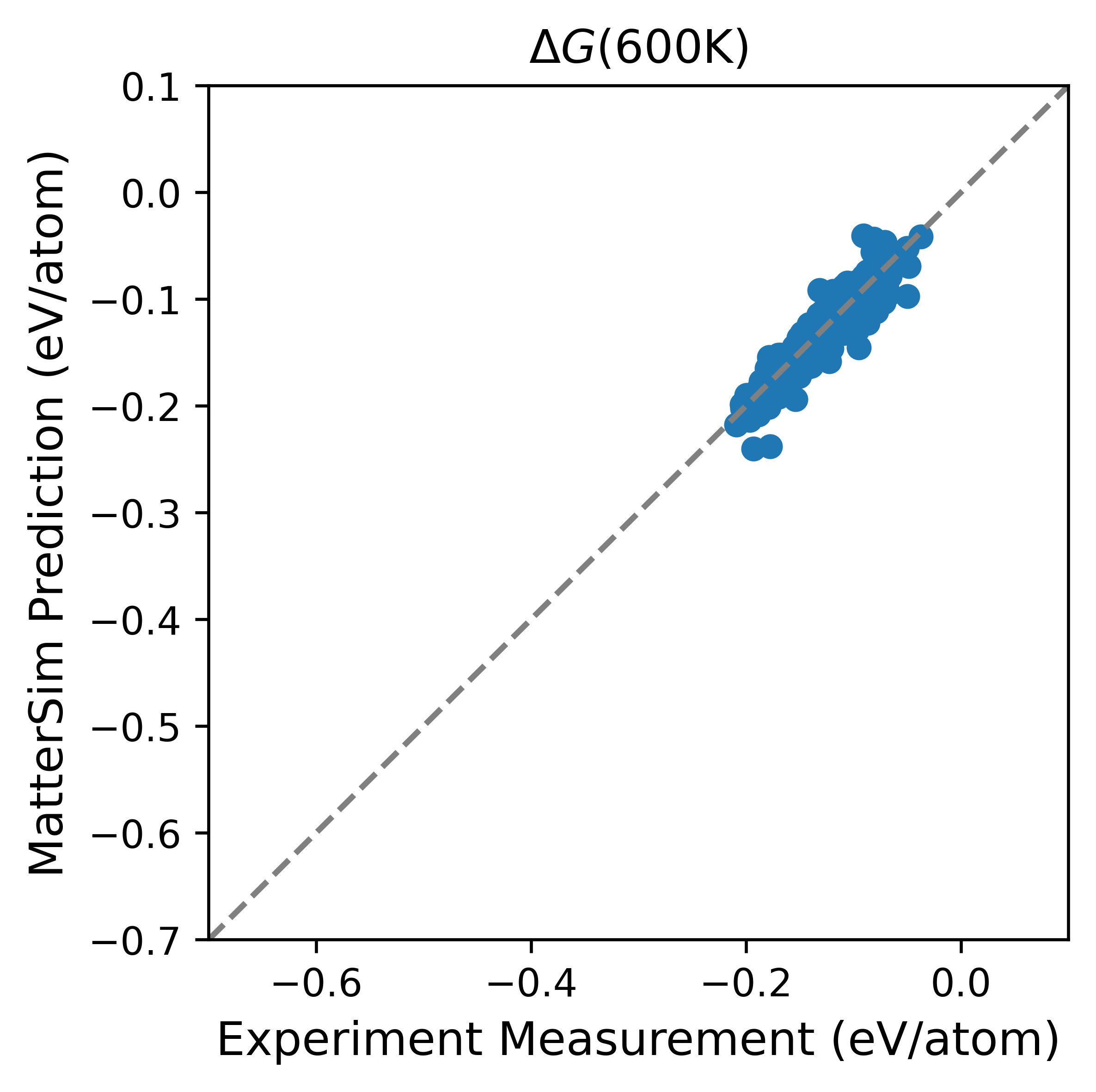}
    \caption{}
    \label{si-fig:free-energy-comparison-vs-expt-600}
    \end{subfigure}
    \begin{subfigure}{0.45\textwidth}
    \includegraphics[width=\textwidth]{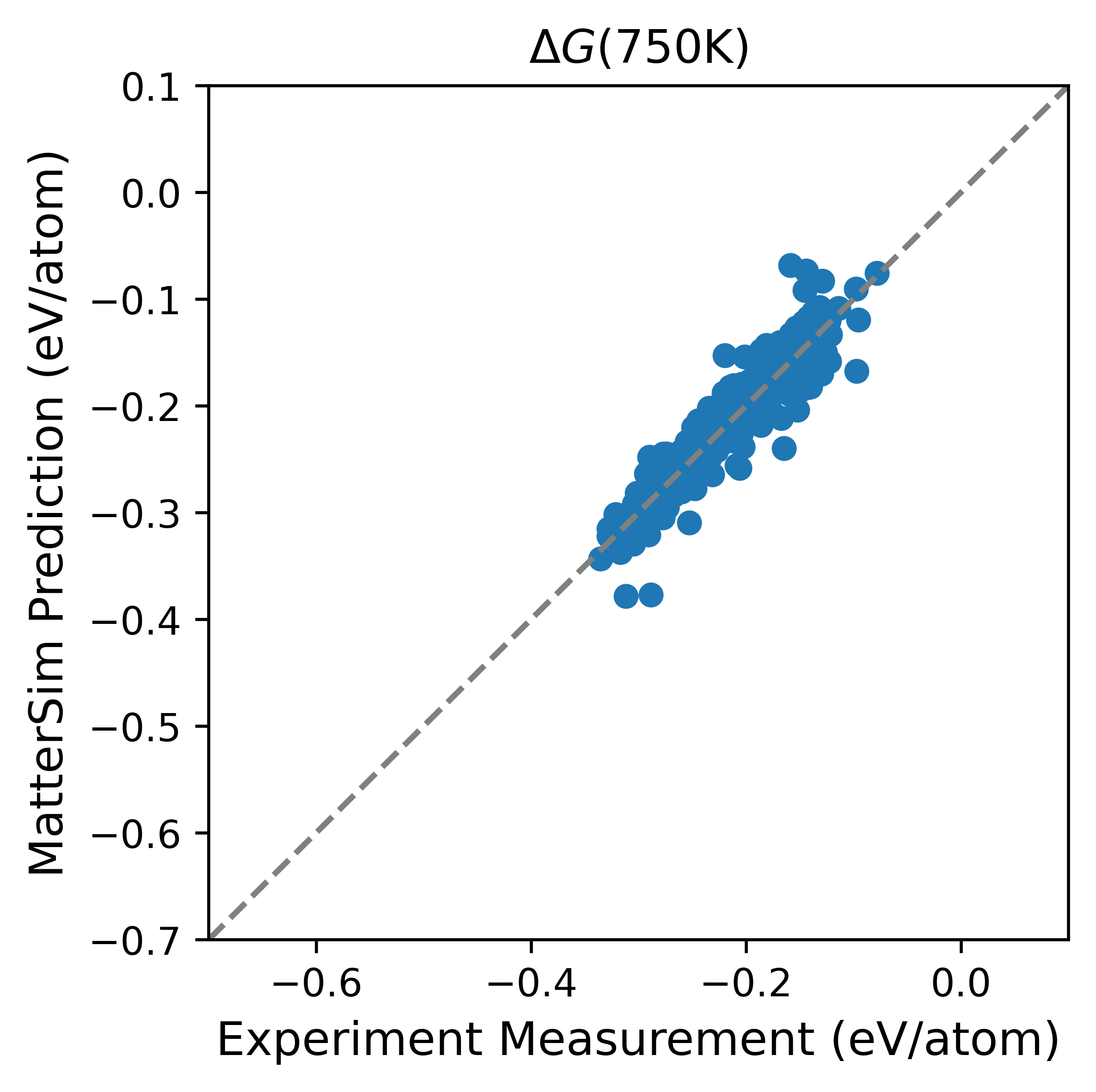}
    \caption{}
    \label{si-fig:free-energy-comparison-vs-expt-750}
    \end{subfigure}
    \begin{subfigure}{0.45\textwidth}
    \includegraphics[width=\textwidth]{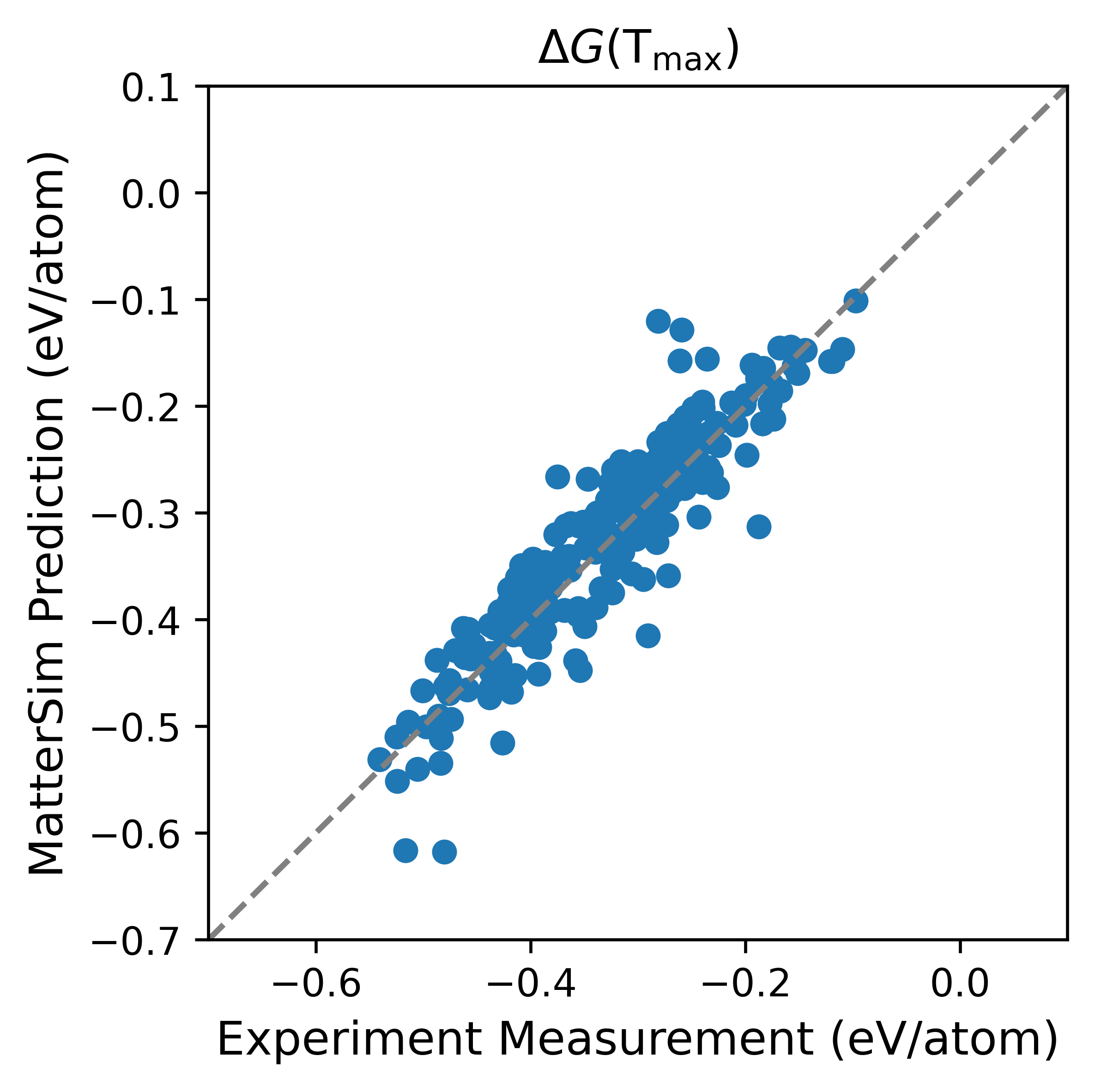}
    \caption{}
    \label{si-fig:free-energy-comparison-vs-expt-tmax}
    \end{subfigure}
    \caption{Parity plots of the predicted $\Delta_\mathrm{MatterSim}G(T)$ and the reference experimental values $\Delta_\mathrm{ref}G(T)$ at 450, 600, \SI{750}{\kelvin}, and $T_\mathrm{max}$, respectively. }
    \label{si-fig:free-energy-comparison-vs-expt-overview}
\end{figure}

We then benchmark the quantitative prediction capability of MatterSim on free energy to experimental measurements. Ref.~\citenum{bartel2018physical} reported the analytical form of the experimental Gibbs free energy of the materials collected from \texttt{FactSage} dataset\cite{bale2016reprint} based feature selection using Sure-Independence Screening and Sparsifying Operator (SISSO) method\cite{ouyang2018sisso, purcell2023recent, bartel2018physical}:
\begin{equation}
    G_{\mathrm{SISSO}}^{\delta}(T)\left[\frac{\mathrm{eV}}{\mathrm{atom}}\right] = \left(-2.48 \times 10^{-4} \, \ln V - 8.94 \times 10^{-5} \times \frac{m}{V} \right) T + 0.181 \times \ln T - 0.882,
\end{equation}
where $V$, $T$ and $m$ are the volume of unit cell, temperature and the mass for each material defined by Ref.~\citenum{bartel2018physical}. Using the Gibbs free energy at \SI{300}{\kelvin} as the common reference point, the Gibbs free energy difference between at given temperatures can thus be inferred from this analytical form by
\begin{equation}
    \Delta_\mathrm{ref} G(T) = G_\mathrm{SISSO}^\delta(T)-G_\mathrm{SISSO}^\delta(300\,\mathrm{K}),
\end{equation}
which will be used as our reference experimental value. The Gibbs free energy difference can also be predicted with MattterSim,
\begin{equation}
    \Delta_\mathrm{MatterSim} G(T) = G_\mathrm{MatterSim}(T) - G_\mathrm{MatterSim}(300\,\mathrm{K}).
\end{equation}
In \autoref{si-fig:free-energy-comparison-vs-expt-overview}, we reported the comparison between MatterSim's prediction of Gibbs free energy at \SI{450}{\kelvin}, \SI{600}{\kelvin}, \SI{750}{\kelvin} and $T_\mathrm{max}$ to the experimental values inferred from the analytical form defined in Ref.~\citenum{bartel2018physical}, where $T_\mathrm{max}$ is the highest temperature for each material release in Ref.~\citenum{bartel2018physical}. At each temperature, the MAEs between MatterSim' prediction and the reference experimental values are 7.1, 11.4, 18.0 and  \SI{28.9}{meV\per atom}, respectively. We also conduct an analysis of the MAEs over the entire temperature range from \SI{300}{\kelvin} to $T_\mathrm{max}$ for each material by integrating their prediction error,
\begin{equation}
    \mathrm{MAE}_{\mathrm{G}} = \frac{1}{T_{\mathrm{max}}-300} \int_{300}^{T_{\mathrm{max}}} \left\vert \Delta_\mathrm{MatterSim} G(t) - \Delta_\mathrm{ref} G(t) \right\vert\,\mathrm{d}t.
\end{equation}
The MAE of Gibbs free energy over the 300--\SI{1000}{\kelvin} temperature range is \SI{15}{meV\per atom}, outperforming  dedicated model trained explicitly on experimental data (MAE of \SI{50}{meV\per atom})  ~\citenum{bartel2018physical}. This underscores the potential of greatly improved accuracy and generaliazability with  machine learning models trained on large-scale materials data supervised by fundamental materials properties generated from first-principles approach.

\subsection{Phase diagram prediction}

\begin{figure} \label{si-fig:Si-free-energy-phase-diagram}
    \centering
    \begin{subfigure}[b]{0.45\textwidth}
    \includegraphics[width=\textwidth]{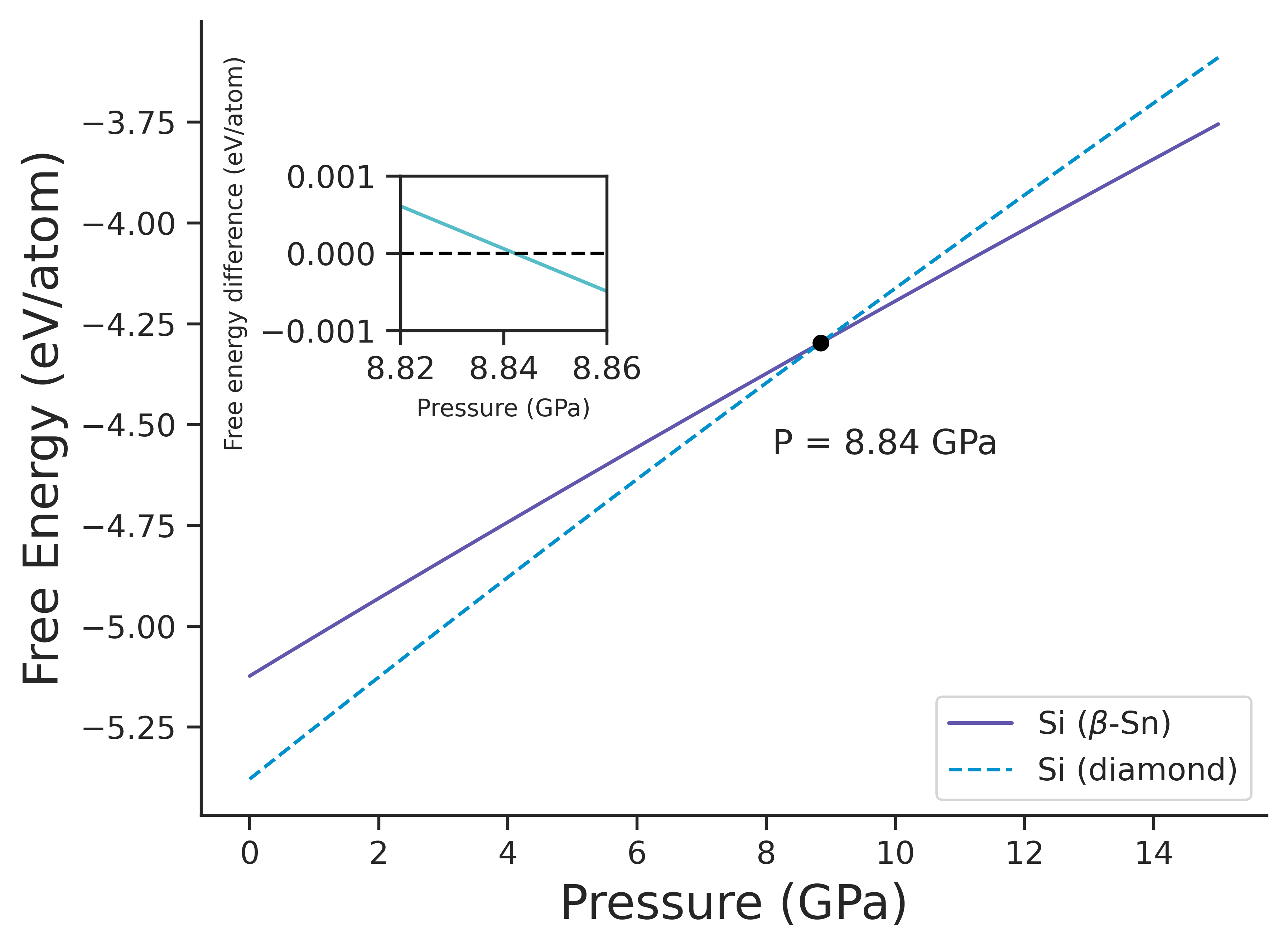}
    \caption{}
    \label{si-fig:Si-free-energy}
    \end{subfigure}
    \begin{subfigure}[b]{0.45\textwidth}
    \includegraphics[width=\textwidth]{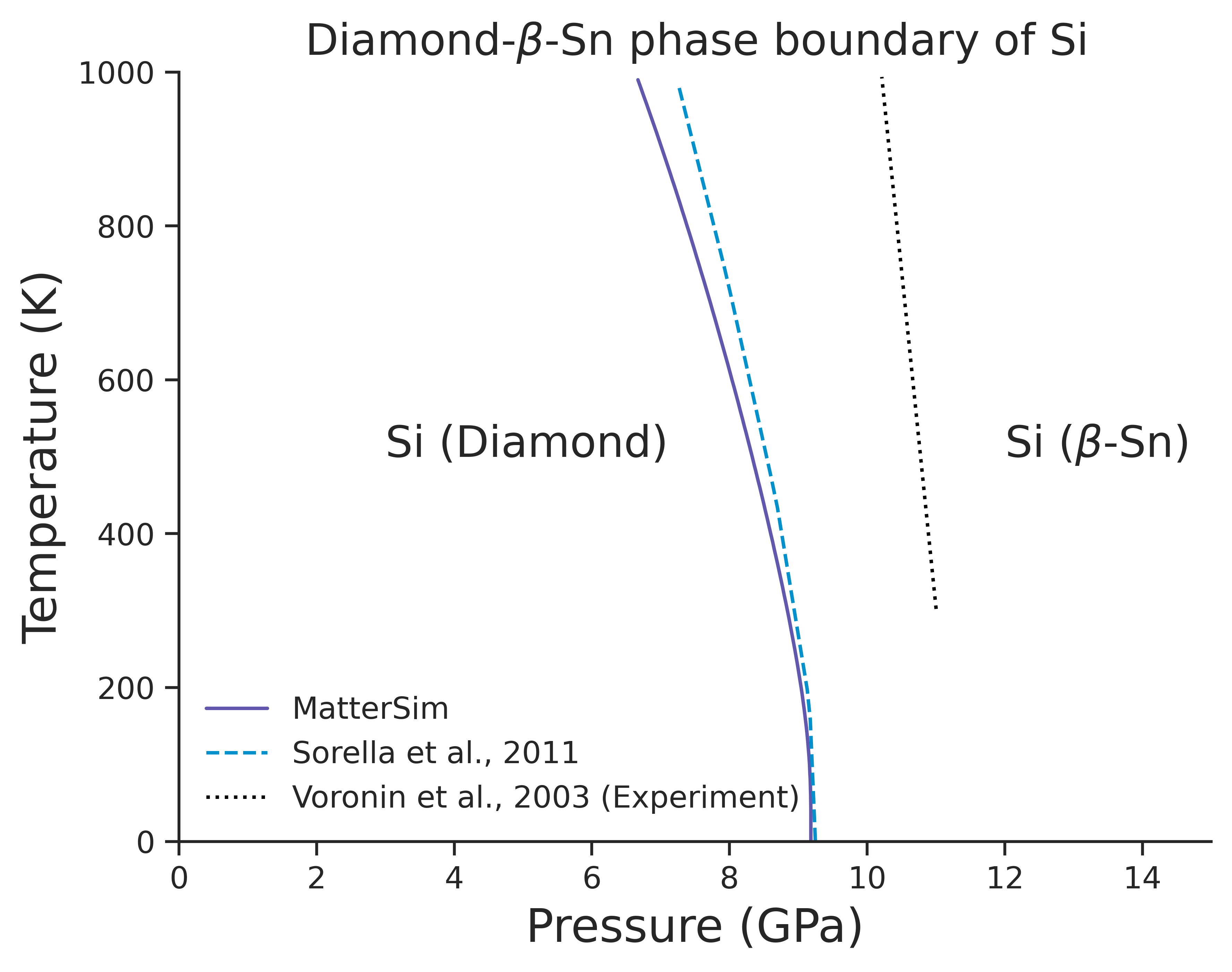}
    \caption{}
    \label{si-fig:Si-phase-diagram}
    \end{subfigure}
    \caption{\textbf{(a)} Pressure dependent Gibbs free energy of the $\beta$-Sn and diamond phases of Si under 300 K; \textbf{(b)} Phase diagrams of Si. Blue solid line: Calculation by MatterSim. Black dashed line: Theoretical calculation by Sorella et al.\cite{sorella2011ab} Black dotted line: Experiment by Voronin et al.\cite{voronin2003situ}}
\end{figure}

We calculate silicon's Gibbs free energy with QHA for the two competing phases, the $\beta$-Sn and the diamond phases, and construct their phase diagram. The results from MatterSim are directly compared with established theoretical predictions\cite{sorella2011ab} and experimental measurements\cite{voronin2003situ}. \autoref{si-fig:Si-free-energy} presents the pressure-dependent Gibbs free energy of silicon in both the $\beta$-Sn and diamond phases, as calculated by MatterSim at 300K. The figure marks the point where the free energies of the two phases intersect, indicating a phase transition. According to our calculations, this transition occurs at a pressure of \SI{8.84}{GPa}. This value demonstrates a remarkable agreement with the theoretical transition pressure of \SI{8.99}{GPa}, substantiating the reliability of MatterSim's prediction power. Further insights are provided by \autoref{si-fig:Si-phase-diagram} that displays the phase diagram of Si, wherein the phase boundary calculated by MatterSim is compared with that obtained from PBE calculation. This comparison reveals an excellent alignment between the phase boundaries derived from both MatterSim and PBE calculations, thereby validating the accuracy of MatterSim's prediction in a wide range of pressure and temperature conditions. While our computational results align very well with first-principles predictions, we still observe that the temperature-dependent phase transition pressures are slightly underestimated in comparison to experimental data, and the possible reason could be the inaccuracy of PBE functional used to generate the training data of our model.

\section{Molecular dynamics simulations} \label{sec:si_md}\label{si-sec:md}

\begin{figure}
    \centering

    \begin{subfigure}[b]{0.3\textwidth}
    \includegraphics[width=\textwidth]{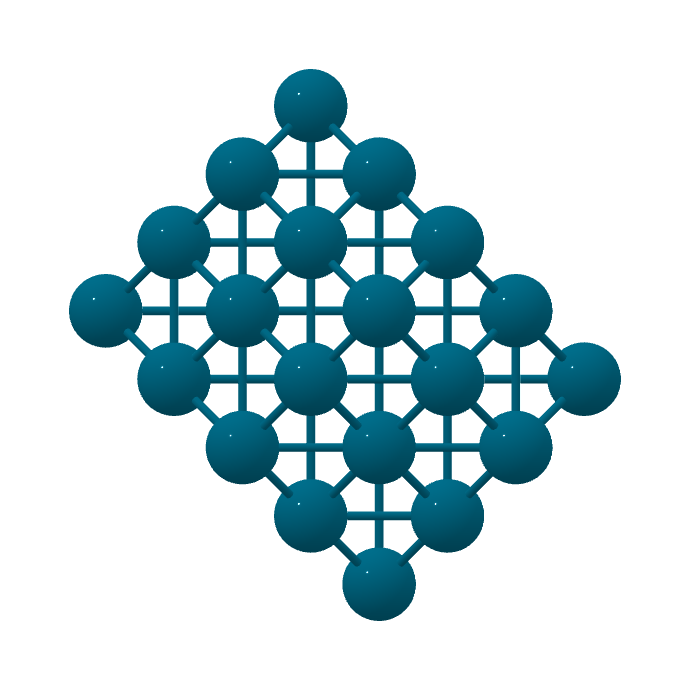}
    \caption{\ce{Pd} (Bulk)}
    \end{subfigure}
    \hfill
    \begin{subfigure}[b]{0.3\textwidth}
    \includegraphics[width=\textwidth]{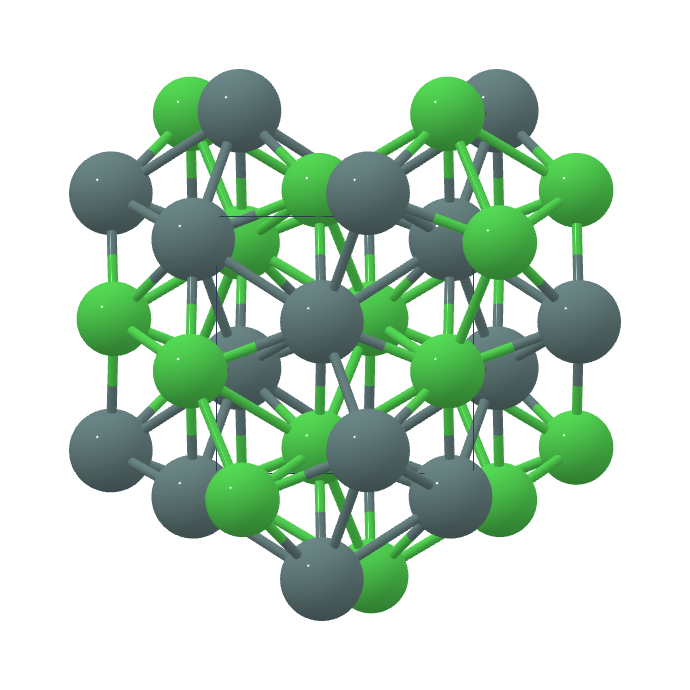}
    \caption{\ce{Ni4Sn4} (Bulk)}
    \end{subfigure}
    \hfill
    \begin{subfigure}[b]{0.3\textwidth}
    \includegraphics[width=\textwidth]{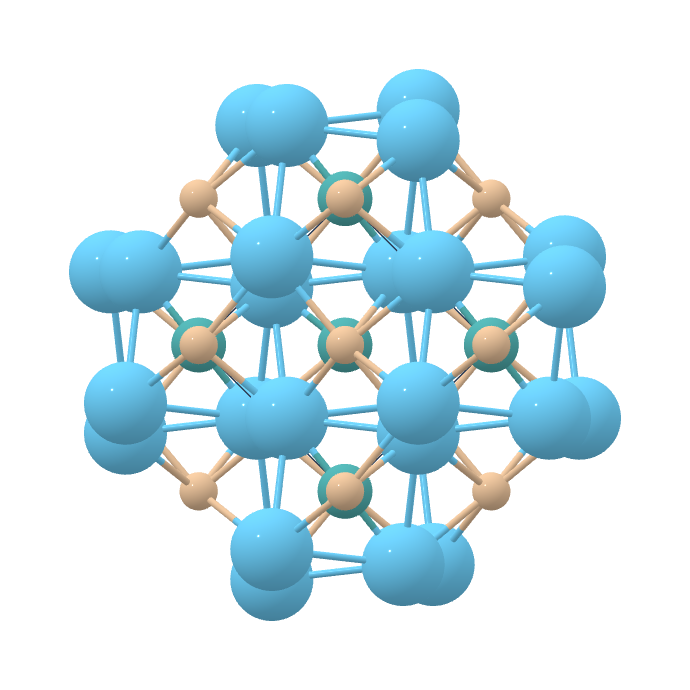}
    \caption{\ce{La4Mo2Si2} (Bulk)}
    \end{subfigure}
    \begin{subfigure}[b]{0.3\textwidth}
    \includegraphics[width=\textwidth]{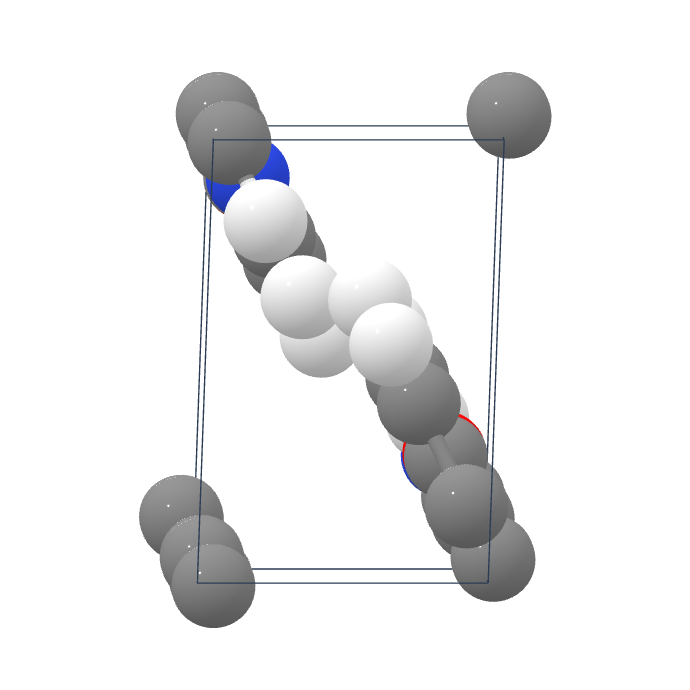}
    \caption{ABPBO (Polymer)}
    \end{subfigure}
    \hfill
    \begin{subfigure}[b]{0.3\textwidth}
    \includegraphics[width=\textwidth]{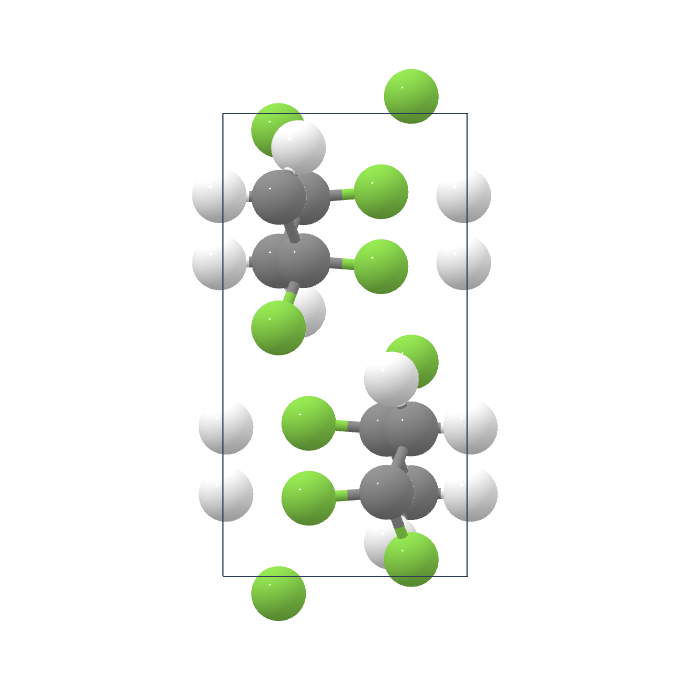}
    \caption{$\alpha$-PVDF (Polymer)}
    \end{subfigure}
    \hfill
    \begin{subfigure}[b]{0.3\textwidth}
    \includegraphics[width=\textwidth]{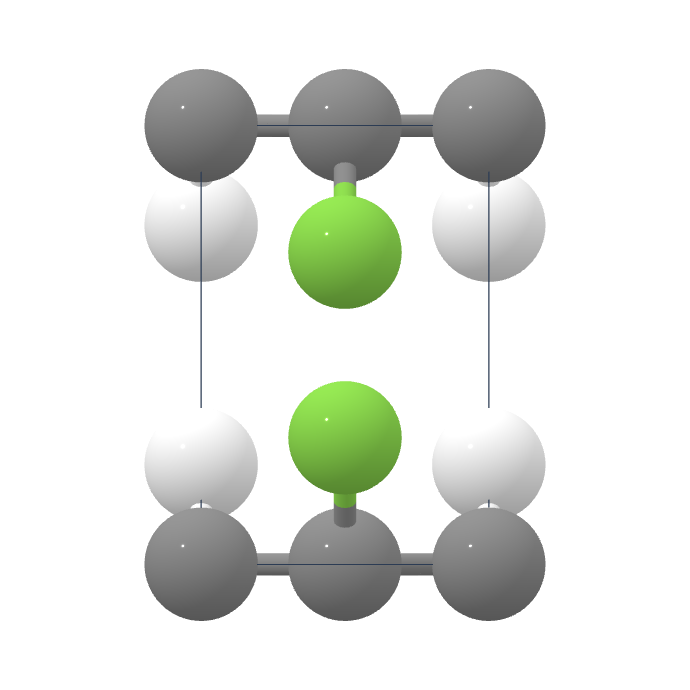}
    \caption{$\beta$-PVDF (Polymer)}
    \end{subfigure}
    \begin{subfigure}[b]{0.3\textwidth}
    \includegraphics[width=\textwidth]{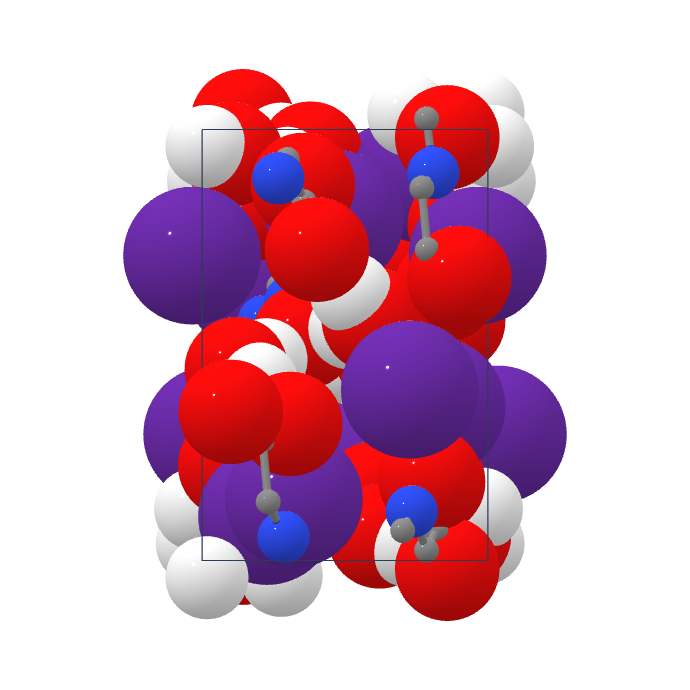}
    \caption{\ce{SnH10C6(BrN)2} (MOF)}
    \end{subfigure}
    \hfill
    \begin{subfigure}[b]{0.3\textwidth}
    \includegraphics[width=\textwidth]{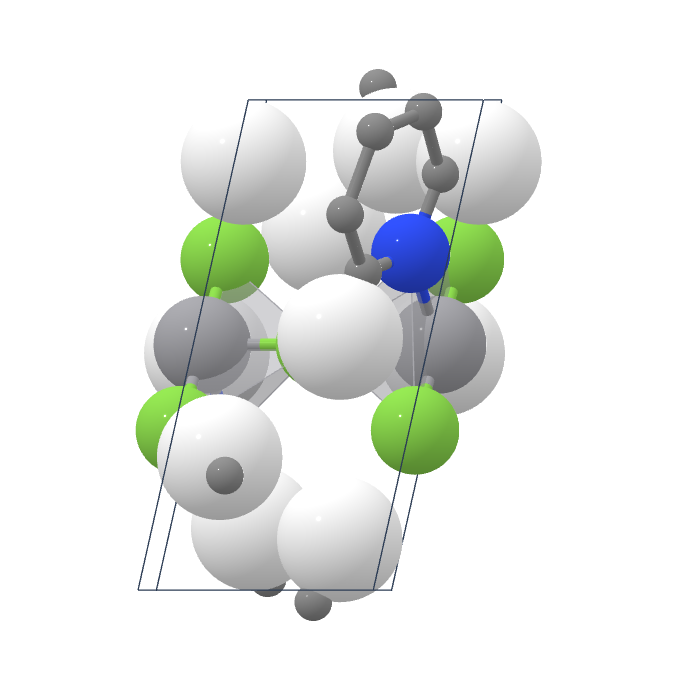}
    \caption{\ce{Zr3H62(C9O2)8} (MOF)}
    \end{subfigure}
    \hfill
    \begin{subfigure}[b]{0.3\textwidth}
    \includegraphics[width=\textwidth]{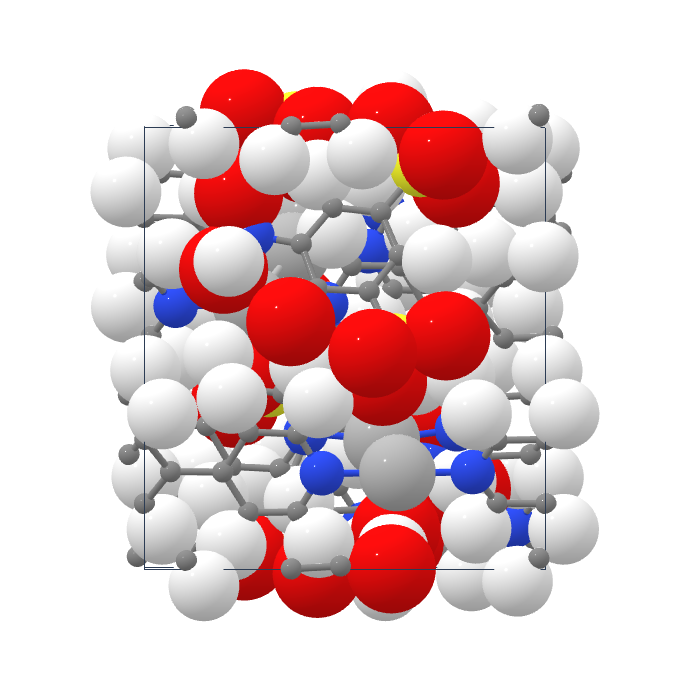}
    \caption{\ce{Mn3H56C66(N4O11)2} (MOF)}
    \end{subfigure}
    \begin{subfigure}[b]{0.3\textwidth}
    \includegraphics[width=\textwidth]{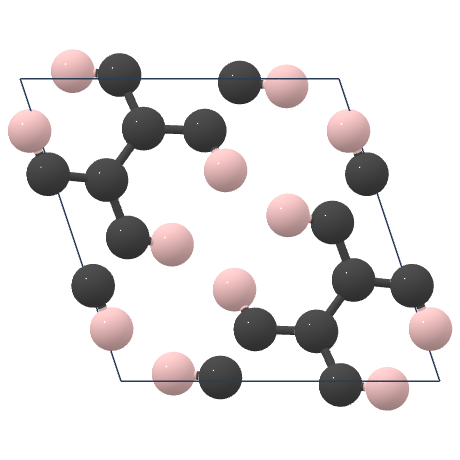}
    \caption{\ce{C3H2} (Molecular crystal)}
    \end{subfigure}
    \hfill
    \begin{subfigure}[b]{0.3\textwidth}
    \includegraphics[width=\textwidth]{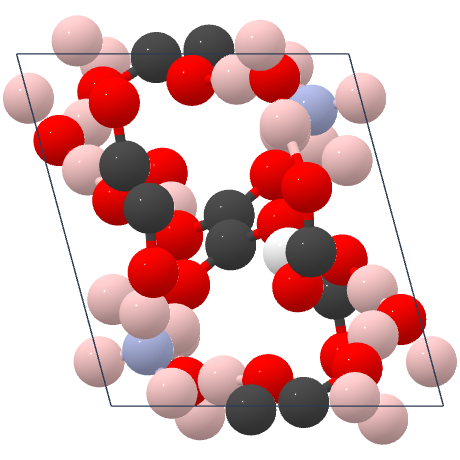}
    \caption{\ce{C4H11NO10} (Molecular crystal)}
    \end{subfigure}
    \hfill
    \begin{subfigure}[b]{0.3\textwidth}
    \includegraphics[width=\textwidth]{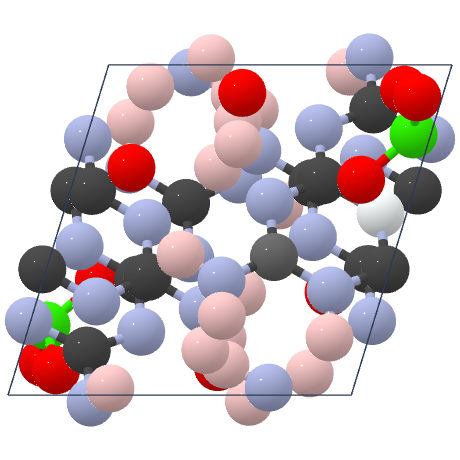}
    \caption{\ce{C6H9N10O5Cl} (Molecular crystal)}
    \end{subfigure}
    \caption{Examples of bulk, polymer, MOF and molecular crystal materials selected in MD simulations.}
    \label{si-fig:examples-of-materials}
\end{figure}

\begin{figure}
    \centering
    \begin{subfigure}[b]{0.3\textwidth}
    \includegraphics[width=\textwidth]{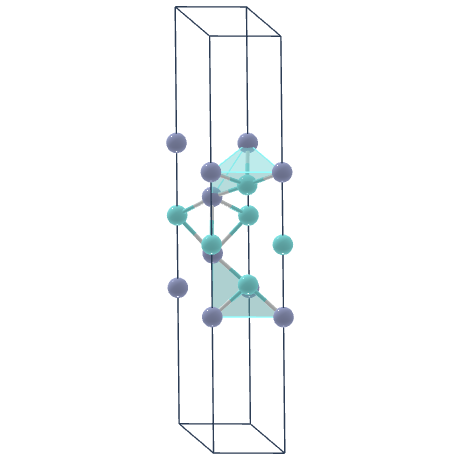}
    \caption{\ce{Y3(NF)2} (2D)}
    \end{subfigure}
    \hfill
    \begin{subfigure}[b]{0.3\textwidth}
    \includegraphics[width=\textwidth]{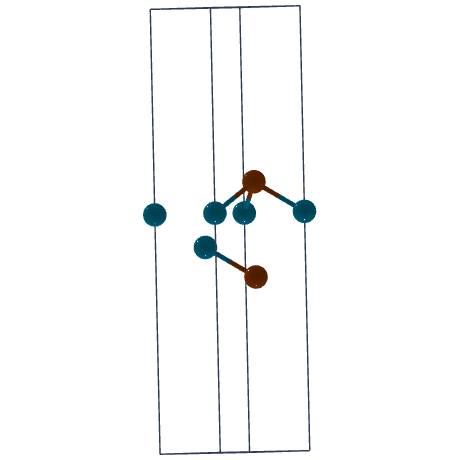}
    \caption{\ce{PdBr} (2D)}
    \end{subfigure}
    \hfill
    \begin{subfigure}[b]{0.3\textwidth}
    \includegraphics[width=\textwidth]{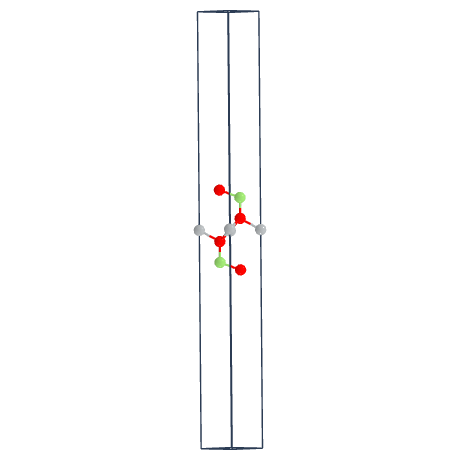}
    \caption{\ce{Ga2NiO4} (2D)}
    \end{subfigure}
    \begin{subfigure}[b]{0.3\textwidth}
    \includegraphics[width=\textwidth]{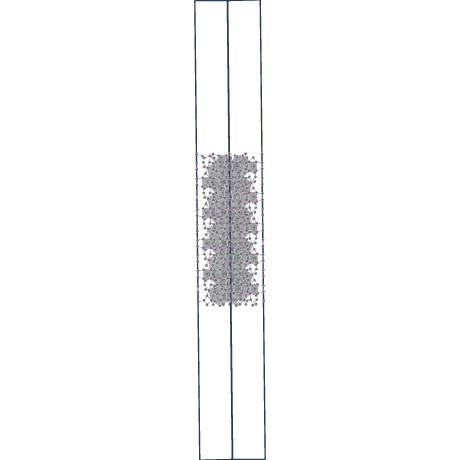}
    \caption{\ce{Ni5P4} (Surface)}
    \end{subfigure}
    \hfill
    \begin{subfigure}[b]{0.3\textwidth}
    \includegraphics[width=\textwidth]{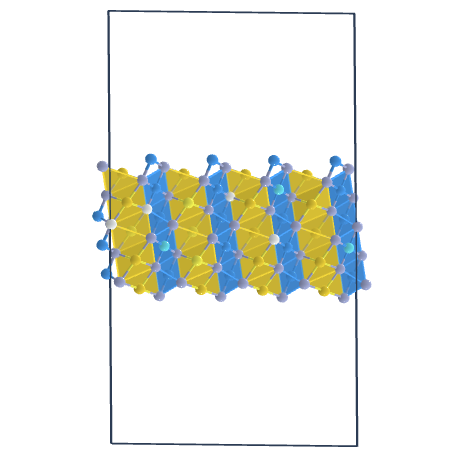}
    \caption{\ce{NaTaN2} (Surface)}
    \end{subfigure}
    \hfill
    \begin{subfigure}[b]{0.3\textwidth}
    \includegraphics[width=\textwidth]{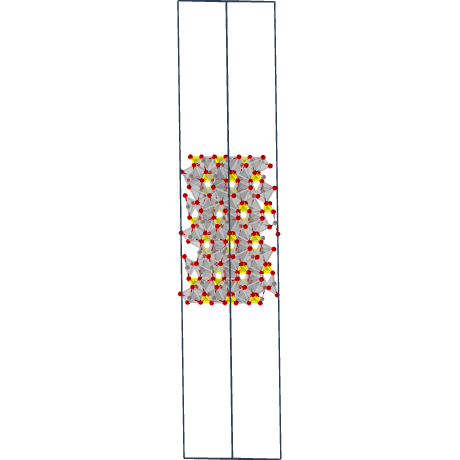}
    \caption{\ce{Ag2SO4} (Surface)}
    \end{subfigure}
    \begin{subfigure}[b]{0.3\textwidth}
    \includegraphics[width=\textwidth]{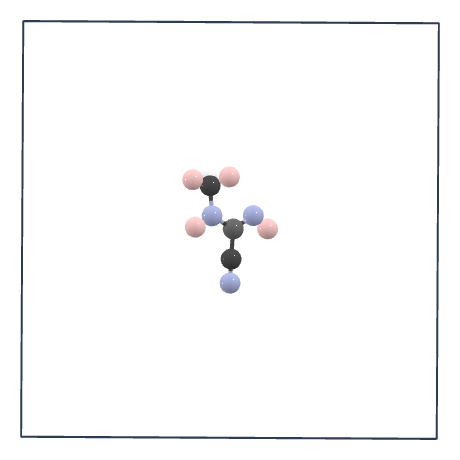}
    \caption{\ce{C3H5N3} (Molecule)}
    \end{subfigure}
    \hfill
    \begin{subfigure}[b]{0.3\textwidth}
    \includegraphics[width=\textwidth]{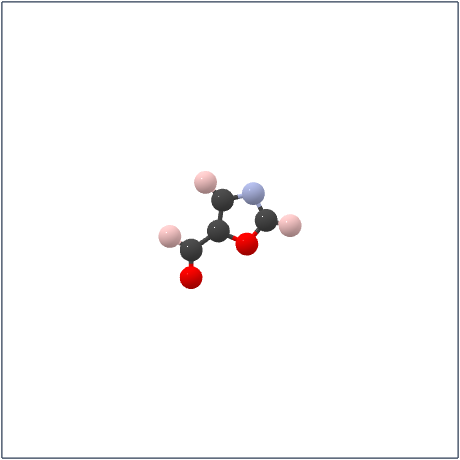}
    \caption{\ce{C4H3NO2} (Molecule)}
    \end{subfigure}
    \hfill
    \begin{subfigure}[b]{0.3\textwidth}
    \includegraphics[width=\textwidth]{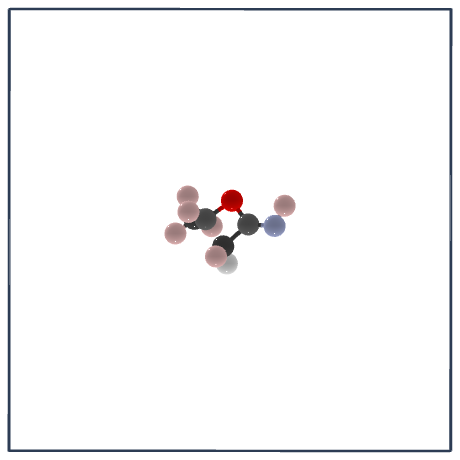}
    \caption{\ce{C4H7NO} (Molecule)}
    \end{subfigure}
    \caption{Examples of 2D, surface and isolated molecule materials selected in MD simulations.}
    \label{si-fig:examples-of-materials2}
\end{figure}

\subsection{System selection}
Representative systems including bulk inorganic materials, molecular crystals, organic polymers, metal-organic frameworks, two-dimensional materials, surfaces, and interfaces are collected by random selection from existing databases as follows:

\textbf{Bulk materials} are selected from \texttt{Alexandria}\cite{schmidt2022dataset,schmidt2022large} by randomly picking 10 structures from elementary, binary, and systems upto 5 elements, totaling 50. Supercells are created so that the number of atoms are larger than 200. See \autoref{si-fig:examples-of-materials} for the systems. Here are the IDs of the selected materials:
\begin{verbatim}
agm002149563, agm002150952, agm003157429, agm004442528, agm003273446
agm002179334, agm002179335, agm002189288, agm002190484, agm002191655
agm001283210, agm003297155, agm003212357, agm003258002, agm001828968
agm003273876, agm002168629, agm001194755, agm002245725, agm002321789
agm001253754, agm003249496, agm003454508, agm002943999, agm002732042
agm001106569, agm003611845, agm002891389, agm003100546, agm002299853
agm001550263, agm001504211, agm001465572, agm001633734, agm001781210
agm001428770, agm001288246, agm001707747, agm001433416, agm001803188
agm002129576, agm002158503, agm002028333, agm003279523, agm002078654
agm002215995, agm003239376, agm002083664, agm002080756, agm003282154
\end{verbatim}

\textbf{Isolated molecules} are selected from the \texttt{QM9} dataset\cite{kim2019energy} randomly totaling 10. See \autoref{si-fig:examples-of-materials2} for the systems. Here are the IDs of the selected molecules:
\begin{verbatim}
dsgdb9nsd_000305, dsgdb9nsd_000347, dsgdb9nsd_000404, dsgdb9nsd_000514, dsgdb9nsd_000608
dsgdb9nsd_000673, dsgdb9nsd_000742, dsgdb9nsd_000952, dsgdb9nsd_000981, dsgdb9nsd_001028
\end{verbatim}

\textbf{Molecular crystals} are chosen from the Materials Project within the C-H-O-N-S-Cl chemical space totaling 10\cite{jain2013commentary}. Supercells are created so that the number of atoms are larger than 200. See \autoref{si-fig:examples-of-materials} for the systems. Here are the IDs of the selected crystals:
\begin{verbatim}
 mp-1195829,   mp-23909,  mp-557379,   mp-560323, mp-866659, mp-995217
mv-15630958, mv-5673042, mv-5675009,  mv-9791995
\end{verbatim}

\textbf{Metal organic frameworks} (MOF) are selected from the QMOF database\cite{rosen2022high,rosen2021machine} on the Materials Project totaling 16.See \autoref{si-fig:examples-of-materials} for the systems. Here are the IDs of the selected materials:
\begin{verbatim}
qmof-ecfd7a0, qmof-49ce9a8, qmof-cbf6511, qmof-d6662a5, qmof-cbf6511
qmof-d6662a5, qmof-2941470, qmof-c3fa563, qmof-2941470, qmof-2a42bc4
qmof-1452981, qmof-2a42bc4, qmof-bb88cf5, qmof-d675ae6, qmof-bb88cf5
qmof-d675ae6
\end{verbatim}

\textbf{Surface systems} are constructed using the \texttt{SlabGenerator} from \texttt{pymatgen}\cite{ong2013python} from randomly selected bulk materials. 11 structures are generated in total by cleaving from their (0,0,1) surfaces with at least 7 layers of atoms in the slab and 20 Angstrom in the vacumm. See \autoref{si-fig:examples-of-materials2} for the systems. Here are the IDs of the selected materials:
\begin{verbatim}
mp-1523, mp-2802, mp-3862, mp-5505,  mp-155
mp-2908,  mp-451, mp-5625, mp-1920, mp-3862
mp-5475
\end{verbatim}

\textbf{Interface structures} are constructed using the \texttt{interface\_master}\cite{xie2022brute,xie2022interface_master} tool. Two interfaces between GaN/Fe and ZnO/Al2O3 are constructed. See \autoref{si-fig:examples-of-materials3} for the systems. The two interface structures are generated with the jupyter notebook following this \url{https://github.com/nmdl-mizo/interface_master/blob/develop/test_files/Tutorial_Two-dimensional%20CSL%20interfaces_graphene_GaN.ipynb}

\textbf{Two-dimensional materials} (2D) are collected from the Computational 2D Materials Database\cite{haastrup2018computational} with random selection, totaling 10 materials. Supercells are created during simulations so that the cell contained at least 100 atoms by replicating along the periodic directions. See \autoref{si-fig:examples-of-materials2} for the systems. Here are the IDs of the selected materials in the dataset.
\begin{verbatim}
  c2db-118,  c2db-1332, c2db-14861, c2db-15102, c2db-15134
c2db-16066, c2db-16132, c2db-16381, c2db-16451,   c2db-628
\end{verbatim}

\textbf{Polymers} are collected from \cite{huan2020polymer} where their experimental crystalline structures are reported. A total of 9 crystalline polymers with different polymorphs are included containing BPBO, PE, PPS, PVDF, and PAN. See \autoref{si-fig:examples-of-materials} for the systems. Here are the IDs of the selected polymers:
\begin{verbatim}
     ABPBO,  ortho-PE,    mono-PE
alpha-PVDF, beta-PVDF, delta-PVDF
       PAN,       PPS,        PVC
\end{verbatim}

\begin{figure}
    \centering
    \begin{subfigure}[b]{0.4\textwidth}
    \includegraphics[width=\textwidth]{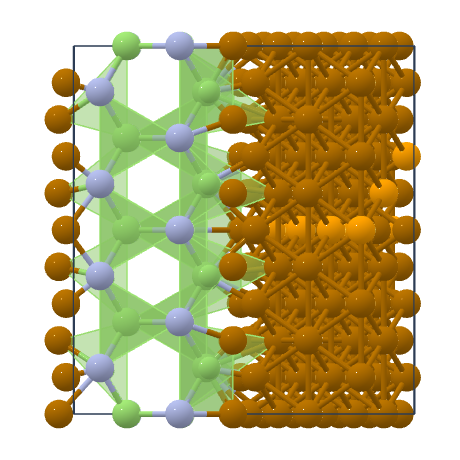}
    \caption{GaN/Fe (Interface)}  
    \end{subfigure}
    \hfill
    \begin{subfigure}[b]{0.4\textwidth}
    \includegraphics[width=\textwidth]{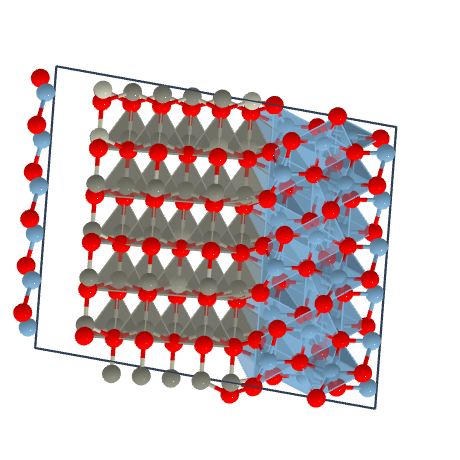}  
    \caption{ZnO/\ce{Al2O3} (Interface)}  
    \end{subfigure}
    \caption{Examples of interface materials selected in MD simulations.}
    \label{si-fig:examples-of-materials3}
\end{figure}

\subsection{MD Setup}
Molecular dynamics simulations are conducted using LAMMPS\cite{thompson2022lammps} via an interface to MatterSim, encompassing both the canonical (NVT) and isothermal-isobaric (NPT) ensembles. The NVT ensemble is employed for all 118 systems under investigation, with a temperature ramp from 0 K to 5000 K over a total simulation duration of 500 ps. In the case of bulk systems, segmented NPT ensemble simulations are performed, where the pressure is initially increased from 0 GPa to 1000 GPa at a constant temperature of 300 K, followed by a temperature ramp from 300 K to 5000 K at a maintained pressure of 1000 GPa. The simulation time of both segments is 500 ps, as shown in the inset of \autoref{fig:md}(c). All time step is set to 1 fs.

\subsection{MD Trajectory}
\begin{figure}
    \centering
    \includegraphics[width=0.9\textwidth]{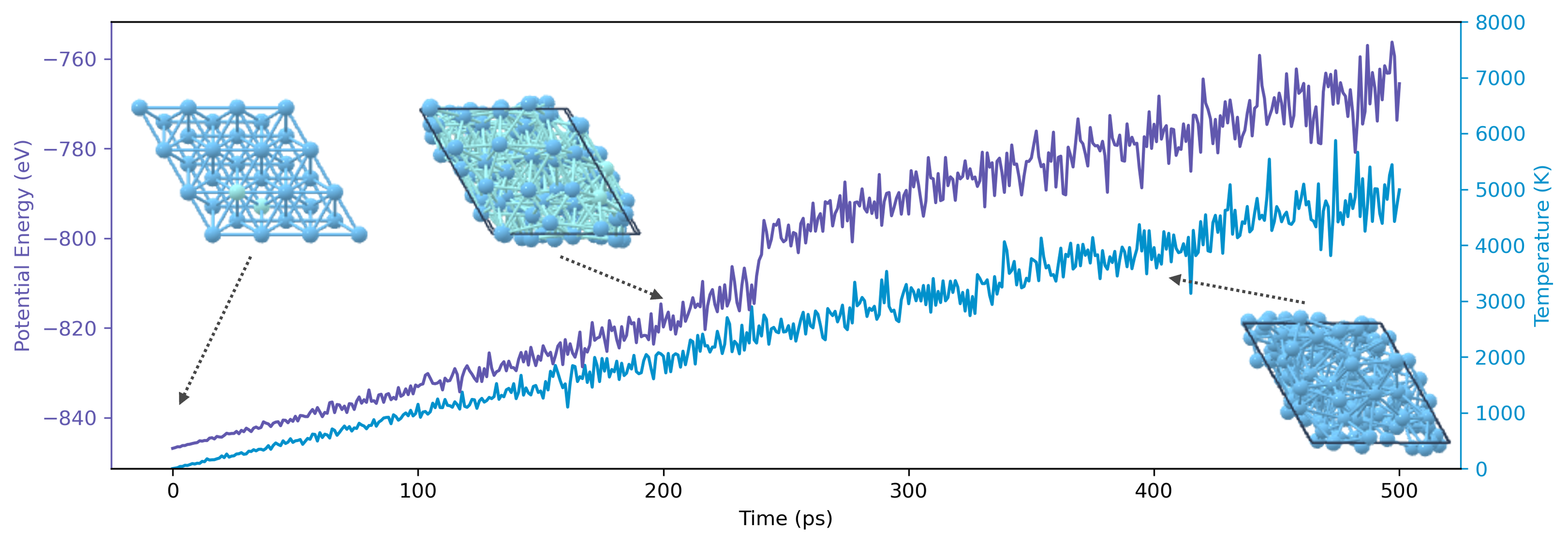}
    \caption{The potential energy of bulk Ti under increasing temperature and NVT ensemble.}\label{fig:bulk_1elements_Ti4_agm002190484_nvt_traj}
\end{figure}

\begin{figure}
    \centering
    \includegraphics[width=0.9\textwidth]{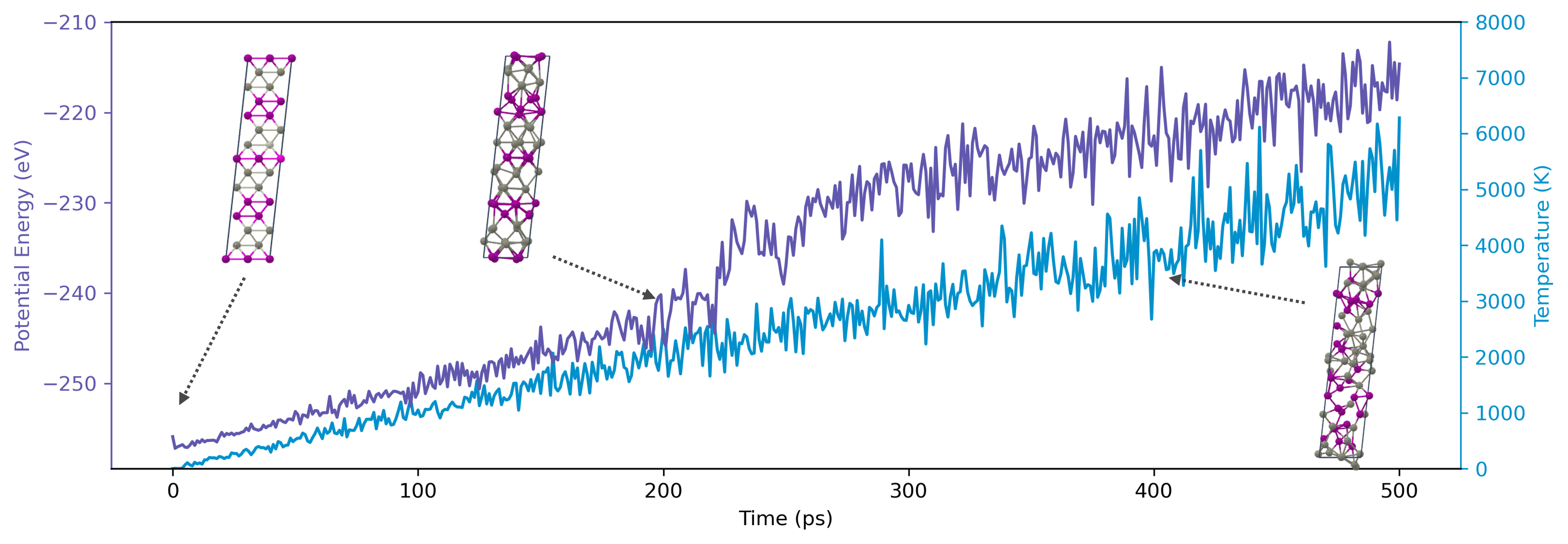}
    \caption{The potential energy of bulk \ce{Mn3Zn4} under increasing temperature and NVT ensemble.}\label{fig:bulk_2elements_Mn3Zn4_agm003258002_nvt_traj}
\end{figure}

\begin{figure}
    \centering
    \includegraphics[width=0.9\textwidth]{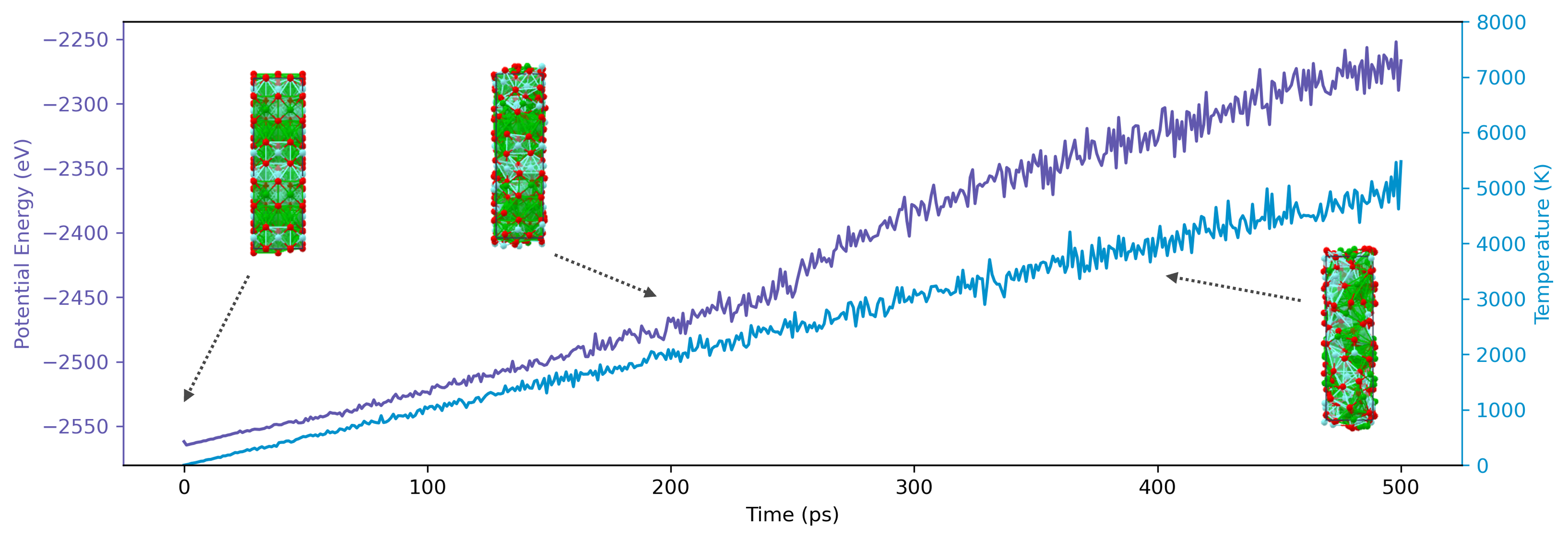}
    \caption{The potential energy of bulk \ce{Ba3O6Y2} under increasing temperature and NVT ensemble.}\label{fig:bulk_3elements_Ba12O24Y8_agm003249496_nvt_traj}
\end{figure}

\begin{figure}
    \centering
    \includegraphics[width=0.9\textwidth]{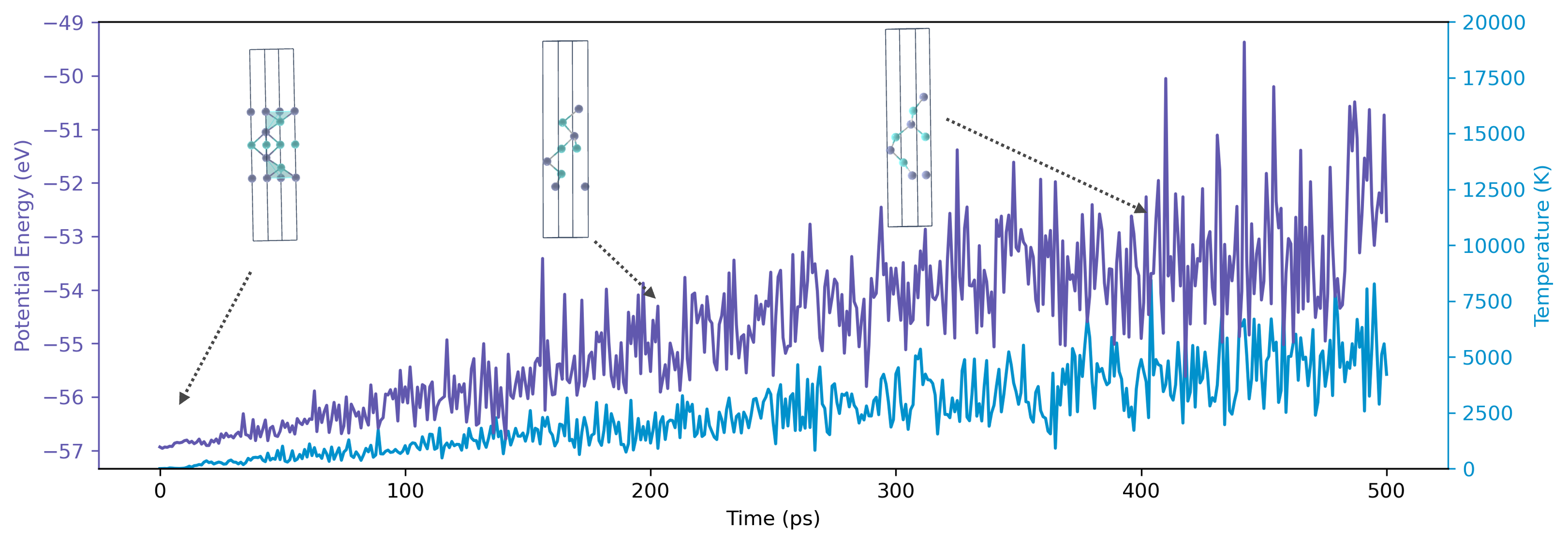}
    \caption{The potential energy of 2D \ce{Y3N2F2} under increasing temperature and NVT ensemble.}\label{fig:2D_c2db-118_nvt_traj}
\end{figure}

\begin{figure}
    \centering
    \includegraphics[width=0.9\textwidth]{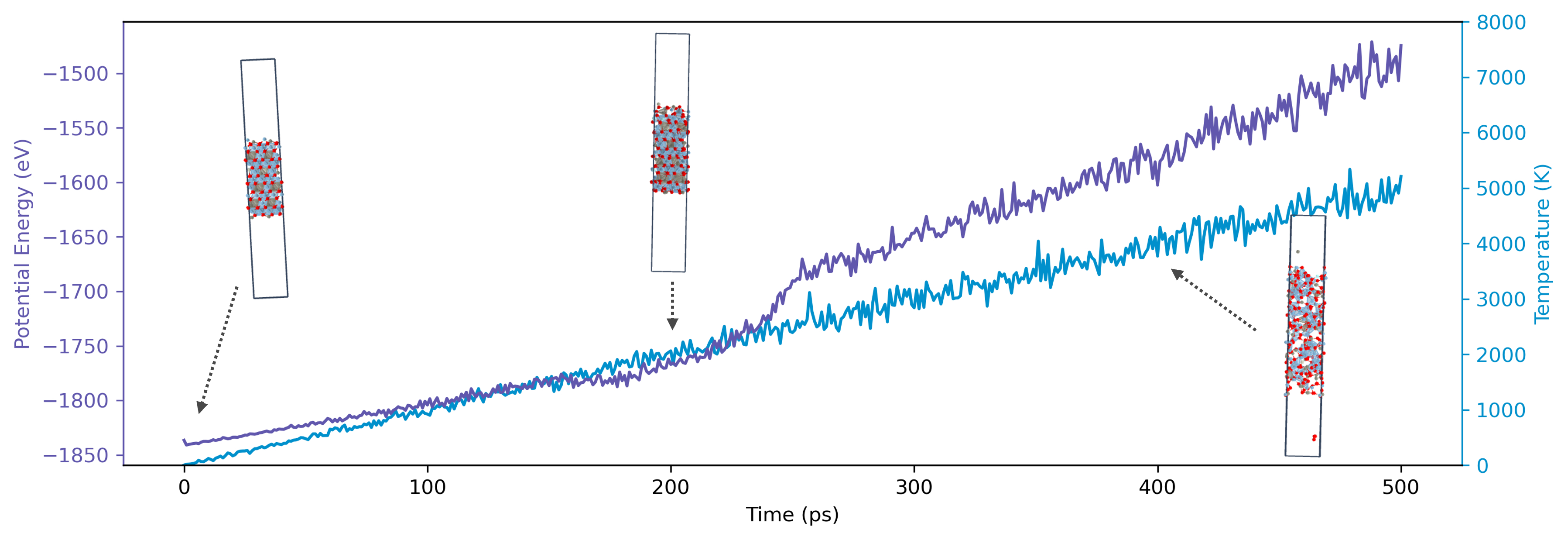}
    \caption{The potential energy of surface \ce{Al2ZnO4} under increasing temperature and NVT ensemble.}\label{fig:Surface_mp-2908_Al2ZnO4_nvt_traj}
\end{figure}

\begin{figure}
    \centering
    \includegraphics[width=0.9\textwidth]{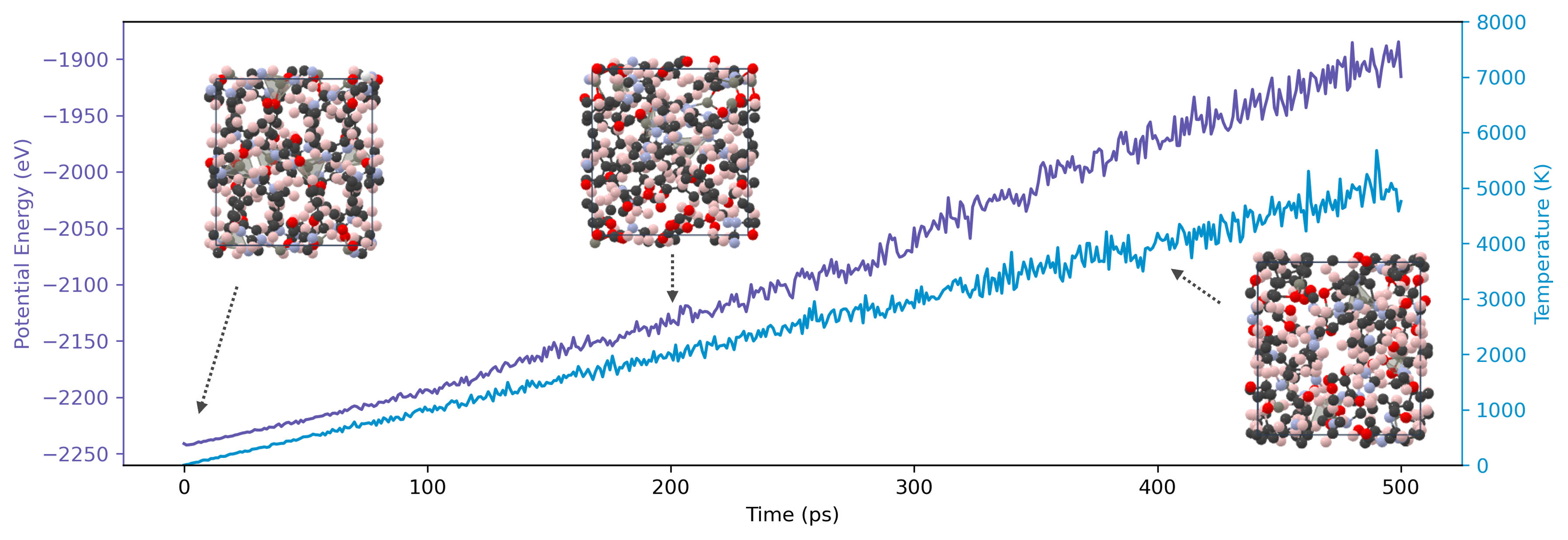}
    \caption{The potential energy of MOF \ce{ZnH16C18NO4} under increasing temperature and NVT ensemble.}\label{fig:MOF_ZnH16C18NO4_qmof-2a42bc4_nvt_traj}
\end{figure}

\begin{figure}
    \centering
    \includegraphics[width=0.9\textwidth]{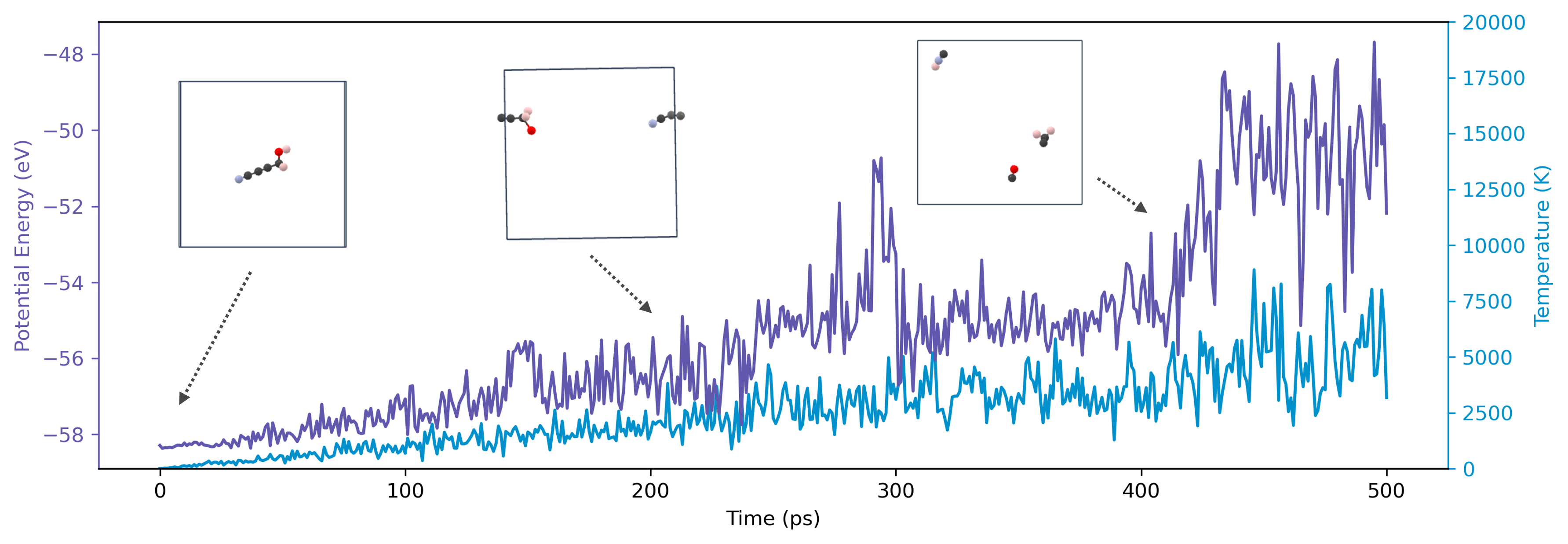}
    \caption{The potential energy of molecule \ce{C4H3NO} under increasing temperature and NVT ensemble.}\label{fig:Molecule_dsgdb9nsd_000514_nvt_traj}
\end{figure}

\begin{figure}
    \centering
    \includegraphics[width=0.9\textwidth]{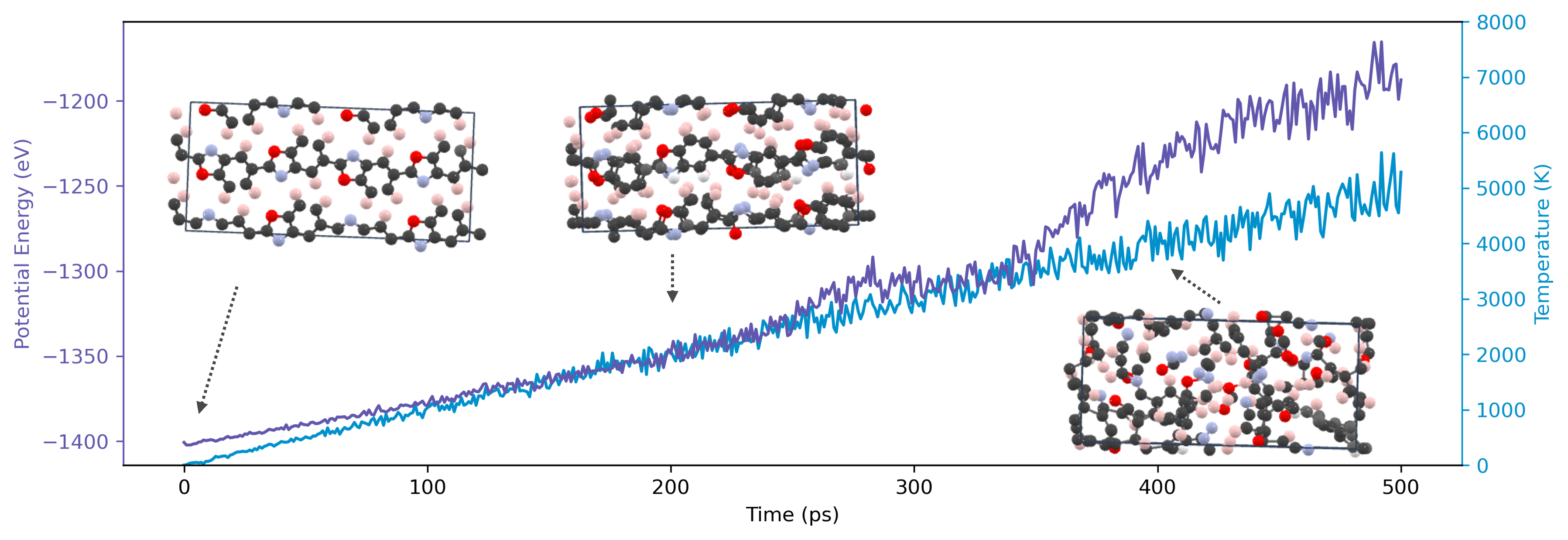}
    \caption{The potential energy of polymer \ce{C7H3NO} under increasing temperature and NVT ensemble.}\label{fig:Polymer_ABPBO_known_nvt_traj}
\end{figure}

\begin{figure}
    \centering
    \includegraphics[width=0.9\textwidth]{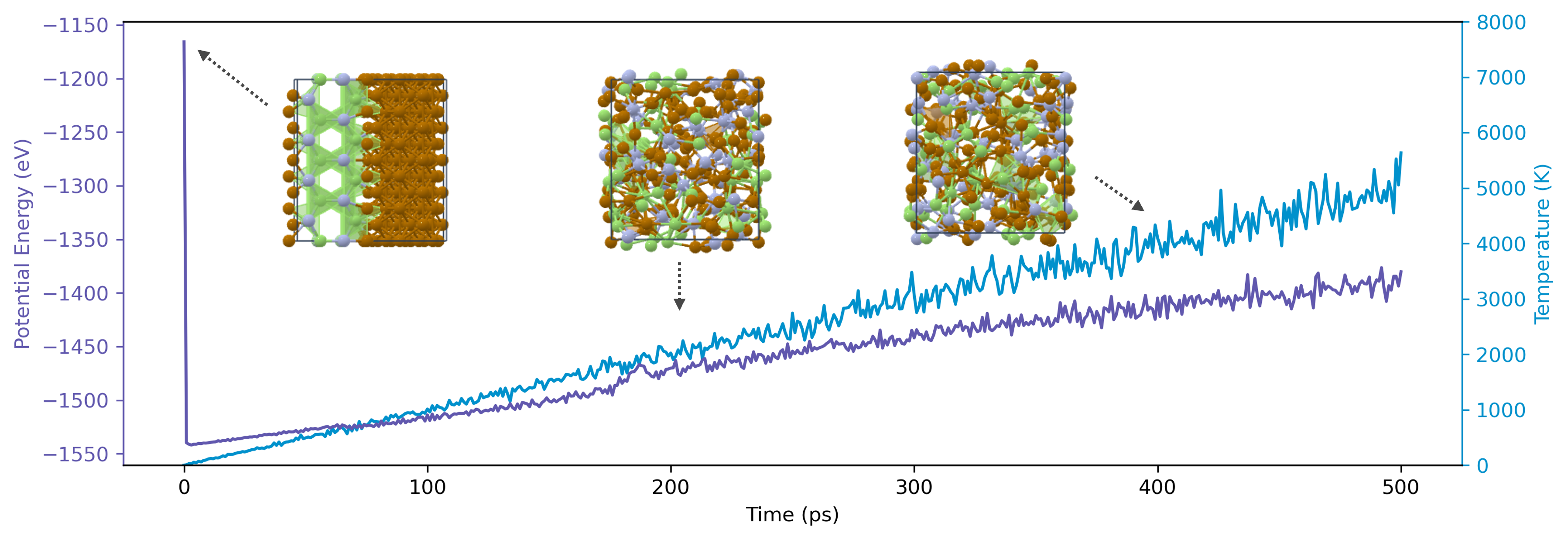}
    \caption{The potential energy of interface GaN/Fe under increasing temperature and NVT ensemble.}\label{fig:Interface_GaN_Fe_nvt_traj}
\end{figure}

\begin{figure}
    \centering
    \begin{subfigure}[b]{0.9\textwidth}
    \includegraphics[width=\textwidth]{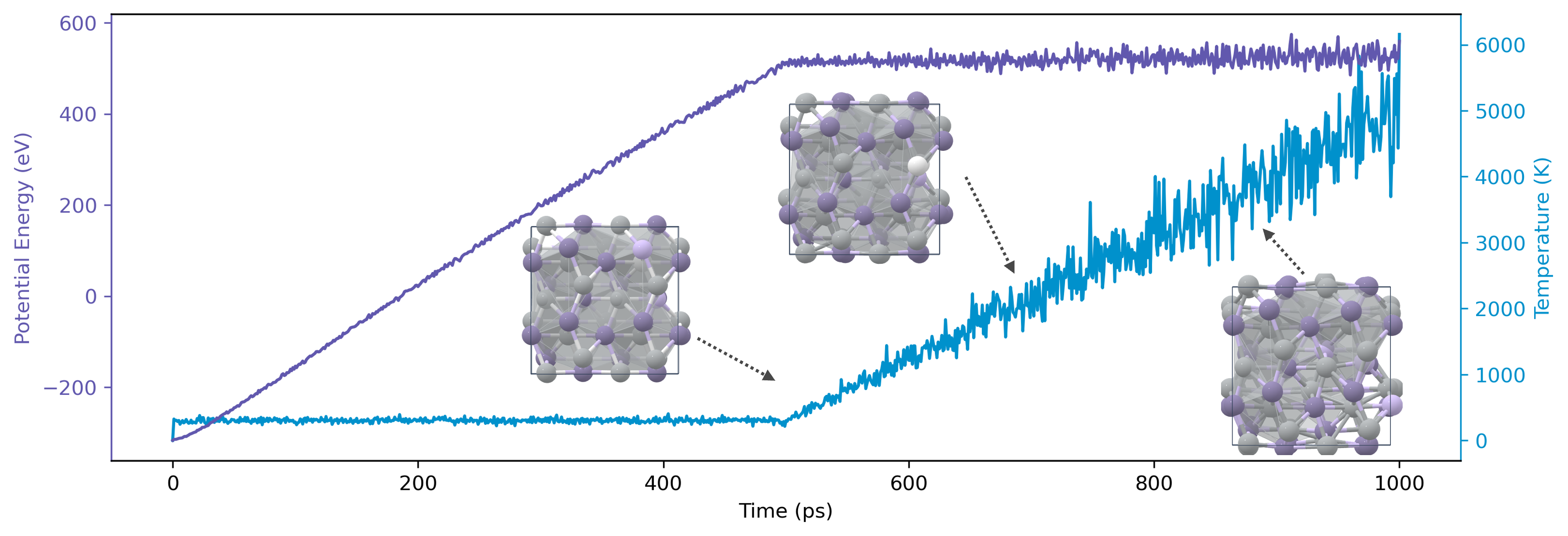}
    \caption{}
\label{fig:md:Bulk_bulk_2elements_Ni4Sn4_agm002168629_npt_traj_press}
    \end{subfigure}
    \hfill
    \begin{subfigure}[b]{0.9\textwidth}
    \includegraphics[width=\textwidth]{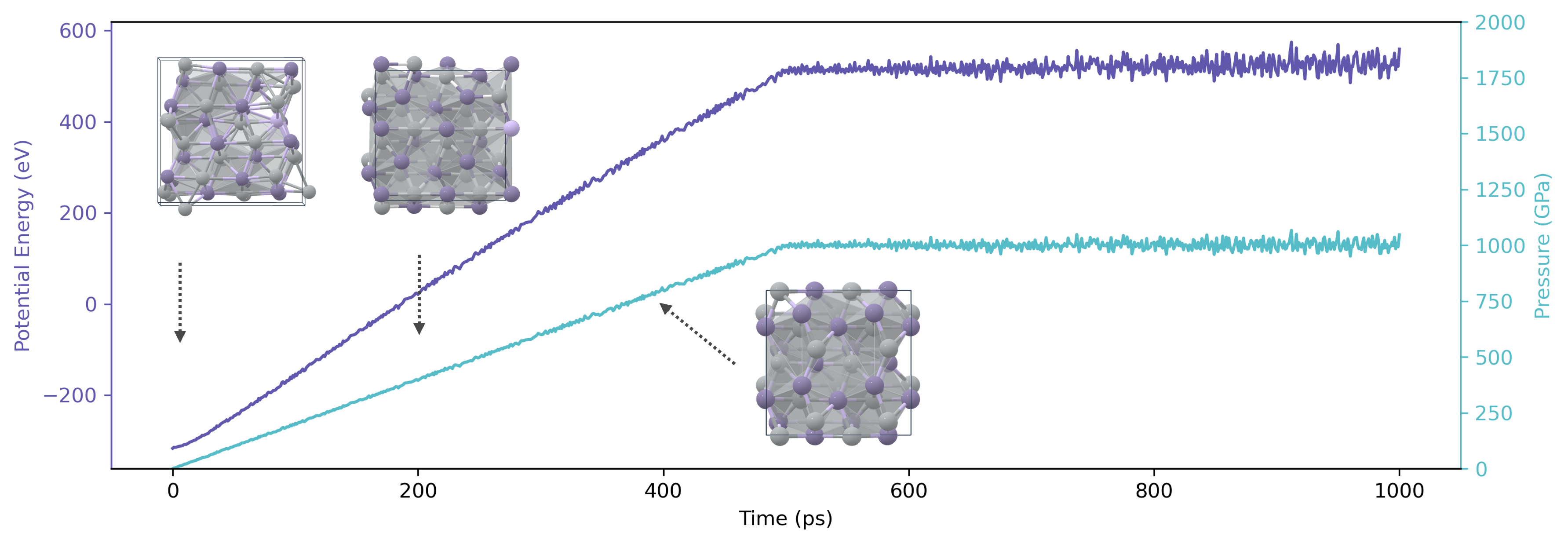}
    \caption{}    \label{fig:md:Bulk_bulk_2elements_Ni4Sn4_agm002168629_npt_traj_temp}
    \end{subfigure}
    \caption{The potential energy, pressure and temperature of bulk \ce{Ni4Sn4} under NPT ensemble.}\label{fig:md:Bulk_bulk_2elements_Ni4Sn4_agm002168629_npt_traj}
\end{figure}

\begin{figure}
    \centering
    \begin{subfigure}[b]{0.9\textwidth}
    \includegraphics[width=\textwidth]{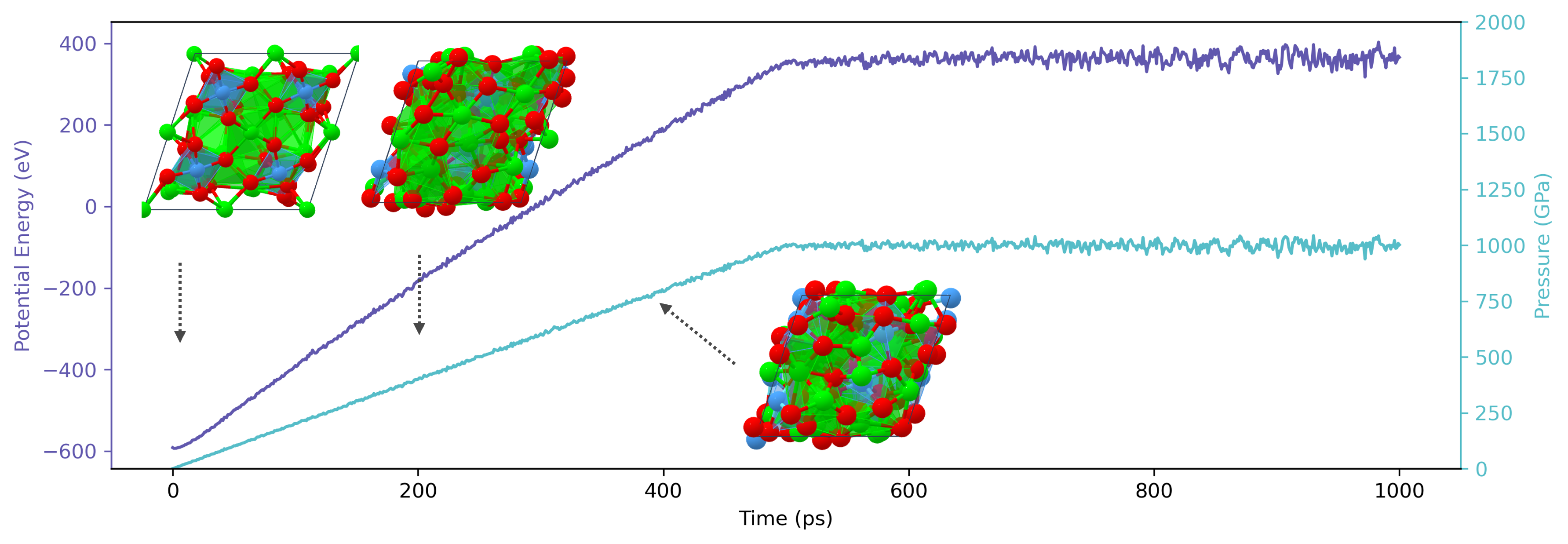}
    \caption{}
\label{fig:md:Bulk_bulk_3elements_O6Sr3Ta_agm002299853_npt_traj_press}
    \end{subfigure}
    \hfill
    \begin{subfigure}[b]{0.9\textwidth}
    \includegraphics[width=\textwidth]{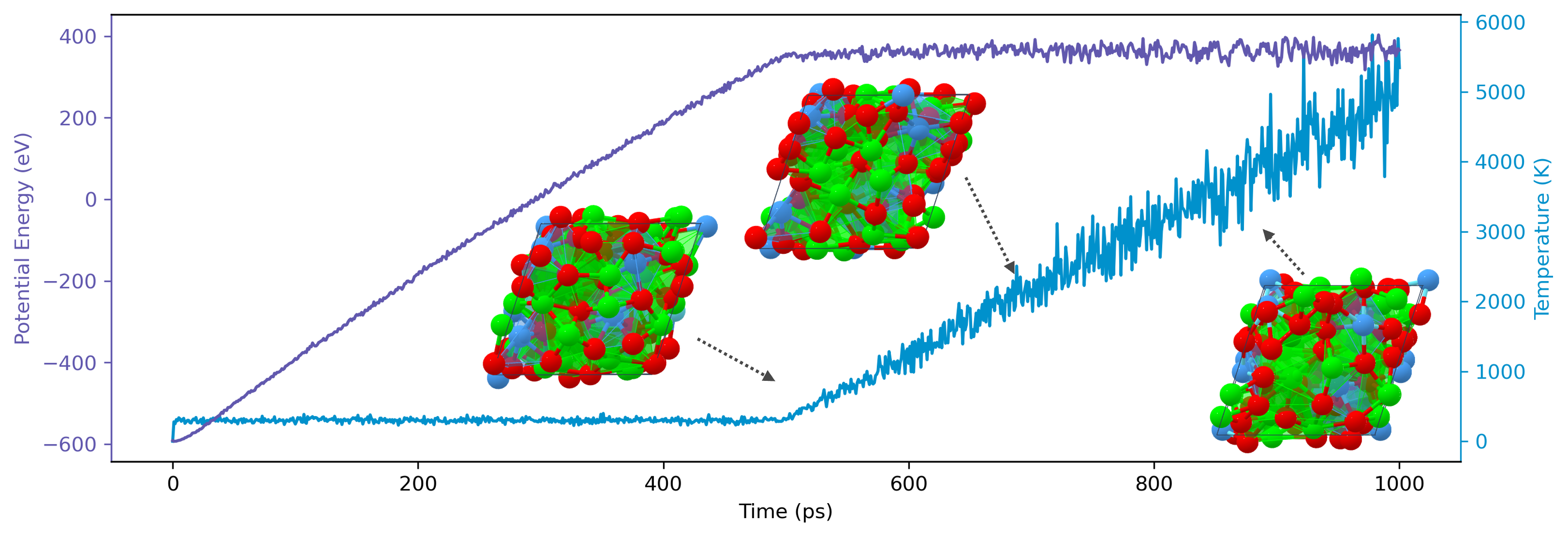}
    \caption{}    \label{fig:md:Bulk_bulk_3elements_O6Sr3Ta_agm002299853_npt_traj_temp}
    \end{subfigure}
    \caption{The potential energy, pressure and temperature of bulk \ce{TaSr3O6} under NPT ensemble.}\label{fig:md:Bulk_bulk_3elements_O6Sr3Ta_agm002299853_npt_traj}
\end{figure}

\begin{figure}
    \centering
    \begin{subfigure}[b]{0.9\textwidth}
    \includegraphics[width=\textwidth]{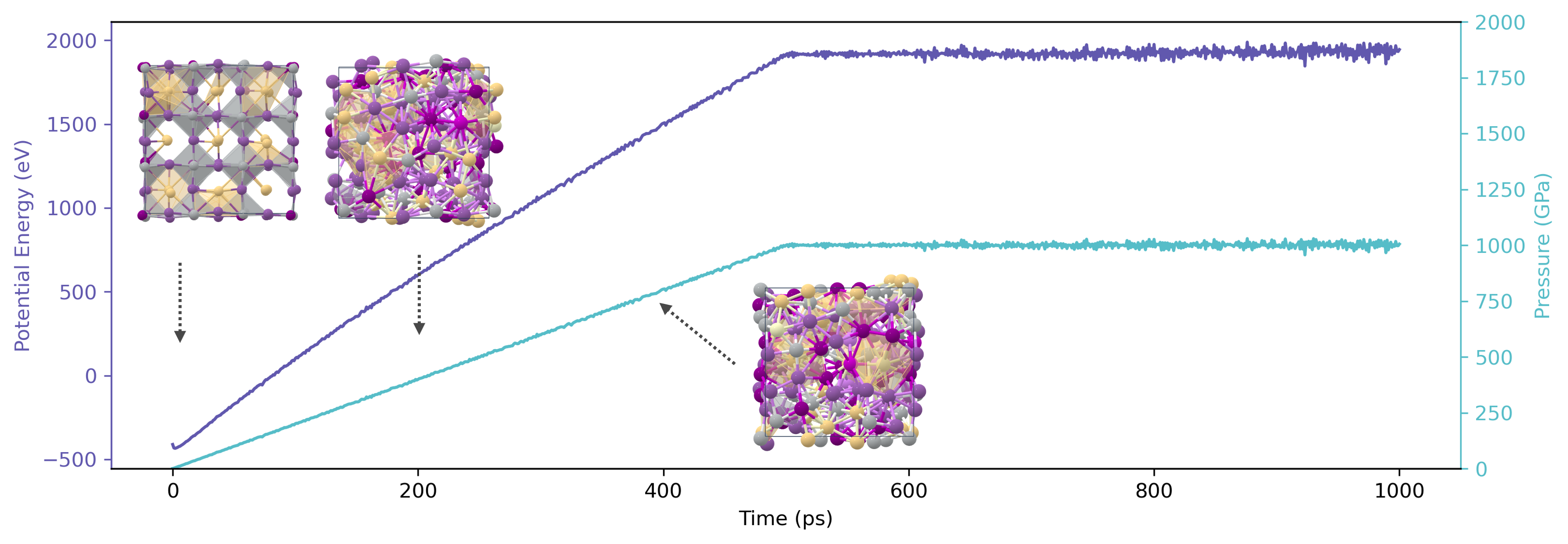}
    \caption{}
\label{fig:md:Bulk_bulk_4elements_CdINiSb2_agm001781210_npt_traj_press}
    \end{subfigure}
    \hfill
    \begin{subfigure}[b]{0.9\textwidth}
    \includegraphics[width=\textwidth]{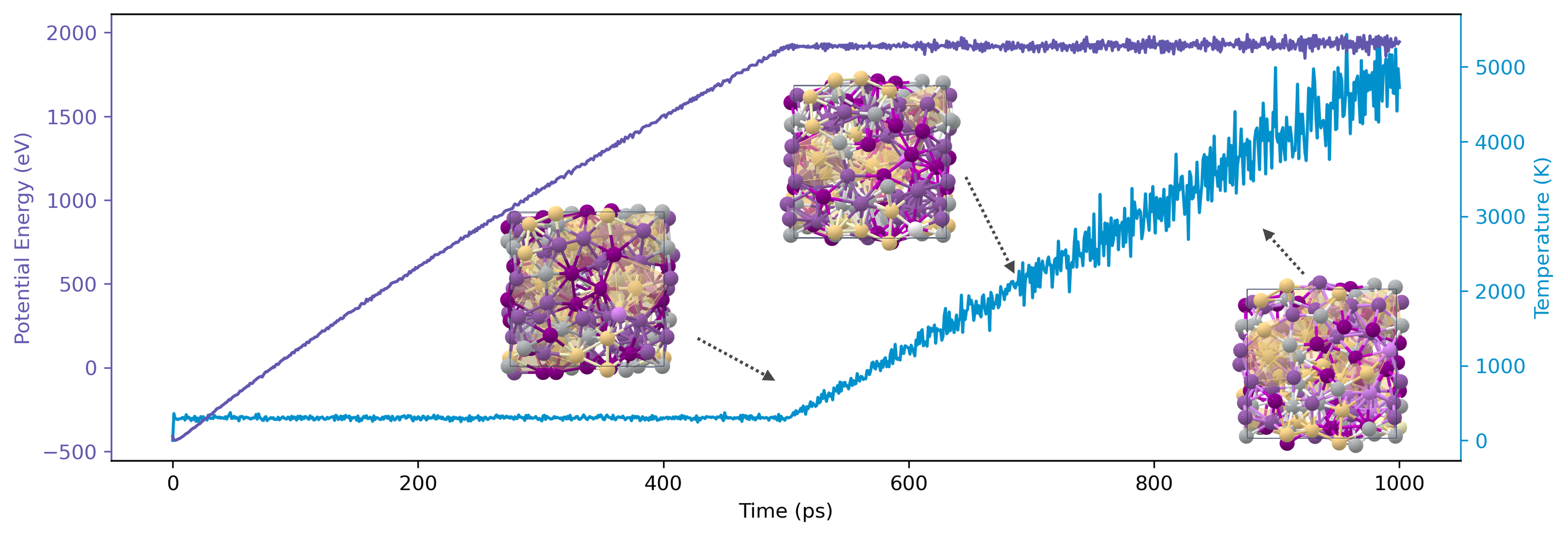}
    \caption{}    \label{fig:md:Bulk_bulk_4elements_CdINiSb2_agm001781210_npt_traj_temp}
    \end{subfigure}
    \caption{The potential energy, pressure and temperature of bulk \ce{CdNiISb2} under NPT ensemble.}\label{fig:md:Bulk_bulk_4elements_CdINiSb2_agm001781210_npt_traj}
\end{figure}

\begin{figure}
    \centering
    \begin{subfigure}[b]{0.9\textwidth}
    \includegraphics[width=\textwidth]{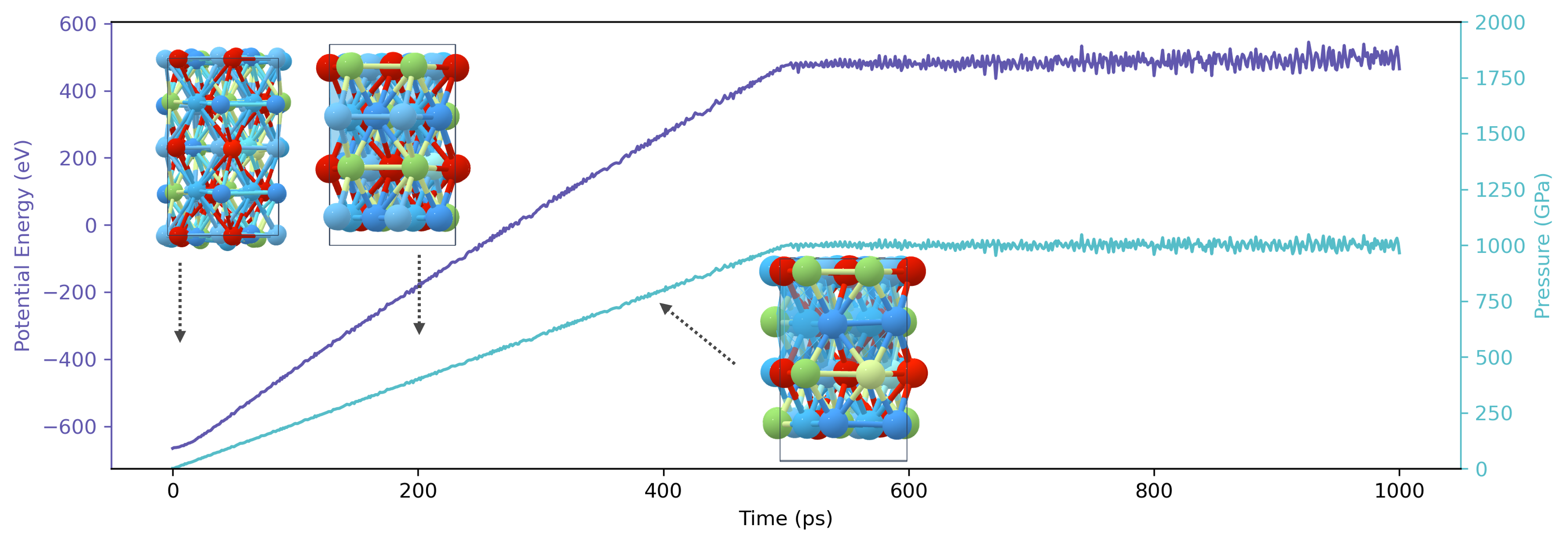}
    \caption{}
\label{fig:md:Bulk_bulk_5elements_Ga2Hf2Ta2Ti2V2_agm002028333_npt_traj_press}
    \end{subfigure}
    \hfill
    \begin{subfigure}[b]{0.9\textwidth}
    \includegraphics[width=\textwidth]{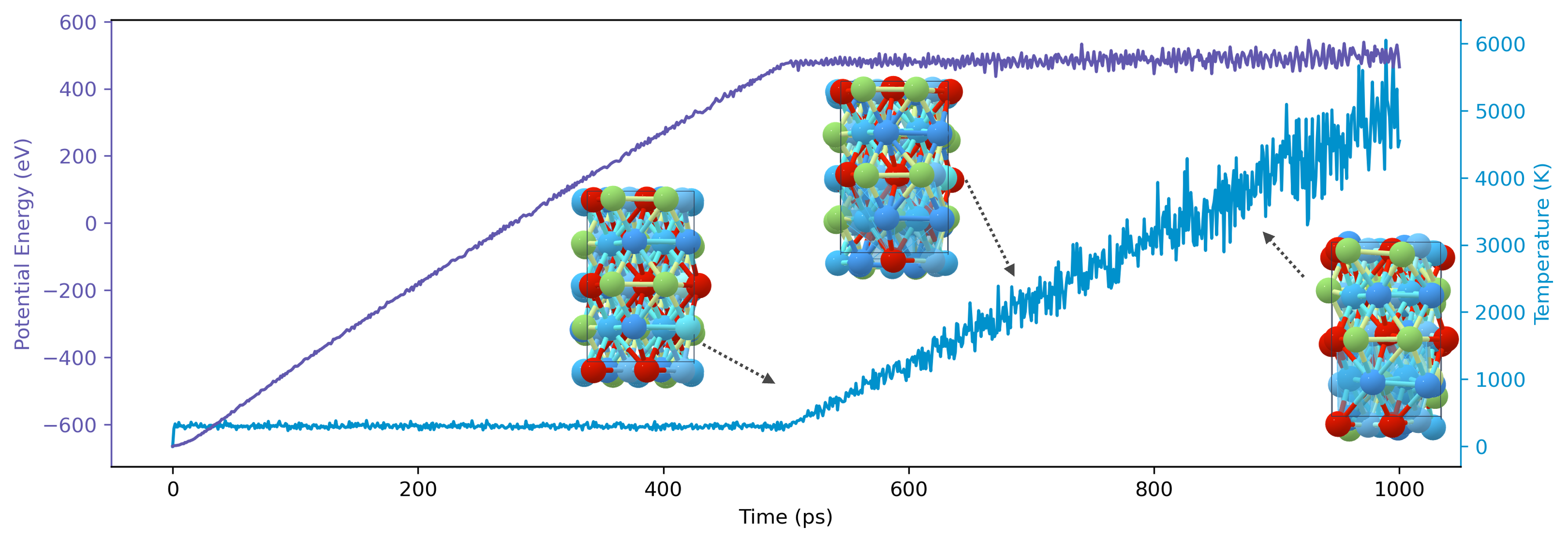}
    \caption{}    \label{fig:md:Bulk_bulk_5elements_Ga2Hf2Ta2Ti2V2_agm002028333_npt_traj_temp}
    \end{subfigure}
    \caption{The potential energy, pressure and temperature of bulk \ce{Ga2Hf2Ta2Ti2V2} under NPT ensemble.}\label{fig:md:Bulk_bulk_5elements_Ga2Hf2Ta2Ti2V2_agm002028333_npt_traj}
\end{figure}

Several example MD trajectories are depicted from \autoref{fig:bulk_1elements_Ti4_agm002190484_nvt_traj} to \autoref{fig:Interface_GaN_Fe_nvt_traj}. From the NVT simulations conducted with MatterSim upon heating, it is evident that the potential energy of all systems increases progressively with the rise in the temperature, leading to their structural transition from ordered to disordered states. The inset in \autoref{fig:md}(d) demonstrates the difference of radial distribution function (RDF) in the heating process, further confirming the melting behavior in the molecular dynamics. For the molecular system \ce{C4H3NO} presented in \autoref{fig:Molecule_dsgdb9nsd_000514_nvt_traj}, by examining the structures at the initial 0 ps, and subsequently at 200 ps and 400 ps, we observe molecular dissociation. In the case of the GaN/Fe interface system shown in \autoref{fig:Interface_GaN_Fe_nvt_traj}, the increase in temperature also results in the transformation of the originally distinct crystal phases into a mixed phase. From \autoref{fig:md:Bulk_bulk_2elements_Ni4Sn4_agm002168629_npt_traj} to \autoref{fig:md:Bulk_bulk_5elements_Ga2Hf2Ta2Ti2V2_agm002028333_npt_traj} illustrate the NPT simulation processes, where MatterSim successfully simulates the effects of pressurization and heating. The RDF shown in \autoref{fig:md}(e) implies the decrease of bond length during pressurization process.

\section{Active learning} \label{sec:si_active_learning}

\subsection{Dataset}
\begin{figure}
    \centering
    \begin{subfigure}[b]{0.3\textwidth}
    \includegraphics[width=\textwidth]{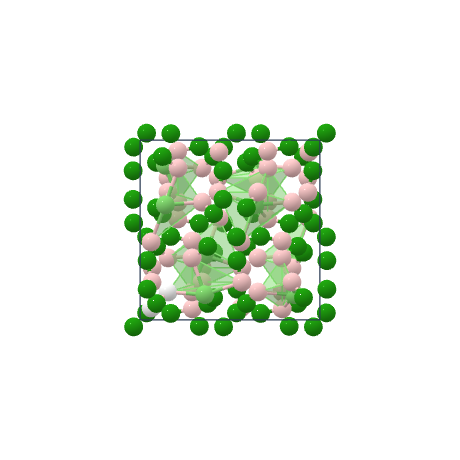}
    \caption{}
    \label{fig:al-and-ft:LiBH_structure1}
    \end{subfigure}
    \begin{subfigure}[b]{0.3\textwidth}
    \includegraphics[width=\textwidth]{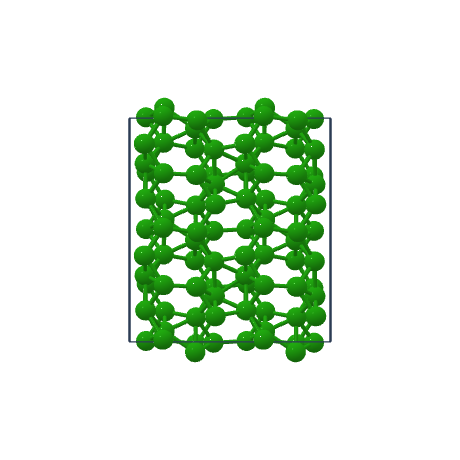}
    \caption{}
    \label{fig:al-and-ft:B_structure1}
    \end{subfigure}
    \begin{subfigure}[b]{0.3\textwidth}
    \includegraphics[width=\textwidth]{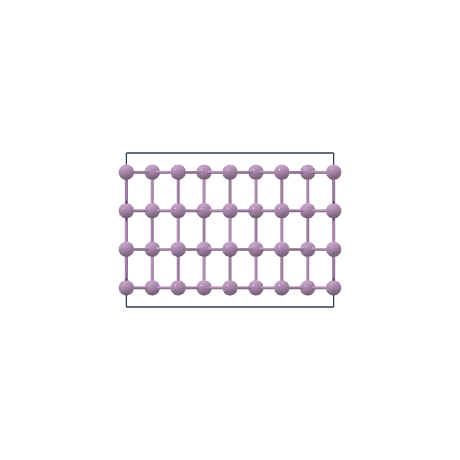}
    \caption{}
    \label{fig:al-and-ft:P_structure1}
    \end{subfigure}
    \hfill
    \begin{subfigure}[b]{0.3\textwidth}
    \includegraphics[width=\textwidth]{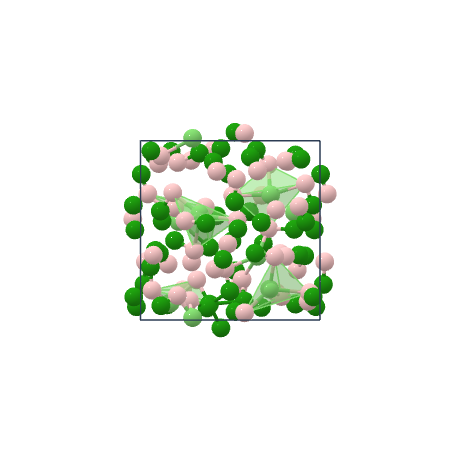}
    \caption{}
    \label{fig:al-and-ft:LiBH_structure2}
    \end{subfigure}
    \begin{subfigure}[b]{0.3\textwidth}
    \includegraphics[width=\textwidth]{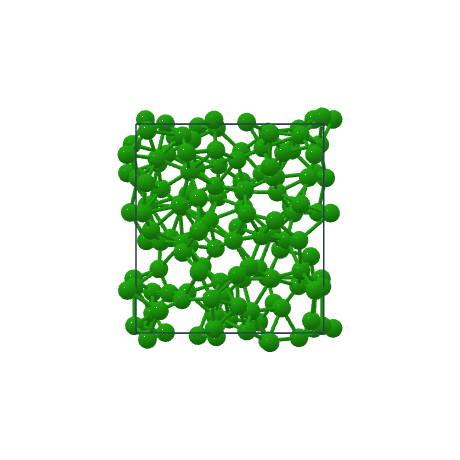}
    \caption{}
    \label{fig:al-and-ft:B_structure2}
    \end{subfigure}
    \begin{subfigure}[b]{0.3\textwidth}
    \includegraphics[width=\textwidth]{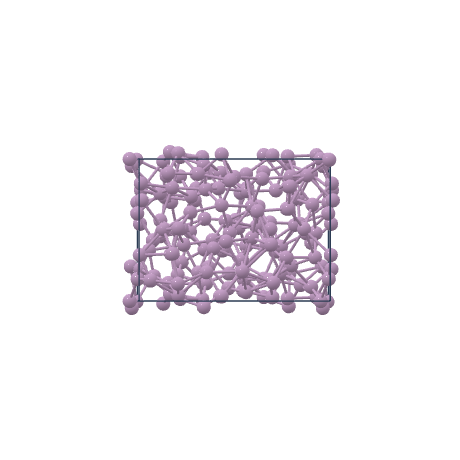}
    \caption{}
    \label{fig:al-and-ft:P_structure2}
    \end{subfigure}
    \caption{Crystal structures of initial configurations (a-c) and molten configurations (d-e) for \ce{Li2B12H12}, boron and phosphorus, respectively.}\label{fig:al_structures}
\end{figure}

As a universal predictive model for material properties, MatterSim may not yield satisfactory accuracy for highly complex systems that have not been previously encountered in its training dataset. Under such circumstances, an active learning approach can be employed to selectively filter data, followed by finetuning of the MatterSim model. This procedure facilitates the rapid development of a sufficiently accurate and operational model. In this work, we present three examples of complex systems, including the ionic superconductor \ce{Li2B12H12}, molten phosphorus and boron, as \autoref{fig:al_structures} shows. To generate the training dataset, we performed NVT simulations on \ce{Li2B12H12} by VASP\cite{kresse1996efficiency,kresse1996efficient} with a time step of 0.5 fs for a total duration of 5.0 ps at a simulation temperature of 2000 K. For molten phosphorus and boron, the simulations were conducted at a temperature of 5000 K, with a time step of 1.0 fs and a total simulation time of 10.0 ps. The test dataset belongs to trajectories that are not included in the training set.

\subsection{Results}
Active learning procedure can improve significantly the accuracy of MLFF by augmenting the data set with data points that exhibit high uncertainty based on the ensemble models. The uncertainty arises from the variability inherent in different pre-trained models, each initialized with a distinct random number seed. This variability is quantified as follows:
\begin{equation}\label{eq:unc}  
\mathrm{unc} = \max \left[ \frac{1}{N} \sum_{i=1}^N  (|F_{ia}| - |\bar{F}_{a}|)^2 \right]
\end{equation}
where $F_{ia}$ denotes the predicted atomic force of the a-th atoms by the i-th zero-shot model and $N$ represents the total number of pretrained models utilized. $\bar{F}_{a}$ signifies the average of the atomic forces predicted by the ensemble models. 

\begin{figure}
    \centering
    \includegraphics[width=0.8\textwidth]{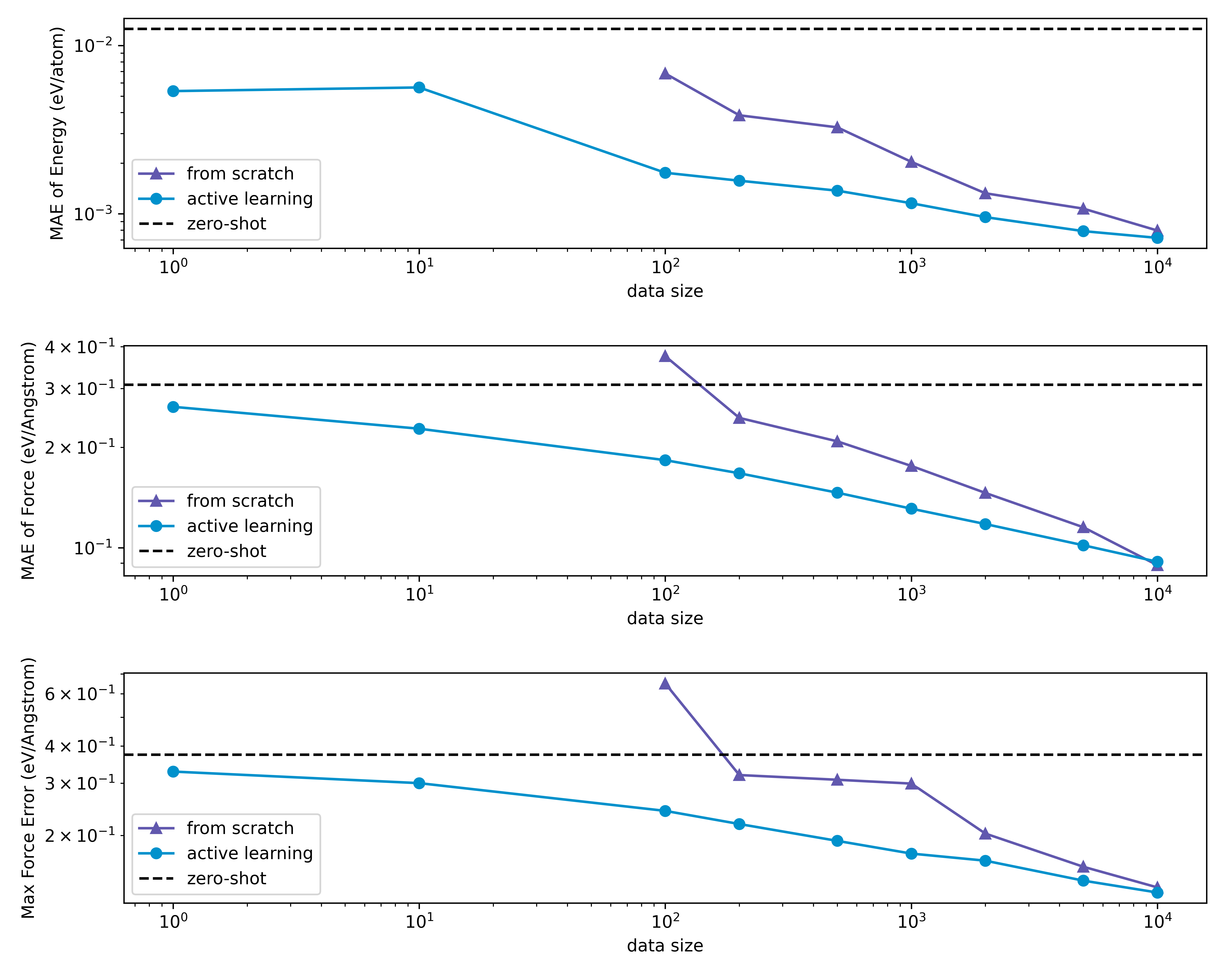}
    \caption{The accuracy by training from scratch and active learning procedure with respect to data size for \ce{Li2B12H12}.}
    \label{fig:si_LiBH_al}
\end{figure}

\begin{figure}
    \centering
    \includegraphics[width=0.8\textwidth]{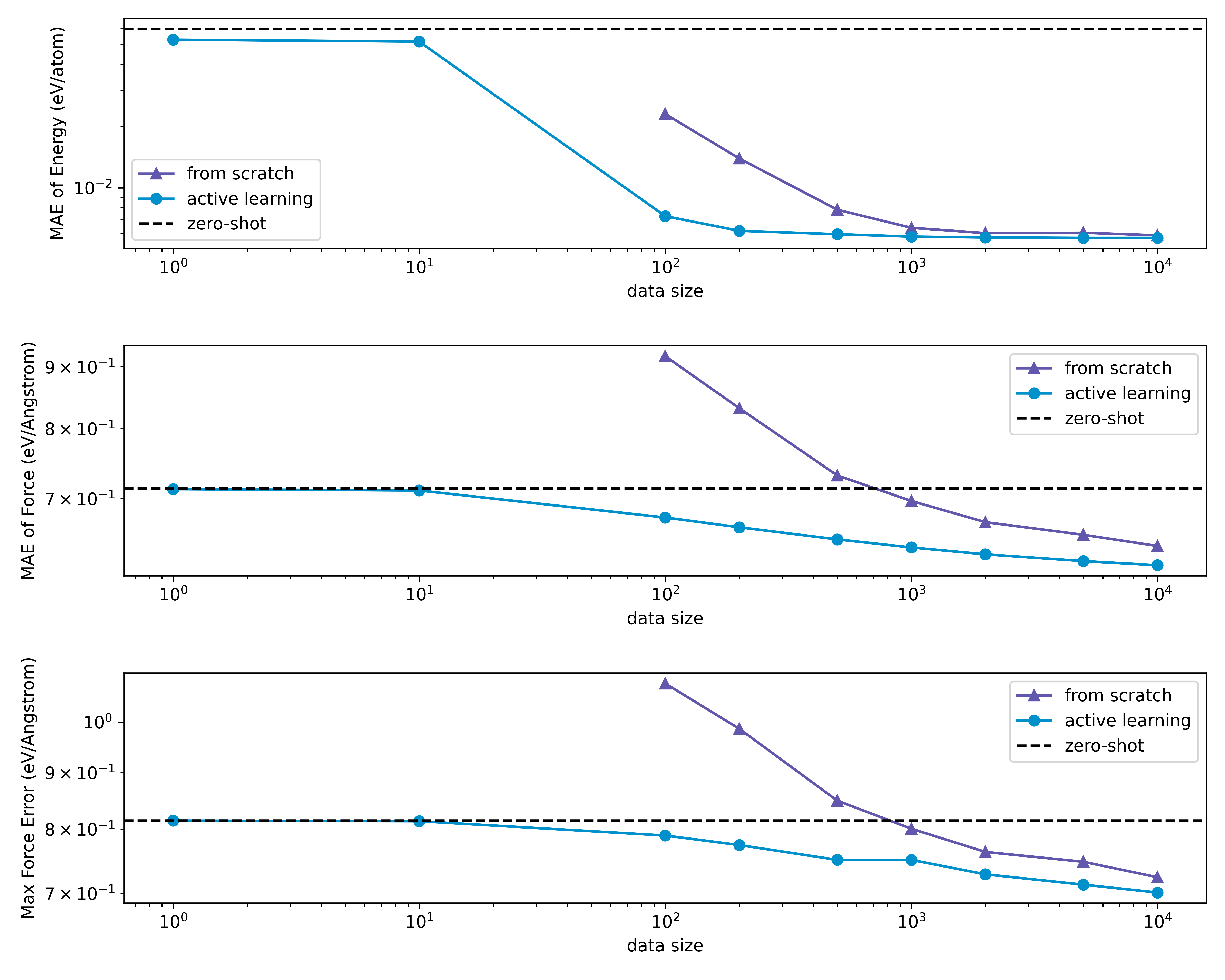}
    \caption{The accuracy by training from scratch and active learning procedure with respect to data size for molten boron.}
    \label{fig:si_B_al}
\end{figure}

\begin{figure}
    \centering
    \includegraphics[width=0.8\textwidth]{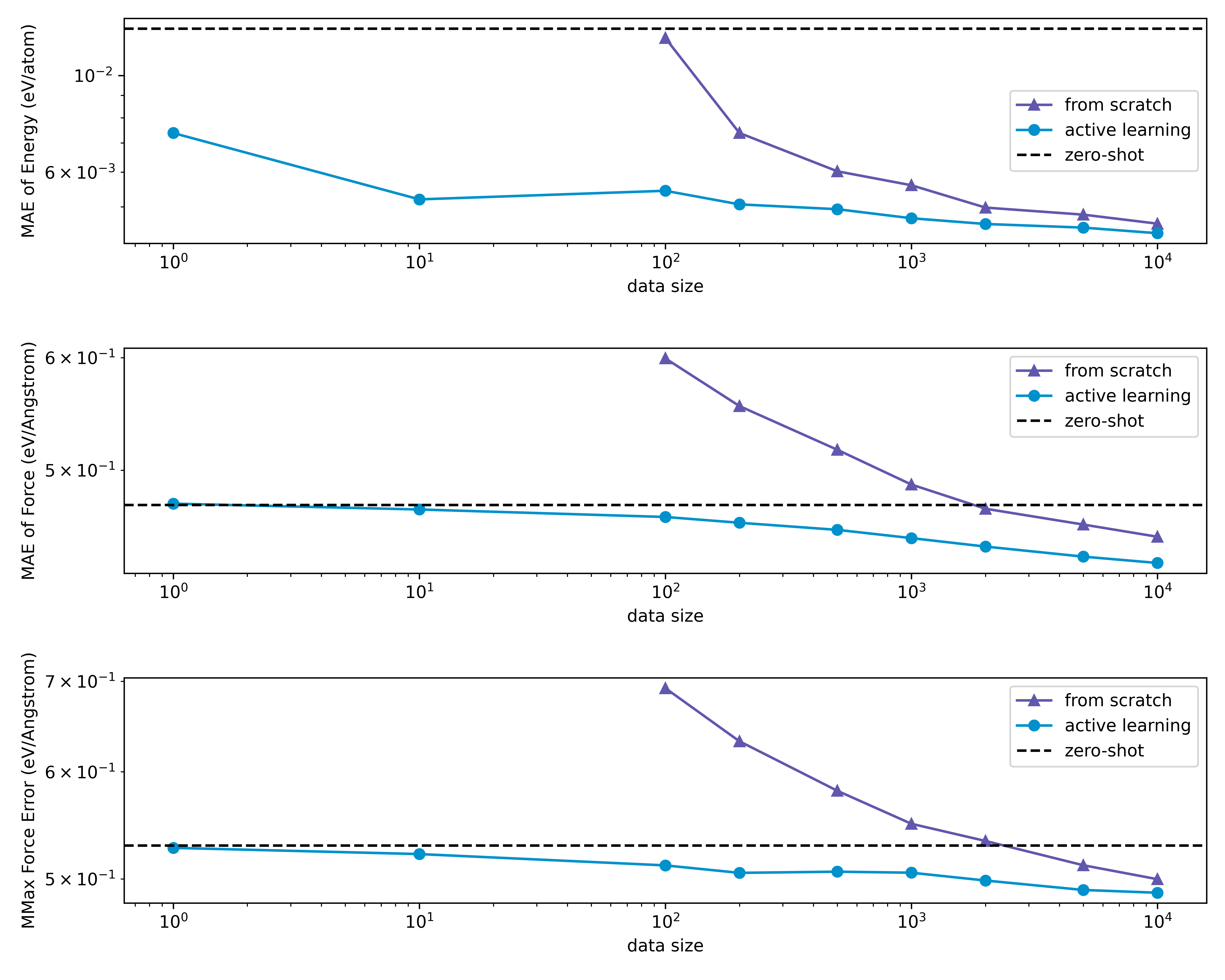}
    \caption{The accuracy by training from scratch and active learning procedure with respect to data size for molten phosphorus.}
    \label{fig:si_P_al}
\end{figure}

\autoref{fig:si_LiBH_al} illustrates the comparative accuracy achieved by training from scratch versus employing supervised finetuning through active learning, building upon the zero-shot model for the crystalline \ce{Li2B12H12}. The learning rate of active learning and training from scratch is set to \SI{1d-4}{} and \SI{1d-3}{}, respectively. The MAE of the atomic forces of the initial MatterSim as a zero-shot model is 30.8 meV/\AA, which may fall short of the desired accuracy for practical use. Nonetheless, with the inclusion of merely 100 additional data points, the MAE is significantly improved to 18.3 meV/\AA. To attain similar accuracy through training from scratch, an order of magnitude larger dataset would be required. This conclusion applies to molten boron and phosphorus as well, as ~\autoref{fig:si_B_al} and ~\autoref{fig:si_P_al} shows.

\section{Finetuning and molecular dynamics on liquid water}\label{si-sec:finetuning-and-molecular-dynamics-on-liquid-water}
In the following section, we provide further information about applying finetuning to simulate different properties for liquid water. To begin with, we detail the parameter settings used in the finetuning process and MD simulations. The latter section explains the post-processing of angular distribution function (ADF) of oxygen-oxygen-oxygen, illustrates the oxygen-hydrogen and hydrogen-hydrogen radial distribution functions (RDF), explores the data efficiency, as well as formulating the diffusion coefficients. For simplicity, we abbreviate three trained models: zero-shot, scratch-900 and finetune-30. The zero-shot model is the model trained with PBE level theory and demonstrated in the main text without any finetuning . 
Scratch-900 is a model trained from scratch using 900 bulk liquid water configurations from Ref. \cite{cheng2019ab, monserrat2020liquid} with rev-PBE0-D3 level of theory. Finetune-30 is a model fine-tuned with only 30 out of those 900 configurations.

\subsection{Parameter settings for finetuning and MD simulations}\label{si-sec:finetune-water-setups} 
\begin{figure}
    \centering
    \includegraphics[width=\textwidth]{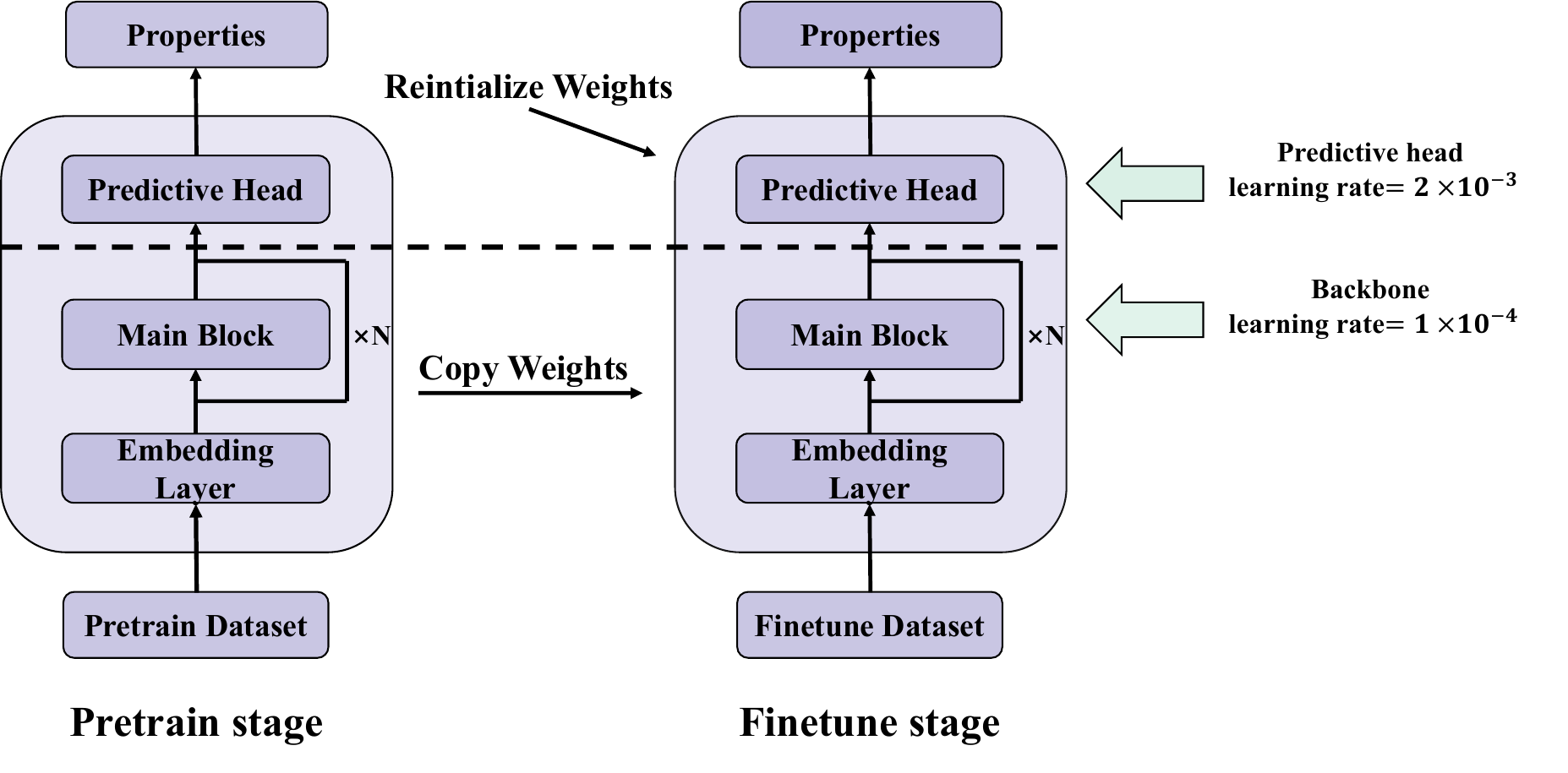}
    \caption{Pretrain--finetuning framework.}
    \label{fig:pretrain-illustration}
\end{figure}

To address the limitations imposed by the level of theory of the training data, we implemented finetuning on the MatterSim model (\autoref{fig:pretrain-illustration}).
Before the finetuning process, we first reset the initial parameters of the predictive head while retaining those of the backbone. During the training process, to maximize the transfer of MatterSim's predictability from PBE to rev-PBE0-D3 for liquid water, we adopted an aggressive learning rate of \SI{2e-3}{} to the head but a relatively lower learning rate of \SI{1e-4}{} to the backbone. We uniformly sampled 100 liquid water configurations based on the total energy as validation data, leaving the remaining 900 liquid water configurations as the potential candidates of training data. Among the 900 available candidates, 30 configurations were selected at random to create various training sets through the alteration of random seeds, which were subsequently employed to fine-tune the MatterSim model. Upon sufficient convergence, the finetuning process early stopped at \SI{151}{epochs} with an MAE of \SI{2.1}{meV/atom} for energies and \SI{58.9}{meV/\AA} for forces. 

\subsection{Simulation settings for molecular dynamics}
In this work, we evaluated the performance of finetuning scheme by probing the structural and dynamical properties of liquid water. MD simulations are conducted up to nanoseconds using the LAMMPS software package.\cite{thompson2022lammps}
The initial structure used for these simulations is composed of a cubic liquid water box containing 512 water molecules with a box length of \SI{24.68}{\AA}. 
Production runs of the MD simulations are carried out in the NVT ensemble at \SI{298}{K} and we regulate the temperature using the Nos{\'e}-Hoover thermostat\cite{parrinello1981polymorphic,nose1984unified, martyna1994constant, capinski1997thermal,shinoda2004rapid} for every 100 steps. The timestep used to propagate the dynamics is chosen to be \SI{0.5}{fs}. Lastly, the first \SI{200}{ps} out of the nanosecond trajectories are discarded for pre-equlibration.
ADF and RDF, along with diffusion coefficients, are analyzed using the General Purpose Trajectory Analyser (GPTA) software tool.\cite{GPTARepo2023}

\subsection{ADF and RDF of Liquid Water \texorpdfstring{$\&$}  \ \ Data Efficiency of Fine-tuneing}

\begin{figure}[ht!]
    \centering
    \includegraphics{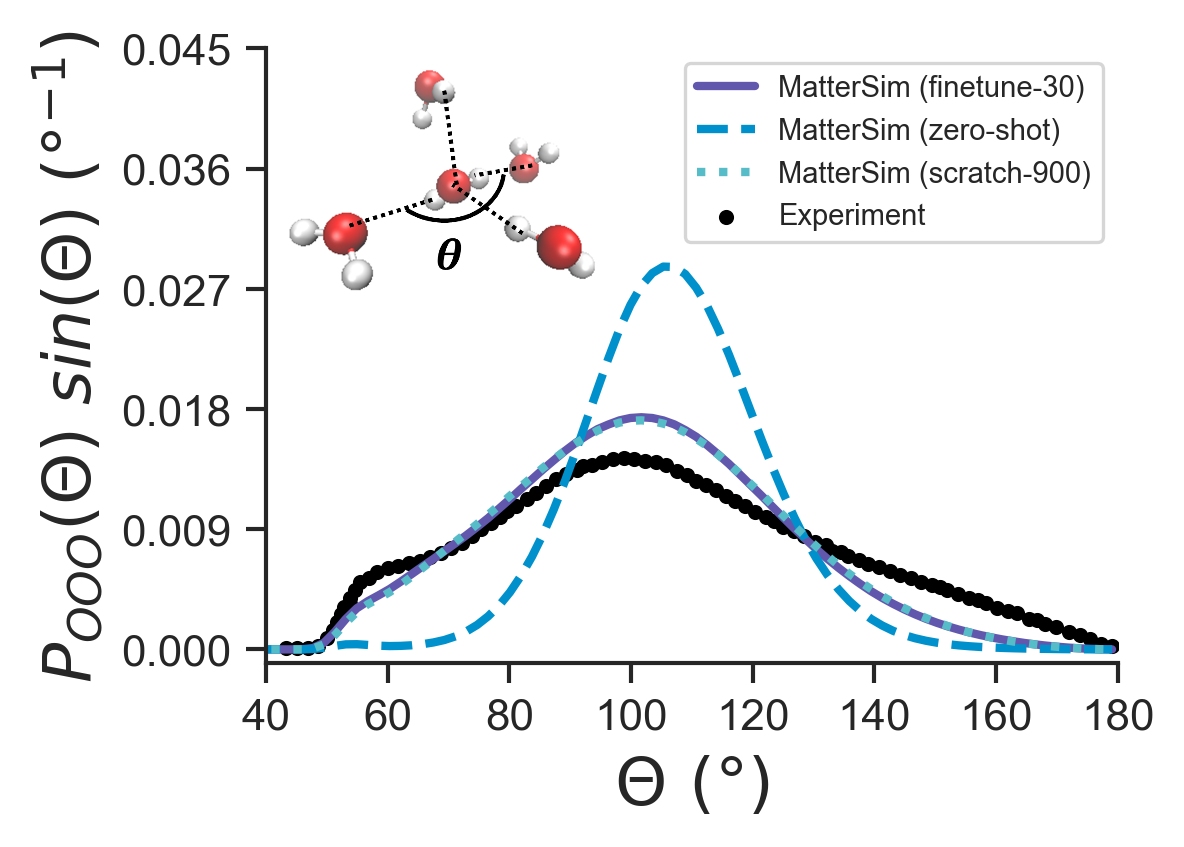}
    \caption{Comparison between $P_{OOO} (\Theta)$ obtained from MD simulations using the three models and the EPSR of joint Xray-Neutron measurements for bulk water at \SI{298}{K}.\cite{soper2008quantum} The inset showcases the $\Theta$ angle used to calculate $P_{OOO} (\Theta)$ and $\sin(\Theta)$. }
    \label{fig:ADF-O-O-O-zoomed-in}
\end{figure}

As illustrated \autoref{fig:ADF-O-O-O-zoomed-in}, we multiple ADF of the oxygen species $\left( P_{OOO}(\Theta)\right)$ by the sine angle composed of the corresponding oxygen triples. Such a representation has been adopted to elucidate the local arrangement of water molecules in condense phase, as $P_{OOO}\left(\Theta\right)\sin\left(\Theta\right)$ allows for a direct comparison with angular distribution extracted from
empirical potential structural refinement (EPSR) based on joint X-ray/neutron scattering measurements.\cite{soper2008quantum}. Following the same procedure outlined in the previous studies\cite{distasio2014individual, gaiduk2018first, zheng2018structural}, $P_{OOO}(\Theta)$ has been normalized such that $\int_{0}^{\pi} \ P_{OOO}(\Theta) \sin(\Theta) \ d\Theta$ goes to unity. In a similar note, the cutoff value applied to identify the oxygen triples is chosen such that oxygen-oxygen coordination number averages around 4.\cite{distasio2014individual} Detailed discussion can be found in the main text regarding \autoref{fig:al-and-ft}(d). 

\begin{figure}[ht!]
    \centering
    
    \begin{subfigure}[b]{0.45\textwidth}
    \includegraphics[width=\textwidth]{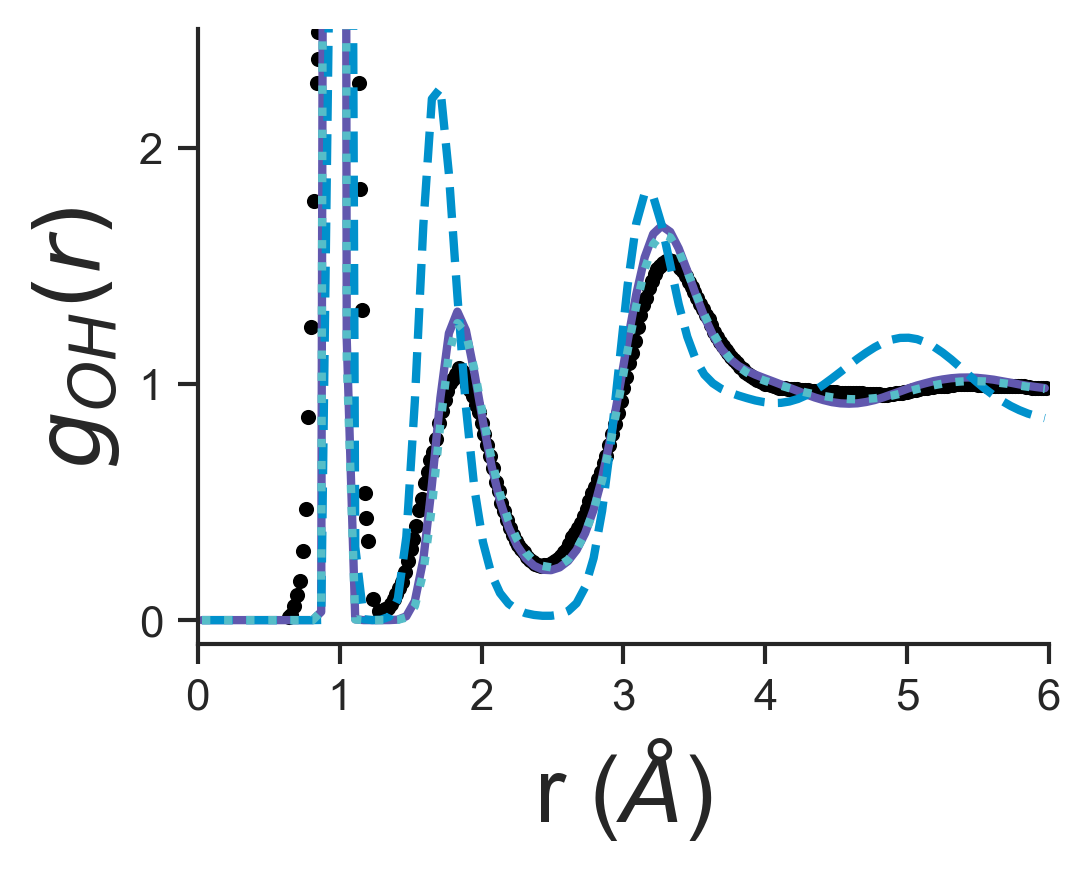}
    \caption{}
    \end{subfigure}
    \begin{subfigure}[b]{0.45\textwidth}
    \includegraphics[width=\textwidth]{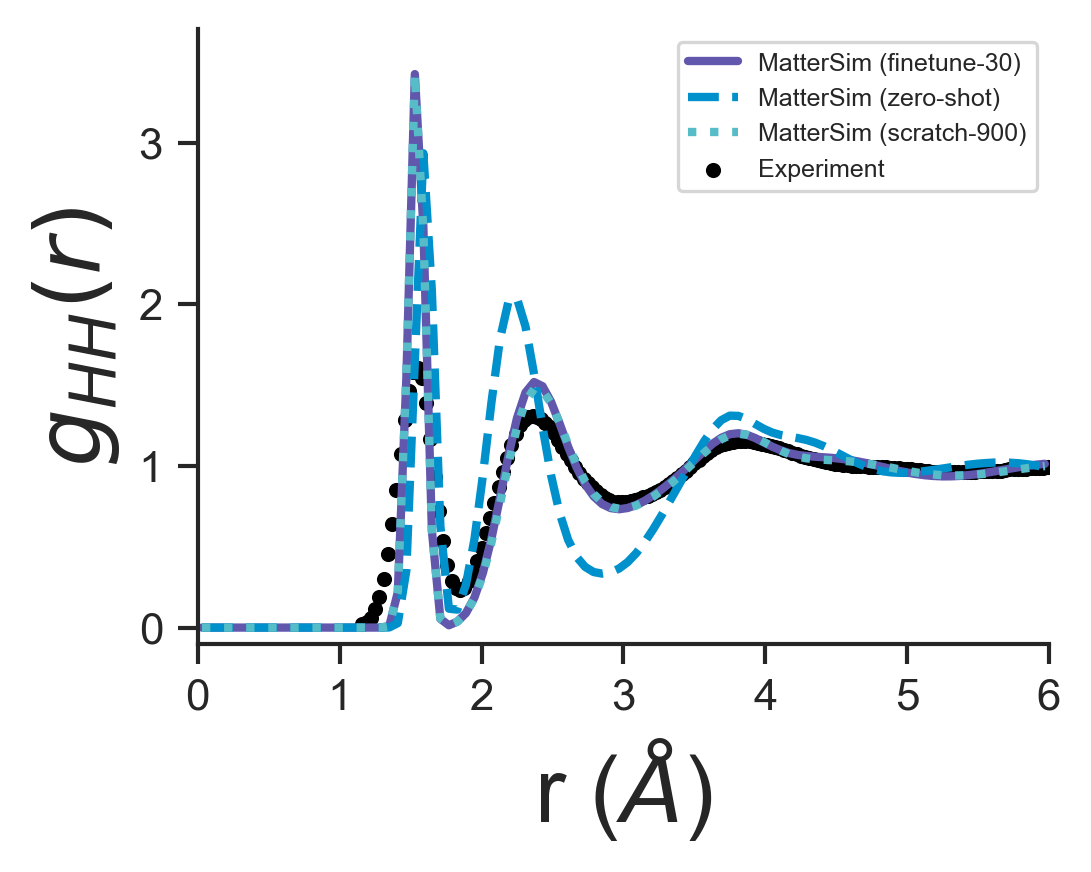}
    \caption{}
    \end{subfigure}
    \caption{Oxygen-hydrogen (a) and hydrogen-hydrogen (b) RDFs obtained from MD simulations performed by the three models. Black dots represent experimental references\cite{skinner2014structure,chen2016ab}}
    \label{fig:rdf-finetune-oh-and-hh}
\end{figure}

\autoref{fig:rdf-finetune-oh-and-hh} presents a comparison between the RDF of $g_{OH}(r)$ and $g_{HH}(r)$ from experiments and those obtained from MD simulations employing each of the three models. Noticeably, RDFs from the three models under-predict the broadening of the first peaks of $g_{OH}(r) \ \& \ g_{HH}(r)$ due to the exclusion of nuclear quantum effects (NQEs), which has been investigated thoroughly in various MD studies with either DFT or specialized machine learning potentials for water and ice\cite{fritsch2014nuclear, michele2016nuclear, cheng2019ab, chen2023thermodynamics}. Yet, exploring and capturing NQEs are beyond the scope of the current work.

\begin{figure}[ht!]
   \centering
   \includegraphics[width=0.5\textwidth]{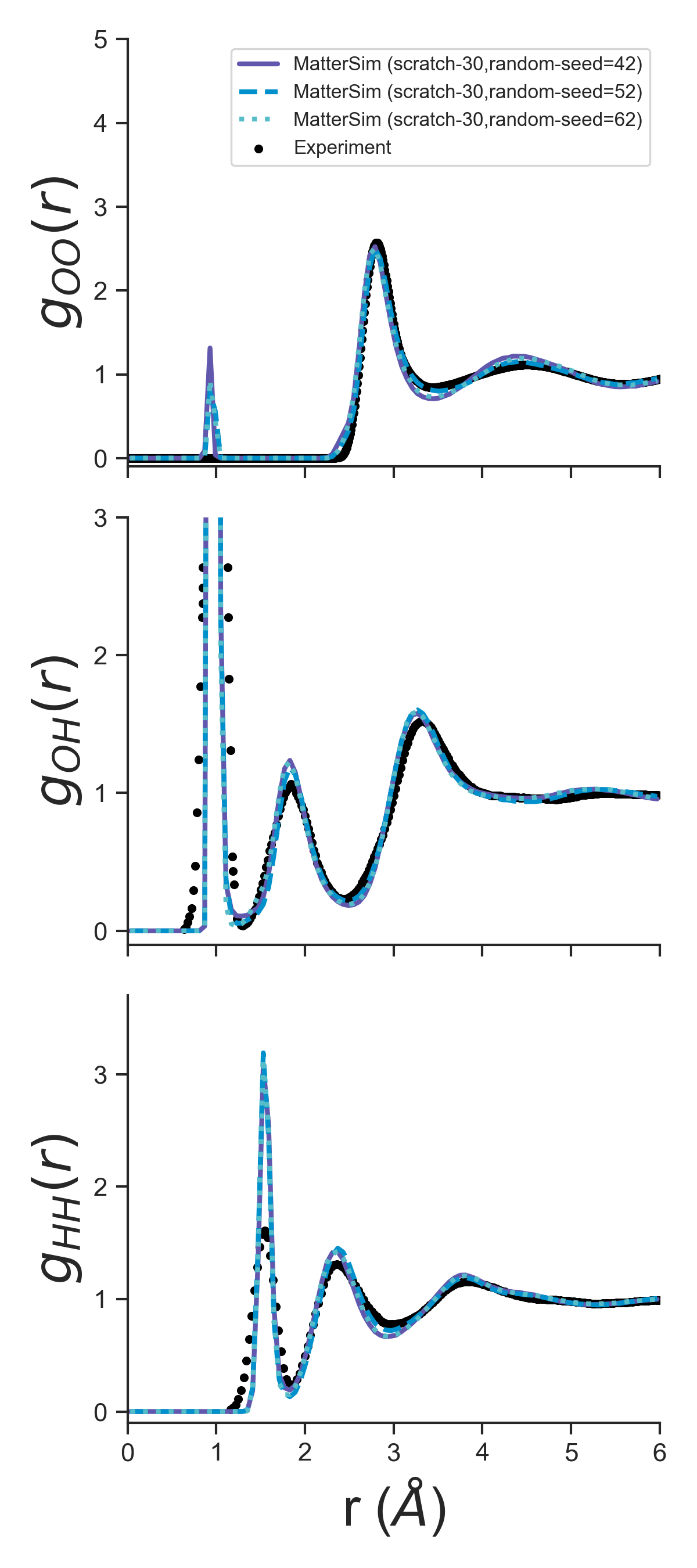}
    \caption{RDF for liquid water at \SI{300}{K} from scratch-30 models in comparison to experimental references.\cite{skinner2014structure,chen2016ab}}
    \label{fig:rdf-rs}
\end{figure}

To explore the efficiency of finetuning process, we here trained the MatterSim model from scratch using the same 30 configurations (here denote as scratch-30) as those used in the finetuning. As shown in  \autoref{fig:rdf-rs}, despite reasonably predicted the RDF of $g_{OH}(r)$ and $g_{HH}(r)$, the scratch-30 model yields nonphysical $g_{OO}(r)$ peak that over-coordinates around \SI{1.1}{\text{\AA}}. Upon training three scratch-30 models with different random seeds, their corresponding $g_{OO}(r)$ RDFs still possess the nonphysical peak at \SI{1.1}{\text{\AA}}, justifying that these over-coordination features are not resulted from the bias of a specific data split. We notice that this nonphysical peak exists even when we train from scratch using 800 configurations and disappear until using all the 900 configurations to train the model from scratch. Conversely, RDFs are indistinguishable when derived from the finetune-30 and scratch-900 model. This trend underscores a significant data efficiency improvement by a factor of 30 through fine-tuning.

\subsection{Deriving Diffusion Coefficients from Mean Squared Displacements}\label{si-section:al-and-ft:liquid-water-dissusion}

Diffusion coefficients $\left( \mathcal{D}\right)$ of liquid water at \SI{300}{K} are determined via the Einstein relation\cite{einstein1905annalen, maginn2019best}:

\begin{equation}\label{eq:Einstein_relation}  
\mathcal{D}_\mathrm{raw} = \frac{1}{6}\lim_{\tau \rightarrow \infty}  \frac{\mathrm{d}\lambda}{\mathrm{d}\tau}, 
\end{equation}

where $\lambda$ stands for the mean squared displacements along a MD trajectory, and $\tau$ represents the correlation time chosen. \autoref{eq:Einstein_relation} suggests that $\mathcal{D}_\mathrm{raw}$ can be obtained by extracting the slope from a linear fit between $\lambda$ and $\tau$. In this work, $\lambda$ is computed along the same trajectory used to compute the RDFs and ADFs. To properly accounting for the finite size effect, a posterior correction\cite{dunweg1993molecular, yeh2004system} can be introduced:

\begin{equation}\label{eq:finite_size_corrections}
\mathcal{D}_\mathrm{corrected} = \mathcal{D}_\mathrm{raw}  + \frac{k_{B}T\varepsilon}{6\pi} \frac{1}{\eta L},  
\end{equation}

where $k_{B}$ and $T$ denote Boltzmann constant and temperature respectively. $\eta$ indicates the shear viscosity. The experimental measurements of $\eta \approx  \SI{0.89}{mPa\cdot s}$ for liquid water at \SI{298}{K}\cite{vega2011simulating} is substituted into \autoref{eq:finite_size_corrections} when obtaining $\mathcal{D}_\mathrm{corrected}$. $\varepsilon$ and $L$ represent the shape factor and length of the simulation box respectively. For a cubic box, $\varepsilon \approx2.873297$.  \autoref{tab:Summary_diffusion_coefficients} summaries $\mathcal{D}$ obtained using the finetune-30 and scratch-900 models, in comparison to the experimental references\cite{mills1973self, krynicki1978pressure, hardy2001isotope}.

\begin{table}[ht!]
    \centering
    \begin{tabular}{c c c }
    \toprule
    Type  & $\mathcal{D} $ (\SI{e-5}{\centi\meter^2\per\second}) & Source \\
    \hline
     $\mathcal{D}_\text{Experiment}$ & $2.3\sim2.4$ & \cite{mills1973self, krynicki1978pressure, hardy2001isotope}\\
     $\mathcal{D}_\text{MatterSim from scratch, raw}$ & $2.117\pm0.036$ &\textbf{This work} \\
     $\mathcal{D}_\text{MatterSim from scratch, corrected}$ & $2.402\pm0.036$ &\textbf{This work} \\
     $\mathcal{D}_\text{MatterSim finetune, raw}$ & $1.576\pm0.032$ &\textbf{This work} \\
     $\mathcal{D}_\text{MatterSim finetune, corrected}$ & $1.862\pm0.032$ &\textbf{This work} \\
    \bottomrule
    \end{tabular}
    \caption{Summary of $\mathcal{D}$ from different approaches. $\mathcal{D}$ presented in the table are from the average and standard errors of $\mathcal{D}$ computed at $\tau=2.5, 5, 25, 50, 100$ and $200$ ps. %
    The \textit{corrected}  and  \textit{raw} subscript indicate $\mathcal{D}$ obtained with and without imposing finite size corrections.} %
    \label{tab:Summary_diffusion_coefficients}
\end{table}

\section{End-to-end property prediction}\label{si-sec:end-to-end-model}
\subsection{Matbench} 

Matbench \cite{dunn2020benchmarking}  is an open leaderboard to test the capability of machine learning models to predict properties of inorganic materials. It contains 13 tasks that cover a wide range of materials' characteristics originating from both experimental measurements and first-principles calculations, including optical, thermal, and mechanical properties. In this section, we focus on the following structure-to-property tasks:

\textbf{MP Gap.} This task involves predicting the electronic band gap in electrovolts (eV) for inorganic compounds. The dataset contains 106,113 materials collected from the Materials Project\cite{jain2013materials}. The dataset has been curated to exclude structures with high formation energy or noble gas elements. 
   
\textbf{log\_gvrh.} This task is to predict the logarithmic of the  Voigt-Reuss-Hill average shear modulus ($G_\mathrm{VRH}$) in gigapascal (GPa). The dataset includes 10,987 materials collected from the Materials Project\cite{jain2013materials}. It has been curated to exclude those with unrealistic mechanical properties or noble gase elements.
   
\textbf{log\_kvrh} This task is to predict the logarithm of the Voigt-Reuss-Hill average bulk modulus ($\log K_\mathrm{VRH}$) in gigapascal (GPa). The dataset contains 10,987 materials collected from the Materials Project\cite{de2015charting}. The dataset has been curated to exclude structures with negative bulk moduli or noble gase elements.  
   
\textbf{Dielectric.} This task is to predict the refractive index of materials. The dataset contains 4,764 entries collected from  the Materials Project\cite{petousis2017high}. The dataset has been curated to exclude materials with low refractive indices or noble gas elements. The refractive index is unitless.
   
\textbf{Phonons.} This task is to predict the frequency of the highest optical phonon mode in wavenumber (\SI{}{\per\centi\meter}). The dataset contains for 1,265 materials collected from Ref.~\citenum{petretto2018high}. This dataset has been curated to exclude materials with high formation energy.

\textbf{jdft2d.} This task is to predict  the exfoliation energy in milli-electrovolt per atom (\SI{}{eV\per atom}) for two-dimentional materials. The dataset contains 636 materials collected from the JARVIS DFT database\cite{choudhary2017high}.

\subsection{Training MatterSim as an end-to-end model}
After the message passing in M3GNet or structure encoder in Graphormer, we obtain a global representation of a structure using scatter operation to aggregate the node feature with the following functions: 
\begin{equation}
\boldsymbol{\kappa}_\mathcal{G} =
\mathrm{Readout}(\mathcal{G})=\left\{
\begin{array}{rl}
\sum_{i\in \mathcal{G}}^N \boldsymbol{v}_i, & \text{if reduction} = \text{summation}; \\
\frac{1}{N}\sum_{i\in \mathcal{G}}^N \boldsymbol{v}_i, & \text{if reduction} = \text{mean}, \\ 
\end{array}\right.
\end{equation}
where $\mathcal{G}$ denotes a material graph consisting of $N$ atoms defined in \autoref{si-fig:mattersim-materials-graph}, $\boldsymbol{v}_i$ is the node feature for atom $i$ in graph $\mathcal{G}$. With two different reduction methods, mean or summation, we obtain the readout vectors of a given material, which will be subsequently sent to a multi-layer perceptron (MLP) to make direct property predictions,
\begin{equation}
\text{M3GNet: }\boldsymbol{\kappa}_\mathcal{G}^\prime = \mathrm{MLP}_1(\boldsymbol{\kappa}_\mathcal{G}) =
\varphi(\boldsymbol{W}_2\rho(\boldsymbol{W}_1\boldsymbol{\kappa}_\mathcal{G}+\boldsymbol{b}_1) + \boldsymbol{b}_2) ,
\end{equation}
and
\begin{equation}
\text{Graphormer: }\boldsymbol{\kappa}_\mathcal{G}^\prime = \mathrm{MLP}_2 (\boldsymbol{\kappa}_\mathcal{G}) = \boldsymbol{W}_2\lambda(\rho(\boldsymbol{W}_1\boldsymbol{\kappa}_\mathcal{G}+\boldsymbol{b}_1)) + \boldsymbol{b}_2,
\end{equation}
where $\rho(\cdot)$ and $\varphi(\cdot)$ are ReLU function and $\lambda(\cdot)$ is the layered normalization, $\boldsymbol{W}$ and $\boldsymbol{b}$ are learnable parameters, and $\boldsymbol{\kappa}_\mathcal{G}^\prime$ is the final output of the model for representing a given material.

The goal of our fine-tuning process is to adjust the parameters in a pre-trained MatterSim, originally developed for calculating structural energies, to make direct predictions on other properties of materials from their representation $\boldsymbol{\kappa}_\mathcal{G}^\prime$. This adjustment can be reformulated as a minimization problem: 
\begin{equation}
\min_{\boldsymbol{\theta}} L\left(\boldsymbol{\theta}; D_{\text {finetuning}}\right).
\end{equation}
Here $\boldsymbol{\theta}$ is the set of parameters in our model, $L$ is the loss function which calculates the error on new fine-tuning data points $D_{\text {finetuning}}$. For every finetuning step $t$, the parameters $\boldsymbol{\theta}$ will be updated using the following equation:
\begin{equation}
\boldsymbol{\theta}^{(t+1)}=\boldsymbol{\theta}^{(t)}-\alpha \nabla_{\boldsymbol{\theta}} L\left(\boldsymbol{\theta}^{(t)} ; D_{\text {finetuning}}\right)
\end{equation}
In this equation, $\alpha$ is the learning rate and $\nabla_{\boldsymbol{\theta}} L$ denotes the gradient of the loss function with respect to parameters. $L$ have been regularized with $L^2$-norm to induce the risk of over-fitting for M3GNet, which can be presented as:
\begin{equation}
L_{\text {regularized }}\left(\boldsymbol{\theta}; D_{\text {finetuning }}\right)=L\left(\boldsymbol{\theta}; D_{\text {finetuning}}\right)+\lambda\norm{\boldsymbol{\theta}}^2
\end{equation}
where $\lambda$ is the regularization coefficient, which controls the trade-off between fitting the training data and imposing smoothness on the parameter estimates. 

\subsection{Performance comparison of M3GNet and Graphormer as end-to-end models}

\begin{table}[htpb]  
    \centering  
    \begin{tabular}{c|ccccc}  
    \toprule  
    \multirow{2}{*}{Property} & M3GNet & M3GNet & Graphormer & Graphomer \\
                             & (From scratch) & (Fine-tuning) & (From scratch) & (Fine-tuning) \\
    \midrule
    MP Gap (eV) & 0.321 & 0.2646 & 0.3031 & 0.1290 \\  
    $\log G_\mathrm{VRH}$ (GPa) & 0.1563 & 0.0959 & 0.0895 & 0.0608 \\  
    $\log K_\mathrm{VRH}$ (GPa) & 0.1464 & 0.0717 & 0.0687 & 0.0488 \\  
    Dielectric (unitless) & 0.4615 & 0.3001 & 0.3823 & 0.2516 \\  
    Phonons (\si{\per\centi\meter}) & 72.3314 & 56.0441 & 65.8220  & 26.0220 \\  
    jdft2d (\si{meV\per atom}) & 77.3612 & 48.1290 & 47.8040 & 32.7620 \\  
    \bottomrule  
    \end{tabular}  
    \caption{Comparison of property prediction performance for M3GNet and Graphormer models.}
    \label{si-tab:comparison-e2e-models}  
\end{table}  

As listed in \autoref{si-tab:comparison-e2e-models}, finetuning from MatterSim, either with M3GNet or Graphormer architecture, always outperforms their counterparts training from scratch. Notably, the model that has been finetuned from the Graphormer architecture has outperformed previous models trained exclusively with domain specific data on all of the 6 tasks, as discussed in \autoref{sec:e2e-model}. This finding underscores the effectiveness of MatterSim to capture the representation of materials, and it can significantly expedite the the future research on materials property prediction and materials discovery. When we are preparing this manuscript, we notice a recent model\cite{shoghi2023molecules} by multi-task pre-training on multiple datasets achieved the best results for all these tasks. This further indicates the power of large-scale pre-training and the advantage of data coverage.

\clearpage
\phantomsection %
\addcontentsline{toc}{part}{References}
\bibliography{ms}

\begin{thebibliography}{100}
\providecommand{\url}[1]{{#1}}
\providecommand{\urlprefix}{URL }
\providecommand{\doi}[1]{\url{https://doi.org/#1}}
\bibcommenthead

\bibitem{fiori2014electronics}
G.~Fiori, F.~Bonaccorso, G.~Iannaccone, T.~Palacios, D.~Neumaier, A.~Seabaugh,
  S.K. Banerjee, L.~Colombo, Electronics based on two-dimensional materials.
\newblock Nature nanotechnology \textbf{9}(10), 768--779 (2014)

\bibitem{li2007electronic}
T.~Li, G.~Galli, Electronic properties of mos2 nanoparticles.
\newblock The Journal of Physical Chemistry C \textbf{111}(44), 16192--16196
  (2007)

\bibitem{mizushima1980lixcoo2}
K.~Mizushima, P.~Jones, P.~Wiseman, J.B. Goodenough, Lixcoo2 (0< x<-1): A new
  cathode material for batteries of high energy density.
\newblock Materials Research Bulletin \textbf{15}(6), 783--789 (1980)

\bibitem{ceder1998identification}
G.~Ceder, Y.M. Chiang, D.~Sadoway, M.~Aydinol, Y.I. Jang, B.~Huang,
  Identification of cathode materials for lithium batteries guided by
  first-principles calculations.
\newblock Nature \textbf{392}(6677), 694--696 (1998)

\bibitem{tibbitt2015progress}
M.W. Tibbitt, C.B. Rodell, J.A. Burdick, K.S. Anseth, Progress in material
  design for biomedical applications.
\newblock Proceedings of the National Academy of Sciences \textbf{112}(47),
  14444--14451 (2015)

\bibitem{li2018review}
X.~Li, J.~Xie, C.~Jiang, J.~Yu, P.~Zhang, Review on design and evaluation of
  environmental photocatalysts.
\newblock Frontiers of Environmental Science \& Engineering \textbf{12}, 1--32
  (2018)

\bibitem{hu2010design}
X.~Hu, G.~Li, J.C. Yu, Design, fabrication, and modification of nanostructured
  semiconductor materials for environmental and energy applications.
\newblock Langmuir \textbf{26}(5), 3031--3039 (2010)

\bibitem{curtarolo2013high}
S.~Curtarolo, G.L. Hart, M.B. Nardelli, N.~Mingo, S.~Sanvito, O.~Levy, The
  high-throughput highway to computational materials design.
\newblock Nature materials \textbf{12}(3), 191--201 (2013)

\bibitem{choudhary2022recent}
K.~Choudhary, B.~DeCost, C.~Chen, A.~Jain, F.~Tavazza, R.~Cohn, C.W. Park,
  A.~Choudhary, A.~Agrawal, S.J. Billinge, et~al., Recent advances and
  applications of deep learning methods in materials science.
\newblock npj Computational Materials \textbf{8}(1), 59 (2022)

\bibitem{xie2018crystal}
T.~Xie, J.C. Grossman, Crystal graph convolutional neural networks for an
  accurate and interpretable prediction of material properties.
\newblock Physical review letters \textbf{120}(14), 145301 (2018)

\bibitem{merchant2023scaling}
A.~Merchant, S.~Batzner, S.S. Schoenholz, M.~Aykol, G.~Cheon, E.D. Cubuk,
  Scaling deep learning for materials discovery.
\newblock Nature pp. 1--6 (2023)

\bibitem{chen2024accelerating}
C.~Chen, D.T. Nguyen, S.J. Lee, N.A. Baker, A.S. Karakoti, L.~Lauw, C.~Owen,
  K.T. Mueller, B.A. Bilodeau, V.~Murugesan, et~al., Accelerating computational
  materials discovery with artificial intelligence and cloud high-performance
  computing: from large-scale screening to experimental validation.
\newblock arXiv preprint arXiv:2401.04070  (2024)

\bibitem{lindsey2017chimes}
R.K. Lindsey, L.E. Fried, N.~Goldman, Chimes: A force matched potential with
  explicit three-body interactions for molten carbon.
\newblock Journal of chemical theory and computation \textbf{13}(12),
  6222--6229 (2017)

\bibitem{schutt2017schnet}
K.~Sch{\"u}tt, P.J. Kindermans, H.E. Sauceda~Felix, S.~Chmiela, A.~Tkatchenko,
  K.R. M{\"u}ller, Schnet: A continuous-filter convolutional neural network for
  modeling quantum interactions.
\newblock Advances in neural information processing systems \textbf{30} (2017)

\bibitem{musaelian2023learning}
A.~Musaelian, S.~Batzner, A.~Johansson, L.~Sun, C.J. Owen, M.~Kornbluth,
  B.~Kozinsky, Learning local equivariant representations for large-scale
  atomistic dynamics.
\newblock Nature Communications \textbf{14}(1), 579 (2023)

\bibitem{batzner20223}
S.~Batzner, A.~Musaelian, L.~Sun, M.~Geiger, J.P. Mailoa, M.~Kornbluth,
  N.~Molinari, T.E. Smidt, B.~Kozinsky, E (3)-equivariant graph neural networks
  for data-efficient and accurate interatomic potentials.
\newblock Nature communications \textbf{13}(1), 2453 (2022)

\bibitem{chen2019graph}
C.~Chen, W.~Ye, Y.~Zuo, C.~Zheng, S.P. Ong, Graph networks as a universal
  machine learning framework for molecules and crystals.
\newblock Chemistry of Materials \textbf{31}(9), 3564--3572 (2019)

\bibitem{choudhary2021atomistic}
K.~Choudhary, B.~DeCost, Atomistic line graph neural network for improved
  materials property predictions.
\newblock npj Computational Materials \textbf{7}(1), 185 (2021)

\bibitem{chen2022universal}
C.~Chen, S.P. Ong, A universal graph deep learning interatomic potential for
  the periodic table.
\newblock Nature Computational Science \textbf{2}(11), 718--728 (2022)

\bibitem{deng2023chgnet}
B.~Deng, P.~Zhong, K.~Jun, J.~Riebesell, K.~Han, C.J. Bartel, G.~Ceder, Chgnet
  as a pretrained universal neural network potential for charge-informed
  atomistic modelling.
\newblock Nature Machine Intelligence \textbf{5}(9), 1031--1041 (2023)

\bibitem{batatia2023foundation}
I.~Batatia, P.~Benner, Y.~Chiang, A.M. Elena, D.P. Kov{\'a}cs, J.~Riebesell,
  X.R. Advincula, M.~Asta, W.J. Baldwin, N.~Bernstein, et~al., A foundation
  model for atomistic materials chemistry.
\newblock arXiv preprint arXiv:2401.00096  (2023)

\bibitem{zhang2023dpa}
D.~Zhang, X.~Liu, X.~Zhang, C.~Zhang, C.~Cai, H.~Bi, Y.~Du, X.~Qin, J.~Huang,
  B.~Li, et~al., Dpa-2: Towards a universal large atomic model for molecular
  and material simulation.
\newblock arXiv preprint arXiv:2312.15492  (2023)

\bibitem{shoghi2023molecules}
N.~Shoghi, A.~Kolluru, J.R. Kitchin, Z.W. Ulissi, C.L. Zitnick, B.M. Wood, From
  molecules to materials: Pre-training large generalizable models for atomic
  property prediction.
\newblock arXiv preprint arXiv:2310.16802  (2023)

\bibitem{kresse1996efficient}
G.~Kresse, J.~Furthm{\"u}ller, Efficient iterative schemes for ab initio
  total-energy calculations using a plane-wave basis set.
\newblock Physical review B \textbf{54}(16), 11169 (1996)

\bibitem{dunn2020benchmarking}
A.~Dunn, Q.~Wang, A.~Ganose, D.~Dopp, A.~Jain, Benchmarking materials property
  prediction methods: the matbench test set and automatminer reference
  algorithm.
\newblock npj Computational Materials \textbf{6}(1), 138 (2020)

\bibitem{kresse1996efficiency}
G.~Kresse, J.~Furthm{\"u}ller, Efficiency of ab-initio total energy
  calculations for metals and semiconductors using a plane-wave basis set.
\newblock Computational materials science \textbf{6}(1), 15--50 (1996)

\bibitem{kohn1965self}
W.~Kohn, L.J. Sham, Self-consistent equations including exchange and
  correlation effects.
\newblock Physical review \textbf{140}(4A), A1133 (1965)

\bibitem{hohenberg1964inhomogeneous}
P.~Hohenberg, W.~Kohn, Inhomogeneous electron gas.
\newblock Physical review \textbf{136}(3B), B864 (1964)

\bibitem{perdew1996generalized}
J.P. Perdew, K.~Burke, M.~Ernzerhof, Generalized gradient approximation made
  simple.
\newblock Physical review letters \textbf{77}(18), 3865 (1996)

\bibitem{anisimov1991band}
V.I. Anisimov, J.~Zaanen, O.K. Andersen, Band theory and mott insulators:
  Hubbard u instead of stoner i.
\newblock Physical Review B \textbf{44}(3), 943 (1991)

\bibitem{jain2013commentary}
A.~Jain, S.P. Ong, G.~Hautier, W.~Chen, W.D. Richards, S.~Dacek, S.~Cholia,
  D.~Gunter, D.~Skinner, G.~Ceder, et~al., Commentary: The materials project: A
  materials genome approach to accelerating materials innovation.
\newblock APL materials \textbf{1}(1) (2013)

\bibitem{saal2013materials}
J.E. Saal, S.~Kirklin, M.~Aykol, B.~Meredig, C.~Wolverton, Materials design and
  discovery with high-throughput density functional theory: the open quantum
  materials database (oqmd).
\newblock Jom \textbf{65}, 1501--1509 (2013)

\bibitem{kirklin2015open}
S.~Kirklin, J.E. Saal, B.~Meredig, A.~Thompson, J.W. Doak, M.~Aykol,
  S.~R{\"u}hl, C.~Wolverton, The open quantum materials database (oqmd):
  assessing the accuracy of dft formation energies.
\newblock npj Computational Materials \textbf{1}(1), 1--15 (2015)

\bibitem{schmidt2023machine}
J.~Schmidt, N.~Hoffmann, H.C. Wang, P.~Borlido, P.J. Carri{\c{c}}o, T.F.
  Cerqueira, S.~Botti, M.A. Marques, Machine-learning-assisted determination of
  the global zero-temperature phase diagram of materials.
\newblock Advanced Materials \textbf{35}(22), 2210788 (2023)

\bibitem{ying2021transformers}
C.~Ying, T.~Cai, S.~Luo, S.~Zheng, G.~Ke, D.~He, Y.~Shen, T.Y. Liu, Do
  transformers really perform badly for graph representation?
\newblock Advances in neural information processing systems \textbf{34},
  28877--28888 (2021)

\bibitem{shi2022benchmarking}
Y.~Shi, S.~Zheng, G.~Ke, Y.~Shen, J.~You, J.~He, S.~Luo, C.~Liu, D.~He, T.Y.
  Liu, Benchmarking graphormer on large-scale molecular modeling datasets.
\newblock arXiv preprint arXiv:2203.04810  (2022)

\bibitem{riebesell_pymatviz_2022}
J.~Riebesell, H.~Yang, R.~Goodall, S.G. Baird.
\newblock Pymatviz: visualization toolkit for materials informatics (2022).
\newblock \doi{10.5281/zenodo.7486816}.
\newblock \urlprefix\url{https://github.com/janosh/pymatviz}.
\newblock 10.5281/zenodo.7486816 - https://github.com/janosh/pymatviz

\bibitem{githubstrict_anionsOption}
M.~Horton.
\newblock Add strict\_anions option to MaterialsProject2020Compatibility by
  mkhorton (2024).
\newblock Accessed May 07, 2024

\bibitem{zeni2023mattergen}
C.~Zeni, R.~Pinsler, D.~Z{\"u}gner, A.~Fowler, M.~Horton, X.~Fu, S.~Shysheya,
  J.~Crabb{\'e}, L.~Sun, J.~Smith, et~al., Mattergen: a generative model for
  inorganic materials design.
\newblock arXiv preprint arXiv:2312.03687  (2023)

\bibitem{xie2021crystal}
T.~Xie, X.~Fu, O.E. Ganea, R.~Barzilay, T.~Jaakkola, Crystal diffusion
  variational autoencoder for periodic material generation.
\newblock arXiv preprint arXiv:2110.06197  (2021)

\bibitem{pickard2011ab}
C.J. Pickard, R.~Needs, Ab initio random structure searching.
\newblock Journal of Physics: Condensed Matter \textbf{23}(5), 053201 (2011)

\bibitem{schmidt2022dataset}
J.~Schmidt, H.C. Wang, T.F. Cerqueira, S.~Botti, M.A. Marques, A dataset of
  175k stable and metastable materials calculated with the pbesol and scan
  functionals.
\newblock Scientific Data \textbf{9}(1), 64 (2022)

\bibitem{schmidt2022large}
J.~Schmidt, N.~Hoffmann, H.C. Wang, P.~Borlido, P.J. Carri{\c{c}}o, T.F.
  Cerqueira, S.~Botti, M.A. Marques, Large-scale machine-learning-assisted
  exploration of the whole materials space.
\newblock arXiv preprint arXiv:2210.00579  (2022)

\bibitem{bergerhoff1987crystallographic}
G.~Bergerhoff, I.~Brown, F.~Allen, et~al., Crystallographic databases.
\newblock International Union of Crystallography, Chester \textbf{360}, 77--95
  (1987)

\bibitem{bloch1929quantenmechanik}
F.~Bloch, {\"U}ber die quantenmechanik der elektronen in kristallgittern.
\newblock Zeitschrift f{\"u}r physik \textbf{52}(7), 555--600 (1929)

\bibitem{baroni2001phonons}
S.~Baroni, S.~De~Gironcoli, A.~Dal~Corso, P.~Giannozzi, Phonons and related
  crystal properties from density-functional perturbation theory.
\newblock Reviews of modern Physics \textbf{73}(2), 515 (2001)

\bibitem{baroni1987green}
S.~Baroni, P.~Giannozzi, A.~Testa, Green’s-function approach to linear
  response in solids.
\newblock Physical review letters \textbf{58}(18), 1861 (1987)

\bibitem{giannozzi1991ab}
P.~Giannozzi, S.~De~Gironcoli, P.~Pavone, S.~Baroni, Ab initio calculation of
  phonon dispersions in semiconductors.
\newblock Physical Review B \textbf{43}(9), 7231 (1991)

\bibitem{kresse1995ab}
G.~Kresse, J.~Furthm{\"u}ller, J.~Hafner, Ab initio force constant approach to
  phonon dispersion relations of diamond and graphite.
\newblock Europhysics Letters \textbf{32}(9), 729 (1995)

\bibitem{yang2021combined}
H.~Yang, M.~Govoni, A.~Kundu, G.~Galli, Combined first-principles calculations
  of electron--electron and electron--phonon self-energies in condensed
  systems.
\newblock Journal of Chemical Theory and Computation \textbf{17}(12),
  7468--7476 (2021)

\bibitem{yang2022computational}
H.~Yang, M.~Govoni, A.~Kundu, G.~Galli, Computational protocol to evaluate
  electron--phonon interactions within density matrix perturbation theory.
\newblock Journal of Chemical Theory and Computation \textbf{18}(10),
  6031--6042 (2022)

\bibitem{fang2024phonon}
S.~Fang, M.~Geiger, J.G. Checkelsky, T.~Smidt.
\newblock Phonon predictions with e(3)-equivariant graph neural networks (2024)

\bibitem{PhononDB}
A.~Togo.
\newblock Atsushi togo.
\newblock \urlprefix\url{https://doi.org/10.48505/nims.4197}

\bibitem{tolborg2022free}
K.~Tolborg, J.~Klarbring, A.M. Ganose, A.~Walsh, Free energy predictions for
  crystal stability and synthesisability.
\newblock Digital Discovery \textbf{1}(5), 586--595 (2022)

\bibitem{bartel2018physical}
C.J. Bartel, S.L. Millican, A.M. Deml, J.R. Rumptz, W.~Tumas, A.W. Weimer,
  S.~Lany, V.~Stevanovi{\'c}, C.B. Musgrave, A.M. Holder, Physical descriptor
  for the gibbs energy of inorganic crystalline solids and
  temperature-dependent materials chemistry.
\newblock Nature communications \textbf{9}(1), 4168 (2018)

\bibitem{dubrovinskaia2019b1}
N.~Dubrovinskaia, S.~Petitgirard, S.~Chariton, R.~Tucoulou, J.~Garrevoet,
  K.~Glazyrin, H.P. Liermann, V.B. Prakapenka, L.~Dubrovinsky, B1-b2 phase
  transition in mgo at ultra-high static pressure.
\newblock arXiv preprint arXiv:1904.00476  (2019)

\bibitem{zhang2023toward}
S.~Zhang, R.~Paul, S.~Hu, M.A. Morales, Toward an accurate equation of state
  and b1-b2 phase boundary for magnesium oxide up to terapascal pressures and
  electron-volt temperatures.
\newblock Physical Review B \textbf{107}(22), 224109 (2023)

\bibitem{mcwilliams2012phase}
R.S. McWilliams, D.K. Spaulding, J.H. Eggert, P.M. Celliers, D.G. Hicks, R.F.
  Smith, G.W. Collins, R.~Jeanloz, Phase transformations and metallization of
  magnesium oxide at high pressure and temperature.
\newblock Science \textbf{338}(6112), 1330--1333 (2012)

\bibitem{unke2021machine}
O.T. Unke, S.~Chmiela, H.E. Sauceda, M.~Gastegger, I.~Poltavsky, K.T.
  Sch{\"u}tt, A.~Tkatchenko, K.R. M{\"u}ller, Machine learning force fields.
\newblock Chemical Reviews \textbf{121}(16), 10142--10186 (2021)

\bibitem{fu2022forces}
X.~Fu, Z.~Wu, W.~Wang, T.~Xie, S.~Keten, R.~Gomez-Bombarelli, T.~Jaakkola,
  Forces are not enough: Benchmark and critical evaluation for machine learning
  force fields with molecular simulations.
\newblock arXiv preprint arXiv:2210.07237  (2022)

\bibitem{deringer2020general}
V.L. Deringer, M.A. Caro, G.~Cs{\'a}nyi, A general-purpose machine-learning
  force field for bulk and nanostructured phosphorus.
\newblock Nature communications \textbf{11}(1), 5461 (2020)

\bibitem{zhou2023structure}
Y.~Zhou, S.R. Elliott, V.L. Deringer, Structure and bonding in amorphous red
  phosphorus.
\newblock Angewandte Chemie International Edition \textbf{62}(24), e202216658
  (2023)

\bibitem{skinner2014structure}
L.B. Skinner, C.~Benmore, J.C. Neuefeind, J.B. Parise, The structure of water
  around the compressibility minimum.
\newblock The Journal of chemical physics \textbf{141}(21) (2014)

\bibitem{chen2016ab}
W.~Chen, F.~Ambrosio, G.~Miceli, A.~Pasquarello, Ab initio electronic structure
  of liquid water.
\newblock Physical review letters \textbf{117}(18), 186401 (2016)

\bibitem{soper2008quantum}
A.~Soper, C.~Benmore, Quantum differences between heavy and light water.
\newblock Physical review letters \textbf{101}(6), 065502 (2008)

\bibitem{distasio2014individual}
R.A. DiStasio, B.~Santra, Z.~Li, X.~Wu, R.~Car, The individual and collective
  effects of exact exchange and dispersion interactions on the ab initio
  structure of liquid water.
\newblock The Journal of chemical physics \textbf{141}(8) (2014)

\bibitem{cheng2019ab}
B.~Cheng, E.A. Engel, J.~Behler, C.~Dellago, M.~Ceriotti, Ab initio
  thermodynamics of liquid and solid water.
\newblock Proceedings of the National Academy of Sciences \textbf{116}(4),
  1110--1115 (2019)

\bibitem{monserrat2020liquid}
B.~Monserrat, J.G. Brandenburg, E.A. Engel, B.~Cheng, Liquid water contains the
  building blocks of diverse ice phases.
\newblock Nature communications \textbf{11}(1), 5757 (2020)

\bibitem{chen2017ab}
M.~Chen, H.Y. Ko, R.C. Remsing, M.F. Calegari~Andrade, B.~Santra, Z.~Sun,
  A.~Selloni, R.~Car, M.L. Klein, J.P. Perdew, et~al., Ab initio theory and
  modeling of water.
\newblock Proceedings of the National Academy of Sciences \textbf{114}(41),
  10846--10851 (2017)

\bibitem{ruff2023connectivity}
R.~Ruff, P.~Reiser, J.~St{\"u}hmer, P.~Friederich, Connectivity optimized
  nested graph networks for crystal structures.
\newblock arXiv preprint arXiv:2302.14102  (2023)

\bibitem{de2021robust}
P.P. De~Breuck, M.L. Evans, G.M. Rignanese, Robust model benchmarking and
  bias-imbalance in data-driven materials science: a case study on modnet.
\newblock Journal of Physics: Condensed Matter \textbf{33}(40), 404002 (2021)

\bibitem{chmiela2023accurate}
S.~Chmiela, V.~Vassilev-Galindo, O.T. Unke, A.~Kabylda, H.E. Sauceda,
  A.~Tkatchenko, K.R. M{\"u}ller, Accurate global machine learning force fields
  for molecules with hundreds of atoms.
\newblock Science Advances \textbf{9}(2), eadf0873 (2023)

\bibitem{ko2023accurate}
T.W. Ko, J.A. Finkler, S.~Goedecker, J.~Behler, Accurate fourth-generation
  machine learning potentials by electrostatic embedding.
\newblock Journal of Chemical Theory and Computation \textbf{19}(12),
  3567--3579 (2023)

\bibitem{Ansel_PyTorch_2_Faster_2024}
J.~Ansel, E.~Yang, H.~He, N.~Gimelshein, A.~Jain, M.~Voznesensky, B.~Bao,
  P.~Bell, D.~Berard, E.~Burovski, G.~Chauhan, A.~Chourdia, W.~Constable,
  A.~Desmaison, Z.~DeVito, E.~Ellison, W.~Feng, J.~Gong, M.~Gschwind, B.~Hirsh,
  S.~Huang, K.~Kalambarkar, L.~Kirsch, M.~Lazos, M.~Lezcano, Y.~Liang,
  J.~Liang, Y.~Lu, C.~Luk, B.~Maher, Y.~Pan, C.~Puhrsch, M.~Reso, M.~Saroufim,
  M.Y. Siraichi, H.~Suk, M.~Suo, P.~Tillet, E.~Wang, X.~Wang, W.~Wen, S.~Zhang,
  X.~Zhao, K.~Zhou, R.~Zou, A.~Mathews, G.~Chanan, P.~Wu, S.~Chintala,
  \emph{{PyTorch 2: Faster Machine Learning Through Dynamic Python Bytecode
  Transformation and Graph Compilation}}, in \emph{29th ACM International
  Conference on Architectural Support for Programming Languages and Operating
  Systems, Volume 2 (ASPLOS '24)} (ACM, 2024).
\newblock \doi{10.1145/3620665.3640366}.
\newblock \urlprefix\url{https://pytorch.org/assets/pytorch2-2.pdf}

\bibitem{Ko_Materials_Graph_Library_2021}
T.W. Ko, M.~Nassar, S.~Miret, E.~Liu, J.~Qi, S.P. Ong.
\newblock {Materials Graph Library} (2021).
\newblock \doi{10.5281/zenodo.8025189}

\bibitem{chen2023geomformer}
T.~Chen, S.~Luo, D.~He, S.~Zheng, T.Y. Liu, L.~Wang, Geomformer: A general
  architecture for geometric molecular representation learning  (2023)

\bibitem{vaswani2017attention}
A.~Vaswani, N.~Shazeer, N.~Parmar, J.~Uszkoreit, L.~Jones, A.N. Gomez,
  {\L}.~Kaiser, I.~Polosukhin, Attention is all you need.
\newblock Advances in neural information processing systems \textbf{30} (2017)

\bibitem{loshchilov2017decoupled}
I.~Loshchilov, F.~Hutter, Decoupled weight decay regularization.
\newblock arXiv preprint arXiv:1711.05101  (2017)

\bibitem{tan2024enhanced}
A.R. Tan, J.C. Dietschreit, R.~Gomez-Bombarelli, Enhanced sampling of robust
  molecular datasets with uncertainty-based collective variables.
\newblock arXiv preprint arXiv:2402.03753  (2024)

\bibitem{krajewski2022extensible}
A.M. Krajewski, J.W. Siegel, J.~Xu, Z.K. Liu, Extensible structure-informed
  prediction of formation energy with improved accuracy and usability employing
  neural networks.
\newblock Computational Materials Science \textbf{208}, 111254 (2022)

\bibitem{thomas2023calibration}
A.~Thomas-Mitchell, G.~Hawe, P.L. Popelier, Calibration of uncertainty in the
  active learning of machine learning force fields.
\newblock Machine Learning: Science and Technology \textbf{4}(4), 045034 (2023)

\bibitem{thaler2023scalable}
S.~Thaler, G.~Doehner, J.~Zavadlav, Scalable bayesian uncertainty
  quantification for neural network potentials: promise and pitfalls.
\newblock Journal of Chemical Theory and Computation \textbf{19}(14),
  4520--4532 (2023)

\bibitem{caldeira2020deeply}
J.~Caldeira, B.~Nord, Deeply uncertain: comparing methods of uncertainty
  quantification in deep learning algorithms.
\newblock Machine Learning: Science and Technology \textbf{2}(1), 015002 (2020)

\bibitem{ong2013python}
S.P. Ong, W.D. Richards, A.~Jain, G.~Hautier, M.~Kocher, S.~Cholia, D.~Gunter,
  V.L. Chevrier, K.A. Persson, G.~Ceder, Python materials genomics (pymatgen):
  A robust, open-source python library for materials analysis.
\newblock Computational Materials Science \textbf{68}, 314--319 (2013)

\bibitem{blochl1994projector}
P.E. Bl{\"o}chl, Projector augmented-wave method.
\newblock Physical review B \textbf{50}(24), 17953 (1994)

\bibitem{wang2021predicting}
H.C. Wang, S.~Botti, M.A. Marques, Predicting stable crystalline compounds
  using chemical similarity.
\newblock npj Computational Materials \textbf{7}(1), 12 (2021)

\bibitem{larsen2017atomic}
A.H. Larsen, J.J. Mortensen, J.~Blomqvist, I.E. Castelli, R.~Christensen,
  M.~Du{\l}ak, J.~Friis, M.N. Groves, B.~Hammer, C.~Hargus, et~al., The atomic
  simulation environment—a python library for working with atoms.
\newblock Journal of Physics: Condensed Matter \textbf{29}(27), 273002 (2017)

\bibitem{riebesell2023matbench}
J.~Riebesell, R.E. Goodall, A.~Jain, P.~Benner, K.A. Persson, A.A. Lee,
  Matbench discovery--an evaluation framework for machine learning crystal
  stability prediction.
\newblock arXiv preprint arXiv:2308.14920  (2023)

\bibitem{pickard2006high}
C.J. Pickard, R.~Needs, High-pressure phases of silane.
\newblock Physical review letters \textbf{97}(4), 045504 (2006)

\bibitem{phonopy-phono3py-JPCM}
A.~Togo, L.~Chaput, T.~Tadano, I.~Tanaka, Implementation strategies in phonopy
  and phono3py.
\newblock J. Phys. Condens. Matter \textbf{35}(35), 353001 (2023).
\newblock \doi{10.1088/1361-648X/acd831}

\bibitem{phonopy-phono3py-JPSJ}
A.~Togo, First-principles phonon calculations with phonopy and phono3py.
\newblock J. Phys. Soc. Jpn. \textbf{92}(1), 012001 (2023).
\newblock \doi{10.7566/JPSJ.92.012001}

\bibitem{perdew2008restoring}
J.P. Perdew, A.~Ruzsinszky, G.I. Csonka, O.A. Vydrov, G.E. Scuseria, L.A.
  Constantin, X.~Zhou, K.~Burke, Restoring the density-gradient expansion for
  exchange in solids and surfaces.
\newblock Physical review letters \textbf{100}(13), 136406 (2008)

\bibitem{Togo2024private}
A.~Togo.
\newblock Private communication (2024)

\bibitem{petretto2018high}
G.~Petretto, S.~Dwaraknath, H.~PC~Miranda, D.~Winston, M.~Giantomassi, M.J.
  Van~Setten, X.~Gonze, K.A. Persson, G.~Hautier, G.M. Rignanese,
  High-throughput density-functional perturbation theory phonons for inorganic
  materials.
\newblock Scientific data \textbf{5}(1), 1--12 (2018)

\bibitem{slack1958thermal}
G.~Slack, R.~Newman, Thermal conductivity of mno and nio.
\newblock Physical Review Letters \textbf{1}(10), 359 (1958)

\bibitem{labotz1963thermal}
R.J. LaBotz, D.R. Mason, The thermal conductivities of mg2si and mg2ge.
\newblock Journal of The Electrochemical Society \textbf{110}(2), 121 (1963)

\bibitem{martin1972thermal}
J.~Martin, Thermal conductivity of mg2si, mg2ge and mg2sn.
\newblock Journal of physics and chemistry of solids \textbf{33}(5), 1139--1148
  (1972)

\bibitem{takahashi1980porosity}
T.~Takahashi, T.~Kikuchi, Porosity dependence on thermal diffusivity and
  thermal conductivity of lithium oxide li2o from 200 to 900° c.
\newblock Journal of Nuclear Materials \textbf{91}(1), 93--102 (1980)

\bibitem{gerlich1982temperature}
D.~Gerlich, P.~Andersson, Temperature and pressure effects on the thermal
  conductivity and heat capacity of cscl, csbr and csi.
\newblock Journal of Physics C: Solid State Physics \textbf{15}(25), 5211
  (1982)

\bibitem{moore1985thermal}
J.~Moore, F.~Weaver, R.~Graves, D.~McElroy, The thermal conductivities of srcl
  2 and srf 2 from 85 to 400 k.
\newblock Thermal Conductivity 18 pp. 115--124 (1985)

\bibitem{morelli1995low}
D.~Morelli, T.~Caillat, J.P. Fleurial, A.~Borshchevsky, J.~Vandersande,
  B.~Chen, C.~Uher, Low-temperature transport properties of p-type cosb 3.
\newblock Physical Review B \textbf{51}(15), 9622 (1995)

\bibitem{hohl1999efficient}
H.~Hohl, A.P. Ramirez, C.~Goldmann, G.~Ernst, B.~W{\"o}lfing, E.~Bucher,
  Efficient dopants for zrnisn-based thermoelectric materials.
\newblock Journal of Physics: Condensed Matter \textbf{11}(7), 1697 (1999)

\bibitem{young2000thermoelectric}
D.~Young, P.~Khalifah, R.J. Cava, A.~Ramirez, Thermoelectric properties of pure
  and doped femsb (m= v, nb).
\newblock Journal of Applied Physics \textbf{87}(1), 317--321 (2000)

\bibitem{inbook}
D.~Morelli, G.~Slack, \emph{High Lattice Thermal Conductivity Solids} (2006),
  pp. 37--68.
\newblock \doi{10.1007/0-387-25100-6_2}

\bibitem{popov2010thermal}
P.~Popov, P.~Fedorov, V.~Osiko, Thermal conductivity of single crystals with a
  fluorite structure: cadmium fluoride.
\newblock Physics of the Solid State \textbf{52}, 504--508 (2010)

\bibitem{mann2010hydrothermal}
M.~Mann, D.~Thompson, K.~Serivalsatit, T.M. Tritt, J.~Ballato, J.~Kolis,
  Hydrothermal growth and thermal property characterization of tho2 single
  crystals.
\newblock Crystal growth \& design \textbf{10}(5), 2146--2151 (2010)

\bibitem{book}
A.~Jha, \emph{Rare Earth Materials: Properties and Applications} (2014), pp.
  1--332.
\newblock \doi{10.1201/b17045}

\bibitem{toberer2011phonon}
E.S. Toberer, A.~Zevalkink, G.J. Snyder, Phonon engineering through crystal
  chemistry.
\newblock Journal of Materials Chemistry \textbf{21}(40), 15843--15852 (2011)

\bibitem{lindsay2013phonon}
L.~Lindsay, D.~Broido, T.~Reinecke, Phonon-isotope scattering and thermal
  conductivity in materials with a large isotope effect: A first-principles
  study.
\newblock Physical review B \textbf{88}(14), 144306 (2013)

\bibitem{xiao2013cubic}
W.~Xiao, D.~Tan, W.~Zhou, J.~Liu, J.~Xu, Cubic perovskite polymorph of
  strontium metasilicate at high pressures.
\newblock American Mineralogist \textbf{98}(11-12), 2096--2104 (2013)

\bibitem{seko2015prediction}
A.~Seko, A.~Togo, H.~Hayashi, K.~Tsuda, L.~Chaput, I.~Tanaka, Prediction of
  low-thermal-conductivity compounds with first-principles anharmonic
  lattice-dynamics calculations and bayesian optimization.
\newblock Physical review letters \textbf{115}(20), 205901 (2015)

\bibitem{togo2015distributions}
A.~Togo, L.~Chaput, I.~Tanaka, Distributions of phonon lifetimes in brillouin
  zones.
\newblock Physical review B \textbf{91}(9), 094306 (2015)

\bibitem{van2016high}
A.~van Roekeghem, J.~Carrete, C.~Oses, S.~Curtarolo, N.~Mingo, High-throughput
  computation of thermal conductivity of high-temperature solid phases: the
  case of oxide and fluoride perovskites.
\newblock Physical Review X \textbf{6}(4), 041061 (2016)

\bibitem{campi2017first}
D.~Campi, L.~Paulatto, G.~Fugallo, F.~Mauri, M.~Bernasconi, First-principles
  calculation of lattice thermal conductivity in crystalline phase change
  materials: Gete, sb 2 te 3, and ge 2 sb 2 te 5.
\newblock Physical Review B \textbf{95}(2), 024311 (2017)

\bibitem{skelton2017lattice}
J.M. Skelton, L.A. Burton, A.J. Jackson, F.~Oba, S.C. Parker, A.~Walsh, Lattice
  dynamics of the tin sulphides sns 2, sns and sn 2 s 3: vibrational spectra
  and thermal transport.
\newblock Physical Chemistry Chemical Physics \textbf{19}(19), 12452--12465
  (2017)

\bibitem{qian2019thermal}
X.~Qian, S.~Peng, X.~Li, Y.~Wei, R.~Yang, Thermal conductivity modeling using
  machine learning potentials: application to crystalline and amorphous
  silicon.
\newblock Materials Today Physics \textbf{10}, 100140 (2019)

\bibitem{xia2020high}
Y.~Xia, V.I. Hegde, K.~Pal, X.~Hua, D.~Gaines, S.~Patel, J.~He, M.~Aykol,
  C.~Wolverton, High-throughput study of lattice thermal conductivity in binary
  rocksalt and zinc blende compounds including higher-order anharmonicity.
\newblock Physical Review X \textbf{10}(4), 041029 (2020)

\bibitem{rakesh2021anomalous}
S.~Rakesh~Roshan, N.~Yedukondalu, R.~Muthaiah, K.~Lavanya, P.~Anees, R.R.
  Kumar, T.V. Rao, L.~Ehm, J.B. Parise, Anomalous lattice thermal conductivity
  in rocksalt iia--via compounds.
\newblock ACS Applied Energy Materials \textbf{5}(1), 882--896 (2021)

\bibitem{zhu2021charting}
T.~Zhu, R.~He, S.~Gong, T.~Xie, P.~Gorai, K.~Nielsch, J.C. Grossman, Charting
  lattice thermal conductivity for inorganic crystals and discovering rare
  earth chalcogenides for thermoelectrics.
\newblock Energy \& Environmental Science \textbf{14}(6), 3559--3566 (2021)

\bibitem{ju2021exploring}
S.~Ju, R.~Yoshida, C.~Liu, S.~Wu, K.~Hongo, T.~Tadano, J.~Shiomi, Exploring
  diamondlike lattice thermal conductivity crystals via feature-based transfer
  learning.
\newblock Physical Review Materials \textbf{5}(5), 053801 (2021)

\bibitem{tranaas2022lattice}
R.~Tran{\aa}s, O.M. L{\o}vvik, O.~Tomic, K.~Berland, Lattice thermal
  conductivity of half-heuslers with density functional theory and machine
  learning: Enhancing predictivity by active sampling with principal component
  analysis.
\newblock Computational Materials Science \textbf{202}, 110938 (2022)

\bibitem{cao2023anomalous}
W.~Cao, J.~Shi, R.~Xiong, L.~Miao, Z.~Wang, Z.~Liu, Anomalous thermal transport
  in mgse with diamond phase under pressure.
\newblock Physical Review B \textbf{107}(23), 235201 (2023)

\bibitem{bale2016reprint}
C.W. Bale, E.~B{\'e}lisle, P.~Chartrand, S.A. Decterov, G.~Eriksson, A.E.
  Gheribi, K.~Hack, I.H. Jung, Y.B. Kang, J.~Melan{\c{c}}on, et~al., Reprint
  of: Factsage thermochemical software and databases, 2010--2016.
\newblock Calphad \textbf{55}, 1--19 (2016)

\bibitem{ouyang2018sisso}
R.~Ouyang, S.~Curtarolo, E.~Ahmetcik, M.~Scheffler, L.M. Ghiringhelli, Sisso: A
  compressed-sensing method for identifying the best low-dimensional descriptor
  in an immensity of offered candidates.
\newblock Physical Review Materials \textbf{2}(8), 083802 (2018)

\bibitem{purcell2023recent}
T.A. Purcell, M.~Scheffler, L.M. Ghiringhelli, Recent advances in the sisso
  method and their implementation in the sisso++ code.
\newblock The Journal of Chemical Physics \textbf{159}(11) (2023)

\bibitem{sorella2011ab}
S.~Sorella, M.~Casula, L.~Spanu, A.~Dal~Corso, Ab initio calculations for the
  $\beta$-tin diamond transition in silicon: Comparing theories with
  experiments.
\newblock Physical Review B \textbf{83}(7), 075119 (2011)

\bibitem{voronin2003situ}
G.~Voronin, C.~Pantea, T.~Zerda, L.~Wang, Y.~Zhao, In situ x-ray diffraction
  study of silicon at pressures up to 15.5 gpa and temperatures up to 1073 k.
\newblock Physical Review B \textbf{68}(2), 020102 (2003)

\bibitem{kim2019energy}
H.~Kim, J.Y. Park, S.~Choi, Energy refinement and analysis of structures in the
  qm9 database via a highly accurate quantum chemical method.
\newblock Scientific data \textbf{6}(1), 109 (2019)

\bibitem{rosen2022high}
A.S. Rosen, V.~Fung, P.~Huck, C.T. O’Donnell, M.K. Horton, D.G. Truhlar, K.A.
  Persson, J.M. Notestein, R.Q. Snurr, High-throughput predictions of
  metal--organic framework electronic properties: theoretical challenges, graph
  neural networks, and data exploration.
\newblock npj Computational Materials \textbf{8}(1), 1--10 (2022)

\bibitem{rosen2021machine}
A.S. Rosen, S.M. Iyer, D.~Ray, Z.~Yao, A.~Aspuru-Guzik, L.~Gagliardi, J.M.
  Notestein, R.Q. Snurr, Machine learning the quantum-chemical properties of
  metal--organic frameworks for accelerated materials discovery.
\newblock Matter \textbf{4}(5), 1578--1597 (2021)

\bibitem{xie2022brute}
Y.~Xie, K.~Shibata, T.~Mizoguchi, A brute-force code searching for cell of
  non-identical displacement for csl grain boundaries and interfaces.
\newblock Computer Physics Communications \textbf{273}, 108260 (2022)

\bibitem{xie2022interface_master}
Y.~Xie, K.~Shibata, T.~Mizoguchi, interface\_master: Python package building
  csl and approximate csl interfaces of any two lattices--an effective tool for
  interface engineers.
\newblock arXiv preprint arXiv:2211.15173  (2022)

\bibitem{haastrup2018computational}
S.~Haastrup, M.~Strange, M.~Pandey, T.~Deilmann, P.S. Schmidt, N.F. Hinsche,
  M.N. Gjerding, D.~Torelli, P.M. Larsen, A.C. Riis-Jensen, et~al., The
  computational 2d materials database: high-throughput modeling and discovery
  of atomically thin crystals.
\newblock 2D Materials \textbf{5}(4), 042002 (2018)

\bibitem{huan2020polymer}
T.D. Huan, R.~Ramprasad, Polymer structure prediction from first principles.
\newblock The Journal of Physical Chemistry Letters \textbf{11}(15), 5823--5829
  (2020)

\bibitem{thompson2022lammps}
A.P. Thompson, H.M. Aktulga, R.~Berger, D.S. Bolintineanu, W.M. Brown, P.S.
  Crozier, P.J. in't Veld, A.~Kohlmeyer, S.G. Moore, T.D. Nguyen, et~al.,
  Lammps-a flexible simulation tool for particle-based materials modeling at
  the atomic, meso, and continuum scales.
\newblock Computer Physics Communications \textbf{271}, 108171 (2022)

\bibitem{parrinello1981polymorphic}
M.~Parrinello, A.~Rahman, Polymorphic transitions in single crystals: A new
  molecular dynamics method.
\newblock Journal of Applied physics \textbf{52}(12), 7182--7190 (1981)

\bibitem{nose1984unified}
S.~Nos{\'e}, A unified formulation of the constant temperature molecular
  dynamics methods.
\newblock The Journal of chemical physics \textbf{81}(1), 511--519 (1984)

\bibitem{martyna1994constant}
G.J. Martyna, D.J. Tobias, M.L. Klein, Constant pressure molecular dynamics
  algorithms.
\newblock The Journal of chemical physics \textbf{101}(5), 4177--4189 (1994)

\bibitem{capinski1997thermal}
W.~Capinski, H.~Maris, E.~Bauser, I.~Silier, M.~Asen-Palmer, T.~Ruf,
  M.~Cardona, E.~Gmelin, Thermal conductivity of isotopically enriched si.
\newblock Applied physics letters \textbf{71}(15), 2109--2111 (1997)

\bibitem{shinoda2004rapid}
W.~Shinoda, M.~Shiga, M.~Mikami, Rapid estimation of elastic constants by
  molecular dynamics simulation under constant stress.
\newblock Physical Review B \textbf{69}(13), 134103 (2004)

\bibitem{GPTARepo2023}
M.~Praiteri.
\newblock {GPTA}: Github repository for gpt assisted tasks.
\newblock \url{https://github.com/praiteri/GPTA} (2023).
\newblock Accessed: 2024-03-18

\bibitem{gaiduk2018first}
A.P. Gaiduk, J.~Gustafson, F.~Gygi, G.~Galli, First-principles simulations of
  liquid water using a dielectric-dependent hybrid functional.
\newblock The journal of physical chemistry letters \textbf{9}(11), 3068--3073
  (2018)

\bibitem{zheng2018structural}
L.~Zheng, M.~Chen, Z.~Sun, H.Y. Ko, B.~Santra, P.~Dhuvad, X.~Wu, Structural,
  electronic, and dynamical properties of liquid water by ab initio molecular
  dynamics based on scan functional within the canonical ensemble.
\newblock The Journal of Chemical Physics \textbf{148}(16) (2018)

\bibitem{fritsch2014nuclear}
S.~Fritsch, R.~Potestio, D.~Donadio, K.~Kremer, Nuclear quantum effects in
  water: A multiscale study.
\newblock Journal of chemical theory and computation \textbf{10}(2), 816--824
  (2014)

\bibitem{michele2016nuclear}
C.~Michele, F.~Wei, M.~Angelos, et~al., Nuclear quantum effects in water and
  aqueous systems: Experiment, theory, and current challenges.
\newblock Chemical Reviews  (2016)

\bibitem{chen2023thermodynamics}
Z.~Chen, M.L. Berrens, K.T. Chan, Z.~Fan, D.~Donadio, Thermodynamics of water
  and ice from a fast and scalable first-principles neuroevolution potential.
\newblock Journal of Chemical \& Engineering Data  (2023)

\bibitem{einstein1905annalen}
A.~Einstein, Annalen der physik.
\newblock Nr \textbf{10}, 891ff (1905)

\bibitem{maginn2019best}
E.J. Maginn, R.A. Messerly, D.J. Carlson, D.R. Roe, J.R. Elliot, Best practices
  for computing transport properties 1. self-diffusivity and viscosity from
  equilibrium molecular dynamics [article v1. 0].
\newblock Living Journal of Computational Molecular Science \textbf{1}(1),
  6324--6324 (2019)

\bibitem{dunweg1993molecular}
B.~D{\"u}nweg, K.~Kremer, Molecular dynamics simulation of a polymer chain in
  solution.
\newblock The Journal of chemical physics \textbf{99}(9), 6983--6997 (1993)

\bibitem{yeh2004system}
I.C. Yeh, G.~Hummer, System-size dependence of diffusion coefficients and
  viscosities from molecular dynamics simulations with periodic boundary
  conditions.
\newblock The Journal of Physical Chemistry B \textbf{108}(40), 15873--15879
  (2004)

\bibitem{vega2011simulating}
C.~Vega, J.L. Abascal, Simulating water with rigid non-polarizable models: a
  general perspective.
\newblock Physical Chemistry Chemical Physics \textbf{13}(44), 19663--19688
  (2011)

\bibitem{mills1973self}
R.~Mills, Self-diffusion in normal and heavy water in the range 1-45. deg.
\newblock The Journal of Physical Chemistry \textbf{77}(5), 685--688 (1973)

\bibitem{krynicki1978pressure}
K.~Krynicki, C.D. Green, D.W. Sawyer, Pressure and temperature dependence of
  self-diffusion in water.
\newblock Faraday Discussions of the Chemical Society \textbf{66}, 199--208
  (1978)

\bibitem{hardy2001isotope}
E.H. Hardy, A.~Zygar, M.D. Zeidler, M.~Holz, F.D. Sacher, Isotope effect on the
  translational and rotational motion in liquid water and ammonia.
\newblock The Journal of Chemical Physics \textbf{114}(7), 3174--3181 (2001)

\bibitem{jain2013materials}
A.~Jain, S.P. Ong, G.~Hautier, W.~Chen, W.D. Richards, S.~Dacek, S.~Cholia,
  D.~Gunter, D.~Skinner, G.~Ceder, et~al., The materials project: A materials
  genome approach to accelerating materials innovation, apl mater  (2013)

\bibitem{de2015charting}
M.~De~Jong, W.~Chen, T.~Angsten, A.~Jain, R.~Notestine, A.~Gamst, M.~Sluiter,
  C.~Krishna~Ande, S.~Van Der~Zwaag, J.J. Plata, et~al., Charting the complete
  elastic properties of inorganic crystalline compounds.
\newblock Scientific data \textbf{2}(1), 1--13 (2015)

\bibitem{petousis2017high}
I.~Petousis, D.~Mrdjenovich, E.~Ballouz, M.~Liu, D.~Winston, W.~Chen, T.~Graf,
  T.D. Schladt, K.A. Persson, F.B. Prinz, High-throughput screening of
  inorganic compounds for the discovery of novel dielectric and optical
  materials.
\newblock Scientific data \textbf{4}(1), 1--12 (2017)

\bibitem{choudhary2017high}
K.~Choudhary, I.~Kalish, R.~Beams, F.~Tavazza, High-throughput identification
  and characterization of two-dimensional materials using density functional
  theory.
\newblock Scientific reports \textbf{7}(1), 5179 (2017)

\end{thebibliography}

\end{document}